\documentclass{aastex6}

\usepackage{natbib}
\usepackage{color}

\usepackage{ulem}
\usepackage{lineno}
\shorttitle{Sustained Solar $\gamma$-Ray Emission}
\shortauthors{Share et al.}

\begin{document}


\title{Characteristics of Sustained $>$100 MeV $\gamma$-Ray Emission Associated with Solar Flares} 


\author{G. H. Share\altaffilmark{1,2}, R. J. Murphy\altaffilmark{3}, A. K. Tolbert\altaffilmark{4,5}, B. R. Dennis\altaffilmark{4}, S. M. White\altaffilmark{6},  R. A. Schwartz\altaffilmark{4,5},  and A. J. Tylka\altaffilmark{4} }

\altaffiltext{1}{Astronomy Department, University of Maryland, College Park, MD 20740, USA} 
\altaffiltext{2}{National Observatory of Athens, Athens, Greece}
\altaffiltext{3}{Space Science Division, Naval Research Laboratory, Washington DC 20375, USA}
\altaffiltext{4}{NASA Goddard Spaceflight Center, Greenbelt, MD 20771, USA}
\altaffiltext{5}{Physics Department, American University, Washington, DC 20016, USA}
\altaffiltext{6}{Kirtland AFB, Albuquerque, NM 87117, USA}

\setcounter {footnote}{0}
\setcounter {mpfootnote}{0}



\begin{abstract}

We characterize and provide a catalog of thirty $>$100 MeV sustained $\gamma$-ray emission (SGRE) events observed by {\it Fermi} LAT.  These events are temporally and spectrally distinct from the associated solar flares.  Their spectra are consistent with decay of pions produced by $>$300 MeV protons and are not consistent with electron bremsstrahlung.  SGRE start times range from CME onset to two hours later. Their durations range from about four minutes to twenty hours and appear to be correlated with durations of $>$100 MeV SEP proton events.  The $>$300 MeV protons producing SGRE have spectra that can be fit with power laws with a mean index of $\sim$4 and RMS spread of 1.8.  $\gamma$-ray line measurements indicate that SGRE proton spectra are steeper above 300 MeV than they are below 300 MeV.    The number of SGRE protons $>$500 MeV is on average about ten times more than the number in the associated flare and about fifty to one hundred times less than the number in the accompanying SEP.  SGRE can extend tens of degrees from the flare site.    Sustained bremsstrahlung from MeV electrons was observed in one SGRE event.  Flare $>$100 keV X-ray emission appears to be associated with SGRE and with intense SEPs.  From this observation, we provide arguments that lead us to propose that sub-MeV to MeV protons escaping from the flare contribute to the seed population that is accelerated by shocks onto open field lines to produce SEPs and onto field lines returning to the Sun to produce SGRE.
 

\end{abstract}



\section{Introduction} \label{sec:intro}

With the advent of the space age it became possible to observe $\gamma$ rays above Earth's absorbing atmosphere, including those emitted from the Sun.  \citet{Dola65} provided the framework for future observations and theoretical studies in their pioneering paper describing the various components of the $\gamma$-ray spectrum  expected to be produced by the interactions of high-energy electrons and protons in the chromosphere and photosphere.  For energies $>$50 MeV, they concluded that radiation from the decay of neutral pions would dominate the spectrum. The pions are produced by protons interacting with solar hydrogen at energies above the $\sim$300 MeV threshold.  They also expected that positrons, produced as a result of neutral and charged pion-decays, would annihilate with ambient electrons to produce the characteristic 511-keV line.     

\begin{figure}
\epsscale{.60}
\plotone{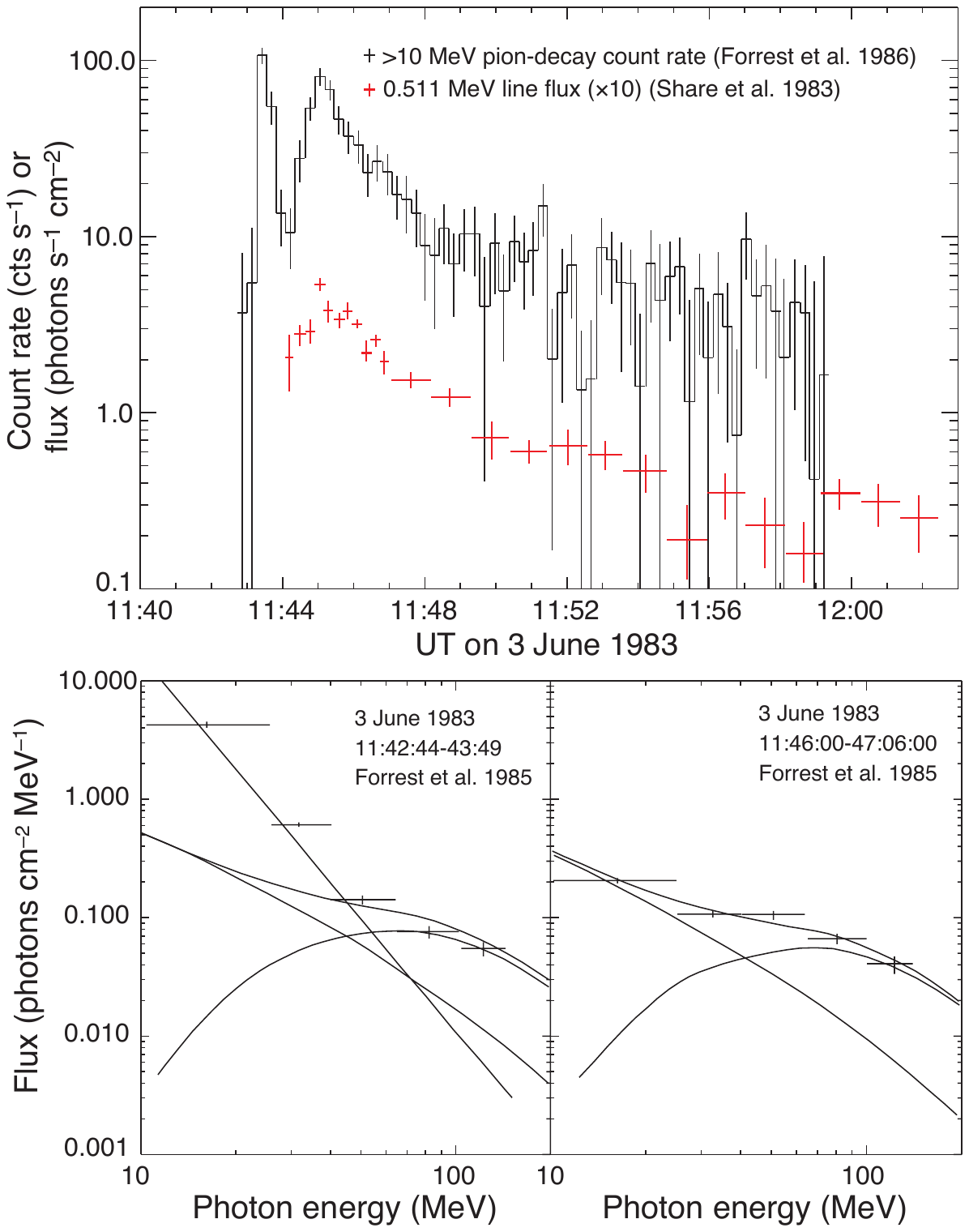}
\caption{Observations of high-energy emission from the 1982 June 3 flare made by the {\it SMM} GRS.  Top Panel:  Time history of the pion-decay $\gamma$-ray count rate revealing two clear phases of emission \citep{forr86}.  Shown in red, and scaled arbitrarily, is the count rate observed in the 511 keV annihilation line \citep{shar83}.  $\gamma$-ray spectra observed during the impulsive phase (lower left panel) and during the second phase (right panel) \citep{forr85}.  The data points with errors show the observations. The curves show the different components of the spectrum, including: bremsstrahlung from primary impulsive-flare electrons (steep power-law spectrum),  bremsstrahlung from charged pion-decay (flat power-law spectrum), and neutral pion-decay (broad peak).}
\label{june3}
\end{figure}

The first unambiguous detection of pion-decay $\gamma$-ray emission from the Sun was made during and after the 1982 June 3 X8.0 {\it GOES}-class flare (using the standard naming convention SOL1982-06-03T11:42 \citep{leib10}) by the Gamma Ray Spectrometer (GRS) on the {\it Solar Maximum Mission} ({\it SMM}) satellite \citep{forr86}.   This emission revealed two distinct phases as shown by the time history of the $>$10 MeV flux in the top panel of Figure \ref{june3}.   The impulsive phase lasted about a minute and was dominated by bremsstrahlung from flare-accelerated electrons, but also contained pion-decay emission.  This can be be seen in the left bottom panel of Figure \ref{june3} taken from \citet{forr85} which distinguishes the bremsstrahlung from flare-accelerated electrons, the bremsstrahlung from charged pion-decay, and the broad peak near 70 MeV from neutral pion decay.  The second phase began within a minute after the peak of the impulsive phase and lasted at least 15 minutes.  As can be seen in the bottom right panel of Figure \ref{june3}, this phase was dominated by pion-decay emission.  Its temporal history plotted in the top panel is similar to the time history, plotted in red, that was observed for the 511 keV positron-annihilation line from decay of positively charged pions \citep{shar83}.  No spectroscopic measurements of the 511 keV line and other nuclear lines were possible during the intense impulsive flare due to saturation effects.  Neutrons from this event were also observed by GRS and by neutron monitors on Earth \citep{chup87}.  \citet{forr85} speculated that the second phase might be associated with the acceleration of solar energetic particles (SEPs) observed in interplanetary space.

\citet{murp87} analyzed the 1982 June 3 flare as part of their comprehensive study of high-energy processes in solar flares using cross sections for proton and $\alpha$-particle reactions.  This resulted in a quantitative understanding of the $\gamma$-ray spectrum from the decay of charged and neutral pions, the yield of 511 keV photons from positron annihilation, and their relationship to nuclear-line emission.   Using the neutron data cited above and nuclear-deexcitation line data provided by \citet{prin83}, \citet{murp87} showed that the two emission phases were produced by distinctly different accelerated particle populations.  The first phase is consistent with a steep ion spectrum from stochastic acceleration. The second phase is consistent with a harder ion spectrum from shock acceleration, similar to that of the accompanying SEPs observed in space.  They estimated that the number of protons $>$30 MeV responsible for the second phase of $\gamma$-ray emission was only about 1\% of the number in the first phase, and about 10\% of the number of SEPs; however, it is important to note that these comparisons are strongly dependent on the energy spectrum of the protons, especially at energies below a few hundred MeV.

There have been several events reported where $\gamma$-ray emission above tens of MeV was observed for extended periods of time after the impulsive phase of the flares.  \citet{ryan00} listed thirteen such events from 1982 until 1991, all of which were associated with intense impulsive X-ray flares with {\it GOES} classifications of X4 or greater.  These high-energy events lasted from 8 minutes to several hours.  The longest duration event, SOL1991-06-11T02:09, was observed $>$50 MeV by the {\it{Compton Gamma-Ray Observatory (CGRO)}} Energetic Gamma Ray Experiment Telescope (EGRET) for at least 8 hours \citep{kanb93,schn96}. The radiation at energies $>$50 MeV observed in most of these events appeared to be due to decay of pions and was observed even when the soft X-ray flux was greatly diminished or no longer detectable.  \citet{ryan00} called a flare associated with time-extended $\gamma$-ray emission a Long Duration Solar Gamma-Ray Flare (LDGRF).  In this paper we prefer to identify these high-energy episodes as Sustained Gamma-Ray Emission (SGRE) events because they have temporal and spectral characteristics that are distinct from the accompanying impulsive flare, and, although they most often extend well past the end of the flare, they are not always of long duration.  The distinct post-flare emission phase of the 1982 June 3 event shown in Figure \ref{june3} is the first example of SGRE.  However, it is different from most of the SGRE events discussed in this paper because $>$100 MeV $\gamma$-ray emission was also detected during the impulsive flare.

\citet{ryan00} discussed four possible origins for the SGRE: 1. delayed precipitation into the chromosphere of high-energy particles accelerated in the impulsive phase of the flare and stored in magnetic structures high in the corona; 2. acceleration by an expanding shock far from the flare site of particles that make their way back to the chromosphere; 3. particles accelerated in the reconnection current sheet behind a Coronal Mass Ejection (CME); and 4. a mixture of the first two where particles are both trapped in high altitude magnetic structures and accelerated continuously.  After considering the data and models, \citet{ryan00} concluded that the SGRE originates from particles in coronal loops accelerated by a relatively constant second-order stochastic process or by DC acceleration associated with a CME.  However, he did not rule out precipitation of particles from a CME-driven shock.  He suggested that a key distinguishing observation would be the detection of SGRE, associated with a CME, but not accompanied by a flare with strong $\gamma$-ray emission.  With the launch of the {\it Fermi} Mission in 2008, it became possible to make the high-sensitivity measurements necessary to detect events of this type.

The {\it Fermi} Large Area Telescope (LAT) \citep{atwo09} has a peak effective area and a field of view about five times and four times larger, respectively, than EGRET.  This provides an order of magnitude improvement in sensitivity to solar events.  LAT first detected solar-flare $\gamma$-ray emission from SOL2010-06-12T00:57 \citep{acke12a, acke12b} (see also Table \ref{tab:full} Event 3).  This flare was a rather modest M2 {\it{GOES}} class event.  It was associated with a relatively slow 500 km s$^{-1}$ CME and a barely detectable SEP event.  The \citet{acke12a, acke12b} papers primarily focused on LAT and the Gamma-ray Burst Monitor (GBM) \citep{meeg09} observations of bremsstrahlung, nuclear $\gamma$-ray, and pion-decay emissions from 200 keV to 1 GeV during the minute-long impulsive phase.  There was no evidence for $>$100 MeV emission in the hours following the flare.  Had the June 12 flare $>$100 MeV emission time history followed that of the 8-hour long 1991 June 11 SGRE it would have been easily detected by LAT.  

After that first {\it Fermi} impulsive flare observation, the LAT Team reported detection of three solar flares that were followed by $>$100 MeV SGRE.  The first reported SGRE observation (event 1 in Table \ref{tab:latlist}) was made just after the impulsive phase of SOL2011-03-07T19:43, a {\it GOES} M3.7 flare \citep{alla11, alla11a} that was also detected in the impulsive flare at energies $>$300 keV by GBM.  The second SGRE event (number 3 in Table \ref{tab:latlist}) followed SOL2011-06-07T06:16, a {\it GOES} M2.5 flare \citep{tana11} detected by both {\it RHESSI (Reuven Ramaty High Energy Solar Spectroscopic Imager)} and GBM at energies $>$100 keV. The third event ((number 9 in Table \ref{tab:latlist}) was associated with SOL2012-01-23T0338, a {\it GOES} M8.7 flare \citep{tana12}, that was also detected by GBM at energies $>$100 keV.  The common characteristics of all three events were that they were associated with fast, $>$1000 km s$^{-1}$ halo CMEs, $>$100 keV hard X-ray emission during the impulsive flare, and SEPs.  Thus, the events all had characteristics that could be explained by one of the origins for SGRE events discussed by \citet{ryan00}. However, they were all associated with flares having relatively modest {\it GOES} soft X-ray emission, and their impulsive emission did not exceed 1 MeV as measured by {\it RHESSI} or GBM. Thus, these three events differ qualitatively from the SGRE events discussed by \citet{ryan00};  the latter were all associated with large X-class flares exhibiting significant $\gamma$-ray emission up to tens of MeV during the impulsive phase. 


In this paper we catalog and analyze 30 SGRE events observed by LAT from 2008 until the end of 2016.  We perform temporal and spectroscopic studies of the events and relate the measurements to the associated solar flares, radio emissions, coronal mass ejections, and solar energetic particle events.  In $\S$\ref{sec:solobs}, we discuss how LAT is used as a solar observatory, and in $\S$\ref{subsec:lightbucket} we describe our method for analyzing LAT data. This method is different from that used by \citet{acke14} and \citet{ajel14} but provides comparable measurements of $\gamma$-ray fluxes and spectra.  A major advantage is that the {\it Fermi} data can be accessed and studied using standard solar data analysis software.   In $\S$\ref{sec:timehist} we list the 30 SGRE events identified and provide an example of our analysis for one of the events, including details of the observation and time histories of the emission.  Similar studies of each of the other 29 events are presented in Appendix \ref{sec:append} where we provide evidence that the SGRE time histories are distinctly different from those of the accompanying impulsive flares.  In $\S$\ref{sec:spect} we describe our methods for analyzing LAT $>$100 MeV photon spectra from the impulsive-flare and SGRE phases of the events and for providing information on the protons producing these emissions.  We also discuss {\it RHESSI} and GBM spectroscopic studies that provide further information on the proton spectra. In $\S$\ref{sec:results} we detail what we have learned about the overall characteristics of SGRE events and their relationship to the accompanying flares, CMEs, and SEPs.  We list the principal findings of our study in $\S$\ref{sec:summary} and discuss their implications in $\S$ \ref{sec:discussion}.  The paper includes five appendices which allow us to focus on the scientific results in the body of the paper.  The appendices are: \ref{sec:characteristics}. a four-year systematic study of 95 eruptive events and their relationship to SGREs, \ref{sec:append}. time histories and detailed discussion of the 30 SGRE events,  \ref{sec:resultstable}. tabulated results from detailed spectroscopic studies of the SGRE events, \ref{sec:sep}. our method for estimating the number protons in SEPs observed in interplanetary space, and \ref{sec:solarradio}. a description of accompanying solar radio observations.

\section{{\it Fermi}/LAT as a Solar Observatory} \label{sec:solobs}

LAT is an electron-positron pair-conversion telescope \citep{atwo09} that is sensitive to $\gamma$ rays from $\sim$20 MeV to $\sim$300 GeV.   It is made up of 16 identical towers, each comprised of a tracker with alternating layers of Silicon Strip Detectors (SSD) and tungsten converter foils, and a calorimeter with logs of CsI arranged in a `hodoscopic' configuration so that the energy deposition can be imaged in three dimensions. The towers are surrounded by a multi-tile plastic scintillator Anti-Coincidence Detector (ACD). Detector events not accompanied by an energy loss equivalent to 45\% of a minimum ionizing singly-charged particle traversing an ACD tile are telemetered to the ground for further analysis.   These events are further processed to produce what is known as ``source-class" data that are used for celestial $\gamma$-ray studies and for studying SGRE after impulsive flares \citep{acke14}.  We used both what the LAT team call `Pass7' and `Pass8' selections of these data in this paper, depending on when the event was processed. During intense solar flares, when there are large energy losses in the ACD from pileup of tens of keV X rays, source-class data are not available because their quality is compromised.   To allow analysis of high-energy $\gamma$-rays during flares, the LAT team developed two special classes of data: LAT Low Energy (LLE) and Pass8 solar-impulsive-class data.  More detailed information on the {\it Fermi}/LAT instrument, operation, and data related to solar flare observations can be found in \citet{acke12a,acke14} and \cite{ajel14}. 

In the standard {\it{Fermi}} sky-survey mode the spacecraft points away from the Earth and rocks $\pm$50$^{\circ}$ from the zenith on alternating 95-minute orbits: therefore, the Sun can typically be observed only for $\sim$20-40 minutes every one or two orbits for an average duty cycle of 15-20\%.  Depending on statistics, data can be accumulated over shorter time intervals.   In response to a `burst' trigger from GBM during a high-energy solar flare, {\it Fermi} can autonomously be pointed at the Sun for up to five hours.  In addition, when the Sun is in an unusually high state of activity, a Target of Opportunity can be declared where the spacecraft remains in the rocking orientation most favorable to solar observations, thus providing $\sim$20-40 minute exposure during every 95 minute orbit.

\subsection{Light Bucket Analysis of {\it Fermi} LAT Data} \label{subsec:lightbucket}

Our method for analyzing LAT data can be described as a `light bucket'.  We accumulated $>$100 MeV photons having measured locations within 10$^{\circ}$ of the Sun; about 95\% of all 200 MeV solar $\gamma$ rays have measured locations within this 10$^{\circ}$ region.  We reduced $\gamma$-ray background from the Earth`s atmosphere by restricting events to those with angles $<$100$^{\circ}$ from the zenith.  The light-bucket approach is less sophisticated but significantly faster than the Maximum Likelihood procedure used by \citet{acke14} and \cite{ajel14} and was implemented with standard SSW IDL software\footnote{\url{http://www.lmsal.com/solarsoft/ssw_whatitis.html}}. It is almost as sensitive for detecting SGRE events and, as we detail in the next section and in Appendix \ref{sec:append}, it provides comparable time histories and spectroscopic results.  Included in the light bucket from the 10$^{\circ}$ region around the Sun are photons from the Galaxy, extra-galactic background, and the quiescent Sun \citep{abdo11}.  As a result, the $>$100 MeV background flux varies from $\sim3 \times 10^{-6}$ $\gamma$ cm$^{-2}$ s$^{-1}$ in March to $\sim4.5 \times 10^{-5}$ $\gamma$ cm$^{-2}$ s$^{-1}$ in mid-December, as the Sun moves along the ecliptic and passes through the Galactic plane.  

 
\subsection{Four-Day Plots of $>$100 MeV Solar Fluxes Observed by {\it Fermi}} \label{subsec:browplots}

\begin{figure}
\epsscale{.70}
\plotone{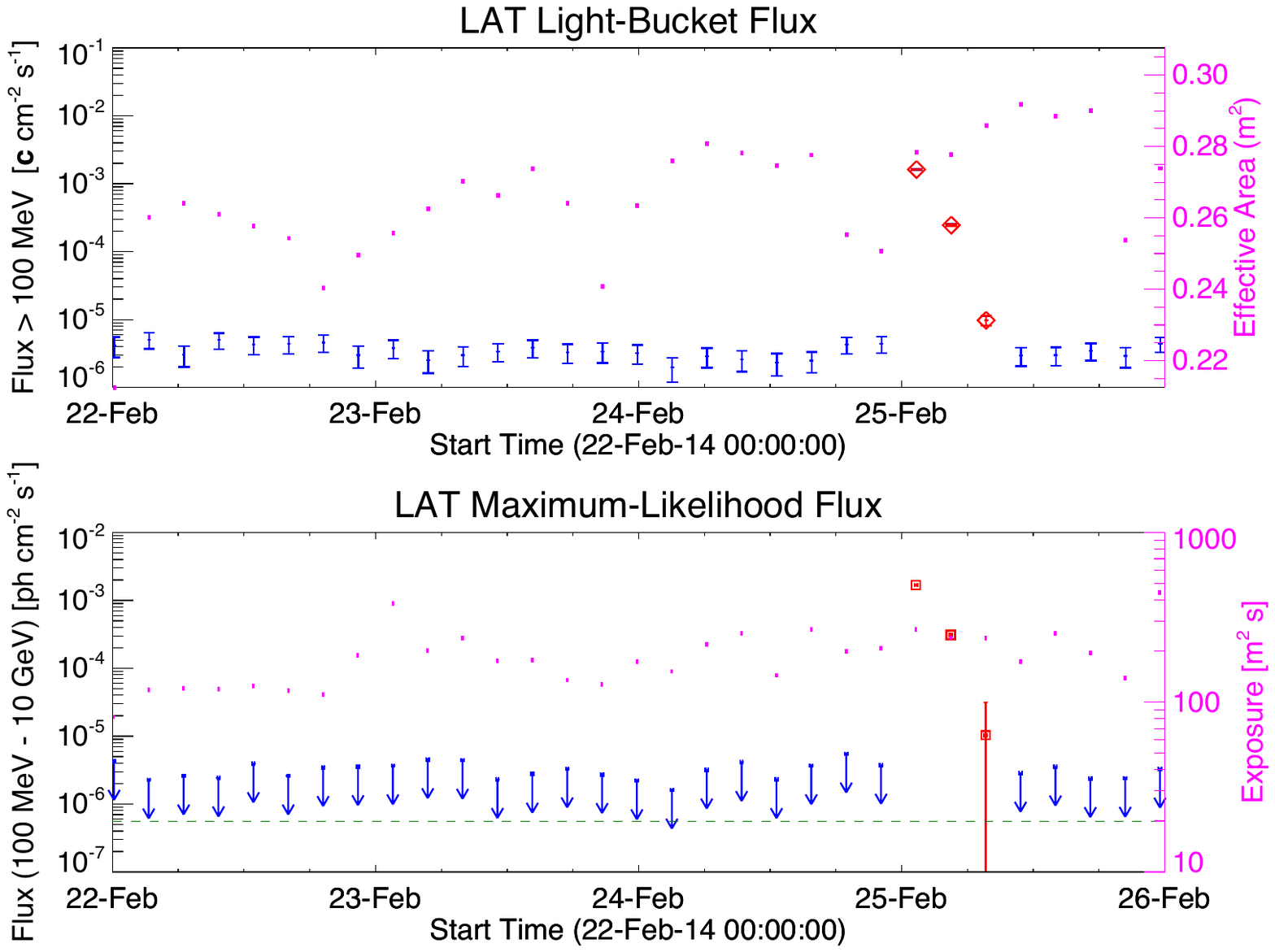}
\caption{Example of {\it RHESSI} Browser 4-day plots of LAT $>$100 MeV fluxes derived from the `Light Bucket' technique used in this paper (upper panel) and the Maximum Likelihood technique developed by \citet{acke14, ajel14} (lower panel).   The light bucket fluxes include background from Galactic, extra-galactic, and quiescent solar sources.  The right ordinate provides the scale for the pink rectangles that give the average effective area (top panel) and effective area $\times$ time (bottom panel).  The time period shown includes the SGRE event observed by {\it Fermi} on 2014 February 25 shown by the three red points with their $\pm 1 \sigma$ uncertainties in both plots.}
\label{browser}
\end{figure}
 
In Figure \ref{browser}, we show an example of the four-day time histories of $>$100 MeV solar fluxes observed by LAT that are available for the entire {\it Fermi} Mission on the {\it RHESSI} Browser\footnote{\url{http://sprg.ssl.berkeley.edu/~tohban/browser}}.  The top panel shows the $>$100 MeV `light-bucket' fluxes plotted as blue data points with $\pm1\sigma$ statistical uncertainties.  When the $>$100 MeV flux exceeds about 4.2 $\sigma$ above background ($\sim 2 \times 10^{-5}$ probability for a random distribution), as it did in three exposures separated by three hours on 2014 February 25, the points are plotted in red and identify possible $>$100 MeV solar events.  In the bottom panel of Figure \ref{browser} we plot the solar gamma-ray flux provided by the {\it FERMI} LAT team using their more sophisticated and sensitive Maximum-Likelihood method \citep{acke14}.  This technique models and removes the background so that only upper limits (in blue) are plotted unless there is a statistically significant solar transient which is plotted in red with $\pm1\sigma$ errors.  The fact that the solar transient fluxes derived using our `light bucket' method are in good agreement with those derived using the Maximum-Likelihood technique gives confidence that our simpler and faster analysis provides adequate accuracy for the study reported in this paper.  Both methods also provide information on the LAT solar exposures shown in pink in each plot; the average effective area is given in the top panel and the effective area $\times$ exposure time in the bottom panel.   More information about these plots can be found at \url{http://hesperia.gsfc.nasa.gov/fermi/lat/qlook/LAT_qlook_plots.htm}.  An advantage of the light-bucket solar data is that they are available in the standard form for spectral analysis using the OSPEX/SSW analysis routines; therefore, any scientist familiar with OSPEX\footnote{\url{https://hesperia.gsfc.nasa.gov/ssw/packages/spex/doc/ospex_explanation.htm}} can study LAT data in the same way that they study {\it RHESSI} and GBM spectral data. 


\section{The Thirty SGRE Events Observed by LAT} \label{sec:timehist}

We have identified 30 SGRE events from 2008 to 2016 from a search of the LAT fluxes plotted in the {\it RHESSI} Browser (see the example in Figure \ref{browser}).    These events are are listed in Table \ref{tab:latlist} and detailed in Appendix \ref{sec:characteristics}.   We do not include the behind-the-limb event on 2014 January 6 discussed by \citet{acke17} because it was not identified in our search the LAT data and was too weak for us to analyze.  There were three other 20--40 minute solar exposures exhibiting fluxes $>$4.2$\sigma$ above the mean 4-day background that we found in our search of the light-bucket plots in the {\it RHESSI} Browser.  Random fluctuations would be expected to produce one to two such events in the 5 $\times 10^4$ exposures of the study.

\begin{deluxetable}{ccccccc}
\tabletypesize{\scriptsize}
\tablecaption{LAT Sustained $>$100 MeV Emission (SGRE) Events from June 2008 to December 2016 \label{tab:latlist}}
\tablehead{
\colhead{Number} & \colhead{Date, Location} & \colhead{{\it GOES} X-Ray}& \colhead{CME}&\colhead{Type II}&\colhead{SEP}&\colhead{Hard X-ray } \\
\colhead{} & \colhead{yyyy/mm/dd, deg} & \colhead{Class, Start-End }&{Speed, km s$^{-1}$}& \colhead{M\tablenotemark{*}, DH} &\colhead{Flux (pfu), Energy (MeV)}&\colhead{Energy (keV)}}
\colnumbers
\startdata
1& 2011/03/07, N30W47 & M3.7, 19:43--20:58 & 2125 & 3?, Y  & 39.6, $>$60 & 300--1000\tablenotemark{d} \\
2 & 2011/06/02, S18E22 & C3.7, 07:22--07:57 & 976 & N, Y  & $\sim$0.1, $<$40\tablenotemark{b}  & --\tablenotemark{{e}}\\
3 & 2011/06/07, S21W54 & M2.5, 06:16--06:59  & 1255 & 2?, Y & 60.5, $>$100 & 300--800 \\
4 & 2011/08/04, N19W46 &  M9.3, 03:41--04:04  & 1315 & 2, Y  & 48.4, $>$100 &  300--1000\tablenotemark{d}  \\
5 &  2011/08/09, N16W70 &  X6.9, 07:48--08:08  & 1610 & 1?, Y & 16.3, $>$10 & 800--7000 \\
6 & 2011/09/06, N14W18 &  X2.1,  22:12--22:24 & 575, $\sim$1000\tablenotemark{a,b,h}  & 2, Y & 5.6, $>$100    &  300--1000\\
7 & 2011/09/07, N18W32&  X1.8, 22:32--22:44 & 792 & 1, N  & $<$1.7, $>$10\tablenotemark{f} & 300--1000\tablenotemark{d}  \\
8 & 2011/09/24, N14E61 & X1.9, 09:21--09:48 & 1936 & 2?, N & $<$77, $>$13\tablenotemark{b,f} & 800--7000  \\ 
9 & 2012/01/23, N33W21 & M8.7, 03:38--04:34 & 2175 & N, Y & 3280, $>$100 & 100--300\tablenotemark{{d,e}} \\
10 &2012/01/27, N35W81 & X1.7, 17:37--18:56 & 2508 & 3, Y &  518, $>$100 & 100--300\tablenotemark{d,e}  \\
11 & 2012/03/05, N16E54 & X1.1, 02:30--04:43  &  1531 & N, Y & $<$33, $>$13\tablenotemark{b,f} & 100--300\tablenotemark{{d,e}} \\
12 & 2012/03/07, N17E27 & X5.4, 00:02--00:40 & 2684 & 2?, Y  &  1800, $>$100 &  $>$1000\tablenotemark{g}  \\
  &    & M3, 01:05--01:23 & 1825 & 2?, Y & 1800, $>$100    &  $>$1000\tablenotemark{g} \\
13 & 2012/03/09,  N16W02 &  M6.3, 03:22--04:18 & 950 & 2, Y & $<$528, $>$10\tablenotemark{f} & 100--300  \\
14 & 2012/03/10, N18W26 &  M8.4, 17:15--18:30 & 1296 & N?, Y  &  $<$115, $>$10\tablenotemark{f} & 100--300\tablenotemark{d}  \\
15 & 2012/05/17, N05W77 & M5.1, 01:25--02:14 & 1582 & 3, Y & 180,  $>$100 & 100--300\tablenotemark{c}  \\ 
16 & 2012/06/03, N15E38 & M3.3, 17:48--17:57 & 605, 892\tablenotemark{b,h} & 2, N & 0.6, $>$60\tablenotemark{b}  &  300--800  \\
17& 2012/07/06, S17W52 & X1.1, 23:01--23:14  & 1828 & 3, Y & 19.1, $>$100  &  --\tablenotemark{e}  \\
18& 2012/10/23, S15E57 & X1.8, 03:13--03:21  & -- & Y, N & $<$0.1, $>$13\tablenotemark{b} & $>$9000 \\
19& 2012/11/27, N05W73 & M1.6, 15:52--16:03 & -- & N, N & $<$0.1, $>$10     &  300--1000\\
20& 2013/04/11, N07E13 & M6.5, 06:55--07:29  & 861 & 3, Y & 184, $>$60\tablenotemark{b}  &  100--300\tablenotemark{d}  \\
21& 2013/05/13, N11E89 & X1.7, 01:53--02:32  & 1270 & 1, Y & 9.3, $>$60\tablenotemark{b}  &  100--300  \\
22& 2013/05/13, N10E80 &  X2.8, 15:48--16:16  & 1850 & 2, Y & 176, $>$60\tablenotemark{b}  &  $>$1000  \\
23& 2013/05/14, N10E77 &  X3.2, 00:00--01:20  & 2625  & 1, Y? & 306, $>$60\tablenotemark{b}  & 300--1000\tablenotemark{d}    \\
24& 2013/05/15, N11E65 &  X1.2, 01:25--01:58  & 1366 & 1, Y & $<$17, $>$13\tablenotemark{b,f}  &  300--1000\\
25& 2013/10/11, N21E103 & M4.9\tablenotemark{i}, 07:01--07:45 & 1182 & 2, Y & 156, $>$60\tablenotemark{b} & -- \tablenotemark{j}\\
26& 2013/10/25, S08E71&  X1.7, 07:53--08:09 & 587 & 2, N & 32.6, $>$60\tablenotemark{b} & 800--7000\tablenotemark{c} \\ 
27&2013/10/28, S14E28& M4.4, 15:07--15:21 & 812 & 2, N & 5.6, $>$13\tablenotemark{b}&100--300\tablenotemark{c} \\
28& 2014/02/25, N00E78 & X4.9, 00:39--01:03 & 2147   & 3, Y & 219\tablenotemark{b}, $>$700 & 1000--10000\\
29& 2014/09/01, N14E126 & X2.1\tablenotemark{i}, 10:58--11:34 & 1901 & Y?, Y  & $\sim$1000, $>$13 &
-- \tablenotemark{j}\\
30& 2015/06/21, N13E16 & M2.6, 02:03--03:15 & 1434 & 2?, Y &$\sim$40, $>$10 & 100-300\tablenotemark{d}\\
\enddata
\tablenotetext{a}{{\it STEREO A}}
\tablenotetext{b}{{\it STEREO B}}
\tablenotetext{c}{{\it RHESSI}}
\tablenotetext{d}{{\it Fermi}/GBM}
\tablenotetext{e}{Missing HXR data due to night time or SAA passage} 
\tablenotetext{f}{Preceding SEP} 
\tablenotetext{g}{\it INTEGRAL}
\tablenotetext{h}{CACTUS}
\tablenotetext{i}{\citet{pesc15c}}
\tablenotetext{j}{flare behind solar limb}
\tablenotetext{*}{1, 2, 3 $\simeq$ $<$50, 50--500, $>$500 $\times 10^{-22}$ W m$^{-2}$ Hz$^{-1}$}

\end{deluxetable}
\

Table \ref{tab:latlist} is in the same format as Table \ref{tab:full} discussed in Appendix \ref{sec:characteristics}.  Much of the information comes from the NOAA solar event reports\footnote{\url{ftp://ftp.ngdc.noaa.gov/STP/swpc_products/daily_reports/solar_event_reports/}}.  Column 1 lists the event number, column 2 lists the date of the event and location of the flare or centroid of the hard X-ray footpoints imaged by {\it RHESSI}, column 3 lists the {\it GOES} soft X-ray class (we used estimates made by \citet{pesc15c} for the location and {\it GOES} class for the two events beyond the limb) and the {\it GOES} start and stop times, column 4 lists the projected CME speed from the {\it SOHO} LASCO catalog\footnote{\url{https://cdaw.gsfc.nasa.gov/CME_list/index.html}} (unless noted), column 5 lists the Type II metric intensity from the solar event reports (``?" means that we could not confirm the detection spectroscopically), and whether decameter-hectometric (DH) Type II emission\footnote{\url{http://secchirh.obspm.fr/select.php},\url{https://ssed.gsfc.nasa.gov/waves/data_products.html}} was observed in space (Y/N), column 6 lists the peak SEP proton flux above 10 MeV in pfu (1 proton cm$^{-2}$ sr$^{-1}$ s$^{-1}$) and highest energy measured at the best magnetically connected spacecraft, and column 7 lists the highest energy channel that the impulsive flare was detected by {\it RHESSI} or GBM in hard X-rays.  Additional information about the sources of the data are provided in Table \ref{tab:full}.  We note that event 12 in Table \ref{tab:latlist} on 2012 March 7 was comprised of two flares and two SGRE events; the second flare, reported as an X1.3 {\it GOES}-class flare, is actually an M7 flare when the tail of the preceding X5.3 flare emission is subtracted.  

\subsection{Time Histories and Details of the SGRE Events} \label{subsec:details}

We present time histories and observational details of each of the 30 SGRE events in Appendix \ref{sec:append}.  In Figure \ref{110307th} we provide an example of one of these time histories.  SOL2011-03-07T19:43, was the first SGRE event detected by LAT \citep{acke14} and lasted about 14 hours.  In the main part of the figure, we plot the $>$100 MeV $\gamma$-ray fluxes (and $\pm 1\sigma$ statistical uncertainties) within the 10$^{\circ}$ accumulation region centered on the Sun\footnote{The flux evaluation is done using the Sun's position at 12 UT on that day}  for several hours before and after the flare.  The flux in each exposure was estimated using the average detector efficiency, assuming that the incident photon spectrum follows a power law with an index of 2 which is consistent with the background spectrum composed of Galactic, extra-galactic, and quiescent solar photons.  Even though solar flare and sustained $>$100 MeV emission spectra are typically softer than the background-subtracted spectrum, we find that our plotted SGRE fluxes are within about 50\% of those derived from detailed spectral fits to background-subtracted data ($\S$\ref{sec:spect}). The dashed vertical lines show the {\it GOES} 1--8{\AA} soft X-ray start and end times given in Table \ref{tab:latlist}.  It is clear from the main plot that the SGRE began around the time of the flare and reached a peak about six hours after the flare, consistent with the time history presented by \citet{acke14}. The six-hour delay to the peak of the SGRE is in stark contrast with the two minute delay observed in the 1982 June 3 event (Figure \ref{june3}).  In addition the associated flare was only an M3.7 {\it GOES}-class event, compared to the X8 flare on June 3.  

In the inset of Figure \ref{110307th} we plot a blowup of data spanning the {\it GOES} soft X-ray event (dashed vertical lines).  The LAT $>$100 MeV fluxes with the magnitudes given by the left-ordinate scale are plotted at 4 minute time resolution. The dashed curve depicts the time profile of the {\it GOES} 1--8{\AA} X-ray flux plotted logarithmically where the magnitude is provided by the right ordinate (e.g., X $= 10^{-4}$ W m$^{-2}$).   The horizontal right- and left-pointing arrow heads show the estimated range of CME onset times using linear and quadratic extrapolations given in the CDAW LASCO CME Catalog\footnote{\url{https://cdaw.gsfc.nasa.gov/CME_list/index.html}}. We also studied {\it SDO} EUV images to determine the CME onset and this is shown by the downward solid arrow.  The dashed downward arrow shows the estimated onset time of Type II radio emission given in the NOAA Solar Event Reports\footnote{\url{ftp://ftp.ngdc.noaa.gov/STP/swpc_products/daily_reports/solar_event_reports/}} or derived from studies of the radio spectra.  Unlike the 1991 June 3 flare where pion-decay $\gamma$-rays were observed during the impulsive flare, GBM and {\it RHESSI} only observed impulsive hard X-rays up to 100--300 keV with no evidence for nuclear line emission.  The combined GBM/{\it RHESSI} 100-300 keV time history is plotted as the solid trace in the inset and the rate peaks after the onsets of CME and Type II radio emission.   LAT began observing the Sun near the end of the impulsive phase.  The $>$100 MeV fluxes plotted at four-minute resolution in the inset clearly reveal that the SGRE began  within minutes of the hard X-ray peak.  We estimated the range of SGRE onset times by extrapolating the best linear fit and $\pm1\sigma$ uncertainties, shown by the dashed lines, to background level.  It is clear from the time history of the SGRE that it is due to a distinct particle acceleration phase and is not just the tail of emission from the impulsive flare.

\begin{figure}
\epsscale{0.8}
\plotone{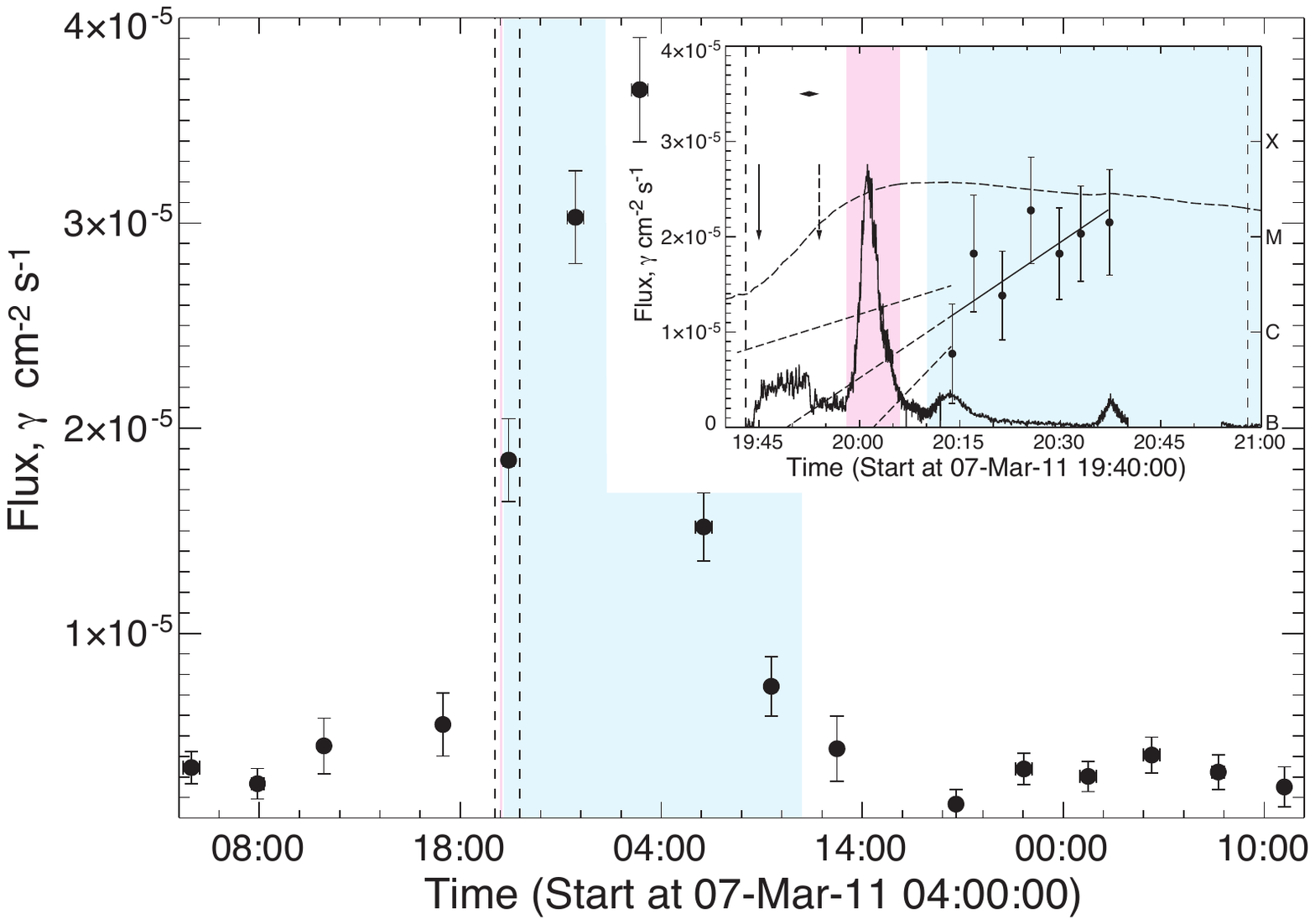}
\caption{The main figure shows the time history of the $>$100 MeV flux from $<$10$^{\circ}$ of the Sun, derived from source-class data, revealing the 2011 March 7 SGRE event.  The fluxes were averaged over the $\sim$20--40 minute solar exposures and the uncertainties are $\pm1\sigma$ statistical errors.  Vertical dashed lines show the {\it GOES} start and end times. The inset shows 4-minute accumulation LAT $>$100 MeV fluxes derived from source-class data. The best linear fit to the rising flux is shown by the solid line and its extrapolation back to background is shown by the dashed line; the other two dashed lines are extrapolations of $\pm 1 \sigma$ deviations from the best fit. The combined 100-300 keV count rate observed by {\it RHESSI} and GBM during the impulsive flare, scaled to the $\gamma$-ray flux, is shown by the solid trace. The dashed curve shows the {\it GOES} 1--8{\AA} time history (logarithmic scale on right ordinate) and the $<->$ symbol shows the range in CME onset times in the CDAW catalog derived for linear and quadratic extrapolations.  The vertical solid arrow depicts our estimate of the CME onset from inspection of {\it SDO}/AIA images and the vertical dashed arrow shows the estimated onset of Type II radio emission.  The blue-shaded region depicts our estimate of the duration of the sustained emission.  The pink-shaded region depicts where we made estimates of the flux of $>$100 MeV impulsive-flare $\gamma$-ray emission.}
\label{110307th}
\end{figure}


\section{Spectroscopic Studies} \label{sec:spect}

Spectroscopic measurements made by LAT, {\it RHESSI}, and GBM are critical to our study of SGRE.  In $\S$\ref{subsec:latspect} we describe the techniques used to analyze the $>$100 MeV solar emissions observed by LAT.  We discuss how we used fits of calculated pion-decay spectra to measured impulsive-flare and SGRE spectra to obtain the $>$100 MeV $\gamma$-ray fluxes, the proton power-law spectral indices, and the numbers of $>$500 MeV protons accelerated at the Sun during these two phases. We include a discussion of fits to a representative SGRE spectrum to  demonstrate that the SGRE can be explained by pion-decay emission but not by bremsstrahlung from primary electrons ($\S$\ref{subsubsec:nobrems}).   In $\S$\ref{subsec:below100MeV} we discuss how we used neutron-capture line measurements from {\it RHESSI} and GBM spectra to estimate of the number of $>$500 MeV protons accelerated in the impulsive flare, especially when there are no LAT data available, and to obtain information about the spectra below 300 MeV of the protons producing the SGRE.  In $\S$\ref{subsec:specfits} we discuss Table \ref{tab:event} that presents the results of our fits to the 30 SGRE events.

\subsection{Spectroscopic Studies of LAT Sustained-Emission Data: Pion-Decay Fits} \label{subsec:latspect}

\begin{figure}
\epsscale{.70}
\plotone{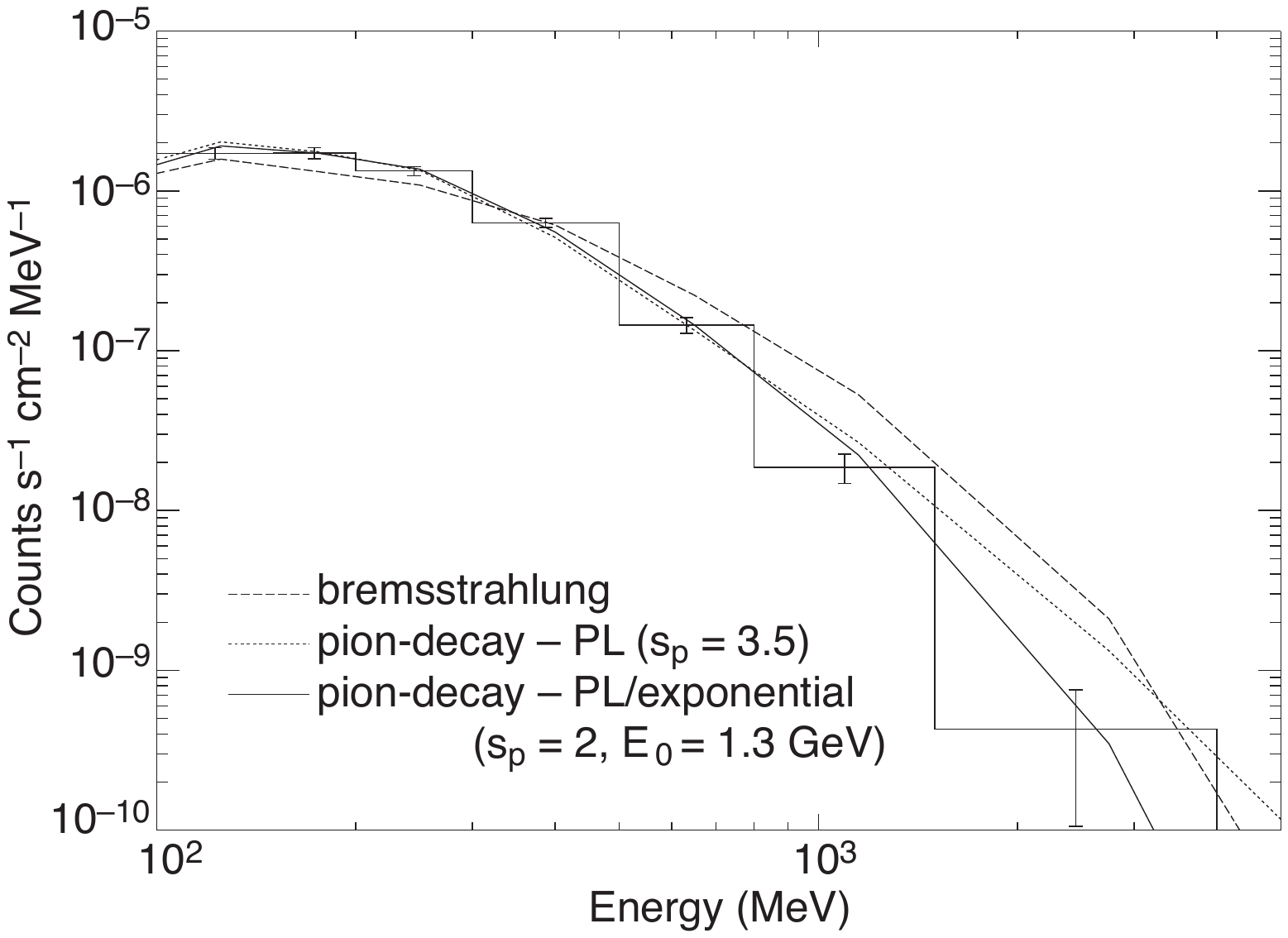}
\caption{Background-subtracted LAT count spectrum with $\pm1\sigma$ statistical uncertainties meaured between 01:13:30 and 01:17:30 UT on 2014 February 25 during LAT's peak exposure to the Sun.  Best fits to the spectrum after passing three different photon spectra through the instrument response are shown: 1) a pion-decay spectrum produced by a power-law spectrum of protons with spectral index, s$_p$ = 3.5 (dotted curve); 2) a pion-decay spectrum produced by a power-law spectrum of protons with spectral index, s$_p$ = 2 and 1.3 GeV exponential cutoff energy ($E_0$) (solid curve); and 3) a bremsstrahlung spectrum produced at a density 10$^{16}$ cm$^{-3}$ from a power-law spectrum of electrons with index, s = 1 and 1 GeV exponential cutoff energy ($E_0$) in a $10^3$ G magnetic field (dashed curve).}
\label{sp140225}
\end{figure}

Our spectroscopic analyses of the LAT data were done using OSPEX software\footnote {\url{http://hesperia.gsfc.nasa.gov/ssw/packages/spex/doc/ospex_explanation.htm}} included in the SolarSoft package\footnote{\url{http://www.lmsal.com/solarsoft/}} developed for the analysis of {\it RHESSI} solar data and operating within IDL\footnote{\url{http://www.harrisgeospatial.com/ProductsandSolutions/GeospatialProducts/IDL.aspx}}.  We converted the publicly available LAT solar data and instrument responses into formats compatible with OSPEX.  The next step in obtaining sustained-emission $\gamma$-ray spectra is subtraction of background.  As can been seen in Figure \ref{browser} in $\S$\ref{subsec:browplots}, the $>$100 MeV background in the LAT source-class data from a 10$^{\circ}$ region around the Sun is relatively constant on timescales of hours, as long as we exclude data $<$12$^{\circ}$ from the Earth's horizon.  Therefore, we chose as background four LAT observation intervals before and/or after the event with solar exposures comparable to the SGRE spectra we were studying.  In Figure \ref{sp140225}, we show the background-subtracted count spectrum on 2014 February 25 (Event 28 Table \ref{tab:latlist}, Appendix \ref{subsec:20140225}) from one of the brightest SGRE events observed by LAT.  The spectrum was accumulated over four minutes (01:13--01:17 UT) during the peak exposure to the Sun to avoid any instrumental issues near the edge of the field of view.

 As discussed in \citet{acke14} and \citet{ajel14}, the photon spectra observed in the SGRE events has the shape of a power-law with an exponential cutoff varying from several tens of MeV to a few hundred MeV.  Such a shape arises naturally from the decay of neutral and charged pions produced when high-energy protons and $\alpha$ particles interact in the solar chromosphere and photosphere \citep{murp87}.  The 1982 June 3 event shown in Figure \ref{june3} and other events discussed by \citet{ryan00} also have $\gamma$-ray spectra $>$10 MeV consistent with this pion-decay shape.  As will be shown below, there is no plausible electron spectrum whose bremsstrahlung can provide an adequate fit to the data. 

To fit LAT spectral data, we use pion-decay spectra calculated with a code developed by \citet{murp87}. The code calculates the yield of neutral and charged pions and the associated $\gamma$-ray spectra from neutral-pion decay, bremsstrahlung of positrons and electrons from charged-pion decay, and annihilation in flight of positrons from positive-pion decay.  The calculations are performed assuming that the particles interact isotropically in a cold thick-target region and use nuclear data and models for the p + p and p + $^4$He reactions (and the inverse reaction  $\alpha$ + H).  Both the accelerated and ambient $^4$He/H abundance ratios are assumed to be 0.1.

Due to the high $>$300-MeV proton threshold for pion production, the pions and photons are produced deep in the solar atmosphere, where scattering effects on the escaping photons are significant, especially for flares observed at heliocentric angles $>$70$^{\circ}$.  We took these heliocentric-dependent scattering effects into account by using an estimated depth distribution for pion production and a Monte Carlo N-Particle (MCNP6) photon-propagation code\footnote{\url{https://mcnp.lanl.gov/}}, employing a spherical geometry model for the Sun, to calculate the spectra of the escaping photons. Details of this work will be discussed in a separate publication. We find that there is only a small difference in the escaping-photon spectral shape from flares at varying heliocentric angles. The primary effect is the reduction in the number of photons that escape. Only $\sim$82\% of the 100 MeV pion-decay photons that escape at disk center can escape at a heliocentric angle of 70$^{\circ}$. This number drops to $\sim$47\% at 85$^{\circ}$ and to $\sim$8\% at 90$^{\circ}$. This has a significant impact on estimates of the derived number of ions required to produce the observed emission.

When fitting LAT spectral data, we use the pion-decay templates derived for the location of the associated active region and vary the proton spectral index to obtain the best fit to the spectrum. The $\chi^2$ statistic was used to determine the proton spectral index and its estimated 1$\sigma$ uncertainty was determined from the values where $\chi^2$ increased by unity. Using the derived $\gamma$-ray flux and proton spectral index, we then applied the results of \citet{murp87} to calculate the number of accelerated protons with energies $>$500 MeV at the Sun during each LAT exposure.  Because of the 300 MeV threshold for neutral pion production, the number of protons $>$500 MeV is relatively insensitive to uncertainties in spectral index. 

Because of the limited duty cycle of the LAT observations, our knowledge of the temporal evolution of the SGRE is limited. For events where only one LAT exposure is available, we generally estimate the number of accelerated protons by assuming that the flux onset occurred at the end of the previous (null) LAT observation and increased linearly to the peak. Similarly, we assume that the flux decayed linearly from the peak to the beginning of the following (null) LAT observation. If there are higher time cadence LAT data available in the exposure, they may be used to estimate the onset time by linear extrapolation (see Figure \ref{110307th}). For events where more than one exposure is available, we estimated the total number of protons by assuming that the flux changed linearly with time between measurements. 

For most of the spectral studies we were able to use the standard source-class data products. However, these products could not be used in studying the spectra of impulsive or sustained emission in periods of intense hard X-ray emission during the flare. The LAT team provides LAT Low Energy (LLE) and Pass8 solar-impulsive-class data products for use during these times that are not sensitive to dead-time effects from high ACD rates. LLE data were used by \citet{acke12a, acke12b} in their study of SOL2010-06-12T00:57. We fit spectra from one or both of these data products during the impulsive flares accompanying 14 of the 30 events to determine the $>$100 MeV $\gamma$-ray fluxes, proton spectral indices, and numbers of protons producing  the impulsive and sustained emissions. As the LLE data products extend to energies below 100 MeV, we were also able to determine the electron bremsstrahlung contribution in a few impulsive phase spectra.

As an illustration of our fits to LAT spectra, we show the LAT count spectrum of the 2014 February 25 event in Figure \ref{sp140225}.  We assumed that the 2014 February 25 emission came from the active region located at a heliocentric angle of 78$^{\circ}$ and fit the count spectrum with pion-decay templates calculated for accelerated proton power-law spectra ($dN/dE \propto E^{-s_p}$) having various indices, s$_p$. The best-fitting spectrum was for an index s$_p$= 3.45 $\pm$ 0.15 (dotted curve; $\chi^2$/dof of 3.7; probability 0.2\%). With the excellent statistics of the measured count spectrum, it is clear that the photon spectrum must steepen above $\sim$1 GeV, implying that the proton spectrum is not a single power-law but rolls over at high energies. We found that a power-law with exponential proton spectrum ($dN/dE \propto E^{-s_p} exp^{-E/E_0}$) having an index, s$_p$ = 2, and an exponential cutoff, $E_0$ = 1.3 GeV, was a good fit to the measured spectral data (solid curve; $\chi^2$/dof of 1.2; probability 31\%). For the statistics available in most events in our study, fits with pion-decay spectra from protons with spectra following a single power-law are adequate. Because of the threshold for pion production the information derived from these fits is only valid for proton energies above 300 MeV.



\subsubsection{Sustained $>$100 MeV $\gamma$-Ray Spectra Cannot be Fit by Electron Bremsstrahlung}\label{subsubsec:nobrems}

\begin{figure}
\epsscale{.80}
\plotone{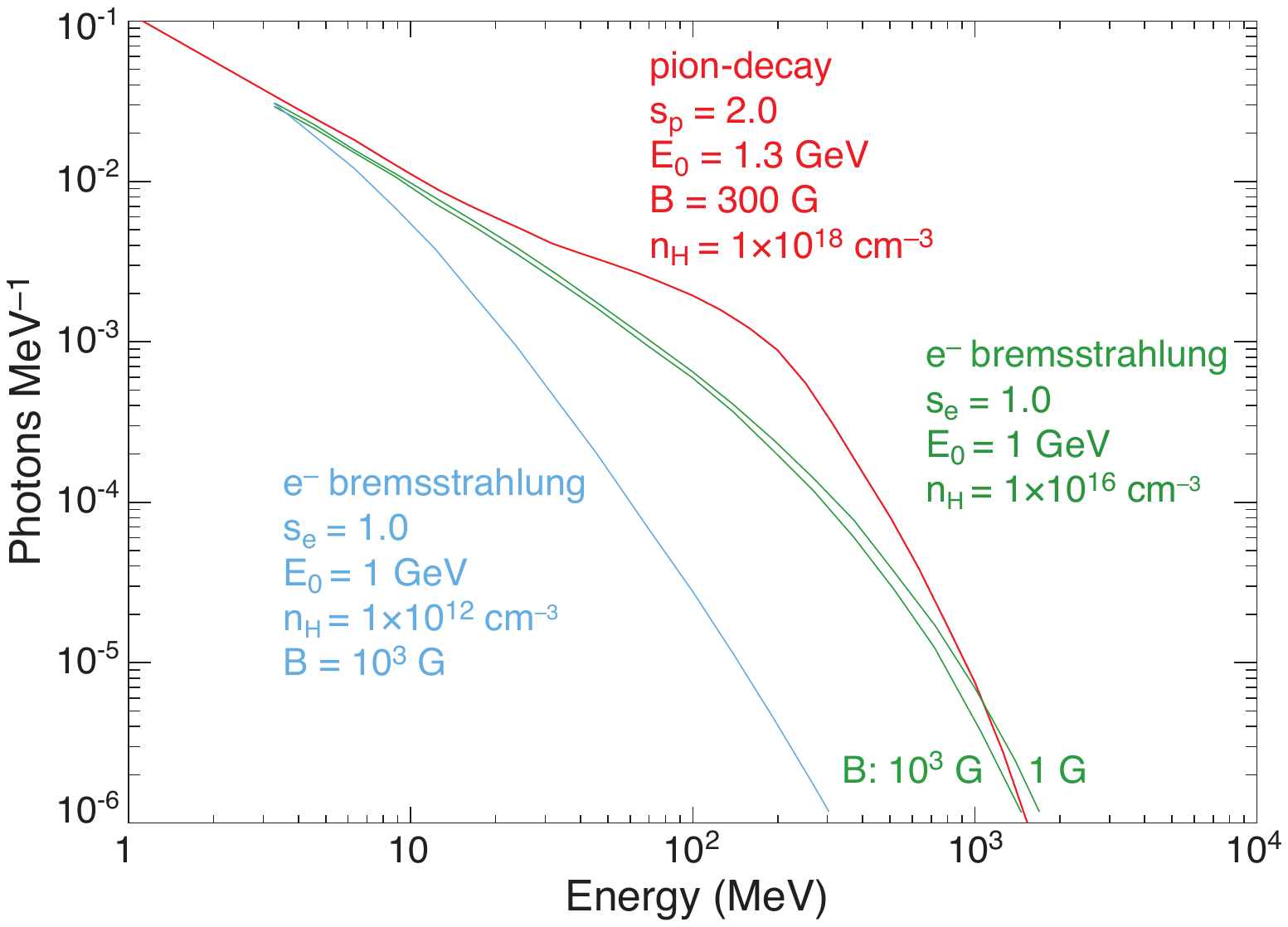}
\caption{Comparison of calculated electron-bremsstrahlung spectra for two different assumed magnetic field strengths (green curves) at a density of 10$^{16}$ cm$^{-3}$ with the pion-decay production spectrum (red curve) that fits the $\gamma$-ray spectrum observed by LAT between 01:13:30 and 01:17:30 UT on 2014 February 25 shown in Figure \ref{sp140225}.  The pion-decay spectrum was produced by protons interacting at a density of 10$^{18}$ cm$^{-3}$ and having a power-law spectrum with 1.3 GeV exponential cutoff.  For comparison we also plot the bremsstrahlung spectrum for the same electron spectrum but for a density of 10$^{12}$ cm$^{-3}$ and 1000 G field (blue curve).}
\label{brem}
\end{figure}

A fundamental question is whether the SGRE spectra are also consistent with bremsstrahlung from accelerated electrons. Based on general arguments, \citet{ajel14} conclude that ``protons seem to be more likely agents of the $\gamma$-ray production."  In Figure \ref{brem} we plot the pion-decay photon spectrum (red curve) produced by interactions of protons with a power-law spectrum with an exponential cutoff, $E_0$ = 1.3 GeV, that provides a good fit to the LAT count spectrum observed at the peak of the SGRE event on 2014 February 25 (Figure \ref{sp140225}).  We attempted to find an electron bremsstrahlung spectrum that approximates the same shape.   In order to duplicate the flat photon spectral shape at low energies, we chose a power-law electron spectrum with index, s$_e$ = 1,.  In order to duplicate the roll-over in the pion-decay spectrum that fits the data well above a few hundred MeV, we introduced an exponential cutoff, $E_0$, in the electron spectrum.  We varied the cutoff energy to provide the closest fit to the pion-decay spectrum and found that this occurred at about one GeV.   

At these high electron energies, it is necessary to take into account the effect that synchrotron losses have on the shape of the resulting bremsstrahlung spectrum.  The solid green curves in Figure \ref{brem} show how the bremsstrahlung spectrum changes with increasing magnetic field when the interaction region has a density of $10^{16}$ cm$^{-3}$.   Electrons with energies $>$100 MeV can penetrate to these chromospheric densities.  At such densities, bremsstrahlung dominates over synchrotron losses so that even with a magnetic field of 1000 G, the photon spectrum changes only marginally as can be seen by the two green curves.  The fit of this optimized bremsstrahlung spectrum to the LAT spectral data is poor (probability of $<10^{-5}$).  If the thick target has a lower density, 10$^{12}$ cm$^{-3}$), the bremsstrahlung spectrum in a 1 G field has a shape similar to the green curve plotted for 1000 G and $10^{16}$ cm$^{-3}$ density.   If we increase the field strength to 1000 G at this lower density, the spectrum rolls over more rapidly at high energies (see blue curve) and is an even worse representation of the best-fit pion-decay spectrum.  It is, therefore, clear that electron bremsstrahlung fails to reproduce the broad peak near 100 MeV resulting from $\pi^{0}$ decay.  This poor fit of electron bremsstrahlung to the LAT count spectrum is shown by the dashed curve in Figure \ref{sp140225}.  We conclude that bremsstrahlung from plausible electron spectra cannot fit the shape of the sustained $\gamma$-ray emission in the intense event on February 25 and that the source of $>$100 MeV $\gamma$-rays is most likely the decay of pions produced by proton interactions. 

We made similar comparisons of pion-decay and bremsstrahlung spectral shapes for sixteen separate LAT exposures where the $>$100 MeV flux exceeded $>$10$^{-4} \gamma$ cm$^{-2}$ s$^{-1}$ in six other SGRE events.  With $>$99.9\% confidence, the pion-decay shape provided a significantly better fit to the $\gamma$-ray spectra in each exposure than did any plausible electron bremsstrahlung spectrum.

\subsection{Using the Neutron-Capture Line for Studying Impulsive Flare Protons and SGRE Protons $<$300 MeV} \label{subsec:below100MeV}

The solar flare $\gamma$-ray spectrum between 300 keV and 8 MeV includes contributions from electron bremsstrahlung, nuclear de-excitation lines such as the 4.43 MeV line from carbon, the 0.511 MeV annihilation line, and the 2.223 MeV neutron-capture line \citep{vilm11}.  The 2.223 MeV line is usually the most prominent feature in the spectrum for flares far from the solar limb and is produced by protons ranging in energy between 2 and several hundred MeV, depending on the steepness of the spectra \citep{murp07}.  Here we describe how we used {\it RHESSI} and GBM measurements of the 2.223 MeV line to estimate the number of $>$500 MeV protons in the impulsive phase of SGRE events, especially when no LAT $>$100 MeV $\gamma$-ray observations were available.

We typically fit background-subtracted spectra between 1.8 and 2.4 MeV with a continuum and Gaussian peak to obtain the measured 2.223 MeV line flux, or its upper limit.  For a few intense events, we fit the full 0.2 to 8 MeV spectra with all the nuclear and bremsstrahlung components.  We then used \citet{murp87} to calculate the number of protons $>$500 MeV accelerated into the thick target solar atmosphere, assuming that the proton spectrum followed a single power law from tens of MeV to several hundred MeV.  Using nuclear-line ratios from a fit the sum of 19 flares observed by SMM, \citet{murp16} found that the average 2--20 MeV proton spectrum had a power-law index of 4.9.  \citet{acke12a, acke12b} provided evidence that the proton spectrum during the impulsive flare on 2010 June 12 flare softened from an index $\sim$4.3 between about 50 and 300 MeV to an index $>$4.5 above 300 MeV.   We therefore assumed a proton power-law index of 4.5 to estimate the number of protons $>$500 MeV during the impulsive flare based on the measured 2.223 MeV line flux.

We also used upper limits on the 2.223 MeV neutron-capture line at times when the $>$100 MeV SGRE was most intense to obtain information on the spectrum of protons below 300 MeV accelerated into the thick target solar atmosphere.  Using \citet{murp87} and 95\% confidence upper limits on the 2.223 MeV/$>$100 MeV $\gamma$-ray flux ratio during each LAT exposure, we obtained upper limits on the 30--300 MeV proton spectral index, s$_p$, (i.e. the spectra could be harder) that we could compare with the spectral index above 300 MeV obtained by fitting the LAT $\gamma$-ray spectrum.  This allowed us to determine whether the SGRE proton spectra softened at energies above several tens of MeV.  We discuss the results of this study in $\S$\ref{subsubsec:LoE_index}.

\subsection{Spectral Fits to 30 SGRE Events} \label{subsec:specfits}

We used the methods described above to obtain spectral information on the protons responsible for the SGRE and for the impulsive flare.   We present the results for each of the 30 SGRE events in Table \ref{tab:event} in Appendix \ref{sec:resultstable}.  Table \ref{tab:example} provides an example of the entries for the two sustained emission eruptions on 2012 March 7, event 12 in Table \ref{tab:event} (see details in Appendix \ref{subsec:20120307}).  We note that \citet{ajel14} treat the first $>$100 MeV eruption as part of the X5.4 flare, but the plotted time history indicates that flux was rising to a peak and that it is not the flare's decay phase (top right inset of Figure \ref{120307th}).  The second increase in $>$100 MeV emission began an hour after the second flare (the M7 flare is listed as X1.3 in the solar event lists, but most of the X-rays came from the tail of the first flare). The first column lists the date of the event and the Table \ref{tab:latlist} event number in parentheses.  In column two we list the type of emission providing the information in each row. There are three emission types: 1) SGRE, 2) impulsive flare designated by its {\it GOES} class, and 3) SEP.  If there was more than one LAT exposure for each event, we put the number of the exposure in parenthesis, e.g. SGRE(2).  As there were two SGREs on 2012 March 7 we distinguish them by designating them as SGRE(A) or SGRE(B).   The third column gives the time intervals of the LAT solar exposures.  The measured $>$100 MeV $\gamma$-ray flux and uncertainty, based on our spectral fits with pion-decay templates, is listed in column 4.\footnote{The fluxes may be different than those plotted in the time history figures in Appendix \ref{sec:append} because the latter included background and were estimated assuming a harder power-law photon spectrum.}  The best-fit spectral index and uncertainty of the protons producing this flux, assuming a differential power-law spectrum at energies above the 300 MeV proton threshold energy, are listed in column 5.  It is clear that the proton spectrum softened with time throughout this 18-hour event.  Using the $>$100 MeV $\gamma$-ray flux, proton spectral index, and heliocentric angle of the flare site and assumptions about the temporal structure of the sustained emission noted above, we show in column 7 the estimated numbers of $>$500 MeV protons at the Sun for each time interval given in column 6.  In the row labelled `SGRE Total' we list our estimate of the total number of $>$500 MeV protons and its uncertainty, based on both statistical errors and our confidence in our knowledge of the time history and duration of the event.  For this event, we estimate that SGRE(B) was produced by three times the number of protons responsible for SGRE(A).

For the 2012 March 7 example event and for other intense SGREs we provide estimates of the proton spectral index between 30 and 300 MeV during the sustained-emission exposures based on a comparison of measured fluxes of the 2.223 MeV neutron-capture line and $>$100 MeV emission, using the procedure outlined in $\S$\ref{subsec:below100MeV}.  This information is provided in column 5 of the rows listed as `SGRE $<$300 MeV'.  There are four such measurements listed in the table and they indicate that the spectrum is harder at lower energies than it is above 300 MeV.  For the rows listing the flare {\it GOES} class, we include in column 7 our estimates for the number of $>$500 MeV protons accelerated during the impulsive phase of the flares, using either available LAT measurements ($\S$\ref{subsec:latspect}) or observations of the 2.223 MeV neutron capture line by {\it RHESSI} or GBM (see $\S$\ref{subsec:below100MeV}).  For this example event, as neither {\it RHESSI} nor GBM observed the X5.4 flare, we used 2.223 MeV line measurements from {\it INTEGRAL} \citep{zhan12}.  We also used {\it INTEGRAL} measurements of the 2.223 MeV line to estimate the number of protons in the M7 flare.  The number of $>$500 MeV protons for the two flares on 2012 March 7 estimated from these line measurements are denoted by the footnote$^b$.  We were also able to estimate an upper limit on the number of impulsive-flare protons in the M7 flare using LAT $>$100 MeV observations.  This limit is provided in the next row of the table.   In column 7, in the rows labelled `SEP' we list estimates of the total, time-integrated number of $>$500 MeV protons in space.  The asymmetric uncertainties are 1$\sigma$.  We summarize our procedure for determining the number of protons in space in Appendix \ref{sec:sep}.

\begin{deluxetable}{ccccccc}
\tablecaption{Spectral Characteristics of SGRE for the 2012 March 7 Event \label{tab:example}}
\tablehead{
\colhead{Date (Event)} & \colhead{Type} & \colhead{Observing} & \colhead{Flux $>$100 MeV}& \colhead{Proton PL Index ,s} &\colhead{Emission} & \colhead{$10^{28}$ Protons } \\
\colhead{yyyy/mm/dd} & \colhead{}  &\colhead{Interval, UT} & \colhead{10$^{-4} \gamma$ cm$^{-2}$ s$^{-1}$ }& $>$300 MeV &\colhead{Interval, UT}  &\colhead{$>$500 MeV}
}
\colnumbers
\startdata
2012/03/07 (11)& SGRE(A) Total & 00:39--01:24 &   28.7 $\pm$ 0.4 & 3.6 +- 0.3  & 00:28--01:24   &     40 $\pm$ 15\\
 &  SGRE(B1)    &02:18--02:48 &  5.8 $\pm$ 0.3  & 3.5 $\pm$ 0.2   &  02:00--02:34   & 3   \\
 & SGRE(B1) $<$300 MeV &   &  & $<$3.3\tablenotemark{c} &  &    \\
 & SGRE(B2)  & 03:50--04:34 &  10.0 $\pm$ 0.2  & 3.85 $\pm$ 0.1   & 02:34--04:12   & 23   \\
  & SGRE(B2) $<$300 MeV  &   &  & $<$3.3\tablenotemark{c}   &  &   \\
 & SGRE(B3)  & 05:34--06:01 &  8.7 $\pm$ 0.4  & 4.25 $\pm$ 0.2    & 04:12--05:46    & 30   \\
 & SGRE(B4)  & 07:02--07:46 &  6.2 $\pm$ 0.2  & 4.5 $\pm$ 0.15  &  05:46-07:24   & 27   \\
    & SGRE(B4) $<$300 MeV  &   &  & $<$3.3\tablenotemark{c} &  &    \\
 & SGRE(B5)  & 08:42--09:12 &  4.1 $\pm$ 0.3  & 4.8 $\pm$ 0.5   & 07:24-08:48    & 18   \\
    & SGRE(B5) $<$300 MeV  &   &  & $<$3.7\tablenotemark{c} &   &    \\
 & SGRE(B6) & 10:33--10:58 &  2.5 $\pm$ 0.2  & 5.2 $\pm$ 0.4   & 8:48--10:46    & 17   \\
    & SGRE(B6) $<$300 MeV  &   &  & $<$3.7\tablenotemark{c} &    &   \\
 & SGRE(B7)  & 13:23--13:33 &  0.6 $\pm$ 0.2  &  & 10:46--13:27   & 9   \\
 & SGRE(B8) & 16:35--16:49 &  0.22 $\pm$ 0.06 &  & 13:27--16:41  &   3 \\
 & SGRE(B9)  & 19:46--20:14 &  0.07 $\pm$ 0.02 &   & 16:41-20:01  &   1 \\
 & SGRE(B) Total &   &   &  & 02:00--20:01 &   131 $\pm$ 15 \\
 & X5.4 flare  &  &  &   & 00:16--00:28   & 1.4\tablenotemark{b}  \\
 & M7 flare &   &  &   & 01:11--01:20  &    1.1\tablenotemark{b} \\
 & M7 flare &   &  &  &  01:12--01:17  &  $<$0.4 \\
 & SEP & & &  &   &   13300 ${^{+31800}_{-9360}}$\\
\enddata
\tablenotetext{b}{from 2.223 MeV line flux assuming protons follow a power-law spectrum with index, s = 4.5 $>$40 MeV (see $\S$\ref{subsec:below100MeV})}
\tablenotetext{c}{95\% confidence limit on index between 30 and 300 MeV based on comparing 2.223 MeV line flux upper limit and $>$100 MeV flux, (see $\S$\ref{subsec:below100MeV})}
\tablenotetext{d}{Better fit E$^{-2}$exp-(E/1300 MeV) }
\tablenotetext{e}{Heliocentric angle of 85$^{\circ}$}

\end{deluxetable}

\section{Results: Characteristics of Sustained $\gamma$-Ray Emission Events} \label{sec:results}

We use the term `Sustained Gamma-Ray Emission' (SGRE) to characterize solar $>$100 MeV $\gamma$-ray emission that has  temporal and spectral characteristics that are distinct from the accompanying impulsive flare and often extends hours after the end of the flare.    Our spectroscopic studies reveal that the SGRE is consistent with pion-decay emission produced by the interaction of high-energy protons and $\alpha$-particles.  We utilized a simple light-bucket approach to study $>$100 MeV $\gamma$-ray emission observed by {\it Fermi}/LAT from the Sun on time scales from minutes to hours ($\S$\ref{subsec:lightbucket}).  We identified 30 SGRE events from 2008 to 2016 and list them in Table \ref{tab:latlist}.  In $\S$\ref{subsec:details} we discussed the $>$100 MeV $\gamma$-ray time history of the 2011 March 7 event (number 1) as an example of plots shown in Appendices \ref{subsec:6nolat} and \ref{sec:append}.   Online four-day long plots of the solar $>$100 MeV $\gamma$-ray flux observed by LAT during the entire mission are provided on the {\it RHESSI} browser\footnote{\url{http://sprg.ssl.berkeley.edu/~tohban/browser}} and discussed in $\S$\ref{subsec:browplots}.   In this section we present the results of our comprehensive study of SGRE events observed by LAT.  In $\S$\ref{subsec:conditions} we discuss the relationship between SGRE events, flares, CMEs, and SEPs and discuss the conditions under which the $\gamma$-ray emission is produced.  We then provide details of the temporal characteristics of the SGRE and compare them with other solar emissions in $\S$\ref{subsec:tempdistinct}.  In $\S$\ref{subsec:spectra} we detail the spectral characteristics of the SGRE and in $\S$\ref{subsec:numprot} compare the numbers of protons in the SGRE event and in the associated flare and SEP.  In the last section we discuss LAT observations of SGRE from two behind-the-limb flares.



\subsection{SGRE Relationship to Flares, CMEs, and SEPs} \label{subsec:conditions}

We identified 95 solar eruptive events between 2008 June and 2012 May having at least one of three characteristics: 1) a broad CME with speeds $\gtrsim$ 800 km s$^{-1}$, 2) an SEP event with proton flux $>$1 pfu, or 3) a hard X-ray flare with energies $>$100 keV.  These were the characteristics shared by the first three SGRE events detected by LAT.  These 95 events and their characteristics are listed in Table \ref{tab:full} which is described in Appendix \ref{sec:characteristics}.  Using these events, we first investigate the relationship between SEP peak proton flux and CME speed, and soft X-ray power and CME speed. We then study how these parameters relate to SGRE events.  

There are 26 events on the visible disk for which CME speed, SEP flux, and flare hard X-rays were all measured.  We plot the measured peak SEP proton flux from the best magnetically aligned spacecraft vs. CME speed for these 26 events\footnote{events without upper limit symbols on SEP flux or open red circle for SGRE event 25 where no hard X-ray measurements were made during the flare because both {\it RHESSI} and GBM were in nighttime} in panel a) and peak soft X-ray power vs. CME speed for the same events in panel b).  The numbers associated with each point are the event numbers in Table \ref{tab:full}.  Both the peak proton flux and soft X-ray power for these events are correlated with CME speed for speeds above 800 km s$^{-1}$.  The correlation between SEP proton flux and CME speed (Spearman rank correlation 0.82 and 99.99\% probability) is consistent with earlier observations \citep{kahl01,gopa03}.  This correlation was used to support the argument that shocks from fast CMEs accelerate the protons observed in gradual SEP events.   We note, however, that there is also a correlation between soft X-ray power and CME speed, although it is weaker (Spearman rank correlation 0.69 and 99.8\% probability).  This latter correlation suggests that the the Big Flare Syndrome\footnote{``energetic flare phenomena are more intense in larger flares regardless of the detailed physics"} (BFS) \citet{kahl82} may be contributing to the observed CME-SEP correlation.


\begin{figure}
\epsscale{.8}
\plotone{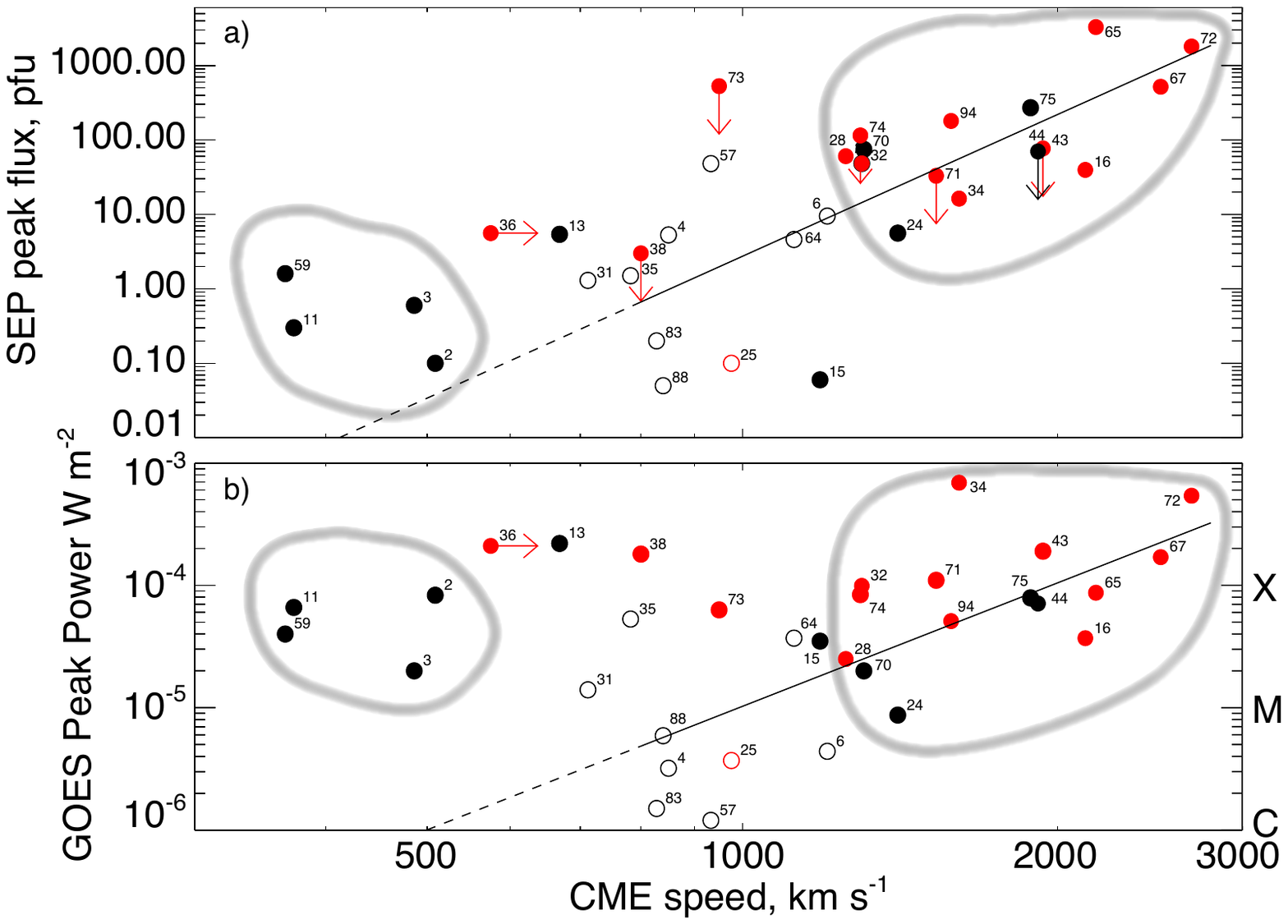}

\caption{Measured SEP flux above $\sim$10 MeV compared with CME speed (panel a) and {\it GOES} peak soft X-ray power compared with CME speed (panel b) for events on the solar disk between June 2008 and the end of May 2012. Events are common to both plots and the associated event numbers in Table \ref{tab:full} (without the preceding `A') are adjacent to the data points. The solid lines show the best linear fits to the data for CME speeds $>$800 km s$^{-1}$ and the dashed lines show the extrapolation to lower speeds.  Open circles show events where the impulsive flare X-ray emission did not exceed 100 keV or there were no measurements.  Filled circles are events where the flare X-ray emission exceeded 100 keV.  Red open or filled circles are SGRE events.  Six events with CME speed $>$800 km s$^{-1}$ and $>$100 keV hard X-ray emission for which only upper limits could be obtained for the SEP fluxes were added to the plot with upper limit symbols, four were SGRE events observed by LAT.  We plotted the CDAW CME speed for event A36 (Appendix \ref{subsec:20110906}) as a lower limit because there is evidence from {\it STEREO} that the speed was higher.  We encircle regions of interest discussed in the text. }
\label{study}
\end{figure}

Events where the accompanying flares emitted hard X-rays that exceeded 100 keV are identified by filled circles.  As shown in Figure \ref{study}b, such energetic hard X-rays are typically emitted in the more intense {\it GOES} soft X-ray events (M-class or higher).  This may just be a manifestation of the the BFS, i.e. the spectra of the most intense flares are also the hardest.  There is a clustering of flares associated with CMEs with speeds between about 700 and 1200 km s$^{-1}$ where the X-ray emission does not exceed 100 keV (black open circles).  These are associated with weaker soft X-ray flares and appear to follow the correlation between {\it GOES} class and CME speed depicted by the solid and dashed line in panel b).  In looking at the SEP proton flux vs. CME speed correlation plot in panel a) we would infer that these events are just weaker CME-shock produced gradual SEP events.  The four events (2, 3, 11, 59) with hard X-ray emission $>$100 keV, $>$M2 {\it GOES} class soft X-rays, and $<$550 km s$^{-1}$ CME speeds are inconsistent with these correlations.  They appear to form a class of relatively intense impulsive flares associated with weak and short duration (less than one day) SEP proton events.  Note that the impulsive 2010 June 12  gamma-ray event \citep{acke12a,acke12b} is one of them.

There are six other events found in the four-year study having CME speeds $\gtrsim$800 km s$^{-1}$ and $>$100 keV X-rays for which there are only upper limits on the SEP proton flux due to a stronger preceding event.  We plot these events as filled circles in both panels and with upper limit symbols on the SEP flux in panel a).  Of the 32 events plotted where hard X-ray measurements were made, 19 (including event 36) had both CME speed $\gtrsim$800 km s$^{-1}$ and flare hard X-ray emission $>$100 keV.  Most of them were associated with intense SEP events and are concentrated in the regions outlined in the upper right hand corner of panels a) and b).   Fourteen of these 19 events had associated SGRE and are plotted as filled red circles in the figure.  The fact that each of the 14 SGRE events was accompanied by both a fast CME and a $>$100 keV X-ray flare suggests that these may be necessary conditions for production of sustained $\gamma$-ray emission.  In $\S$\ref{subsec:6nolat} we discuss reasons why SGRE may not have been detected in the five other events meeting these two conditions.

If the SGRE is related to gradual SEP events, we can understand why they might be associated with fast CMEs.  What is not clear is why they also appear to be associated with flares observed above 100 keV.  We therefore studied a complete sample  
of 32 events between 2008 June and 2012 May with CME speeds $\gtrsim$800 km s$^{-1}$ where hard X-ray measurements of the accompanying flare were made.  This adds eight events, with flares that did not exceed 100 keV and for which there were only upper limits on the peak SEP proton flux, to those plotted in Figure \ref{study}.  Four of them were preceded by stronger proton events (events 37, 45, 66, 81) and couldn't be measured.  The other four (9, 22, 85, and 93) were not detected at a flux above 0.5 pfu even though background conditions were good and the flare site was well connected to Earth.  In Table \ref{tab:bfs} we list the characteristics of the 32 events depending on whether the flare's X-ray energy exceeded 100 keV or not.  The numbers of events in the two categories are comparable, but the characteristics are distinctly different.  In addition to not producing SGRE, the mean peak soft X-ray power for $<$100 keV flares is fifteen times smaller, the mean CME speed is 40\% less and the number of events with metric Type II emission is 3-times smaller.   There are 22 events where the SEP proton flux was measured without a preceding strong SEP.  The mean/median proton fluxes are about two orders of magnitude larger when the flares emit hard X-rays $>$100 keV.

  
 
\begin{deluxetable}{lll}
\tabletypesize{\scriptsize}
\tablecaption{Characteristics of Events in Table \ref{tab:full} Where Maximum HXR Energy is $<$100 keV and $>$100 keV \label{tab:bfs}}
\tablehead{
\colhead{ }  & \colhead{Maximum HXR Energy} & \colhead{ }  \\
\colhead{ } & \colhead{$<$100 keV} &  \colhead{$>$100 keV}}
\startdata
 Number & 14 & 18  \\
 Number with SGRE & 0 & 13\\
 Mean {\it GOES} Class & C8  & X1.4\\
 Mean CME speed, km s$^{-1}$ & 964 & 1635\\
 Number With Type II radio & 4 &13\\
 Mean/Median SEP proton flux, pfu & 6.8/ 0.2\tablenotemark{a}& 525/75\tablenotemark{b}\\
 \enddata
\tablenotetext{a}{10 measured fluxes}
\tablenotetext{b}{12 measured fluxes}
\end{deluxetable}
 
These striking differences in the characteristics of the two groups of events may simply be a manifestation of the BFS. One way to explore this possibility is to look for systematic differences in the X-ray spectra of the two groups.   Of the fourteen events listed in the table with hard X-ray emission that did not exceed 100 keV, only two had emission exceeding even 50 keV: the M3 flare on 2011 September 24 (event number 45 in Table \ref{tab:full}) and the C1.5 flare on 2012 April 5 (event number 83). The best fitting electron spectra had power-law indices of 4.8 and 6.0, respectively.   These compare with power-law indices close to 3.5 for the three weakest {\it GOES} class flares observed above 100 keV (events 15, 24 and 70).  Thus, the flares with hard X-ray emission that does not exceed 100 keV have significantly steeper non-thermal electron energy spectra than the flares where the emission exceeds 100 keV.  This suggests that the BFS is not the primary reason why detection of impulsive $>$100 keV emission is a necessary condition for production of SGRE and why the associated SEP proton flux is so high.  Instead, we suggest that there is a threshold for efficient particle acceleration where both electrons and ions reach energies in excess of 100 keV.  If there is a sufficient number of upwardly directed particles with these energies, then they could constitute a seed population that is further accelerated by a CME shock, contributing to both electrons and protons observed in space as SEPs and those that return to the Sun and interact in the solar atmosphere producing SGRE events.  We develop this line of thought in $\S$\ref{sec:discussion} with implications for both SGRE and SEPs.   

\subsubsection{Necessary and Sufficient Conditions for Sustained $>$100 MeV $\gamma$-ray Emission} \label{subsubsec:conditions}

When we include the 15 SGRE events observed after 2012 May, listed in Table \ref{tab:latlist} with numbers 16--30, we find that 26 of the 30 SGRE events were associated with impulsive flares having electron bremsstrahlung above 100 keV.  There were no hard X-ray measurements for the other four events: the 2011 June 2 SGRE event discussed above, the 2012 July 6 event, and the two behind-the-limb flares on 2013 October 11 (Appendix \ref{subsec:20131011}, Figure \ref{131011th}) and 2014 September 1 (Appendix \ref{subsec:20140901}, Figure \ref{140901th}).  Thus, flare acceleration of electrons to energies above 100 keV appears to be a necessary condition for production of SGRE.

With the inclusion of the 15 SGRE events identified after 2012 May, it appears that association with a fast CME can no longer be considered a necessary condition for the production of SGRE.  Neither the event on 2012 October 23 (event 17; Figure \ref{121023th}, Appendix \ref{subsec:20121023})\footnote{It is of interest that the 2012 October 23 event was accompanied by the largest helio-seismic event of Cycle 24 that was believed to be produced by the energy release of electrons in the solar atmosphere during the impulsive flare \citep{shary17}.} nor the event on 2012 November 27 (event 18; Figure \ref{121127th}, Appendix \ref{subsec:20121127}) had an accompanying CME or detectable SEP.  It is not surprising that CMEs were not observed because the {\it GOES} soft X-ray durations, eight and eleven minutes, were two of the three shortest in our sample of 30 SGRE events.  \citet{shee83} first showed that such short-duration soft X-ray flares are infrequently accompanied by CMEs.   

Although no CMEs were observed, AIA movies reveal evidence for the eruption of a magnetic loop in both events, at the times denoted by the downward arrows in the inset of Figures \ref{121023th} and \ref{121127th}.  In addition, Type II metric radio emission, indicative of shock formation, was detected following the magnetic eruption on 2012 October 23.  This suggests that the two SGRE events may have been accompanied by failed CMEs \citep{ji03} that accelerated protons only onto low lying field lines only reaching back to the Sun.  Thus, a magnetic eruption with an accompanying shock may still be a necessary condition for production of SGRE, even if a CME is not observed. 

As can be seen in Figure \ref{study}, no SGRE was detected in five of the 19 events with CME speeds above 800 km s$^{-1}$ and flare emission above 100 keV.  Thus, eruption of a fast CME accompanied by $>$100 keV flare emission is not a sufficient condition for production of SGRE.   LAT's limited duty cycle for solar observations (see $\S$\ref{sec:solobs}), 20-40 minute every 90 or 180 minutes, is one possible reason why sustained emission was not observed in these five events (see Appendix \ref{subsec:6nolat}).   The proximity of some of the five of events to the solar limb provides another possible explanation because some of the $\gamma$-rays could have been emitted beyond the solar limb.

\subsubsection{Information on SGRE Location and Spatial Distribution} \label{subsubsec:location}

Of the 71 eruptive events in Table \ref{tab:full} from 2008 to 2012 that were associated with CME velocities $\gtrsim$800 km s$^{-1}$, 36 were on the visible disk and 35 were distributed relatively uniformly in heliocentric longitude beyond the limb of the Sun.  Because there were no SGRE events associated with the 35 flares beyond the limb, the protons producing the $\gamma$ rays could not have interacted globally on the Sun, such as might be the case when protons stored in a reservoir behind the receding CME shock return to the Sun via open field lines in coronal holes \citep{huds89}.  \citet{ajel14} reported that the time-integrated centroid location of the sustained emission associated with SOL2012-03-07T00:02/T01:05 (event 12) was $\lesssim$10$^{\circ}$ from the flare site and that the location of the centroid may have moved to the east late in the event.  

With the detection of sustained emission from flares beyond the limb \citep{pesc15,acke17}, it is clear that the protons producing the SGRE can interact tens of degrees from the flare site. There is also evidence that the protons were accelerated by CME-shocks along magnetic field lines returning to the Sun \citep{plot17}.  We discuss these events in Appendices \ref{subsec:20131011} and \ref{subsec:20140901}.  In $\S$\ref{subsec:electrons} we provide evidence that bremsstrahlung observed by GBM during the behind-the-limb event on 2014 September 1 was produced by electrons with energies reaching up to 10 MeV that were likely accelerated in the same process that produced the SGRE.  After emerging from the SAA radiation belts at 11:11 UT, {\it RHESSI} detected $>$20 keV X rays from the Sun having the same spectral characteristics as those observed by GBM. There were a sufficient number of detected events during the exposure to yield an image.  However, no image was produced and the only modulation observed was by taking the differences in the fields of view parallel and perpendicular to the slit axis.  This localized the source in the northeast quadrant of the Sun, consistent with LAT centroid \citep{acke17}.  As there was no rotating modulation collimator (RMC) signal in {\it RHESSI} detector 9, which has the coarsest grids, the source size had to be larger than 300 arc seconds.  This indicates that the hard X-ray emission was distributed over a broad region and not confined to the loop top source imaged between 6 and 12 keV by \citet{acke17}.  If the association of this bremsstrahlung with the SGRE is valid, then these measurements provide additional evidence that the protons producing the emission can be distributed over tens of heliographic degrees.

\subsection{Temporal Characteristics of Sustained Gamma-Ray Emission} \label{subsec:tempdistinct}

We defined SGRE as $>$100 MeV solar $\gamma$ radiation having temporal and spectral characteristics that are distinct from the accompanying impulsive flare.  The 2011 March 7 event provided the first example of such an event ($\S$\ref{subsec:details}).  The SGRE in that event lasted about 14 hours and began during or just after the flare. It was clearly not the tail of the impulsive flare.  In Appendix \ref{sec:append} we provide the histories on minute and hour time scales, as well as the details, of all 30 SGRE events.  Because the 2012 March 7 event (Table \ref{tab:latlist} number 12, Appendix \ref{subsec:20120307}, Figure \ref{120307th}), reveals two separate SGRE eruptions there are 31 cases to study.  For each event we start by posing and answering the question: ``Is the SGRE time history different from that of the impulsive flare?".   

In 17 of the 31 SGRE eruptions, the $>$100 MeV emission began after the impulsive flare hard X-rays and lasted longer.   In four other events (numbers 5, 16, 19, and 30), the sustained $>$100 MeV emission began during or perhaps before the impulsive hard X-ray peaks, but it had a different time history from than that of the X-ray emission.  Each of the remaining ten SGRE events were observed in only one 20--40 minute LAT solar exposure well after the flare.  In seven of these events there is no evidence for temporal variation during the exposure, suggesting that the measurement was made near the peak of the sustained emission.  In two of the events (2013 October 25 and 28, events 26 and 27) there is evidence at the 80--90\% confidence level that the emission was falling in time and in one event (2012 July 06, event 17) the evidence for a falling flux is even stronger.  We could not determine if the emission in these three events came from the falling phase of a short SGRE event or simply from the tail of the impulsive flare.  However, in no event did we find clear temporal evidence that the emission was the tail of $>$100 MeV $\gamma$-rays produced in the impulsive phase. 

The conclusion from the timing studies that the SGRE comes from a second distinct population of accelerated particles is supported by considerations in $\S$\ref{subsubsec:imp_sus} below, where we typically find that the numbers of protons needed to explain the SGRE are significantly larger than those inferred from measurements of the impulsive flare.  Thus, a source of energy distinct from the impulsive flare is needed to account for SGRE.  Particle acceleration at the shock formed by a fast CME is the likely source. In the following sections we: 1) compare CME, Type II radio, and SGRE onset times;  2) study the SGRE onset times and durations;  3) search for timing correlations; and 4) discuss whether the SGRE is due to a series of episodic events.



\subsubsection{Comparative CME, Radio, and SGRE Timing} \label{subsubsec:timing} 

\citet{zhan01} discussed the temporal relationship between the CME and flare soft X-ray emission. They showed that the CME typically has a 30 minute to 2 hour initiation phase with a slow magnetic field rise rise prior to the rapid acceleration that we refer to here as the ``CME onset".  They also showed that the rapid expansion occurs close in time to the rise in flare soft X-rays. 
The solid vertical downward arrows in the insets of the time histories of the 30 SGRE events plotted in Appendix \ref{sec:append} 
show our estimates of the CME onset times from inspection of {\it SDO} AIA  movies.   The horizontal arrows in the insets show the range in CME onset times derived in the CDAW catalog from linear and quadratic extrapolations of the measured CME height vs. time relationship.   In a separate study of 28 eruptive events, including some in which SGRE was observed, we find that the CME onset times estimated from the EUV movies agree well with the rise in {\it GOES} 0.5--4{\AA} emission; the mean difference is less then 0.5 minutes and they differ by at most four minutes.  In contrast, the averages of the CME onset times from linear and quadratic extrapolations of data in the CDAW catalog differ by as much as 20 minutes from our CME onset time estimates based on the the EUV movies, with an average difference of just over a minute.  We therefore use the CME onset times derived from the EUV movies as the reference time in determining the delay in the formation of the shock (Type II radio emission) and the rise of the SGRE.   We list these CME onset times for the 30 SGRE events in column 3 of Table \ref{tab:onsets}.  There were no CMEs reported for the 2012 October 23 (event 17; Appendix \ref{subsec:20121023}, Figure \ref{121023th}) and 2012 November 27 (event 18, Appendix \ref{subsec:20121127}, Figure \ref{121127th}) events, but there is clear evidence for the eruption of magnetic loops in AIA images at 03:15 UT and 15:56 UT, respectively, along with material moving away from the flare region.  We, therefore, use these times as the eruption onsets for these two events. 

Type II radio emission, discussed in Appendix \ref{sec:solarradio} along with Type III radio emission, is attributed to electrons accelerated at a shock and radiating in its vicinity as it moves through the corona. Its onset time is believed to mark the formation of the shock and can be compared with the onset times of the CME and the SGRE.  In column 4 of Table \ref{tab:onsets}, we list the Type II onset times that were primarily derived from metric observations; where these metric measurements are not available we list in italics the times derived from interplanetary decametric/hectometric (DH) observations.  These Type II onset times are shown by the dashed vertical arrows in the insets of the time history plots in Appendix \ref{sec:append}.  No Type II radio emission was observed from the 2012 November 27 event that was one of the two SGRE events not accompanied by a CME.   On average we find that metric Type II emission begins about six minutes after the CME onset times.  Using these delays and the CME linear speeds from the CDAW catalog for the events in the study, we estimate that the shock, on average, was formed at a heliocentric distance of $\sim$1.7 R$_{\odot}$.  Using a more sophisticated orthogonal analysis of CMEs, \citet{gopa13} found that the shock formation locations ranged from 1.20 to 1.93 R$_{\odot}$, with mean and median values of 1.43 and 1.38 R$_{\odot}$, respectively.  These timing studies indicate that the CME shocks can form low in the corona.


\begin{deluxetable}{ccccc}
\tablecaption{Onset Times of CMEs, Type II Radio, and SGREs \label{tab:onsets}}
\tablehead{
\colhead{Number} & \colhead{Date} & \colhead{CME Onset} & \colhead{Type II Onset}& \colhead{SGRE Onset} \\
\colhead{} & \colhead{} & \colhead{UT} & \colhead{UT}& \colhead{UT}
 }
\colnumbers
\startdata
1& 2011/03/07 & 19:45 & 19:54 & 19:40 ${_{-00:20}^{+00:22}}$ \\
2 & 2011/06/02 & 07:43 & {\it 08:00}\tablenotemark{a}  & $<$09:40  \\
3 & 2011/06/07 &  06:16 & 06:25 & $<$07:48 \\
4 & 2011/08/04 &  03:46  & 03:54  & $<$04:56  \\
5 &  2011/08/09 &  08:02  & 08:01 &  08:02:40 $\pm$ 00:00:40\\
6 & 2011/09/06 &  22:18 & 22:19  & 22:20:00 $\pm$ 00:00:40    \\
7 & 2011/09/07 &  22:36 & 22:38 & $<$23:38:00 \\
8 & 2011/09/24 &  09:34 & 09:35 & 09:40 $\pm$ 00:01\\
9 & 2012/01/23 & 03:40 & {\it 04:00}\tablenotemark{a} & 04:17 $\pm$ 00:01 \\
10 &2012/01/27 & 18:07 & 18:10 & 19:31 ${_{-00:22}^{+00:06}}$  \\
11 & 2012/03/05 & 03:32  &  {\it 04:00}\tablenotemark{a} & 04:30 $\pm$ 00:03  \\
12 & 2012/03/07 &00:07 & 00:17  &  00:20 ${_{-00:20}^{+00:08}}$  \\
  &    & 01:01 & 01:09 &  02:01 ${_{-00:13}^{+00:07}}$ \\
13 & 2012/03/09 &  03:38 & 03:43 & 04:00 ${_{-01:10}^{+00:30}}$  \\
14 & 2012/03/10 &  17:24 & {\it 17:55}\tablenotemark{a}  & 20:00 $\pm$ 01:20  \\
15 & 2012/05/17 & 01:27 & 01:31 & $<$02:15  \\ 
16 & 2012/06/03 & 17:53 & 17:59 &  17:55 ${_{-00:04}^{+00:01}}$  \\
17& 2012/07/06 & 23:04  & 23:09 &  $<$23:26  \\
18& 2012/10/23 & 03:15\tablenotemark{b} & 03:17 & $<$04:16 \\
19& 2012/11/27 & 15:56\tablenotemark{b} & --- & 15:55 $\pm$ 00:01 \\
20& 2013/04/11 & 06:53 & 07:02 &  07:11 $\pm$ 00:01  \\
21& 2013/05/13 & 01:58 & 02:10  & $<$04:30  \\
22& 2013/05/13 &  15:48 & 15:57 & 17:07 ${_{-00:09}^{+00:07}}$  \\
23& 2013/05/14 &  01:04 & 01:07  & 01:20 $\pm$ 00:03   \\
24& 2013/05/15, &  01:29  & 01:37  &  03:00 $\pm$ 01:00 \\
25& 2013/10/11& 07:08 & 07:11 & 07:15 $\pm$ 00:01 \\
26& 2013/10/25 &  07:57 & 07:59 & $<$08:17 \\ 
27&2013/10/28 & 15:00 &15:10 & $<$15:45 \\
28& 2014/02/25 & 00:43 & 00:56 & 00:55 ${_{-00:16}^{+00:06}}$\\
29& 2014/09/01 & 10:57 & {\it 11:12}\tablenotemark{a}  & 11:02 $\pm$ 00:01\\
30& 2015/06/21 & 02:10 & 02:24 & 02:23 $\pm$ 00:03\\
\enddata

\tablenotetext{a}{DH Type II start time in the absence of metric report}
\tablenotetext{b}{No CME observed; onset time of magnetic eruption}
\end{deluxetable}

\subsubsection{Study of SGRE Onset Delays and Durations} \label{subsubsec:deldur}

The SGRE onset times were derived by visual inspection of the $>$100 MeV fluxes plotted for the 30 events in Appendix \ref{sec:append}, each of which is accompanied by a detailed discussion.  In some cases the onset can be clearly distinguished with an accuracy of one or two minutes.  In other cases we fit the time history to estimate the onset and its uncertainty, as we did for the 2011 March 7 SGRE plotted in Figure \ref{110307th} discussed in $\S$\ref{subsec:details}.  For events for which there was only one solar exposure with detectable $>$100 MeV emission and no evidence for short-term variability, we used the start time of the exposure as an upper limit on the onset time. The duration of each event used to estimate the number of protons producing the SGRE is given in Table \ref{tab:event} column 6 in the row labelled `SGRE Total'.   As LAT had limited exposure to the Sun, our estimates of the durations are, in many cases, upper limits based on the time of the first LAT solar exposure with no detectable $>$100 MeV emission. This uncertainty does not impact the coarse comparisons that we are making.  Our best estimates of the SGRE onset times and their uncertainties are listed in column 5 of Table \ref{tab:onsets}.

\begin{figure}
\epsscale{.70}
\plotone{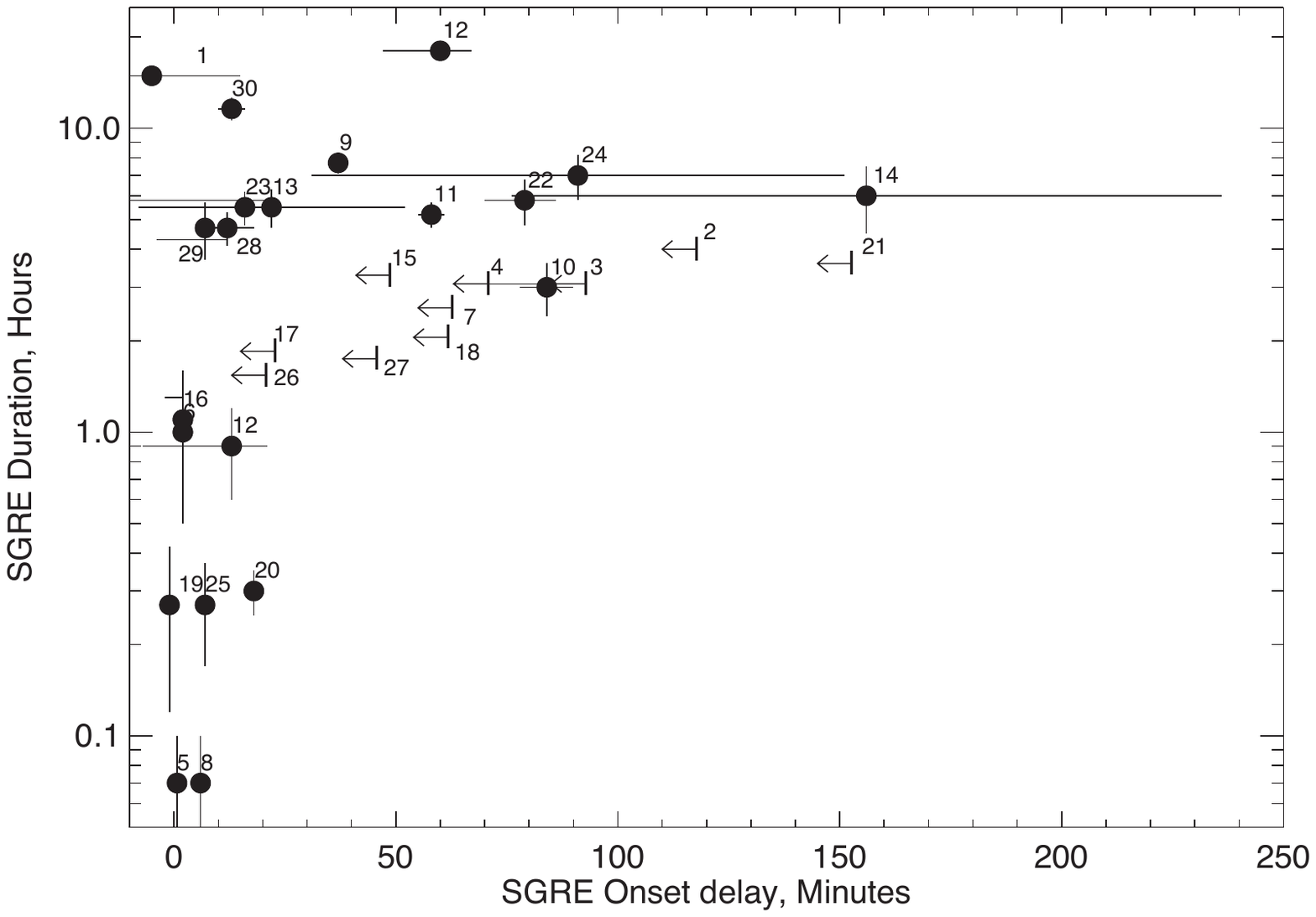}
\caption{Estimated duration of  $>$100 MeV SGRE plotted against the delay in its start time from the onset of the  CME determined from {\it SDO} AIA movies.   When an onset time cannot be estimated, we place a horizontal upper limit symbol at the start time of the exposure in which SGRE was first detected. The event numbers in Table \ref{tab:latlist} are printed adjacent to the data points.}
\label{del_dur}
\end{figure}

We define the SGRE onset delay as the difference between the SGRE onset time and the CME onset times listed in column 3.  In Figure \ref{del_dur}, we plot the duration of the SGRE events versus their onset delays.  We include, separately, the delays and durations of the two post-flare SGRE events on 2012 March 7, each identified by number 12 in the figure (see the event time history in the inset of Figure \ref{120307th} in Appendix \ref{subsec:20120307} ).   Thus, there are 31 individual points in the plot.   We were able to estimate the onset delays based on multiple flux observations for 21 of them (shown by filled circles).  For ten of the events we only have upper limits on the delays.  The measured SGRE onset delays ranged from minutes to an hour or more.  Sixteen of the 31 plotted events have onset delays shorter than 30 minutes.  Assuming that the protons producing these 16 SGRE events were accelerated by a CME shock moving at 2000 km s$^{-1}$, the CME would have been at most 5 R$_{\Sun}$ from the Sun when the protons began interacting in the solar atmosphere.   Four well measured delays exceeded one hour indicating that the CME may have been as far away as 15 R$_{\Sun}$ from the Sun when the SGRE began.  The estimated SGRE duration was as short as about 10 minutes and as long as $\sim$20 hours, with a mean duration of about 4.5 hours.  As some of the longest lasting SGRE events also had large onset delays, the durations and delays of the measured events (filled circles) exhibit some correlation (Spearman rank correlation 0.47, correlation probability 97\%).  

It is interesting to discuss the timing of some of the SGRE events with the shortest onset times in more detail.   For the events on 2013 October 11 (event 25; Appendix \ref{subsec:20131011}) and 2014 September 1 (event 29; Appendix \ref{subsec:20140901}), where the active regions were beyond the limb of the Sun, the SGRE began about seven minutes after the estimated CME onset time.  Therefore, any shocks accelerating the particles producing the $\gamma$ rays for these events had to develop relatively low in the solar corona.  Based on the metric Type II radio emission, the shock on 2013 October 11 formed about three minutes after the CME onset time.   Because both flares occurred beyond the solar limb, the particles had to be accelerated onto low coronal magnetic field lines that reached the visible disk (see $\S$\ref{subsec:electrons}).  We note that even though the onset delays for these two events are similar, their durations appear to be significantly different, 15 minutes and 4.5 hours, respectively.  However, this difference may may be just be due to sensitivity issues because of the ten times more intense $\gamma$-ray flux on September 1.  

The shortest duration SGRE event occurred on 2011 September 24 (event  8; Appendix \ref{subsec:20110924}).  It began about five minutes after the shock formation, as monitored by the Type II emission, and lasted about five minutes.  The associated solar  flare lasted less than 1 minute, began within a minute of the radio emission,  and was observed up to about 100 MeV.  The tens of MeV emission was dominated by electron bremsstrahlung and delayed by about 8 seconds from the 100--300 keV hard X-rays.  The one-hour long SGRE event on 2011 September 6 (event 6; Appendix \ref{subsec:20110906}) began within about a minute of CME onset time raising questions whether it could have been energized by the CME related shock.  However, the Type II radio emission also appears to have begun within a minute of the CME.   The dynamic radio spectrum is not clear, but there appears to be high-frequency emission starting as early at 22:19 UT that looks as though it drifts down in frequency and reveals itself as Type II radiation at longer wavelengths. Thus, shock-related Type II radio emission also started about a minute after the CME onset, suggesting that the $>$100 MeV $\gamma$ rays were produced by particles accelerated at altitudes below 2 R$_{\odot}$.    The $>$100 MeV $\gamma$ rays in the short SGRE event on 2011 August 9 (event 5, Appendix \ref{subsec:20110809}) were also detected within about a minute of the CME onset and associated Type II radio emission, again indicating rapid formation of the shock and particle acceleration low in the corona. 

The 2012 June 3 SGRE (event 16; Appendix \ref{subsec:20120603}) has some puzzling features.   There is no clear onset of the SGRE although there is evidence (3$\sigma$ confidence) that it may have begun at about 17:51 UT, two minutes before the CME onset time inferred from the AIA images (see Figure \ref{120603th}).   The {\it GOES} 0.5--4{\AA} emission and the apparent restructuring of field lines in the AIA images began at about 17:51 UT.  Emission with Type-III-like radio properties also commenced around that time, but Type II radio emission did not begin until 17:59 UT.  This is the only SGRE event where the Type II radio emission clearly began after the SGRE.   It is, therefore, difficult to attribute the $\gamma$-ray emission before the start of Type II emission to protons from the CME shock.   However, there are two arguments against attributing the emission observed after 17:54 UT to the impulsive flare: 1)  the number $>500$ MeV protons that produced the SGRE is about four times the number accelerated in the flare (Table \ref{tab:event}) and 2) the spectrum of protons that produced the flare emission followed a power law with a steep index of 6.5 $\pm$ 1.0 while the protons producing the SGRE followed a much harder index of 4.3 $\pm$ 0.7.   

The low-coronal acceleration of particles necessitated by the short SGRE onset delays is consistent with the evidence for shock acceleration low in the corona \citep{gopa13}.  In addition, the systematic charge-to-mass ratio (Q/M) dependence observed in gradual SEP events \citep{desa16} is also readily explained by such low-corona shock acceleration \citep{schw15}.

The SEP release time at the Sun was determined for the GLE on 2012 May 17 \citep{roui16} (event 14).  It occurred at about 01:37 UT, just before the impulsive hard X-ray peak shown in Figure \ref{120517th}.  Thus, the particle release occurred about nine minutes after the CME onset and about six minutes after shock formation inferred from detection of Type II radio emission.   The first Fermi/LAT solar exposure began about 40 minutes after SEP particle release and, by that time, the proton flux interacting with the solar atmosphere appeared to have reached its peak.  Thus, we cannot compare the release times for the particles interacting at the Sun with those escaping to interplanetary space for this event.

\subsubsection{Other SGRE Temporal Correlations} \label{subsubsec:correlations}

\begin{figure}
\epsscale{.70}
\plotone{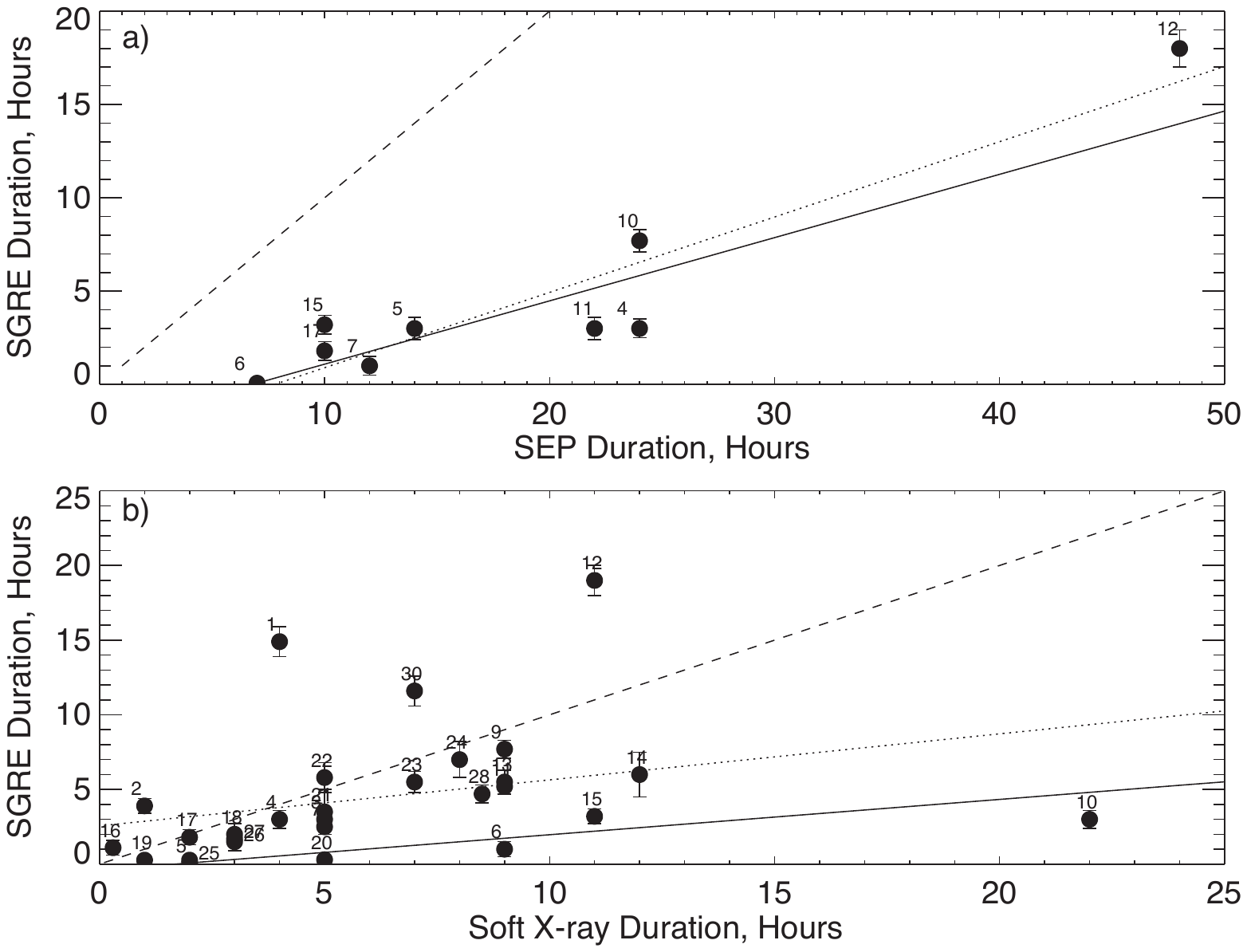}
\caption{a) Estimated durations of SGRE events plotted against durations of ten well-connected $>$100 MeV proton events observed by {\it GOES}. b) Estimated durations of SGRE events plotted against durations of SXR emission observed by {\it GOES}.  Solid lines show linear fits to the statistically weighted data; dotted lines show linear fits to the unweighted data; and dashed lines show where the SGRE and SEP or SXR durations are equal.  }
\label{grsepgoesdur}
\end{figure}

If $>$100 MeV SGRE events are associated with the CME shocks producing SEPs, we might expect some correlations between the $\gamma$-ray and the CME and SEP parameters.  We found no correlation between CME speed and SGRE onset delay and only a weak correlation, 95\% probability, between CME speed and SGRE duration, that increased to to 98\% for durations greater than one hour.    We also compared the durations of ten $>$100 MeV SEP proton events observed by {\it GOES}, that were magnetically well-connected to the flare site, with the durations of the associated SGRE events. The SEP duration was estimated from the time of onset to the time that the flux dropped to $\sim$10\% of its peak value.  As plotted in Figure \ref{grsepgoesdur}a), the SEP and SGRE durations appear to be well correlated.  The correlation coefficient is 0.93 and the Spearman rank correlation coefficient is 0.82 with better than 98\% confidence for a correlation. This correlation is reflected in the solid line that fits the weighted data and contributes to the growing evidence that the SGRE events are produced by protons accelerated by shocks of fast moving CMEs that are also responsible for the high-energy SEP protons observed in interplanetary space.  It is clear from the figure, however, that the SGRE durations are only fractions of the SEP durations; the mean SGRE/SEP duration ratio is $\sim$0.20.

\citet{ryan00} suggested that SGRE might be produced by protons accelerated in the current sheet behind a CME.   If this were true, then there might be some other observable manifestation of this energy source.  \citet{ryan13} and others discuss extended heating of the corona following a flare that may be associated with such an energy source.  The heating is reflected in the long duration tail often observed in soft X-rays (SXRs).  For this purpose, we defined the duration of the SXR emission to be from the time of the peak {\it GOES} 1--8{\AA} flux to the time at which the flux had fallen back to the pre-flare background level.  Small flares that occurred during this decay did not affect this study because of the relatively smooth exponential decay of the SXR emission from the flare of interest. In Figure \ref{grsepgoesdur}b) we compare the SXR and SGRE durations. They appear, on average, to be comparable and to exhibit evidence for a correlation.  The correlation coefficient is 0.55 and the Spearman rank correlation coefficient is 0.39 with better than 98\% confidence for a correlation.  There appears to be large scatter around the best fit to the data shown by the solid line, however.  This is reflected in the significantly different fits to the statistically weighted (solid line) and unweighted data (dotted line).

\subsubsection{Episodic Events Ruled Out as Origin for SGRE} \label{subsec:episodic}

It is possible that the hours-long SGRE events are due to episodic accelerations of protons to energies $>$300 MeV.  We searched source-class LAT data at 4-minute temporal resolution for evidence of impulsive events that could have produced what looked to be hours-long emission.  In all of the individual orbits studied, we found no evidence for such episodic emission.  This study was best done for the long-duration SGRE events on 2011 March 7 (event 1) and 2012 March 7 (event 11).   Any temporal variations that we did find were instrumental and occurred as the Sun was leaving the field of view of the LAT, where the solar exposure was not well determined.
 
\subsection{Spectroscopic Studies of SGRE} \label{subsec:spectra}

In this section we discuss  {\it Fermi}/LAT spectroscopic measurements of $>$100 MeV $\gamma$-rays and related {\it RHESSI} and GBM observations in the nuclear-line energy range.  These include observations made both during the associated impulsive  flare and the SGRE event.  Details of the analysis techniques were presented in $\S$\ref{sec:spect}.    We summarize the results of these spectroscopic studies in Table \ref{tab:event} and describe the entries in the table in $\S$\ref{subsec:specfits} and in Appendix \ref{sec:resultstable}. 



\subsubsection{Spectral Indices of $>$300 MeV Protons Producing SGRE} \label{subsubsec:heindex}

\begin{figure}
\epsscale{.70}
\plotone{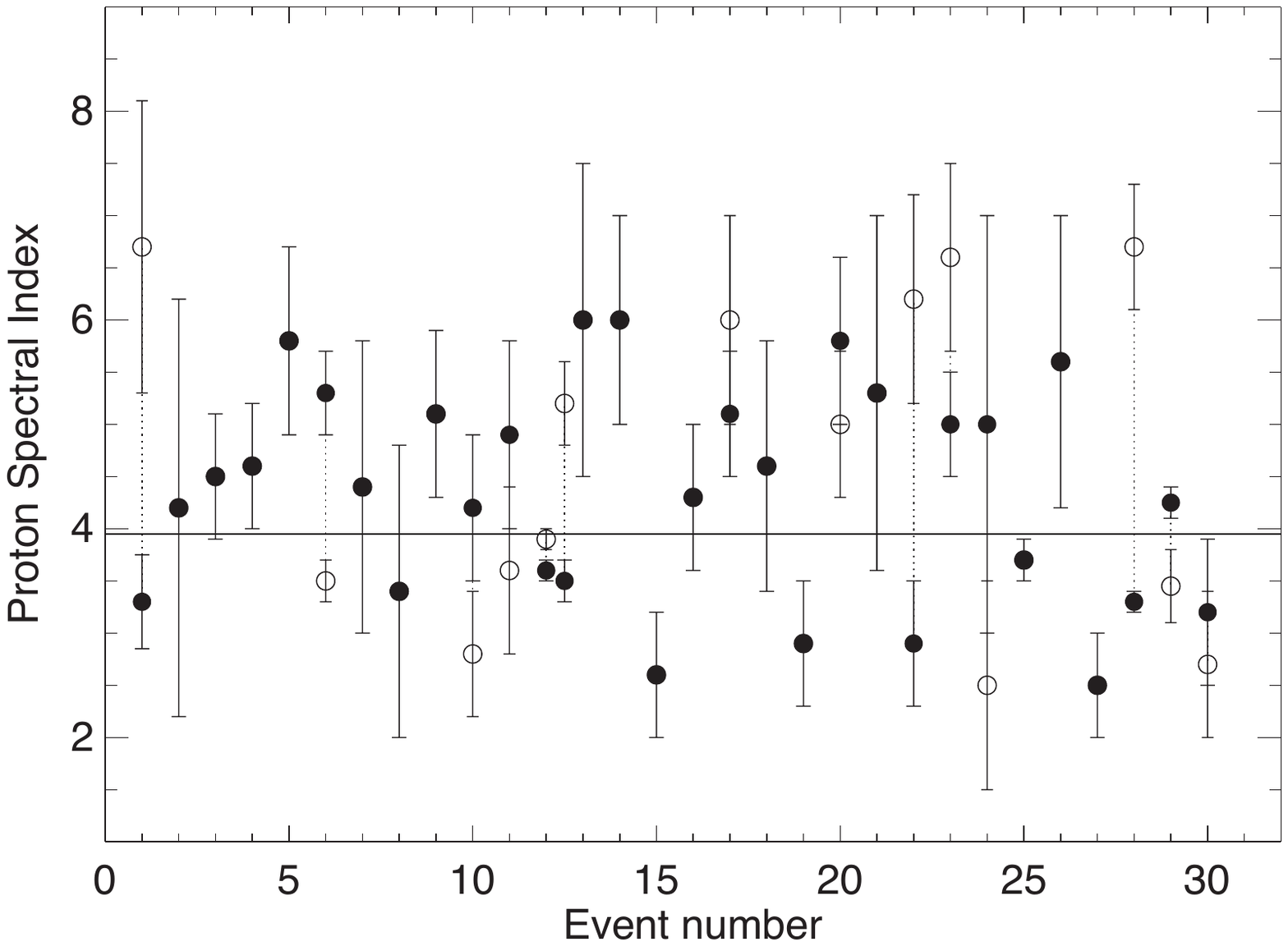}
\caption{Estimated $>$300 MeV proton spectral indices and $\pm$1$\sigma$ statistical uncertainties for the SGRE events.  Filled circles are the measured indices when there is only a single measurement or if the index does not vary significantly through the event.  When the spectral index varies through the event the filled circle gives the first index and the open circle gives the last index.  The weighted mean spectral index is 3.95 (solid line) and the RMS scatter is 1.8.}
\label{index}
\end{figure}

In this section we study the spectra and variability of the $>$300 MeV protons producing the sustained emission under the assumption that the  $>$100 MeV $\gamma$-ray emission comes from the decay of neutral and charged pions.  In most cases we assume that the $>$300 MeV protons interact in a thick target and follow a single power-law spectrum, even though there is evidence that the spectra in the most intense events steepen above 1 GeV.  The derived power-law spectral indices and uncertainties are listed for each LAT solar exposure in column 5 of Table \ref{tab:event} and plotted in Figure \ref{index} by event number.  Recall that there are two SGRE events associated with the two flares on 2012 March 7 (event 12).   For events where there is more than one LAT solar exposure, and statistics are sufficient, we plot the derived spectral index and error for both the first (filled circle) and last (unfilled circles) exposures.   The event-averaged proton power-law indices range from $\sim$2.5 to 6.0 with a mean index weighted by their uncertainties of 3.95 $\pm$ 0.05.  However, the RMS scatter around this mean index is large and there is $<10^{-6}$ probability that the data are randomly distributed about the mean.  

We compared our spectral index measurements with those reported by \citet{acke14} and \citet{ajel14} for the SGRE events on 2011 March 7, 2011 June 7, and 2012 March 7.  When averaged over the same time intervals, we obtained a mean index of 4.4 compared with 4.2 in the two papers prepared by the LAT team authors cited above.  Thus, there is good agreement for flares occurring on the solar disk.  In contrast, we find that our estimated proton indices for the two behind-the-limb flares are somewhat harder than those given in \citet{acke17}: 3.8 $\pm$ 0.2 versus 4.2 $\pm$ 0.2 for the 2013 October 11 event and 4.05 $\pm$ 0.2 versus 4.65 $\pm$ 0.2 for the 2014 September 1 event.   The emission from these two events is likely to have come from close to the solar limb so the difference may be due to atmospheric attenuation that we take into account but believe that the LAT team did not.  

We note that one of the hardest SGRE proton spectrum measured is associated with the 2012 May 17 GLE (event 14).  The proton spectral index measured between 02:10 and 02:48 UT had a value of 2.6 $\pm$ 0.6.  Over a range in energies from 0.3 to 1 GeV, this differential power-law index in energy is equivalent to a differential rigidity index of 3.4.  Neutron monitors recorded rigidity indices ranging from 2.1 to 3.8 from 01:40 to 03:30 UT \citep{plai14}, consistent with the $\gamma$-ray measurement. 

There are six events in which there appears to be significant variation in proton spectral index with time.  Four of these showed spectral softening with time: 2011 March 7 (event 1), 2012 March 7 (2$^{nd}$ point event 12), 2013 May 13 (event 22), and 2014 February 25 (event 28).  All of these events had durations longer than four hours.  As can be seen in Table \ref{tab:event} the spectrum of the second 2012 March 7 event gradually softened from an index of $\sim$3.5 to an index of $\sim$5.2 over a nine-hour period.  This is consistent with the observations of \cite{acke14}.  The second event on 2013 May 13 (event 21), associated with the flare at 16 UT, also showed clear evidence of softening between 17:50 UT, when the power-law index was $\sim$3, and 20:50 UT, when the index was $\sim$6.  A similar spectral softening from a power-law index of $\sim$3 to an index of $\sim$6 also occurred over a three hour period in the event on 2014 February 25.  

There are two events that showed evidence for spectral hardening: 2011 September 6 (event 6) and 2014 September 1 (event 29).  These two events were of shorter-duration.  The September 6 event had a duration of less than an hour and had a much softer proton spectrum, power-law index 5.3, during the rise to maximum than it did during the decaying phase, when the index was 3.5.  A similar, but less significant spectral variation was observed during the behind-the-limb 2014 September 1 event, where the proton spectrum hardened from an index of 4.25 $\pm$ 0.15  during the six minute rising phase to 3.85 $\pm$ 0.1 during the eight minute falling phase.   The protons producing weak emission an hour later also appeared to have a harder index,  3.45 $\pm$ 0.35. 

We studied other SGRE events with durations shorter than an hour in a search for spectral hardening with time.   As shown in Figure \ref{aug9th} panel d), there are two 20 second LAT exposures at the peak of the 6--10 minutes long $>$100 MeV SGRE event on 2011 August 9 (event 5).  The spectrum of the first exposure appears to be softer than the second one.  The spectral evolution of the first SGRE event on 2012 March 7 (1$^{st}$ point; event 12 in Figure \ref{index}), following the X5.7 flare (see inset of Figure \ref{120307th}) is complicated because there were only a few measurements before the peak. The proton spectrum from 00:39 -- 00:44 UT, in what appears to be the rising phase of the emission, is harder than the spectrum during the falling phase before the M7 flare, but the statistical significance is not compelling (see also Table 1 in \citet{ajel14}).  The 2012 November 27 event (19) exhibits a clear rise and fall, but the measured spectra during those times are both consistent with power laws with indices of $\sim$3.  The event on 2013 April 11 (event 20) also showed a clear rise and fall and there is a suggestion that the proton spectrum hardened after the peak flux, but it is not statistically significant.   There was no evidence for spectral hardening from the rise to the fall of the behind-the-limb event on 2013 October 11 (event 25), with the power-law indices both being consistent with an index of 3.7 $\pm$ 0.2.  Thus, there is no clear pattern of spectral hardening with time in SGRE events with durations shorter than about one hour.






\subsubsection{Information on SGRE Proton Spectra Below 300 MeV} \label{subsubsec:LoE_index} 

Due to the threshold for pion production, LAT only provides spectral information on protons with energy above 300 MeV.  As discussed in $\S$\ref{subsec:below100MeV}, spectral information on lower energy protons can be obtained by comparing {\it RHESSI} and GBM flux measurements of de-excitation lines and the 2.223 MeV neutron capture line with the $>$100 MeV $\gamma$-ray fluxes measured by LAT.  Because these instruments are much less sensitive in their energy domains than LAT is above 100 MeV, constraining spectral information can only be obtained for the seven most intense SGRE events with peak $>$100 MeV fluxes $\geq1 \times 10^{-4} \gamma$ cm$^{-2}$ s$^{-1}$: 2011 August 9 (event 5); 2011 September 6 (event 6); 2012 March 7 (event 12); 2013 April 11 (event 20); 2013 October 11 (event 25); 2014 February 25 (event 28); and 2014 September 1 (event 29).  These events are discussed below.

GBM detected de-excitation and 2.223 MeV neutron-capture line emission during the short SGRE event on 2011 August 9 (Appendix \ref{subsec:20110809}, Figures \ref{110809th} and \ref{aug9th}).  After subtracting the contribution from the impulsive flare, we found that the relative 2.223 MeV line and $>$100 MeV fluxes were consistent with sustained-emission produced by  20--300 MeV protons with a spectral index 4.0 $\pm$ 0.3.  This value is listed in the second row for the event in Table \ref{tab:event} and denoted by the footnote `c'.  For comparison, the proton spectrum measured above 300 MeV, using our pion-decay fits, had an index of 5.8 $\pm$ 0.9.  Although GBM detected the 2.223 MeV line during the impulsive flare on 2011 September 6, it did not detect the line at any time during the SGRE event.  We list 95\% confidence limits on the 20--300 MeV proton spectral indices (denoted by the footnote `c') at three different times during the event in Table \ref{tab:event}.  Only during the rise of the sustained emission, when the $>$300 MeV proton spectrum had a power-law index of $\sim$5, is it clear that the lower energy spectrum was significantly harder: index $<$4.0.   The 20-300 MeV sustained-emission proton spectrum was also significantly harder than the $>$300 MeV spectrum in the six hours after 04 UT on 2012 March 7; between 07:10 and 07:30 UT on 2013 April 11; between 04:21 and 04:40 UT on 2014 February 25; and between 11:06 and 11:20 UT on 2014 September 1 (this comparison was made assuming that the protons interacted at a heliocentric angle of 85$^{\circ}$, although the same conclusion holds for smaller heliocentric angles).  Thus, there is compelling evidence that the spectrum of protons producing the SGRE is not an unbroken power law above 20 MeV and that the spectrum softens between energies of tens of MeV and several hundred MeV.

\subsection{Comparisons of Numbers of $>$500 MeV Protons at the Sun and in Space} \label{subsec:numprot}

\begin{figure}
\epsscale{.70}
\plotone{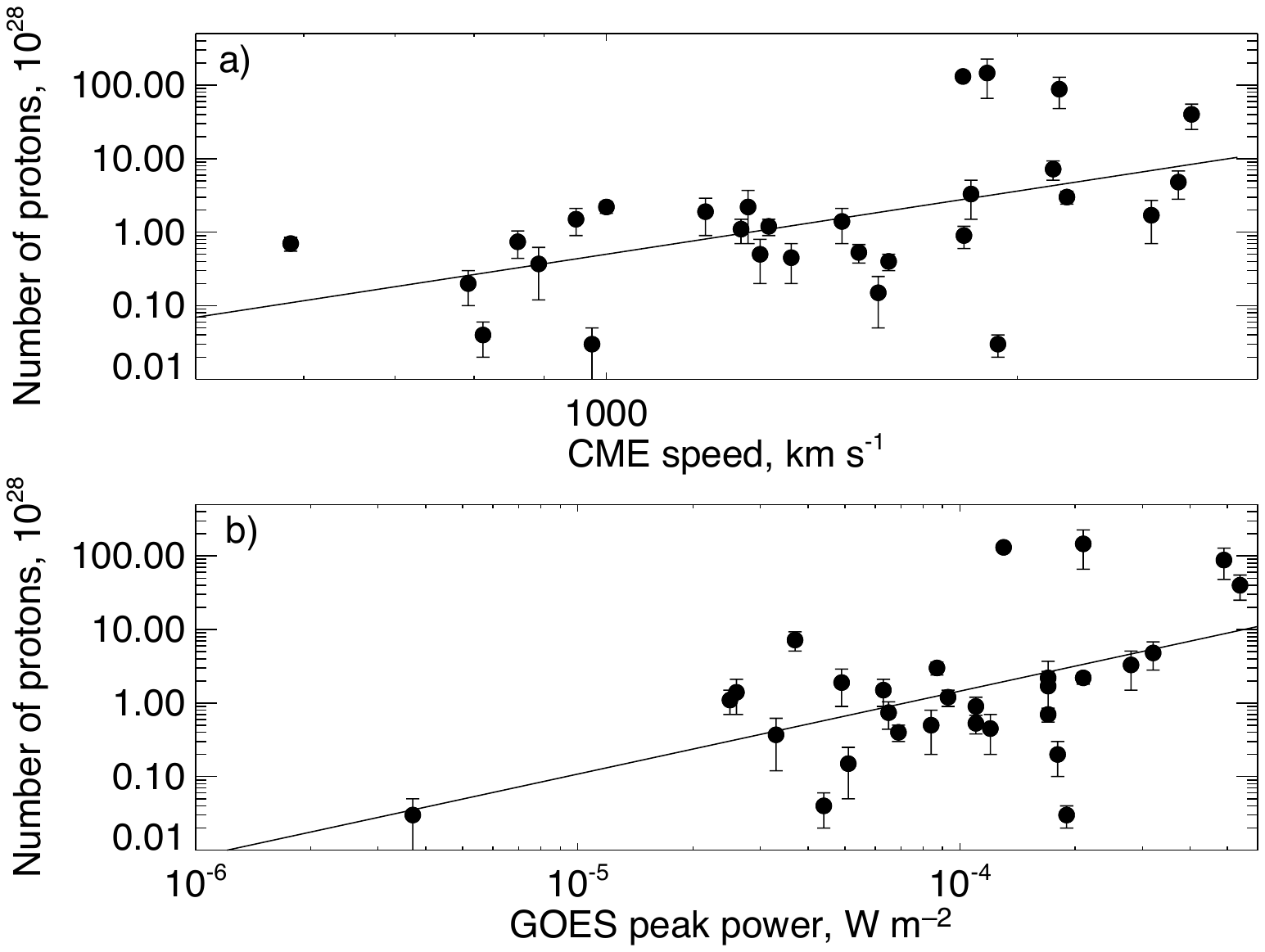}
\caption{Number of protons $>$500 MeV producing the SGRE plotted against CME speed (panel a) and against peak soft X-ray power (panel b).  For reference a {\it GOES} X-class flare has a peak power of 10$^{-4}$ W m$^{-2}$. Solid lines show best linear fits to the data. The RMS scatter in the measured number of protons about the best fit line for the correlation with {\it GOES} peak power is twice as large as that for CME speed.}
\label{prosus_cme_goes}
\end{figure}

Understanding the origin of SGRE and its relationship to both solar flares and SEPs is abetted by comparing the numbers of protons producing these emissions.  We estimated the number of $>$500 MeV protons in the SGRE events and flares using techniques discussed in $\S$\ref{subsec:latspect} and $\S$\ref{subsec:below100MeV}.  We list the number of $>$500 MeV protons accelerated at the Sun producing the SGRE in each 20--40 minute LAT exposure for the 30 events in column 7 of Table \ref{tab:event}.  The total number of protons at the Sun responsible for the SGRE events ranged from $\sim2 \times 10^{26} - 1.5 \times 10^{30}$.  We plot these numbers against the associated CME speeds and peak {\it GOES} soft X-ray powers in Figure \ref{prosus_cme_goes} panels a) and b).  The number of SGRE protons is correlated with CME speed with a confidence exceeding 99.99\% using either Spearman's  ($\rho$) or Kendall's ($\tau$) rank correlation test.  The confidence that the number is correlated with {\it GOES} class is an order of magnitude smaller, but still strong, primarily due to the barely detected LAT event on 2011 June 2 that was associated with weak C3.7 flare.  If we remove this flare from the sample, the confidence that the SGRE proton number and {\it GOES} class are correlated drops to 98\%, while the confidence that the SGRE proton number and CME speed are correlated is still $>$99.9\%.  This provides additional support for the contention that the protons producing the SGRE have a physical relationship with the accompanying CME.


\subsubsection{Comparison of Numbers of Protons in the SGRE and Associated Impulsive Flare} \label{subsubsec:imp_sus}

\begin{figure}
\epsscale{.70}
\plotone{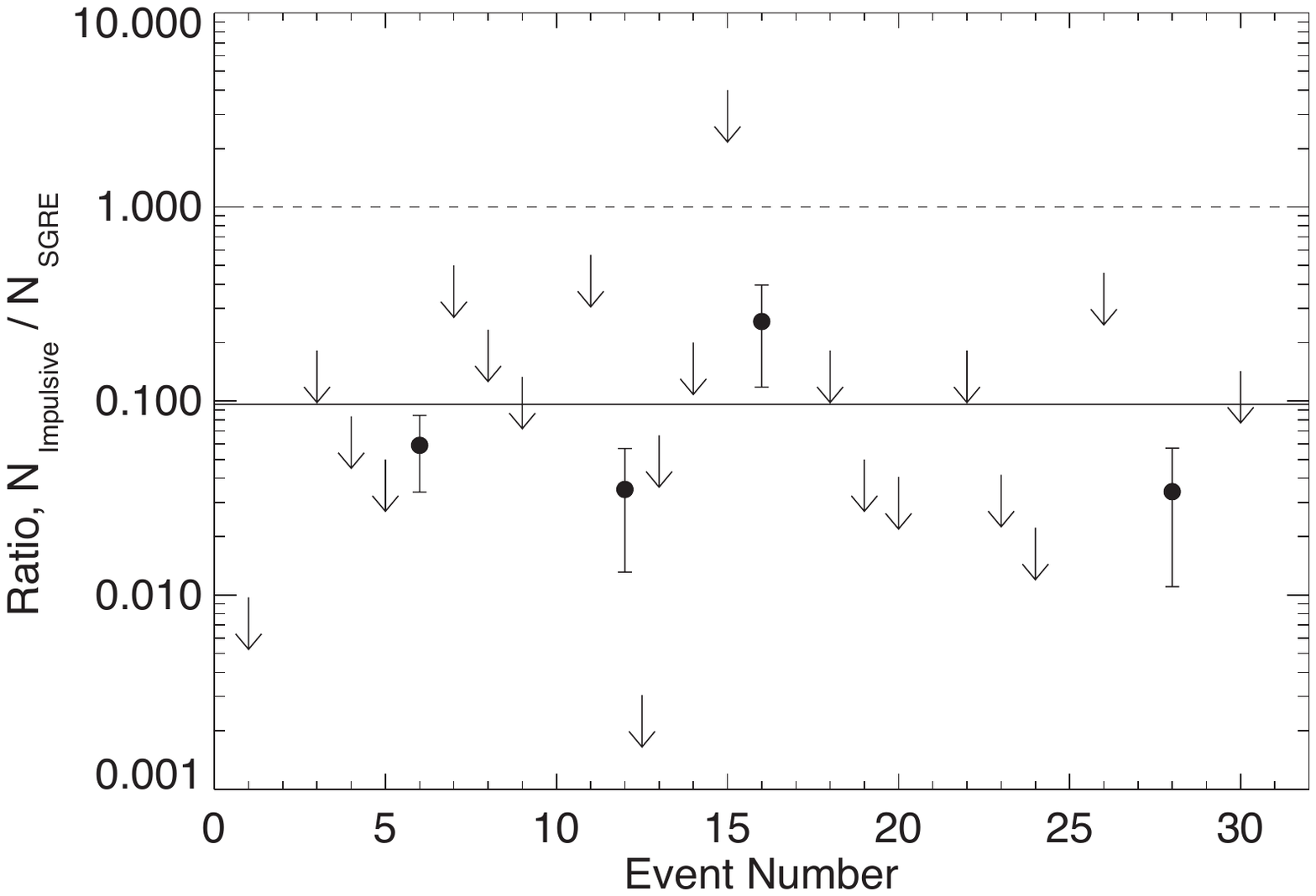}
\caption{Ratio of number of $>$500 MeV protons in the impulsive flare to the number in the SGRE for the events where there were LAT or 2.223 MeV line observations during the flare.  Filled circles give the ratios where the number of protons (not upper limits) were estimated: 20110906 (event 6) using LAT, 20110307 (first eruption of event 12) using {\it INTEGRAL}, 20120603 (event 16) using LAT, and 20140225 (event 28) using GBM. The solid line shows the mean of these four measured ratios. Upper limit symbols provide 95\% confidence limits on the ratios.  The dashed line shows where the ratio is unity.}
\label{ratio_nimp_nsus}
\end{figure}

In this section we compare estimates of the number of $>$500 MeV protons that produce SGRE with the number in the associated impulsive flare.   In making this comparison we assume that the protons interact in a thick target producing pion-decay radiation.  Our study only includes the 24 impulsive flares for which there were either LAT exposures or 2.223 MeV line measurements.  The measured flare proton numbers or upper limits are given in column 7 of Table \ref{tab:event} in the row listing the {\it GOES} X-ray class. LAT had at least partial exposures to 14 impulsive flares associated with SGRE events but detected $>$100 MeV $\gamma$ rays in only two: 2011 September 6 (event 6; Appendix \ref{subsec:20110906}) and 2012 June 3 (event 16; Appendix \ref{subsec:20120603}).  The ratios of the number of impulsive-phase protons to the number of SGRE protons for these two events are plotted as filled circles with uncertainties in Figure \ref{ratio_nimp_nsus} at event numbers 6 and 16.   The SGRE in the hour following the flare on 2011 September 6  was produced by about fifteen times the number of protons responsible for the impulsive $>$100 MeV emission.  The number of protons producing SGRE on June 3 was about four times higher than the number in the impulsive phase.

The 2.223 MeV neutron-capture line was detected during the two impulsive flares on 2012 March 7 (event  12, the X5.4 flare)  Appendix  \ref{subsec:20120307}, Figure \ref{120307th}) by the SPI detector on {\it INTEGRAL} \citep{zhan12}.  From these measurements, we used the technique described in $\S$\ref{subsec:below100MeV} to estimate that there were 1.4 and 1.1 $\times 10^{28}$ protons with energies $>$500 MeV in the X5.4 and M7 flares on 2012 March 7, respectively.  There was a 45-minute SGRE event associated with the X5.4 flare and an 18-hour SGRE event associated with the M7 flare, or possibly from both flares.  In obtaining the impulsive/sustained-emission proton ratios plotted in Figure \ref{ratio_nimp_nsus}, we associated the 18-hour event with the M7 flare.  We plot the ratio for the X5.4 flare, 0.035, as a filled circle with uncertainties at event location 12.  Because we had a more constraining 95\% confidence upper limit on the ratio for the M7 flare using the LAT measurements, we plotted a limit symbol, $<$0.003 for the second flare at event location 12, in lieu of the estimated ratio from the {\it INTEGRAL} 2.223 MeV line observation. The neutron-capture line was detected by GBM during the impulsive flare on 2014 February 25 (event  28,  Appendix  \ref{subsec:20140225}). The measured ratio for this event has a value of 0.034 and is plotted as a filled circle with uncertainties at event location 28.

The mean impulsive/sustained-emission proton ratio for these four events (6, 12, 16, and 28) where we have measurements of the number of protons during the impulsive flare is 0.096.  This is plotted by the solid line in Figure \ref{ratio_nimp_nsus}.  The weighted mean is even smaller, 0.044. 

LAT provided the most constraining upper limits on the ratios for events numbered 5, 8,12 (the M7 flare, plotted at 12.5), 13, 14, 19, 20, 24, and 30 and they are plotted in the figure with upper-limit symbols.   The mean 95\% confidence upper limit on the impulsive/SGRE $>$500 MeV proton-number ratio for these eight events is 0.09; the median upper limit is 0.05.  There were 12 other events for which we obtained 95\% confidence upper limits on the impulsive/sustained proton number ratios using limits on the neutron-capture line.  The mean and median ratios are $<$0.55 and $<$0.18 with 95\% confidence, respectively.  The high mean value is dominated by the high upper limit on the ratio for the 2012 May 17 event (15).  This high limit may be due to observational effects.  As we discuss in $\S$\ref{subsubsec:sus_sep} and Appendix \ref{subsec:20120517}, the SGRE flux from the GLE on the 2012 May 17 was anomalously low, possibly due to the flare's location near the western solar limb.  If we exclude this event, the mean 95\% confidence limit on the impulsive/sustained proton number ratios drops to 0.21. 

In summary, we find that the average number of $>$500 MeV protons producing SGRE is at least ten times larger than the number in the impulsive flare.   Thus, the impulsive flare contains insufficient energy in high-energy protons to be the prime source of the sustained $\gamma$-ray emission and another energy source is required.

\subsubsection{Comparison of Numbers of Protons in the SGRE and Associated SEP } \label{subsubsec:sus_sep}

It is becoming clear from our discussion to this point that most of the SGRE events are associated with magnetic eruptions and CMEs, deriving energy from them to accelerate protons to energies in excess of 300 MeV.  As CME shocks are responsible for accelerating the ions observed in large gradual SEPs, it is important to compare the number of $>$500 MeV protons producing the sustained emission with the number in the associated SEP.   However, most SEP measurements are made at energies below 100 MeV.  Satellite experiments such as {\it PAMELA} \citep{adri15} and {\it AMS} \citep{agui13} provide sensitive observations $>$100 MeV, but results on the total number of protons in space are not yet available.  Such a determination requires knowledge about the spatial distribution of high-energy SEPs.  Using data from the {\it GOES} HEPAD experiment and from neutron monitors, \citet{tylk14} developed a method to estimate the total number of $>$500 MeV SEP protons in space for some of the SGRE events.  We summarize this study in Appendix \ref{sec:sep} and compare below the derived number of SEP protons in space with the number of SGRE protons at the Sun.

\begin{figure}
\epsscale{.70}
\plotone{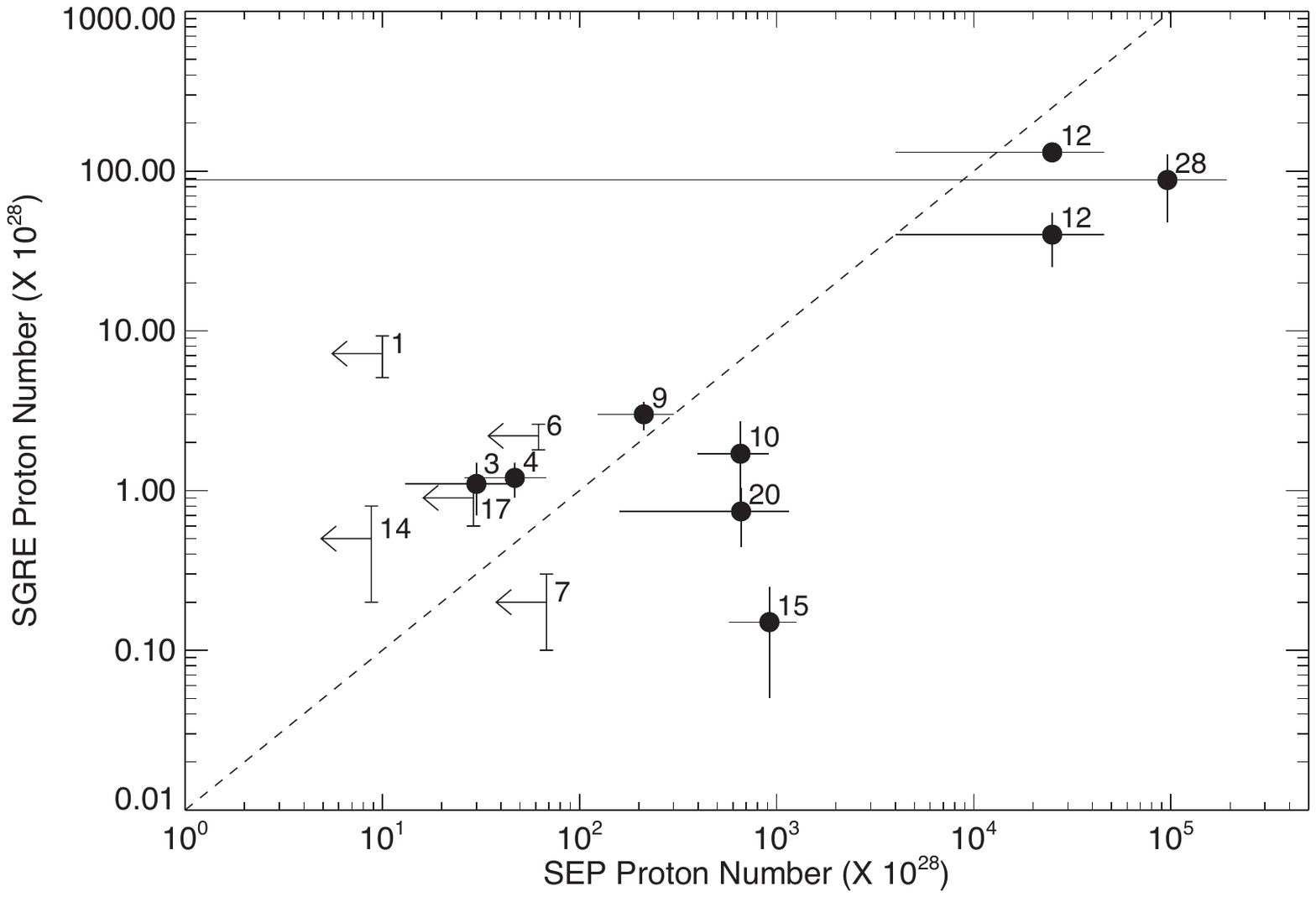}
\caption{Number of $>$500 MeV protons producing the sustained $\gamma$-ray emission plotted vs. the number in the associated SEP event.   The nine events with estimated numbers of SEP protons are plotted as filled circles at the mean values with $\pm1\sigma$ uncertainties.  The five events with upward limits on the number of SEP protons are plotted as left-pointed arrows. The dashed line is where the number of SGRE protons are 0.01 $\times$ the number of SEP protons.}
\label{corr_nsus_nsep}
\end{figure}

There are nine SGRE events for which we made estimates of the number of $>$500 MeV protons in space and there are five events where upper limits on the number were obtained.  The SEP proton numbers for these 14 events are given in column 7 of the rows labelled ``SEP" in Table \ref{tab:event}.  We plot the number of $>$500 MeV SGRE protons vs. the number observed in the accompanying SEP for the nine measured SEP events as filled circles in Figure \ref{corr_nsus_nsep}.   There is weak evidence for a correlation between the numbers of protons interacting at the Sun to produce the SGRE and the number of SEPs observed in space; the Spearman's rank correlation coefficient is close to 0.63 with only 87\% confidence that the numbers are correlated.  The ratio SGRE/SEP proton ratio ranges from 1.7 $\times$ 10$^{-4}$ to 0.04.   There are reasons for such a wide range in ratios.  For example, the small ratio measured for the GLE on 2012 May 17 (event 15) may be explained by the active region's location near the  western solar limb and the inferred few tens of degrees angular distribution of SGRE ($\S$\ref{subsubsec:location}).  The dashed line is the locus of points where the number of SGRE protons are 0.01 $\times$ the number of SEP protons.  The average SGRE/SEP proton number ratio is dependent on how the data are weighted by their uncertainties.  If we take a weighted mean of the nine measured ratios we obtain 0.025  $\pm$ 0.008; however if we first obtain weighted means of the numbers of sustained-emission and SEP protons and divide the former by the latter the ratio is 0.009 $\pm$ 0.003.  So on average the number of $>$500 MeV protons producing the SGRE is between $\sim$40 and $\sim$100 times smaller than the number of SEPs measured in space.   With the exception of the event on 2011 March 7 (event 1), the lower limits on the solar/SEP proton ratio are consistent with the range in ratios discussed above.  We discuss explanations for this ratio in $\S$\ref{sec:discussion}.  

We conclude that the number of protons interacting in a thick target solar atmosphere to produce the SGRE is, on average, close to two orders of magnitude smaller than the number of protons observed in space in the accompanying SEP.  If the SGRE protons and SEP protons have the same CME-shock origin then, on average, only $\sim$2\% of the shock accelerated protons return to the Sun and interact to produce $>$100 MeV SGRE.


\subsection{ Evidence for Sustained-Bremsstrahlung in a Behind-the-Limb Event}  \label{subsec:electrons}

Detection of SGRE from behind-the-limb events on 2013 October 11 (event 25) and 2014 September 1 (event 29) \citep{pesc15, pesc15c, acke17} definitively confirmed the suggestions discussed in $\S$\ref{subsubsec:location} that the $>$100 MeV $\gamma$-ray emission was produced by protons interacting within tens of degrees of the flare site, and neither just from the active region itself nor globally on the Sun.  \citet{plot17} detailed the characteristics of these events and showed that the onsets of the SGRE emission are consistent with the times when CME shock-accelerated protons interacted at the locations  visible from Earth after transport on magnetic fields returning to the Sun. 

We plot the time histories of these two behind-the-limb events in Figures \ref{131011th} and \ref{140901th} and discuss the detailed observations in Appendix \ref{subsec:20131011} and Appendix \ref{subsec:20140901}.  The NaI detectors on GBM observed hard X-rays from high coronal sources associated with both flares after they appeared over the solar limb.  Direct on-disk observations of the flares themselves were made by the Solar Assembly for X-rays (SAX) instrument on {\it MESSENGER} \citep{schl07}.   The SGRE on 2013 October 11 began about seven minutes after the 50--100 keV coronal source rose above the solar limb, lasted 15 minutes, and showed no evidence for spectral evolution.  The $>$100 MeV $\gamma$-ray spectrum is consistent with the decay of pions produced by $>$300 MeV protons with a power-law spectral index, s = 3.8 $\pm$ 0.2 throughout the observation (Table \ref{tab:event}), assuming that the interactions took place at a heliocentric angle of 85$^{\circ}$.
 
The most intense portion of the SGRE on 2014 September 1 lasted about 30 minutes and its flux was about ten times larger at its peak than the 2013 October 11 event.  It was one of the most intense SGRE events observed by LAT, even though the active region was more than 30$^{\circ}$ beyond the solar limb.  Our studies summarized in Table \ref{tab:event} and in $\S$\ref{subsubsec:heindex} provide evidence that the proton spectrum producing the pions hardened from a power-law index $\sim$4.3 to $\sim$3.3 over that time period.  During the peak emission time, we also found that the proton power-law spectrum steepened from one with an index harder than 3.4 between 20 and 300 MeV to one with an index of $\sim$4 above 300 MeV.

What is significantly different about the 2014 September 1 event is that the $>$100 MeV SGRE was accompanied by hard X-ray emission extending to energies above 10 MeV that began within a minute of the $>$100 MeV emission and peaked about five minutes earlier (see Figure \ref{140901th} in Appendix \ref{subsec:20140901}).   This hard X-ray emission could not have come from the flare site behind the limb.  The spectrum of the hard X-rays measured by the GBM NaI detectors from onset to the peak of the sustained $\gamma$-ray emission was fit acceptably by thick-target bremsstrahlung from a power-law electron spectrum with index 3.2 $\pm$ 0.1 and a low-energy electron cutoff of $\sim$130 keV (Appendix \ref{subsec:20140901}.  There is no evidence for variation of the spectrum with time.  The spectrum could not be fit acceptably by thin target bremsstrahlung, suggesting that the electrons interacted in the low corona and the chromosphere.  Additional evidence for the thick target nature of the emission comes from fits to the GBM BGO spectrum from 200 keV to 30 MeV over the same time interval.  Our spectral fits to these high-energy data yielded the same 3.2 $\pm$ 0.1 electron power-law index observed at lower energy.   \citet{acke17} showed that the hard X-ray and microwave time histories match one another and that the electrons producing the $>$1 GHz emission had a power-law index of $\sim$3, the same value we found for electrons producing the bremsstrahlung.  Although \citet{acke17} focus most of their discussion on a thin target origin from a source high in the corona, they leave open the possibility that the emission could be due to bremsstrahlung produced by electrons radiating in the chromosphere, as we suggest.   

The fact that both the protons producing the SGRE and the electrons producing bremsstrahlung interacted in a thick target on the visible disk and that both emissions commenced near the same time, suggests that they had a common origin, probably the CME shock that produced the intense electron and proton SEP event observed by {\it STEREO}.   We might, therefore, expect the relative numbers of the protons and electrons at the Sun and in space to be comparable.  The  {\it STEREO} 0.7--4 MeV electron and 13--100 MeV proton SEP fluxes each rose to a peak after the flare and the electron-to-proton flux ratio during these peaks was about 4.5 as measured in these energy bands.   From our fits to the bremsstrahlung spectrum, we estimate that about 3 $\times 10^{33}$ electrons with energies between 0.7--4.0 MeV interacted in the solar atmosphere.   Assuming that the proton spectral index had a value of 3.0 from 10 MeV to 200 MeV and softened to 4.0 above 200 MeV, as suggested by our spectral measurements, we estimate that there were about 5 $\times 10^{33}$ protons with energies between 13--100 MeV that interacted at the Sun.  From this admittedly uncertain estimate we find that the electron-to-proton ration at the Sun is about 0.6.  Thus, the measured electron/proton ratios at the Sun and in space are at least within a factor of ten of another.  This reasonable agreement, with the uncertainties involved, provides additional evidence that the electrons producing the sustained bremsstrahlung and protons producing the SGRE had a common origin with the SEP electrons and protons observed in interplanetary space.

We did not observe sustained bremsstrahlung from the occulted SGRE event on 2013 October 11 event where the peak pion-decay $\gamma$-ray flux was about 20 times lower than that on 2014 September 1.  The SEP 0.7-4.0 MeV electron flux on October 11 was between 50 and 100 times lower than it was on September 1.  Bremsstrahlung from an electron flux two orders of magnitude smaller than observed on September 1 would not have been detectable over background by GBM.  The SGRE events on 2012 March 7 (Appendix \ref{subsec:20120307}) and on 2014 February 25 (Appendix \ref{subsec:20140225}) were of comparable intensity to the 2014 September 1 event, but we did not detect any evidence for sustained bremsstrahlung emission. However, flare-related and instrumental background could have affected the GBM sensitivity to sustained bremsstrahlung.  We estimate the expected sustained bremsstrahlung flux from these events by assuming that the (peak SGRE)/(peak-sustained-bremsstrahlung) ratio was the same as that measured in the 2014 September event, modified by the relative SEP proton/electron ratio measured in space.   For the first SGRE event on 2012 March 7 the estimated sustained bremsstrahlung flux was about a factor six lower than the upper limit that could be set from the background at that time.  For the second event on March 7 the estimated bremsstrahlung flux was about the same as the upper limit based on the background.  In contrast, we estimate that sustained bremsstrahlung during the peak of the SGRE on 2014 February 24 should have exceeded the upper limit set by background by a factor of five.  Thus, we have evidence for sustained bremsstrahlung emission only during the peak of the SGRE on 2014 September 1, and there is one event where we might have expected to detect the bremsstrahlung, but did not.  

. 


\section{Summary} \label{sec:summary}

Below we list the salient points made in the paper.

\begin{itemize}

\item We use the term `sustained gamma-ray emission' (SGRE) to characterize $>$100 MeV $\gamma$-ray emission that has temporal and spectral characteristics that are distinct from the accompanying impulsive flare.  The 1982 June 3 flare observed by the {\it SMM} gamma ray spectrometer was first to exhibit this classic time history for such high-energy emission. 

\item We utilize a light-bucket approach to study SGRE events observed by {\it Fermi}/LAT from the Sun on time scales from minutes to hours ($\S$\ref{subsec:details}).  Online four-day plots of the solar $>$100 MeV $\gamma$-ray flux observed by LAT over the entire mission are provided on the {\it RHESSI} browser\footnote{\url{http://sprg.ssl.berkeley.edu/~tohban/browser}} ($\S$ \ref{subsec:browplots}).  We performed spectroscopic studies of LAT solar data using OSPEX software\footnote {\url{http://hesperia.gsfc.nasa.gov/ssw/packages/spex/doc/ospex_explanation.htm}} available for use by all researchers. 

\item We list the 30 SGRE events identified in our study from 2008 to 2016, their characteristics, and related solar observations in Tables \ref{tab:latlist}, \ref{tab:onsets}, and \ref{tab:event} and provide their time histories and observational details in Appendix \ref{sec:append}. 

\item The $>$100 MeV SGRE can be fit by the $\gamma$-ray spectrum from decay of neutral and charged pions produced by protons having a power-law spectrum in energy interacting in a thick target ($\S$\ref{subsec:latspect}).  However, there is evidence in the most intense events that the proton spectrum actually steepens from a simple power-law at high energies.  Bremsstrahlung from high-energy electrons with synchrotron losses does not fit the SGRE ($\S$\ref{subsubsec:nobrems}) spectrum.

\item In all 26 SGRE events for which there were available hard X-ray measurements of the accompanying impulsive flare, the emission exceeded 100 keV ($\S$\ref{subsec:conditions}).  This suggests that $>$100 keV emission from the accompanying flare is a necessary condition for SGRE.     

\item In a separate four-year study of 22 flares on the visible disk with CMEs having speeds $>$800 km s$^{-1}$ and measured SEP proton fluxes, the average proton flux was about two orders of magnitude higher when the hard X-ray emission from the flare exceeded 100 keV ($\S$\ref{subsec:conditions}). 

\item SGRE was observed in all 28 events having fast CMEs but also in two events without an accompanying CME.  Thus a fast CME is not a necessary condition for SGRE.

\item Neither an accompanying flare with $>$100 keV nor an accompanying fast CME is a sufficient condition for LAT detection of a SGRE. 

\item The time histories of 21 of 31 SGRE eruptions (two on 2012 March 7) are clearly distinct from that of the impulsive phase emission.  The remaining ten SGRE events extend in time well beyond the impulsive flare and none of them has a time history clearly arising from decay of the flare emission ($\S$\ref{subsec:tempdistinct}).  There is no evidence that the SGRE is due to a series of episodic high-energy outbursts ($\S$\ref{subsec:episodic}).  The SGRE begins as earlier as the onset of the CME and as late as about two hours after onset ($\S$\ref{subsubsec:deldur}).  The duration of the emission ranges from about five minutes to twenty hours.

\item The SGRE durations are well correlated with the durations of the accompanying $>$100 MeV SEP proton events but the SEP durations are on average five-times longer.  The SGRE durations display a weaker correlation with the durations of the associated {\it GOES} soft X-ray flares where late emission measurements were possible ($\S$\ref{subsubsec:correlations}). 

\item The average power-law index of $>$300 MeV protons producing the SGRE is $\sim$4.0 with an RMS scatter in index of 1.8 ($\S$\ref{subsubsec:heindex}).  The proton spectra of six SGRE events varied with time, four softened and two hardened.   Analysis of MeV lines in the $\gamma$-ray spectra of SGRE events indicates that the proton spectra between 20 and 300 MeV are flatter than they are above 300 MeV, often with a change in power-law index of unity or larger ($\S$\ref{subsubsec:LoE_index}).   

\item The number of $>$500 MeV protons producing the SGRE is typically about an order of magnitude larger than the number producing the impulsive $\gamma$-ray emission during the associated flare.   This suggests that the flare is not the primary source of energy for the SGRE ($\S$\ref{subsubsec:imp_sus}).

\item In all 23 SGRE events for which we have measurements of the associated impulsive flare, the distinctly different impulsive flare and SGRE time histories and/or their relative energetics implies that the flare was not the primary energy source for the SGRE.  Therefore, a separate process is required to accelerate the protons to the high energies necessary to produce SGRE. The shock formed by a fast CME or magnetic eruption is the likely candidate.

\item Type II metric or DH radio emission, indicating the presence of a shock, accompanied all of the SGRE events except the 2012 November 27 event which did not have an associated CME (see Appendix \ref{sec:solarradio}).

\item Type III metric or DH radio emission, indicating that flare electrons reached open field lines, accompanied all the SGRE events except the two events which did not have associated CMEs (see Appendix \ref{sec:solarradio}).

\item SEP protons were observed from all SGRE events that were accompanied by a CME and for which the proton flux was not masked by the remnants of a previous SEP.

\item The number of $>$500 MeV protons producing the SGRE is on average only 1--2\% of the number of protons observed in SEPs in interplanetary space ($\S$ \ref{subsubsec:sus_sep}).  This percentage ranged from 0.02 to 4.0\% in individual events.  

\item Behind-the-limb events \citep{acke17} and other observations indicate that SGRE can extend up to a few tens of degrees from the active region but is not globally distributed over the visible solar disk ($\S$\ref{subsubsec:location}).  

\item Sustained bremsstrahlung from electrons with energies in excess of 10 MeV accompanied the SGRE associated with the 2014 September 1 behind-the-limb flare ($\S$\ref{subsec:electrons}).  {\it RHESSI} imaging indicated that the bremsstrahlung was distributed over tens of degrees in heliographic longitude ($\S$\ref{subsubsec:location}) suggesting that the SGRE was similarly distributed. 
\end{itemize}

\section{Discussion} \label{sec:discussion}

We can use the characteristics of sustained $>$100 MeV $\gamma$-ray emission (SGRE) summarized in the previous section to set constraints on its origin.  Our spectroscopic studies provide compelling evidence that the SGRE is from pion-decay produced by $>$300 MeV protons and $>$200 MeV $\alpha$ particles, and not from electron bremsstrahlung.  In his summary of sustained-emission events associated with large solar flares, \citet{ryan00} discussed the possibility that the time-extended emission arises from delayed precipitation of high-energy protons accelerated in the flares and stored in magnetic structures high in the corona.  He pointed out that such an origin could be tested if the sustained emission were observed when the accompanying flare was much weaker.  This test was made possible with the launch of {\it Fermi} and its Large Area Telescope (LAT), which has observed SGRE associated with C- and M-class flares. Observations made by LAT appear to rule out precipitation of stored flare protons as the primary source because: 1) the time histories of the sustained $\gamma$-ray emission indicate that the particles are accelerated in a separate process, mostly delayed from the impulsive phase of the flare, and not consistent with a decaying impulsive phase flux,  2) the numbers of protons responsible for the sustained emission are on average over an order of magnitude higher than the numbers accelerated in the impulsive phase, and 3) the interaction site of the sustained-emission protons can be tens of heliographic degrees from the active region.  Thus, another more energetic acceleration process is required that can distribute the particles over broader spatial regions.  

Because CMEs typically have ten times the energy of flare-accelerated particles \citep{emsl12}, \citet{ryan00} suggested that an accompanying CME can provide sufficient energy for accelerating SGRE particles to high energies through either the associated shock or magnetic reconnection in the current sheet behind the CME.  Particles accelerated to $>$300 MeV energies in the current sheet would likely interact near the active region and would not easily explain SGRE from flares behind the solar limb.  Transport to locations far from the active region requires the acceleration of protons onto magnetic field lines that reach back to the Sun.  This would seem to favor acceleration by the accompanying CME shock that also produces the SEP particles that reach interplanetary space.  Indeed, Type II radio emission indicating the presence of such a coronal shock was observed in all but one of the 30 SGRE events (the 2012 November 27 event without an accompanying CME).  Thus, a plausible scenario for production of SGRE is shock acceleration of a seed population onto magnetic field lines that reach the visible disk and precipitation of these particles into the chromosphere.  

CME-shock acceleration of a seed population onto field lines open to interplanetary space also explains the characteristics of gradual SEP events.  Indeed, SEPs were detected in all of the SGRE events that were accompanied by fast CMEs, except for those where the SEPs were masked by radiation from a previous event. The presence of such open field lines was confirmed in 27 of the 30 SGRE events by detection of Type III radio emission produced by electrons streaming outward from the active region.  This suggests a common origin for both gradual SEPs and the particles producing the SGRE. The fact that we find a correlation between the duration of SGRE events and the duration of SEPs observed above 100 MeV also suggests a common CME-shock origin, even though the SGRE durations are on average about five times shorter than those of the SEPs.   Where measurements are available, our studies indicate that the average number of protons with energies above 500 MeV producing the sustained $\gamma$-ray emission is 1--2\% of the number in the associated SEP event and that individual measured ratios range from 0.02--4\%.  \citet{koch15} presented a shock-wave model to estimate the ratio of the number protons that return to the Sun and interact to the number that escape into interplanetary space.  They find that the ratio for $>$100 MeV protons is about 0.2\%, ten times less than the average we determined from LAT observations.  However, the calculated ratio increases to about 2\% if stochastic re-acceleration from turbulence downstream of the shock direction is included.   

There is another interesting feature observed during the 2014 September 1 event related to the connection between SEPs and SGRE events.  We find evidence for sustained bremsstrahlung emission produced by electrons with energies up to a few tens of MeV that had the same onset time as that of the $>$100 MeV SGRE implying that the electrons were also accelerated in the CME shock.  Detection of sustained bremsstrahlung was made possible because of the intensity of the event and the lack of hard X-ray emission from the behind-the-limb flare.  We found comparable electron/proton ratios measured over the same energy bands in the SGRE and in SEPs for this event, which is again consistent with a common CME-shock origin. 

The CME-shock scenario provides for a wide range in particle spectra at different locations depending on the shock strength and where and at what angle the particles are accelerated onto magnetic field lines.   This can explain how $>$300 MeV protons interacted at the Sun to produce the first LAT SGRE event on 2011 March 7 while the energies of the SEP protons in space barely exceeded 100 MeV.   The scenario can also explain the broad spatial distribution of SGRE implied by detection of SGRE from behind-the-limb flares on 2013 October 11 and 2014 September 1 and by imaging of the sustained bremsstrahlung on September 1.  Such a broad distribution can also explain the weak SGRE associated with the 2012 May 17 GLE from the flare on the visible disk, but within 13$^{\circ}$ of the solar limb.  In this case, most of the particles could have been accelerated onto field lines that return to the Sun at locations beyond the limb as viewed from Earth. 

The fact that two of the sustained-emission events were observed without accompanying CMEs presents a problem for the CME-shock magnetic-field scenario.  We note, however, that metric Type II emission was observed two minutes after a magnetic eruption associated with the 2012 October 23 SGRE event.  This suggests that $>$300 MeV protons could have been accelerated onto low-lying magnetic loops by a shock formed by the magnetic eruption, and stored for the two hour duration of the SGRE, during which time some precipitated and interacted in the chromosphere.  It is not likely that the flare itself produced these trapped protons because we estimate that the number of protons required to produce the SGRE was five times the upper limit on the number of protons accelerated during the flare.  The SGRE event on 2012 November 27 that lasted only 15 minutes also provides a challenge for the scenario because there is no evidence for the presence of a shock.  However, there was evidence for a magnetic eruption.  The impulsive flare could not have produced the emission because its time history was distinctly different from that of the SGRE (see Appendix \ref{subsec:20121127}) and the number of impulsive flare protons was a factor of twenty lower than the number producing the SGRE. 

The smooth time histories of the SGRE events without evidence for episodic eruptions also suggests acceleration of particles by an expanding CME shock onto field lines returning to the Sun.  This model may explain SGRE events lasting less than a few hours but it has difficulty explaining events with emission lasting several hours because it implies that the shock accelerated protons to energies $>$300 MeV at several tenths of an AU from the Sun and that these particles were able to return to the Sun.  Such long-duration events are more likely produced by precipitation of shock-accelerated particles that are magnetically trapped in a reservoir behind the shock.  Particles accelerated onto large coronal loops by shocks can precipitate into the chromosphere on different time scales depending on the amount of magnetic turbulence \citet{ryan15} in the loop.  There are questions about whether such trapping is consistent with the softening of the spectra found in the long-duration SGRE events because it is expected that lower energy ions would precipitate first \citep{mand92}.  It would, however, explain the hardening of the spectra observed in the 2011 September 6 (event 6) and 2014 September 1 (event 28) SGRE events that have durations shorter than a few hours.  Hudson (2017) recently proposed a related ``lasso" scenario in which particles are accelerated onto closed field lines extending out to several R$_{\odot}$, that retract, further accelerating them while moving them closer to the Sun.

There is one feature of the SGRE events that needs to be understood in the context of the CME-shock magnetic-field scenario.  We find that impulsive hard X-rays reached energies greater than 100 keV in every SGRE event for which there were flare observations.  This appears to be a necessary condition for producing sustained $\gamma$-ray emission.  Although it may just be a manifestation of the Big-Flare Syndrome \citep{kahl82}, our spectroscopic studies suggest otherwise and that a threshold for acceleration of hundred keV electrons, and perhaps ions, was reached in the impulsive flares associated with the SGRE.  If this is the case, then sub-MeV flare-accelerated protons could be a seed population for further acceleration by a CME shock if they escape from the flare site.  Flare-accelerated electrons have been shown to be the source of heating of a CME \citep{gles13} and have been found to be present in radio plumes containing both open and closed field lines \citet{flei17}.  In addition, acceleration of flare-produced electrons by CME shocks was suggested by \citet{petr16} to explain the harder electron spectra in the associated SEP events.   The presence of electrons on open field lines is also inferred in 27 of the 30 SGRE events by the accompanying Type III radio emission.  Thus it is possible that hundreds of flare electrons accelerated to MeV energies by the CME shock onto field lines returning to the visible disk produced the observed sustained $\sim$ MeV bremsstrahlung emission in the 2014 September 1 behind-the-limb flare.   

The key question is whether the presence of hundreds of keV electrons in flares implies the acceleration of sub-MeV protons that can form the seed population accelerated by the CME shock to contribute to both the SGRE and gradual SEPs.  Nuclear de-excitation line $\gamma$-rays have a high probability of being produced in flares when the accompanying electron bremsstrahlung reaches energies above 300 keV (e.g. \citet{vest99,shar00,shih09}).  This implies that ions with energy in excess of the 1 MeV nucl$^{-1}$ threshold energy necessary for producing $\gamma$-ray lines are present in flares when electrons with energies of several hundred keV are present.  If this is true, then sub-MeV nucl$^{-1}$ ions, which do not produce $\gamma$ rays and are not easily detected by other means, probably are accelerated in flares along with the electrons that produce the bremsstrahlung above 100 keV.  If sub-MeV ions are also accelerated in the upward direction, as found for flare electrons \citep{gles13}, they could contribute to the seed population accelerated by the CME to produce the SGRE and the SEPs.  

Early in the CME eruption, acceleration of particles by shocks onto field lines returning to the Sun occurs more favorably at the flanks of the CME where the magnetic-field geometry is quasi-perpendicular to the shock \citep{schw15,plot17}.   \citet{zank06} demonstrated that the seed-population of protons needs to have energies above $\sim$50 keV to be accelerated in such a quasi-perpendicular shock geometry at 0.12 AU.  Such quasi-perpendicular geometries low in the corona can produce hard proton spectra \citep{tylk06b} observed in SEPs.  The suggestion of a population of $>$50 keV protons that escape from the flare site and are accelerated by the CME shock is admittedly speculative, but it can explain why detection of $>$100 keV bremsstrahlung in the impulsive flare is a necessary condition for detection of SGRE.   It also can explain why the median SEP flux is about two orders of magnitude higher when $>$100 keV bremsstrahlung is detected in the accompanying solar flare than when it is not ($\S$\ref{subsec:conditions} and Table \ref{tab:bfs}).  \citet{li05} calculated the spectrum of particles in space from a mixed population of flare-accelerated and CME-shock accelerated particles for a shock starting at $5 R_{\odot}$.  This calculation also allowed for further acceleration of flare particles with energies above 1 MeV by the shock. 

Such a seed population of sub-MeV ions associated with the flare may also in part explain the delay in the onset of the SGRE because of the time it would take them to catch up to the CME shock where they can be further accelerated.  If we assume that the protons are ejected into space at the time of the mid-point of flare 100--300 keV X-ray emission, we can estimate the energy they would require in order to reach the shock by the time of the onset of the SGRE.  From a comparison of CME and SGRE onsets and hard X-ray time profiles in 18 events where the SGRE onset was determined, we estimate that these protons have a mean energy of $\sim$ 75 keV.  This energy is consistent with our above discussion.  The energy would be even higher if the SGRE onset delay is in part due to the time it takes for the shock accelerated particles to reach a magnetic field line returning to the Sun and the time it takes to accelerate the protons to energies $>$300 MeV.   \citet{plot17} calculated the times when high-energy protons could be accelerated onto loops reaching the visible solar disk from the behind-the-limb flares on 2013 October 11 and 2014 September 1.  Their calculated times are consistent with the start times of SGRE observed by LAT.


In early September 2017, during solar minimum, Active Region 12673 grew in size and produced 26 M-class and four X-class flares.  Two of the X-class flares were associated with SGRE events.  The X9.3 solar flare began at 11:53 UT on 2017 September 6 and emitted hard X-rays up to 400 keV.  It was associated with a 1200 km s$^{-1}$ CME and an SEP event with protons $>$100 MeV.  We found no evidence for either $>$100 MeV impulsive flare or SGRE from 12:10 to 12:35 UT when the hard X-rays were observed.  LAT first observed SGRE during its next exposure beginning at 12:35 UT, indicating that the SGRE began between 12:35 and 13:25 UT.  The SGRE lasted for $\sim$18 hours, about the same duration as the $>$100 MeV SEP proton event.   The spectrum of the SGRE emission is consistent with the decay of pions produced by $>$300 MeV protons with power-law spectral indices between 5 and 7.  The second SGRE event was associated with an X8.2 flare, located near the west solar limb, that emitted bremsstrahlung, nuclear de-excitation lines, and pion-decay emission.  It was associated with a $\sim$ 2600 km s$^{-1}$ CME and a high energy SEP event that was observed above 500 MeV by neutron monitors at sea level. The $>$100 MeV $\gamma$-ray flux during the impulsive flare was about five times higher than any solar flux previously measured by LAT.  The SGRE began near the time of the flare, reached a peak flux 30 times smaller than that of the flare, and lasted about 18 hours.    The characteristics of these two SGRE events are consistent with those of the 30 events discussed in this paper.  Details of these events will be provided in a future publication.

\acknowledgments
G. Share conducted this research with financial support from NSF Grant 1156092, NASA Fermi/GI grant GSFC \#71080, the EU's Horizon 2020 research and innovation program under grant agreement No 637324 (HESPERIA), a PRAXIS sub-contract, and the U.S. Civil Service Retirement System.  R. Murphy was supported by a NASA Fermi/GI DPR and by the Chief of Naval Research.  We acknowledge stimulating conversations with Dr. Eric Grove and the hospitality of the NRL Space Science Division, the Solar Physics Laboratory (Code 671) of NASA GSFC, and the Department of Astronomy of the University Maryland where much of this work was performed.  This work would not have been possible without the efforts of the {\it Fermi} teams that built the highly successful LAT and GBM detectors and provided the data products and calibrations. We especially appreciate the efforts of Nicola Omodei and Melissa Pesce-Rollins.  We also acknowledge our use of data from NASA's {\it Solar Dynamics Observatory} and the NRL LASCO instrument on ESA's and NASA's {\it Solar Heliospheric Observatory}. 

\appendix

\section{Systematic Study of of High-Energy Solar Eruptive Events from 2008 to 2012} \label{sec:characteristics}

As a step in understanding the origin of SGRE, we identified high-energy eruptive events, limited to the time interval between June 2008 and May 2012, having at least one of the characteristics found in the three originally-reported LAT events.    Specifically, we searched for: (1) CMEs with projected speeds $\gtrsim$ 800 km s$^{-1}$ and widths $>$90$^{\circ}$ in the {\it SOHO} LASCO CME Catalog\footnote{\url{http://cdaw.gsfc.nasa.gov/CME_list/index.html}}, (2) SEP events with proton flux $>$1 pfu at energies $\gtrsim$10 MeV measured at the best magnetically connected spacecraft, or (3) hard X-ray flares with energies $>$100 keV.  We manually identified the SEP events from the {\it GOES} $>$10 MeV proton plots for the four years of the study and performed an automated search of proton fluxes $>$13 MeV in the IMPACT instrument \citep{luhm08} on {\it STEREO} from 2010 to 2012, when the two spacecraft were at large angular separations from Earth.  We identified the flares with $>$50 keV hard X-ray emission observed by {\it RHESSI} and GBM using the flare lists \footnote{\url{http://hesperia.gsfc.nasa.gov/rhessi2/}} and events that triggered the GBM onboard burst mode.  We then visually inspected the RHESSI Browser \footnote{\url{http://sprg.ssl.berkeley.edu/~tohban/browser}} plots to determine the highest-energy band detected.  


\subsection{List of 95 High-Energy Solar Eruptive Events from 2008 June to 2012 May} \label{subsec:95list}

In Table \ref{tab:full} we list the 95 events\footnote{Event 72 contained an X- and M-class flare, each with distinct CME emissions but unresolved SEP contributions} from June 2008 until the end of May  2012 that met at least one of the three criteria listed above.  The first column of the table gives the sequential event number.  The second column contains the date of the associated event and location of the flare from the NOAA Solar Event Reports\footnote{\url{ftp://ftp.ngdc.noaa.gov/STP/swpc_products/daily_reports/solar_event_reports/}}.  Where {\it RHESSI} data are available, we list the centroid of the highest-energy quick-look hard X-ray image in the {\it RHESSI} Browser.  Where the emission appears above the limb and there is no evidence for footpoints, we list the longitude as E91 or W91.  For backside events, we list flare locations from the {\it STEREO} EUVI catalog developed by \citet{asch14}. The third column lists the {\it GOES} soft X-ray class and its start and end times  as given in the NOAA Solar Event Reports. For behind-the-limb events we list the estimated {\it GOES}-class range based on {\it STEREO A/B} data, where available \citep{nitt13}. When the flare occurred behind the solar limb, and there is no associated {\it GOES} X-ray event, we list the linearly extrapolated CME onset time given in the {\it SOHO} LASCO CME Catalog or from the EUVI flare image in {\it STEREO A/B}.

The fourth column lists linear speeds from the {\it SOHO} LASCO CME Catalog.  The uncertainty in speeds is a few hundred km s$^{-1}$ based on a comparative study of different CME catalogs by \citet{rich15}.  In the fifth column we list the relative strength of any observed metric Type II slow drift $\sim$20 to 200 MHz (1.5m to 15m) radio emission in the corona observed by ground-based radio observatories as reported in the NOAA Solar Event Reports.  If we cannot confirm the Type II emission in our studies of the radio spectra, we add a ``?" next to the entry.   We also enter a ``Y" when decameter-hectometric (DH) Type II emission was observed by {\it Wind} and {\it STEREO} spacecraft, as given in the Wind/WAVES type II bursts and CME catalog\footnote{\url{https://cdaw.gsfc.nasa.gov/CME_list/radio/waves_type2.html}}. The sixth column lists the estimated peak SEP proton flux ($>$10 MeV for {\it GOES} and $>$13 MeV for {\it STEREO} ) in protons cm$^{-2}$ s$^{-1}$ (pfu) followed by the highest energy or energy range where protons were observed.  We list the measured flux from the spacecraft that was best magnetically connected to the flare site.  Fluxes are from {\it GOES} unless otherwise specified. These peak fluxes were obtained after background subtraction and do not include energetic particle fluxes from local shocks. We identify events for which only flux limits can be obtained because they were preceded by a much stronger SEP (footnote 'f').  The seventh column lists the maximum hard X-ray energy band detected by the {\it RHESSI} and GBM hard X-ray detectors based on the {\it RHESSI} Browser plots.  A dash indicates that no hard X-rays were detected, either because the active region was beyond the limb of the Sun or hard X-ray data were missing.

Entries in bold print denote the nineteen events for which the CME speed was $\gtrsim$ 800 km s$^{-1}$ and flare hard X-rays reached energies above 100 keV.  The fourteen events listed in the Table with their event numbers in red, and with a double asterisks, had accompanying $>$100 MeV SGRE detected by LAT.  For reference we list the associated event number in Table \ref{tab:latlist} in parenthesis in column 1.  The SGRE event A25 on 2011 June 2 was not in bold print because there were no  observations available to determine if the emission exceeded 100 keV.

\begin{deluxetable}{ccccccc}
\tabletypesize{\scriptsize}
\tablecaption{Solar Eruptive Events from June 2008 to May 2012 \label{tab:full}}
\tablehead{
\colhead{Number} & \colhead{Date, Location} & \colhead{{\it GOES} X-Ray}& \colhead{CME}&\colhead{Type II}&\colhead{SEP}&\colhead{Hard X-ray } \\
\colhead{} & \colhead{yyyy/mm/dd, deg} & \colhead{Class, Start-End }&{Speed, km s$^{-1}$}& \colhead{M\tablenotemark{*}, DH} &\colhead{Flux (pfu), Energy (MeV)}&\colhead{Energy (keV)}}
\colnumbers
\startdata
A1 & 2010/02/08, N22W02  &  M4.0, 07:36--07:46 & N & N, N  &  $<$0.2, $<$10   & 100--300\tablenotemark{d} \\
A2 & 2010/02/12, N25E11 &  M8.3, 11:19--11:28 & 509 & N, N &  0.1, $<$35\tablenotemark{b}; $<$0.2, $<$10& 300--800\tablenotemark{c} \\
A3 & 2010/06/12, N22W57&  M2.0, 00:30--01:02  & 486 &  2, N &  0.6, $<$60   & 1000--40000   \\
A4 & 2010/08/01, N13E21  & C3.2, 07:55--09:35&  850 & N, Y  & 5.3, $<$60\tablenotemark{b}  &  12--25 \\ 
A5 & 2010/08/07, N12E31&  M1.0, 17:55--18:47 & 871  & 2, Y &  5.0, $>$60\tablenotemark{b}   & 12--25\tablenotemark{e}   \\
A6 & 2010/08/14, N11W65&  C4.4, 09:38--10:31  & 1205  & 1, N &  9.5, $>$100   & 25--50 \\
A7 & 2010/08/18, N19W97 &  C4.5, 04:45--06:51 & 1471 & 1, Y &  2.5, $<$60   & 12--25\tablenotemark{c} \\
A8 & 2010/08/31, S22W146  &  M8.4--X2.5,  20:41--? & 1304  & N, Y & 0.6, $>$60\tablenotemark{{a}}  & -- \\
A9 & 2010/12/14, N20W56 & C2.3, 15:03--16:55    & 835 &  N, N & $<$0.2, $<$10 & 6--12 \\
A10 & 2011/01/28, N17W91 & M1.3, 00:44--01:10    &  606 & 1, Y &  1.9, $>$100  & 25--50\tablenotemark{c} \\
A11 & 2011/02/13, S20E04 & M6.6, 17:28--17:47  & 373 & 1, Y & 0.3, $<$60\tablenotemark{b} & 100--300\tablenotemark{d} \\ 
A12 & 2011/02/14, S19W05 & M2.2, 17:20--17:32  & 326 & 2, N & $<$0.2, $<$10 & 100--300\tablenotemark{d} \\
A13 & 2011/02/15, S20W12 & X2.2, 01:44--02:06 & 669 & 2, Y &  5.4, $>$60\tablenotemark{b} & 100--300 \\
A14 & 2011/02/18, S20W55 & M6.6, 09:55--10:15 & N & N,N & $<$0.2, $<$10 & 100--300\tablenotemark{c} \\
\bf{A15} & \bf{2011/02/24, N15E84} & \bf{M3.5, 07:23--07:42}  & \bf{1186} & \bf{2, Y?} & \bf{0.06, $<$35\tablenotemark{b}; $<$0.2,$<$10} & \bf{800--7000} \\
\bf{\color{red}A16\tablenotemark{**}(1)} & \bf{2011/03/07, N30W47} & \bf{M3.7, 19:43--20:58} &\bf{ 2125} &\bf{ 3?, Y}  & \bf{39.6, $>$60} & \bf{300--1000\tablenotemark{d}} \\
A17 & 2011/03/09, N09W11 & X1.5, 23:13--23:29 & 332 & N, N & $<$14.7, $>$10\tablenotemark{f} & 100--300\tablenotemark{d} \\
A18 & 2011/03/14, N15W49 & M4.2, 19:30--19:54 & 512 & N, N & $<$0.2, $<$10 & 100--300\tablenotemark{d} \\
A19 & 2011/03/21, N20W128 & M1.3--X1.3, 02:11--? & 1341 & N, Y & 702, $>$60\tablenotemark{a} & -- \\
A20 & 2011/03/27, N19E101 & ?, $\sim$ 05:16  & 877 & 1, N & $<$0.2, $<$10 & -- \\
A21 & 2011/03/29, N21E115? & ?, $\sim$20:14    &1264 & N, N & 1.3, $>$60\tablenotemark{{b}} & -- \\
A22 & 2011/04/27, N19E59 & C2.0, 02:26--03:01 & 924 & 2, N & $<$0.2, $<$10 & 25--50   \\
A23 & 2011/05/09, N19,E91 & C5.4, 20:42--21:19 & 1318 & N, Y & 0.3, $<$41\tablenotemark{b} & 25--50  \\
\bf{A24} & \bf{2011/05/29, S18E75} & \bf{C8.7, 21:04--21:45} &\bf{1407} & {2?, Y} &\bf {5.6, $<$35\tablenotemark{b}}  &\bf{100--300\tablenotemark{d}} \\
\color{red} A25\tablenotemark{**}(2) & 2011/06/02, S18E22 & C3.7, 07:22--07:57 & 976 & N, Y  & 0.1, $<$40\tablenotemark{b}; $<$0.2, $<$10  & --\tablenotemark{e}\\
A26 & 2011/06/04, N15W140 & M5.2--X1.6, 07:06--? & 1407 & N, Y & 55.8, $>$60\tablenotemark{a} & -- \\
A27 & 2011/06/04, N17W148 & X4--X12 ,21:51--?  & 2425 & N, Y  &  2060, $>$60\tablenotemark{a} & -- \\
\color{red} \bf{A28\tablenotemark{**}(3)} & \bf{2011/06/07, S21W54} & \bf{M2.5, 06:16--06:59}  & \bf{1255} & \bf{2?, Y}& \bf{60.5, $>$100} & \bf{300--800} \\
A29 & 2011/06/13, S18E131 & ?, $\sim$ 03:55 & 957 & N, Y& $<$3.6, $>$10\tablenotemark{f}  & --  \\
A30 & 2011/07/30, N14E34 & M9.3, 02:04--02:12 & N & N, N  & $<$0.2, $<$10  &  100--300 \\
A31 & 2011/08/02, N19W11 & M1.4, 05:19--06:48 &  712  & 2, Y & 1.3, $>$100 & 25--50\tablenotemark{c}  \\
\color{red} \bf{A32\tablenotemark{**}(4)} & \bf{2011/08/04, N19W46} &  \bf{M9.3, 03:41--04:04}  & \bf{1315} & \bf{2, Y}  & \bf{48.4, $>$100} &  \bf{300--1000\tablenotemark{d}}  \\
A33 & 2011/08/08, N15W64 & M3.5, 18:00--18:18 & 1343 & 1, Y & 1.5, $>$100 & 50-100\tablenotemark{d,e} \\
\color{red}\bf{A34\tablenotemark{**}(5)} & \bf{2011/08/09, N16W70} &  \bf{X6.9, 07:48--08:08}  & \bf{1610} & \bf{1?, Y} & \bf{16.3, $>$10} & \bf{800--7000} \\
A35 & 2011/09/06, N14W07 &  M5.3, 01:35--02:05  & 782  & 3, Y & 1.5, $>$100   & 50--100\tablenotemark{d} \\
\color{red} \bf{A36\tablenotemark{**}(6)} & \bf{2011/09/06, N14W18} & \bf{X2.1,  22:12--22:24} & \bf{575, $\sim$1000\tablenotemark{a,b,h}}    & \bf{2, Y} & \bf{5.6, $>$100}    &  \bf{300--1000}\\
A37 & 2011/09/07,  N22E66     & B9.1, 18:24--18:33 & 924 &  N, Y & $<$1.6, $>$10\tablenotemark{f}& 25--50\tablenotemark{c}  \\
\color{red} \bf{A38\tablenotemark{**}(7)} & \bf{2011/09/07, N18W32}&  \bf{X1.8, 22:32--22:44} & \bf{792} & \bf{1, N}  & \bf{$<$1.7, $>$10\tablenotemark{f}} & \bf{300--1000\tablenotemark{d}}  \\
A39 & 2011/09/08, N14W41 &  M6.7, 15:32--15:52 & 351& N, N & $<$0.4, $>$10\tablenotemark{f} & 100--300 \\
A40 & 2011/09/08, N19W134 &  ?, $\sim$ 21:50 & 983& N, Y & $<$0.4, $>$10\tablenotemark{f}  & -- \\
A41 & 2011/09/21, N19W114 & ?,$\sim$ 22:00 & 1007 & N, N  &  0.2, $<$60\tablenotemark{a} & -- \\
A42 & 2011/09/22, N10E91 & X1.4, 10:29--11:44 & 1905 & 2, Y & 1220, $>$60\tablenotemark{b} & 25-50\tablenotemark{c,e}  \\
\color{red} \bf{A43\tablenotemark{**}(8)} & \bf{2011/09/24, N14E61} & \bf{X1.9, 09:21--09:48} & \bf{1936} &\bf{2?, N} &\bf{$<$77, $>$13\tablenotemark{b,f}} & \bf{800--7000}  \\
\bf{A44} & \bf{2011/09/24, N11E61} &  \bf{M7.1, 12:33--14:10} & \bf{1915}& \bf{N, Y}& \bf{$<$70, $>$13\tablenotemark{b,f}}  & \bf{100--300\tablenotemark{d}}\\ 
A45 & 2011/09/24, N12E42  & M3.0, 19:09--19:41 & 972 & 2, Y & $<$127, $>$13\tablenotemark{b,f}  & 50--100 \\
A46 & 2011/09/25, S27W67 &  M4.4, 02:27--02:37 & 613 & N, N  &$<$118, $>$13\tablenotemark{b,f} & 100--300\tablenotemark{d} \\
A47 & 2011/09/25, N08E71 &  M7.4, 04:31--05:05 & 788 & N, Y  & $<$180, $>$13\tablenotemark{b,f} & 50--100  \\
A48 & 2011/09/26, N13E33 &  M4.0, 05:06--05:13 &  N & N, N  & $<$27, $>$10\tablenotemark{f} & 100--300\tablenotemark{c} \\
A49 & 2011/09/28, N15W01 &  C9.3, 12:26--12:38 & 562 & N, N & $<$2.5, $>$10\tablenotemark{f} & 100--300\tablenotemark{d}  \\
A50 & 2011/10/01, N20E169 & ?, $\sim$ 20:30 & 1238 & N, Y  & $<$0.2, $<$10 & -- \\
A51 & 2011/10/04, N23E146 & ?, $\sim$ 12:40   &  1101 & N, N  & 12.4, $>$60\tablenotemark{b} & -- \\
A52 & 2011/10/14, N12E133 & ?, $\sim$ 12:00 & 814 & N, N & $<$0.20, $<$10 & -- \\
A53 & 2011/10/20, N19,W91 &  M1.6, 03:10--03:44 &893 & N, N & $<$0.20, $<$10 & 50--100 \\
A54 & 2011/10/22, N29W91 &  M1.3, 10:00--13:09 & 1005 & N, N? & 8.1, $<$60 & 12--25  \\
A55 & 2011/11/03, N08E156 & M4.7--X1.4,  22:41--? &991 & N, Y & 127, $>$60\tablenotemark{a}   & -- \\
A56 & 2011/11/17, N11E102 & ?, $\sim$ 20:15 & 1041 & N, N  & 1.2, $>$60\tablenotemark{b} &-- \\
A57 & 2011/11/26, N11W48 &  C1.2, 06:09--07:56 & 933 & N, N & 48.1, $>$100  & 25--50\tablenotemark{{d,e}} \\
A58 & 2011/12/21, S19E164 & ?, $\sim$ 03:10  & 1064 & N, Y  & $<$0.2, $>$13\tablenotemark{a,f} & -- \\
A59 & 2011/12/25, S22W25 & M4.0, 18:11--18:20 & 366 & 2, Y & 1.6, $>$100 & 100--300\tablenotemark{d}   \\
A60 & 2011/12/27, S16E32 &  C8.9, 04:11--04:31 & 147 &  N, N  & $<$0.2, $<$10 & 100--300 \\
A61 & 2012/01/02, N11W104 &  C2.4, 14:31--16:04 & 1138 & N, Y  & 0.5, $<$60  & 12-25 \\
A62 & 2012/01/12, N20E115 &  C2.5, 07:54--12:16 & 814 & N, N & $<$0.2, $<$10 & 6--12  \\ 
A63 & 2012/01/16, N09E143  &  C6.5, 02:36--06:46 & 1060& N, N & 0.4, $<$60\tablenotemark{b} & 12--25 \\
A64 & 2012/01/19, N32E22&  M3.2, 13:44--17:50 & 1120 & N, Y  & 4.6, $>$60\tablenotemark{b} & 25--50\tablenotemark{d} \\
\color{red} \bf{A65\tablenotemark{**}(9)} &\bf{ 2012/01/23, N33W21} & \bf{M8.7, 03:38--04:34} &\bf{ 2175} & \bf{N, Y} & \bf{3280, $>$100} & \bf{100--300\tablenotemark{{d,e}} }\\
A66 & 2012/01/26, N26W73 & C6.4, 03:58--07:03  & 1194 & N, N & $<$41, $>$10\tablenotemark{f} & 12--25\tablenotemark{d} \\
\color{red} \bf{A67\tablenotemark{**}(10)} &\bf{2012/01/27, N35W81} & \bf{X1.7, 17:37--18:56} & \bf{2508} & \bf{3, Y} &  \bf{518, $>$100} & \bf{100--300\tablenotemark{{d,e}}}  \\
A68 & 2012/02/24, S16W165  & ?, $\sim$03:15 & 800& N, N & 2.3, $<$50 & -- \\
A69 & 2012/03/03, N18E91  & C1.9, 18:13--20:46  & 1078 & N, N  &  0.1, $>$24\tablenotemark{b}  ; $<$0.2, $<$10 & 12--25 \\
\bf{A70} & \bf{2012/03/04, N17E68}  &  \bf{M2.0, 10:29--12:16} & \bf{1306} & \bf{N, Y} & \bf{74.7, $>$60\tablenotemark{b}} & \bf{100--300} \\
\color{red} \bf{A71\tablenotemark{**}(11)} & \bf{2012/03/05, N16E54} & \bf{X1.1, 02:30--04:43}  & \bf{1531} & \bf{N, Y} & \bf{$<$33, $>$13\tablenotemark{b,f}} & \bf{100--300\tablenotemark{{d,e}} }\\
\color{red} \bf{A72\tablenotemark{**}(12)} & \bf{2012/03/07, N17E27} & \bf{X5.4, 00:02--00:40} &\bf{2684} & \bf{2?, Y}  &  \bf{1800, $>$100} & \bf{$>$1000\tablenotemark{g}}  \\
  &    & \bf{M7, 01:05--01:23} & \bf{1825} &\bf{2?, Y} & \bf{1800, $>$100}    & \bf{$>$1000\tablenotemark{g}} \\
\color{red} \bf{A73\tablenotemark{**}(13)} & \bf{2012/03/09,  N16W0} &  \bf{M6.3, 03:22--04:18} & \bf{950} &\bf{2, Y} &\bf{$<$528, $>$10\tablenotemark{f}} & \bf{100--300}  \\
\color{red} \bf{A74\tablenotemark{**}(14)} &\bf{2012/03/10, N18W26} &  \bf{M8.4, 17:15--18:30} & \bf{1296} & \bf{N?, Y}  &  \bf{$<$115, $>$10\tablenotemark{f}} & \bf{100--300\tablenotemark{d}}  \\
\bf{A75} & \bf{2012/03/13, N19W59} &  \bf{M7.9, 17:12--18:25}  &\bf{1884} & \bf{3?, Y} & \bf{271, $>$100} &\bf{300--1000\tablenotemark{d}}\\ 
A76 & 2012/03/16, N20W104 & ?, $\sim$ 20:30 & 862 &  N, N & $<$0.5, $>$10\tablenotemark{f}  & -- \\
A77 & 2012/03/18, N20W110 & ?,  $<$00:24 & 1210 & N, Y  & $<$0.2, $<$10 & --  \\
A78 & 2012/03/21, N21W152 & ?, $\sim$07:20 & 1178 & N, Y  & 38.5, $>$60\tablenotemark{a} & -- \\
A79 &  2012/03/24, N12E154  & ?, $\sim$ 00:00    & 1152 & N, Y & 71, $>$60\tablenotemark{a} & -- \\
A80 & 2012/03/26, N18E123 & M8.2--X2.5, 22:16--? & 1390 & N, Y & $<$45, $>$13\tablenotemark{{b,f}} & -- \\
A81 & 2012/03/27, N21W17 & C5.3, 02:50--03:22 & 1148 & N, N & $<$45, $>$13\tablenotemark{{b,f}} & 12-25\tablenotemark{d} \\
A82 & 2012/03/28, N19E106 & ?, $\sim$ 01:25 & 1033 & N, Y& $<$26, $>$13\tablenotemark{{b,f}} & -- \\
A83 & 2012/04/05, N18W29 & C1.5, 20:49--21:57 & 828 & N, N & 0.2, $<$50& 50-100\tablenotemark{d}  \\
A84 & 2012/04/07, N13W152 & ?, $\sim$16:15 & 765 & N, Y &  3.2, $>$60\tablenotemark{a} & -- \\
A85 & 2012/04/09, N20W67 & C3.9, 12:12--13:08 &  921 & 1, Y  & $<$0.2, $<$10 & 12--25\tablenotemark{c}  \\
A86 & 2012/04/15, N10E108 &  C1.7, 02:16--02:45 & 1220 & 1, Y & 0.2, $<$60\tablenotemark{b} & 12--25\tablenotemark{c} \\
A87 & 2012/04/16, N12E91 & M1.7, 17:24--18:00  & 1348 & N, N& $<$0.2, $<$10 & 25-50\tablenotemark{c}\\
A88 & 2012/04/18, S26W31 & C5.9, 14:51--15:18 & 840 &  N, N & 0.05, $<$35\tablenotemark{a}; $<$0.2, $<$10 & 25--50\tablenotemark{d} \\
A89 & 2012/04/30, S16W91 & C3.9, 06:56--08:19,  & 992 &  1, N &$<$0.2, $<$10 & 25--50  \\
A90 & 2012/05/09, N13E31 & M4.7, 12:21--12:36 & N\tablenotemark{h} & N, N  &$<$0.2, $<$10& 100--300\tablenotemark{e} \\
A91 & 2012/05/09, N08E19 & M4.1, 21:01--21:09 &N\tablenotemark{h} & N, N &$<$0.2, $<$10 & 100--300\tablenotemark{d}   \\
A92 & 2012/05/10, N12E23 & M5.7, 04:11--04:23 & N\tablenotemark{h} & N, N & $<$0.2, $<$10& 100--300\tablenotemark{c} \\
A93 & 2012/05/11, N05W13 & C3.2, 23:02--00:33 & 805 & N, N & $<$0.2, $<$10 & 12-25 \\
\color{red}\bf{A94\tablenotemark{**}(14)} & \bf{2012/05/17, N05W77} & \bf{M5.1, 01:25--02:14} & \bf{1582 }& \bf{3, Y }& \bf{180,  $>$100} & \bf{100--300\tablenotemark{c}} \\ 
A95 & 2012/05/26, N12W118 & ?, $\sim$ 20:35 & 1966 & 1, Y & 8.5, $<$60 & -- \\ 
\enddata
\tablenotetext{*}{1, 2, 3 $\simeq$ $<$50, 50--500, $>$500 $\times 10^{-22}$ W m$^{-2}$ Hz$^{-1}$}
\tablenotetext{\color{red}{**}}{ Fermi/LAT sustained-emission event}
\tablenotetext{a}{{\it STEREO A}}
\tablenotetext{b}{{\it STEREO B}}
\tablenotetext{c}{{\it RHESSI}}
\tablenotetext{d}{{\it Fermi}/GBM}
\tablenotetext{e}{Missing Data} 
\tablenotetext{f}{Preceding SEP} 
\tablenotetext{g}{\it INTEGRAL}
\tablenotetext{h}{No LASCO data; no clear associated CME in {\it STEREO A/B}}

\end{deluxetable}

Of the 95 events presented in Table \ref{tab:full} 70 are associated with broad CMEs $\gtrsim$ 800 km s$^{-1}$, 38 with measured peak SEP proton fluxes $>$1 pfu, and 38 with flare X-rays $>$100 keV; these are the selection criteria used in the study.   In addition, 58 of the 95 are on the visible disk.  We discuss results from a study of the 95 events in $\S$\ref{subsec:conditions}.  One of the findings is that all of the SGRE events appear to be accompanied by flares with hard X-ray emission $>$100 keV and fast CMEs.  However, there are five events meeting these criteria that were not SGRE events.  Below, we discuss the characteristics of these events and reasons why SGRE might not have been detected.

\subsection{Characteristics of the Five Events Identified in the Study with no Detectable SGRE}  \label{subsec:6nolat}

If the dual requirements, $\gtrsim$800 km s$^{-1}$ CME speed and impulsive $>$100 keV hard X-ray emission, are physically related to the production of SGRE, we need to explain the five events (black numbers with bold print: A15, A24, A44, A70, A75) in Table \ref{tab:full}; events, meeting these requirements that did not exhibit SGRE. The first question to address is whether there is any difference between the average characteristics of the 14 LAT events (red numbers with bold print) with sustained emission and these five events.  We use median values to compare these characteristics because of the small number of events in the samples.  Events associated with SGRE had a higher median {\it GOES} X-ray class (X1.1 vs. M3.5), comparable median CME speeds (1582 vs. 1407 km s$^{-1}$) and peak SEP proton fluxes (61 vs 75 pfu), and median heliographic longitudes closer to disk center (47$^{\circ}$ vs 68$^{\circ}$) than events with no SGRE.  The higher {\it GOES} class for SGRE events might be just a manifestation of the Big Flare Syndrome.  

The higher heliographic longitude of the five events with no detectable SGRE can be explained by detectability of $\gamma$ rays near the solar limb.  The transmission of $>$100 MeV $\gamma$ rays from the lower chromosphere and photosphere, through the overlying solar atmosphere decreases from about 95\% at 45$^{\circ}$ to 82\% at 70$^{\circ}$ and to 15\% at 89$^{\circ}$ relative to disk center.   In addition, if the SGRE does not all come from the flare site, but is distributed over tens of degrees in heliocentric angle, as suggested in $\S$\ref{subsubsec:location} and shown to be the case for two behind-the-limb events \citep{acke17}, then events at large heliocentric angles should be weaker due to attenuation and/or missed emission from behind the solar limb. 

The lack of a LAT detection of $>$100 MeV emission in the five events might also just be due to the limited duty cycle for good LAT solar exposures: 20--40 minutes every 90 or 180 minutes.  LAT would have the highest probability of missing a short 30-minute, transient than it would one that lasted several hours.  Below we plot and discuss the LAT, GBM, and {\it RHESSI} observations of the five events to determine whether these duty-cycle considerations could have played a role in the failure to detect SGRE.  

\vspace{5mm}

{\bf SOL2011-02-24T07:23, Event A15} 

{\bf Why might the SGRE have been missed?}   LAT had good solar exposure that ended just before the impulsive phase and no $>$100 MeV emission was detected at that time.   Any sustained emission with duration less than one hour following the impulsive phase would not have been detected because the next solar exposure began at 08:35 UT.  The flare was near the solar limb (E84), thus some of the emission might have been beyond the solar limb or attenuated by at least a factor of two in the overlying atmosphere.
 
\begin{figure}
\epsscale{.80}
\plotone{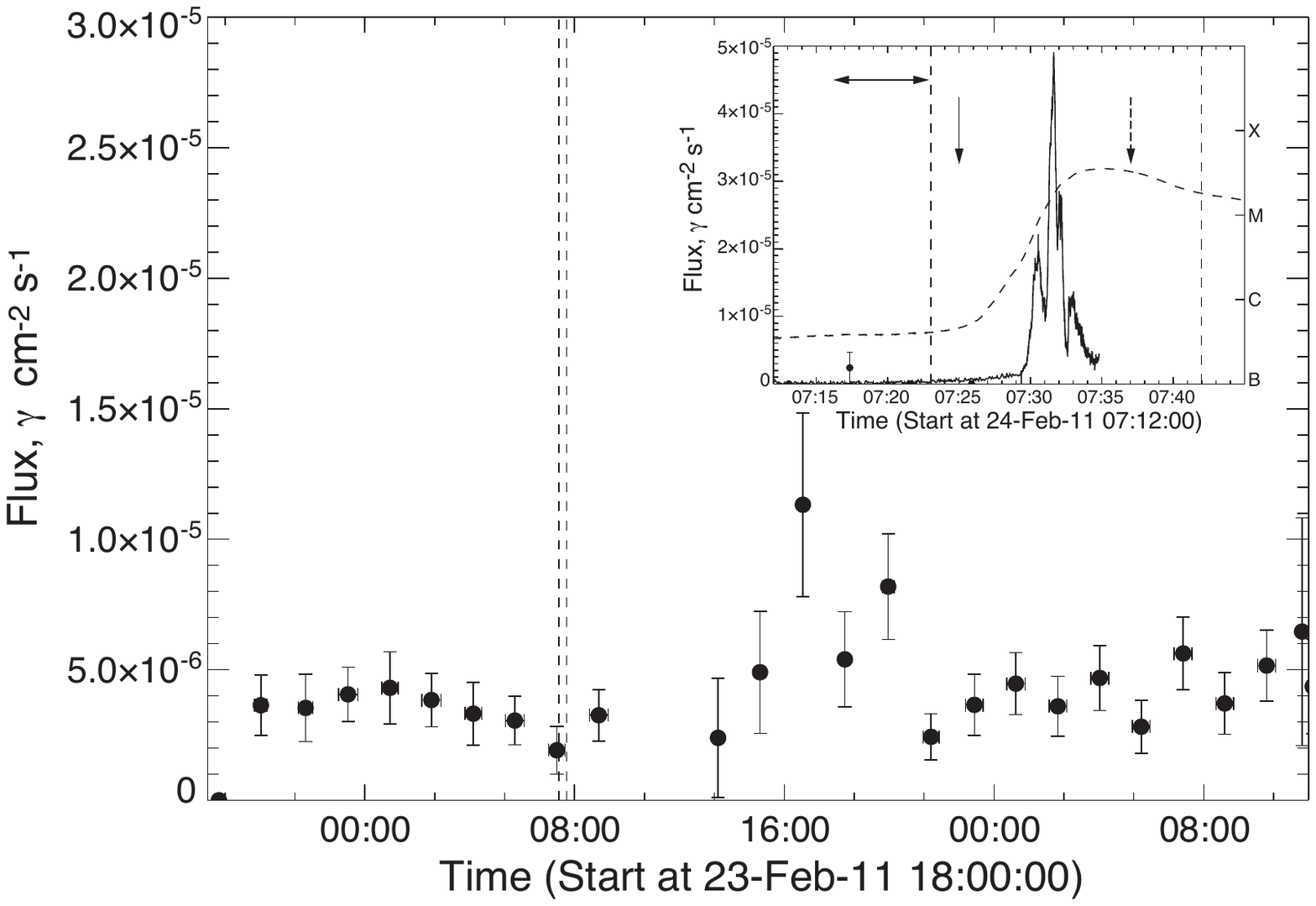}
\caption{Time history of 2011 February 24 event with no SGRE detected by LAT.  Main figure displays the time history of $>$100 MeV flux (data points and uncertainties) accumulated in 10 to 40 minute LAT exposures $\leq$10$^{\circ}$ of the Sun using source-class data.  Vertical dashed lines show the {\it GOES} start and end times. The inset shows an expanded plot around the impulsive phase of the flare with LAT source class data plotted at 4-minute resolution.  The solid trace is the GBM 100-300 keV count rate scaled to the LAT flux.  The vertical dashed lines show the {\it GOES} start and end times and the dashed curves show the {\it GOES} 1--8{\AA} time history; the {\it GOES} scale on right ordinate.  The $<->$ symbol shows the range in CME onset times in the CDAW catalog derived for linear and quadratic extrapolations.  The solid vertical arrow shows our estimate of the CME onset time estimated from {\it SDO}/AIA movies and the dashed vertical arrow shows the onset of Type II radio emission.}
\label{110224th}
\end{figure}

{\bf Details}  Event A15 is plotted in Figure \ref{110224th}.  M3.5 class flare at E84 (limb flare) lasting $\sim$20 minutes;  $\sim$1200 km s$^{-1}$ CME with onset $\sim$5 minutes before the 100--300 keV X-ray onset; M (metric) and DH (decameter-hectometric) Type II emissions observed with the M onset about 12 minutes after the CME onset;  marginal evidence for solar energetic protons; impulsive $\gamma$-ray emission up to 1 MeV was observed by GBM (100--300 keV rates plotted in inset) and {\it RHESSI} with no evidence for 2.223 MeV neutron-capture line, as is expected for such a location near the limb.  Good LAT solar exposure 07:05 -- 07:35 UT, that overlaps the early part of the impulsive phase; however, LAT solar exposure ends just before the rise of the hard X-rays.  No evidence for $>$100 MeV emission 08:35 -- 09:10 UT, one hour after impulsive flare.  No additional LAT solar exposure until 14 UT.  Both 2.223 MeV line and $>$100 MeV emissions would be strongly absorbed if produced at flare site.

\vspace{5mm}

{\bf SOL2011-05-29T21:04, Event A24}

\begin{figure}
\epsscale{.80}
\plotone{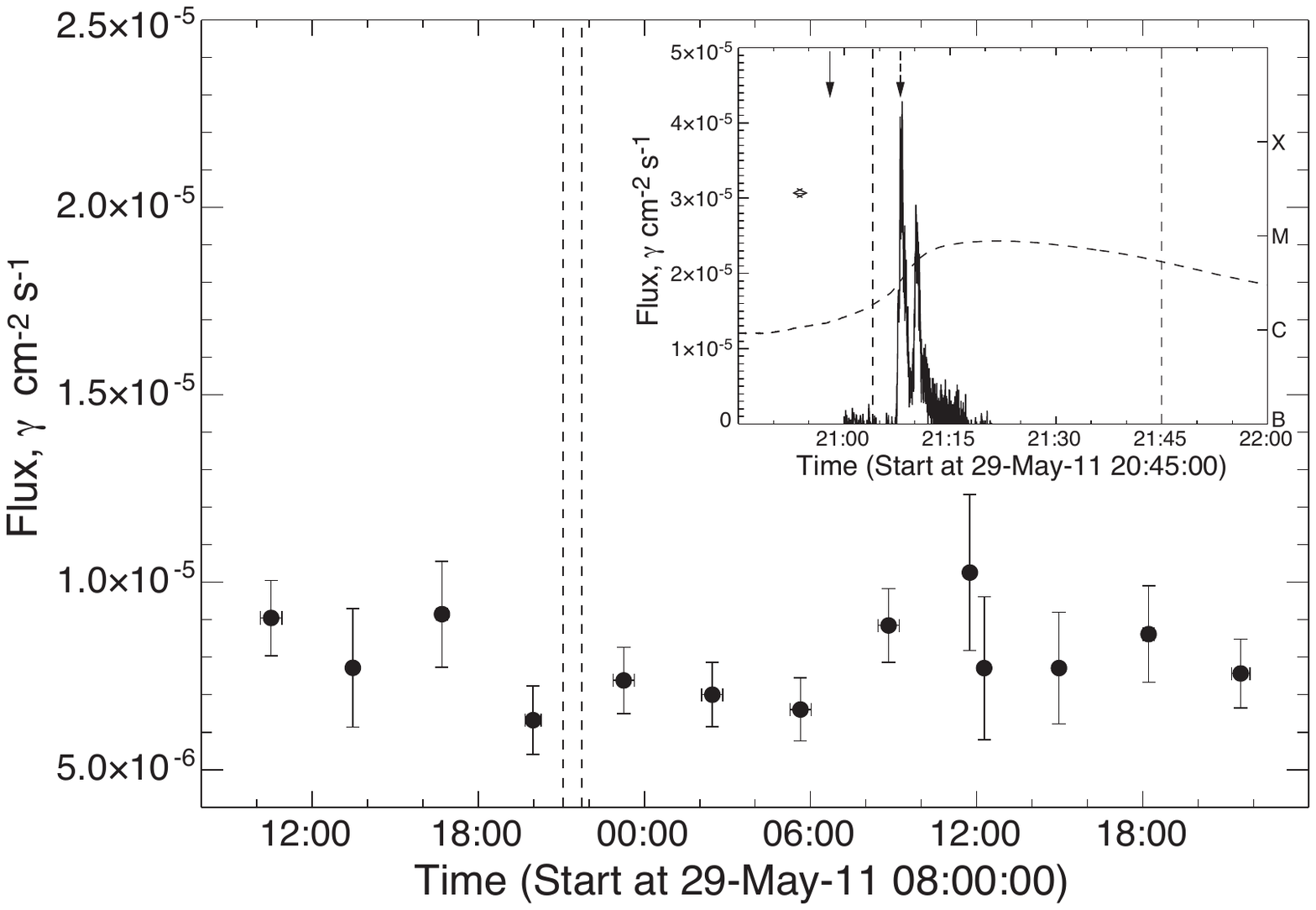}
\caption{Time history of the 2011 May 29 event with no SGRE detected by LAT.  Arbitrarily scaled GBM 100-300 keV rates are shown by solid trace in the inset.  Otherwise the same caption as Figure \ref{110224th}}
\label{110529th}
\end{figure}

{\bf Why might the SGRE have been missed?}  LAT had poor solar exposure to the impulsive phase of the flare and the first good {\it Fermi} solar exposure was at 22:54 UT (see Figure \ref{110529th}; therefore we have no information about the presence of $>$100 MeV emission from about 21:04--22:54 UT, for about two hours after the impulsive peak.  The active region was also relatively close to the solar limb (E75), with consequences for attenuation and missed radiation behind the solar limb if the sustained $\gamma$ rays were broadly distributed.  We also note that there was only a weak (6 pfu) SEP observed by magnetically well-connected {\it STEREO B}.
  
{\bf Details}  Event A24 is plotted in Figure \ref{110529th}.  C8.7 class flare at E75 lasting $\sim$39 minutes;  $\sim$1400 km s$^{-1}$ CME with estimated onset within minutes of the soft X-ray rise; GBM entered sunlight at 21:07 revealing 100--300 keV emission plotted in inset; M (metric) and DH (decameter-hectometric) Type II emissions observed with the M onset $\sim$10 minutes after the CME launch; small flux of solar energetic protons up to energies $>$40 MeV at {\it STEREO B} for which the flare site was well connected.  Poor LAT solar exposure 21:14--21:40 UT after the hard X-ray peak with no evidence for $>$100 MeV emission.  Good LAT exposure 22:54--23:36 UT but no evidence for $>$100 MeV emission.  Cannot rule out $>$100 MeV emission up until 22:54 UT.

\vspace{5mm}

{\bf SOL2011-09-24T12:33, Event A44}

{\bf Why might the SGRE have been missed?}  LAT had good solar exposure at the beginning of the impulsive flare between  12:37--13:00 UT and poor exposure near the end between 14:10--14:30 UT.   There is no evidence for sustained emission in both source class and solar-impulsive-class data in these time intervals.  The next good solar exposure was an hour later and no SGRE was observed. LAT would have missed any SGRE during the flare from 13:10--14:10 UT and for about an hour after the flare.

\begin{figure}
\epsscale{.80}
\plotone{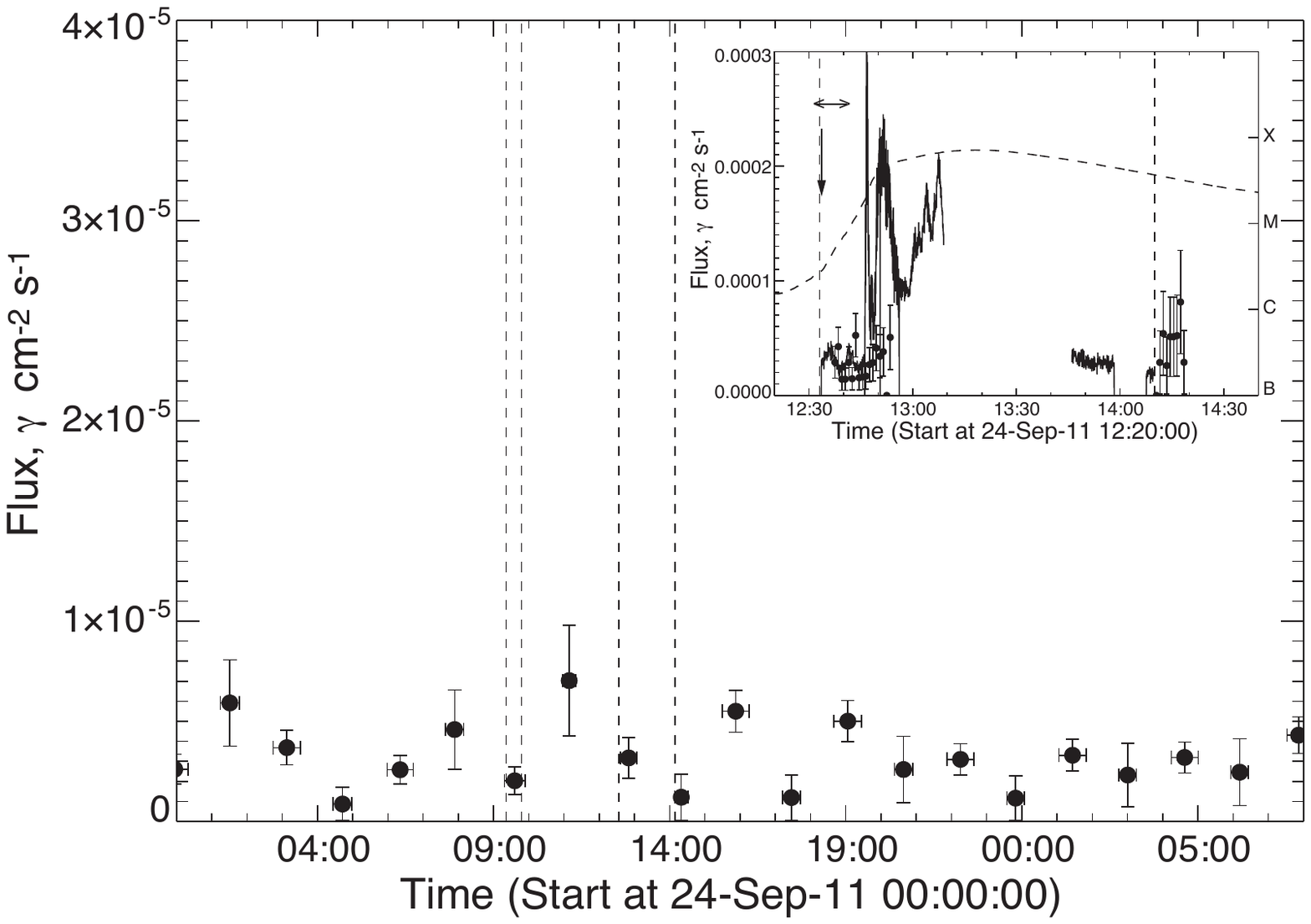}
\caption{Time history of LAT $>$ 100 MeV fluxes on 2011 September 24.  SGRE was observed during the first flare beginning at 09:21 UT (see Appendix \ref{subsec:20110307}). No SGRE was detected by LAT during and after the flare beginning at 12:33 UT.  The inset plots GBM 100--300 keV count rates on the same scale as $>$100 MeV fluxes derived from solar-impulsive-class data.   Otherwise the same caption as Figure \ref{110224th}}
\label{110924ath}
\end{figure} 

{\bf Details}  Event A44 is plotted in Figure \ref{110924ath}.  M7.1 class flare at E61 lasting $\sim$37 minutes; the CME speed was  $\sim$1936 km s$^{-1}$ with estimated onset about 14 minutes before the first intense 100--300 keV X-ray peak; only DH Type II emission;  only an upper limit on the SEP proton flux was obtained because of the large event on September 22.   GBM observed emission up to 100--300 keV from the beginning of the flare until nighttime at 13:05 (time history plotted in the inset).  LAT had good solar exposure 12:37--13:00 UT, poor exposure 14:10--14:30 UT, and good exposure 15:30--16:10 UT.  There is no evidence for $>$100 MeV emission in the source-class data in any of these intervals.  The source-class data in the first two exposures were compromised because high ACE rates.   We therefore plot the solar-impulsive-class $>$100 MeV fluxes during these time periods in the inset.   There is no evidence for SGRE in any of the exposures.

\vspace{5mm}

{\bf SOL2012-03-04T10:29, Event A70}

{\bf Why might the SGRE have been missed?}  LAT had limited exposure during the impulsive flare, ending near the rise of the impulsive X-ray peak and there is no evidence for impulsive emission in our plots of source-class and LLE data. The ability to search for $>$100 MeV emission following the impulsive phase was compromised by the shortened solar exposure split by an SAA passage beginning about an hour after the impulsive phase, and again during the next orbit.  


{\bf Details}  Event A70 is plotted in Figure \ref{120305th} in Appendix \ref{subsec:20120305} where both the full plot and inset used Pass8 source-class data in hopes of improving sensitivity and detecting weak SGRE; there was no significant improvement over Pass7.  M2.0 class flare at W68 lasting $\sim$107 minutes; $\sim$1300 km s$^{-1}$ CME with onset coincident with GBM 50-100 keV X-ray rise (plotted in inset in lieu of the 100--300 keV band); no Type II radio emissions;  moderate solar energetic proton event with emission observed $>$60 MeV; impulsive hard X-ray emission up to 100-300 keV ({\it RHESSI}, GBM) with no evidence for the 2.223 MeV neutron-capture line.  LAT peak solar exposure from 10:40--11:10 UT only 50\% of the exposure of the adjacent orbits. Exposure very small at the time of the 100--300 keV impulsive peak at 11:05 UT.   No reported impulsive $>$100 MeV \citep{acke14} or observed in our study of the Pass8 solar-impulsive-class data.  LAT exposures in the next orbits truncated by SAAs: 12:10--12:20, 12:40--12:50 UT; 13:50--13:57 UT causing large uncertainties in the flux.  No clear evidence for SGRE during these exposures.

\vspace{5mm}

 {\bf SOL2012-03-13T17:12, Event A75} 
 
 {\bf Why might the SGRE have been missed?}   This event was the most likely of the five to have produced sustained emission at energies $>$100 MeV following the flare because of its association with a 1900 km s$^{-1}$ CME and 271 pfu SEP event.  Unfortunately, the first good LAT solar exposure occurred about 90 minutes after the impulsive phase.  Thus, it is possible that any SGRE lasted less than 90 minutes.  There is an alternative explanation, assuming that the particles responsible for the SGRE are produced along with SEPs.  Because of different shock and magnetic field geometries \citep{roui12}), it is possible that the protons on field lines reaching the Sun may not have been accelerated to energies sufficient to produce pions even though the protons observed in space reached energies $>$100 MeV.  This same process might also explain why $>$300 MeV protons reached the Sun during the 2011 March 7 LAT event (number 1 in Table \ref{tab:latlist}, A16 in Table \ref{tab:full} while few SEPs were observed in space with energies $>$100 MeV. 
 
 \begin{figure}
\epsscale{.80}
\plotone{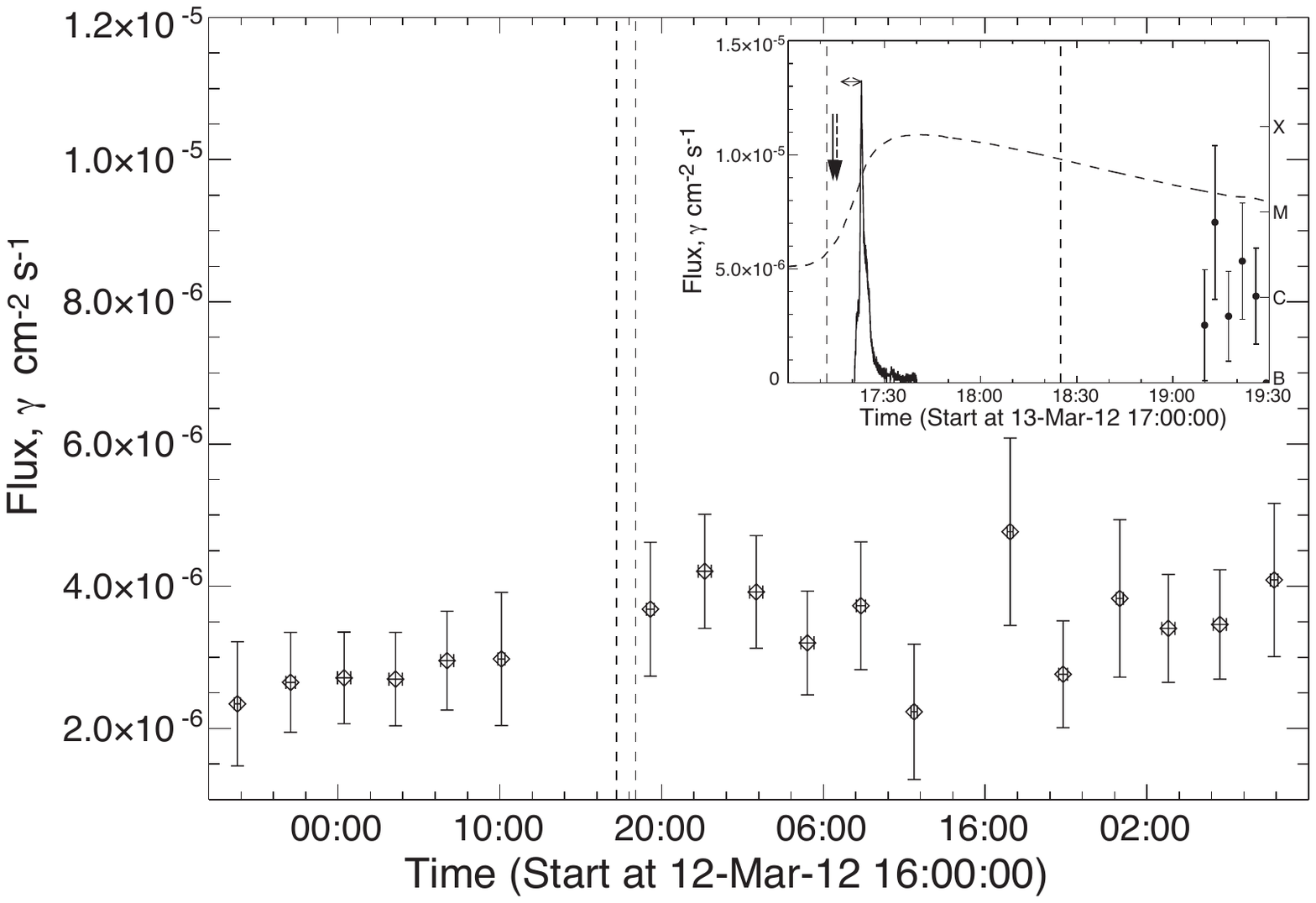}
\caption{Time history of the 2012 March 13 event with no SGRE detected by LAT.   Otherwise same caption as Figure \ref{110224th}}
\label{120313th}
\end{figure} 
 
{\bf Details}  Event A75 is plotted in Figure \ref{120313th}.  M7.9 class flare at W59 lasting $\sim$73 minutes;  $\sim$1600 km s$^{-1}$ CME with estimated onset $\sim$2 minutes after the {\it GOES} start time and six minutes before GBM hard X-ray data become available; M (metric) and DH (decameter-hectometric) Type II emissions observed with the M onset within one minute of the CME launch; strong SEP event with emission $>$100 MeV that appeared to steepen at higher energy because it was not observed in the {\it GOES} HEPAD data; impulsive hard X-ray emission observed up to 1 MeV by GBM (100--300 keV time history plotted in inset).  No LAT solar exposure during the flare.  Good LAT solar exposures only made every other orbit: 19:08--19:30, 22:17--23:00, and 01:28--02:12. No evidence for SGRE at a flux level in excess of 3 $\times 10^{-6} \gamma$ cm$^{-2}$ s$^{-1}$ about 90 minutes after the flare and in the ensuing hours.   However, it is possible that there was time-extended $>$100 MeV emission during the 90 minutes between the time of the hard X-ray peak and first LAT observation at 19:08 UT.  GBM spectrum during the impulsive hard X-ray peak can be fit by bremsstrahlung from the sum of two power-law populations of electrons $>$200 keV: one containing about 2.6 $\times 10^{35}$ electrons and having an index of 5.3 and the second containing about 1.4 $\times 10^{34}$ electrons having an index of 3.2.  The 5.3 index of the soft component is the same as derived by \citet{kauf13} in their paper discussing the discovery of 30 THz emission that has a time history somewhat broader than observed in the hard X-rays.  \citet{kauf13} point out that the softer spectrum of electrons cannot penetrate to depths where the THz emission can be produced.  The fact that there may be a harder spectrum of electrons that is suggested by our fits may provide a source that could penetrate to the depths needed.  However, the time history of the 30 THz emission is significantly more extended in time than the $\gamma$-ray emission observed by GBM.  There is weak evidence (95\% statistical confidence) for the 2.223 MeV neutron capture line in GBM impulsive-phase spectrum, but this could be due to the strong background line.  There is also structure in the spectrum that \citet{trot15a} attribute to nuclear lines, but the features do not appear at the energies of the expected nuclear lines .

\section{Details of the Thirty SGRE Events} \label{sec:append}

In this Appendix we plot and discuss the time histories of the 30 SGRE events observed by {\it Fermi} LAT from 2008 to 2016 and listed in Table \ref{tab:latlist}. These plots lead us to conclude that the sustained emission is temporally distinct from the accompanying impulsive flare in at least 67\% of the events.  For each event, we also provide details of the LAT, GBM, and RHESSI observations, as well as a brief summary of related solar and heliospheric measurements.  

The main plot accompanying each event shows the time history of $>$100 MeV flux from hours before to hours after the associated solar eruption, including the entire duration of any sustained emission.   The plotted fluxes, derived from source-class data, are averages integrated over the LAT solar exposures and the $\pm1\sigma$ uncertainties are statistical.  The solar exposures were made either each orbit or every other orbit, depending the viewing geometries of the spacecraft's two rocking positions.  There are times when LAT solar observations can be made every orbit, e.g., when the orbital procession provides comparable exposures for both rocking positions, as a targeted observation for a few hours after an Autonomous Repoint Request (ARR) following a cosmic gamma-ray/solar flare trigger from GBM, or during a Target of Opportunity (ToO) when the favorable rocking position for solar observations is maintained for days due to high solar activity.  The plotted vertical dashed lines are the {\it GOES} soft X-ray start and stop times.  

The insets in the figures either show blowups of the {\it GOES} flare region (vertical dashed lines) or later solar exposures exhibiting temporal variation of the $>$100 MeV emission.  The dashed curves within the flare interval depict the {\it GOES} 1--8{\AA} X-ray flux histories with its logarithmic scale on the right ordinate.  The horizontal arrow heads designate the range in CME onset times given in the CDAW catalog estimated from linear and quadratic extrapolations to the surface of the Sun (1 $R_{\odot}$).  We have also visually studied movies of SDO 171{\AA}, 193{\AA}, and 211{\AA} images to provide an alternate estimates of the CME onset times.  These times are based on motions of field lines from which the CMEs were released and are shown by the vertical arrows; these times are subjective and are accurate to about two minutes.  The onsets derived using the two methods are not always consistent.  There is an alternative method for estimating the CME onset that uses the {\it GOES} soft X-ray data \citep{zhan01}; these times are are in good agreement with what we found using the {\it SDO} data.   The dashed vertical arrow depicts the onset of Type II radio emission.  The solid traces typically plot arbitrarily scaled rates in the 100-300 keV band from {\it RHESSI} or GBM.  When they are not available, we plot a representative hard X-ray time history.  Where LAT source-class, LLE data, or solar-impulsive-class data are available during the flare, we plot the $>$100 MeV fluxes accumulated over either one or four minute intervals; the uncertainties are statistical.

\subsection{SOL2011-03-07T19:43} \label{subsec:20110307}

\begin{figure}
\epsscale{0.8}
\plotone{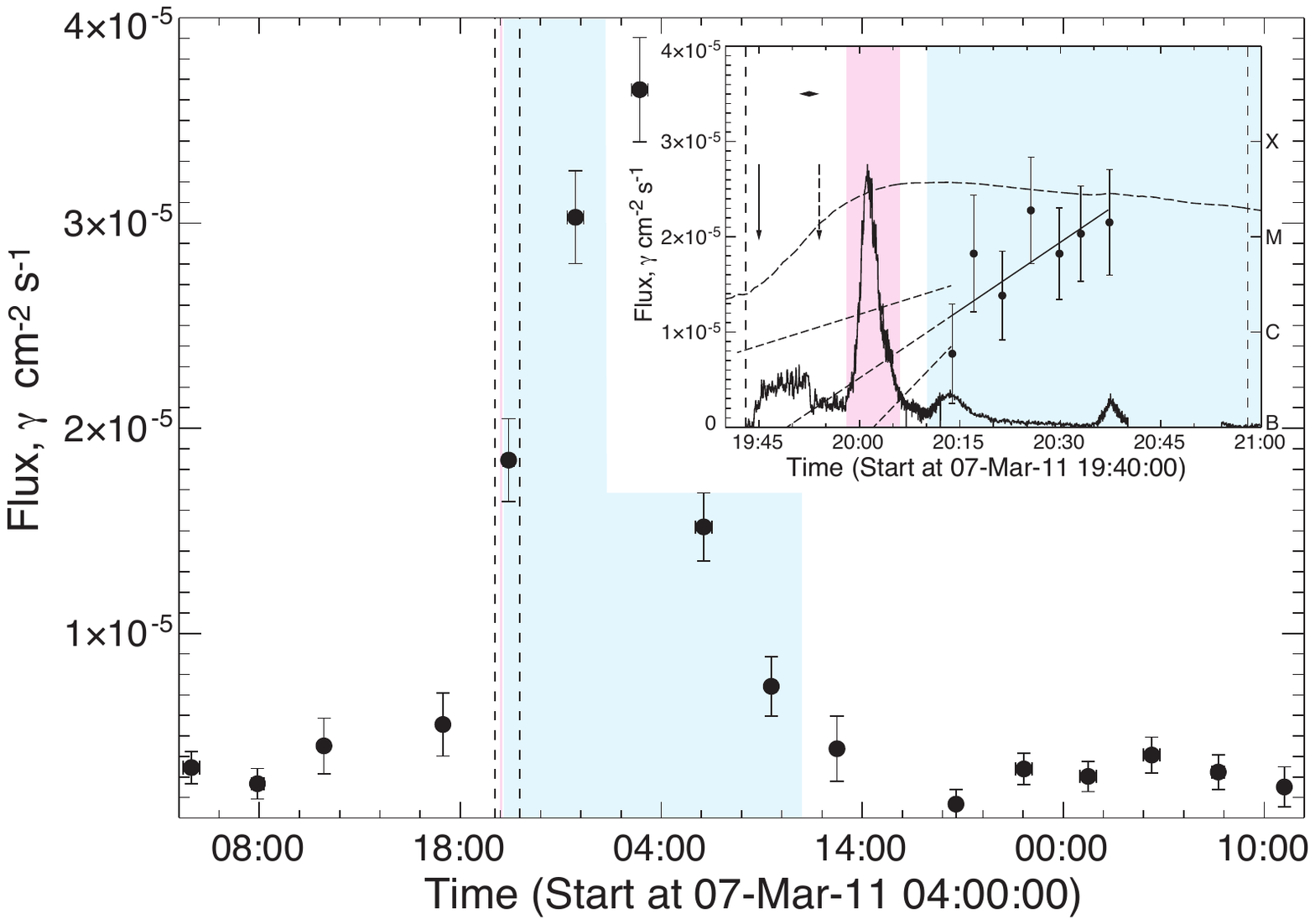}
\caption{The main figure shows the time history of the $>$100 MeV flux from $\leq$10$^{\circ}$ of the Sun, derived from source-class data, revealing the 2011 March 7 SGRE event.  The fluxes were averaged over the $\sim$20-40 minute solar exposures and the uncertainties are $\pm1\sigma$ statistical errors.  Vertical dashed lines show the {\it GOES} start and end times. The inset shows 4-minute accumulation LAT $>$100 MeV fluxes derived from source-class data. The best fit to the onset of the $>$100 MeV emission is shown by the solid line and its extrapolation to zero flux is shown by the dashed line; the other two dashed lines are extrapolations of $\pm1\sigma$ deviations from the best fit. The combined 100-300 keV count rate observed by {\it RHESSI} and GBM during the impulsive flare, scaled to the $\gamma$-ray flux, is shown by the solid trace. The dashed curve shows the {\it GOES} 1--8{\AA} time history (scale on right ordinate) and the $<->$ symbol shows the range in CME onset times in the CDAW catalog\footnote{\url{https://cdaw.gsfc.nasa.gov/CME_list/index.html}} derived for linear and quadratic extrapolations.  The vertical solid arrow depicts our estimate of the CME onset from inspection of {\it SDO}/AIA images and the vertical dashed arrow shows the estimated onset of Type II radio emission.  The solid time history is the arbitrarily scaled 100-300 keV count rates from {\it RHESSI} and GBM. The blue-shaded region depicts our estimate of the duration of the sustained emission.  The pink-shaded regions depict where we made estimates of the flux of $>$100 MeV impulsive-phase $\gamma$-ray emission. The dotted trace shows the high-energy $\gamma$-ray timer history of the 1991 June 11 flare normalized to the first LAT 4-minute accumulation point at 20:14 UT.  We do not show this curve in time histories of other LAT events.}
\label{110307Ath}
\end{figure}
{\bf Is the SGRE time history distinct from that of the impulsive flare? YES.}  The $>$100 MeV flux measured $\sim$10 minutes after flare's hard X-ray peak appears to rise (90\% confidence). 

{\bf Details of the observation}
Plotted in Figure \ref{110307Ath}.  This was the first sustained emission event detected by the LAT.  Additional details in $\S$\ref{sec:timehist} and \citet{acke14}).  M3.7 class flare at W47 lasting 15 minutes; $\sim$2100 km s$^{-1}$ CME with onset from {\it SDO}/AIA 171{\AA} and 211{\AA} images consistent with the rise in 100-300 keV X-rays and $\sim$15 minutes before its peak; M (metric) and DH (decameter-hectometric) Type II emissions observed with the M onset about nine minutes after the CME launch;  moderate SEP event with proton energies barely exceeding 100 MeV; impulsive $\gamma$-ray emission up to 1 MeV with no evidence for 2.223 MeV neutron-capture line (limit on line flux is used to estimate a limit on number of $>$500 MeV protons during impulsive phase of flare).  Note that the CME onset time derived from AIA is about seven minutes earlier than the CDAW onset times.  In plotting the 100-300 keV time history, we used {\it RHESSI} data from the onset of the flare until 20:08, when the satellite entered the radiation belts of the South Atlantic Anomaly (SAA) and normalized GBM 100--300 keV fluxes after $\sim$20:02 UT when {\it Fermi} entered sunlight. 

There was good LAT solar exposure between 20:10--20:40 UT following the impulsive phase;  $>$100 MeV emission appears to increase (90\% confidence) after about 20:14 UT and continues rising to a peak flux of $\sim 3.5 \times 10^{-5} \gamma$ cm$^{-2}$ s$^{-1}$ near 03:00 UT (consistent with \citet{acke14}) and was observable until about 13:00 UT on March 8.  The fluxes plotted in the figure and inset used source-class data; we confirmed the fluxes measured just after the flare using solar-impulsive-class data.  The celestial background level $\sim 3 \times 10^{-6} \gamma$ cm$^{-2}$ s$^{-1}$ in March was low and is about 50\% higher than the quiescent solar flux from cosmic-ray interactions \citep{abdo11}.  From our extrapolations, the SGRE onset appears to have occurred before 20:05 UT.  Best fitting pion-decay spectra suggest that the $>$300 MeV proton spectra softens during the duration of the event (Table \ref{tab:event}).  Number of protons $>$500 MeV in the sustained emission phase exceeds the number in the impulsive phase, determined by the upper limit on the 2.223 MeV line flux, by at least a factor of ten, but it appears to be comparable to the upper limit on the number of $>$500 MeV protons observed in space  (Table \ref{tab:event}).  This result strongly contrasts with other events where the numbers of protons in space appear to be at least an order of magnitude higher than the number at the Sun.  It is interesting that the peak SEP flux is about a factor of 10 smaller than what we might expect for such a fast CME which was  preceded by a broad CME a few hours earlier \citep{gopa04}.  In addition the SEP proton energies only reach 100 MeV, significantly below the threshold for producing the pion-decay photons observed by LAT.  See $\S$\ref{subsec:details} for additional details.


.

\subsection{SOL2011-06-02T07:22} \label{subsec:20110602}

{\bf  Is the SGRE time history distinct from that of the impulsive flare? Uncertain.}  $>$100 MeV $\gamma$ rays were only weakly observed in one spit exposure two hours after the flare; there was no evidence for flux variability.


\begin{figure}
\epsscale{.80}
\plotone{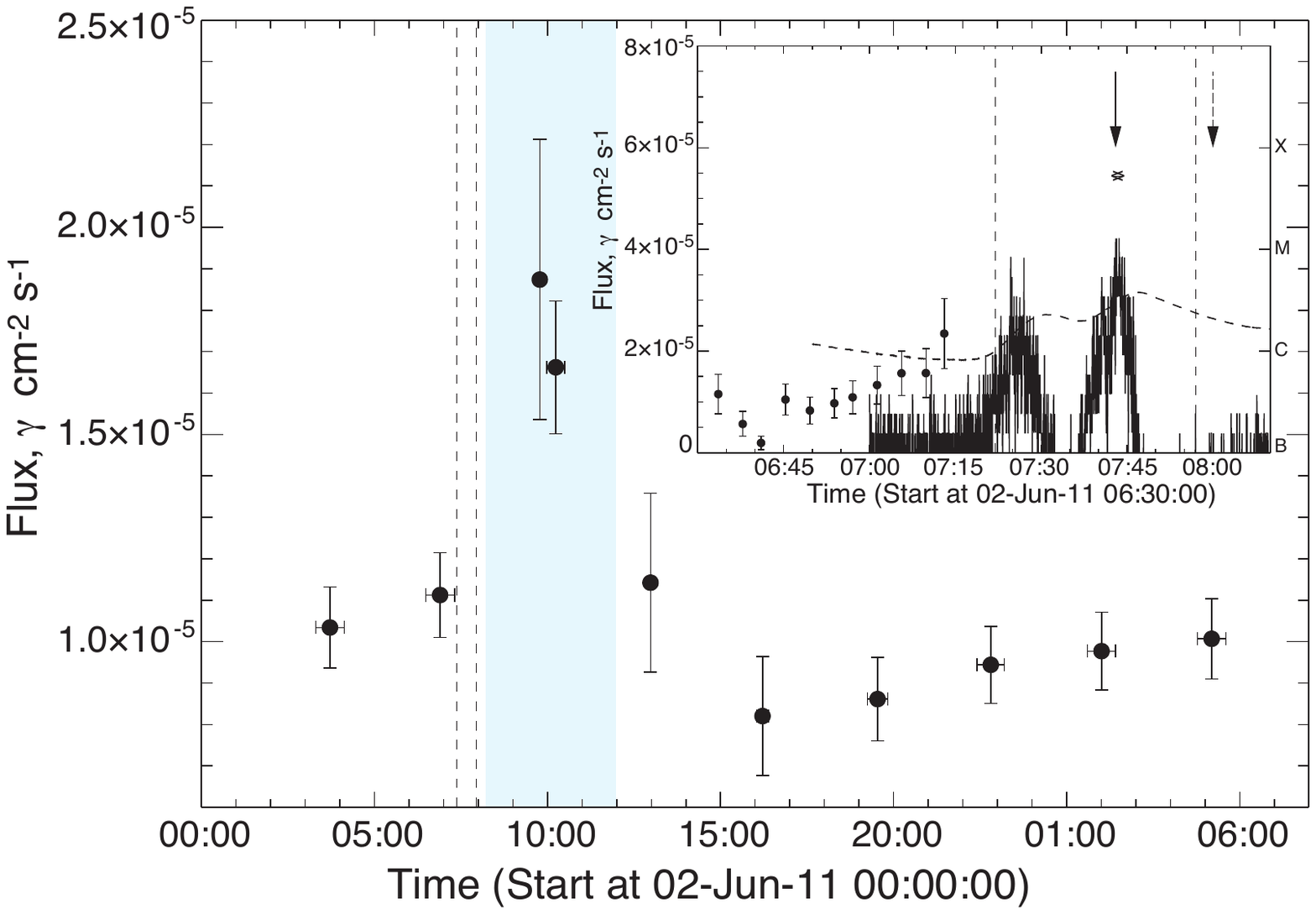}
\caption{Time profile of the 2011 June 2 SGRE event observed by LAT.  The inset shows a blowup of the region around the flare.  $>$100 MeV fluxes at 4-minute resolution with $\pm1\sigma$ statistical errors are shown just before the flare. As neither {\it RHESSI} nor GBM had significant exposure to the flare, we plot an arbitrarily scaled derivative of the {\it GOES} 1--8{\AA} power that is a proxy for the hard X-ray time history. Plotted by the dashed downward arrow is the DH Type II onset time; there were no metric Type II observations.  See caption for Figure \ref{110307Ath} for more details.}
\label{110602th}
\end{figure} 

{\bf Details} Plotted in Figure \ref{110529th}. C3.7 class flare at E22 lasting $\sim$35 minutes.;  $\sim$1000 km s$^{-1}$ CME with CDAW projected onset at $\sim$07:43 UT consistent with measurements of {\it SDO}/AIA movies, and a sharp peak in soft X-ray emission; only DH (decameter-hectometric) Type II emission;  barely detectable flux of solar energetic protons observed up to 35 MeV; GBM did not observe the flare and {\it RHESSI} only observed the flare up until 07:36 UT and thus missed the peak of the soft X-ray emission and time of CME onset.  This is confirmed by plotting the derivative of the soft X-ray emission that is a proxy for the hard X-ray emission.  The derivative shows two peaks, one from about 07:20--07:33 UT and another one from 07:33--07:48 UT.  The latter one peaks at about 07:43 UT, the time of the CME onset.  It is likely then that any high-energy impulsive hard X-ray emission occurred during the second peak, where there were no available {\it RHESSI} observations.  This is a weak LAT event only identified at just over 3$\sigma$ significance in the Pass8 source-class data, but not in our earlier search of Pass7 data. It was also listed as an event in \citet{acke14}. Weak $>$100 MeV emission was only observed in the first LAT exposure from 09:40 to 10:25 UT, two hours after the impulsive phase.  There is no evidence for time variability and thus no information on when the SGRE began.  The flare would not have been identified as having the potential for $>$100 MeV emission because no hard X-ray measurements were available during most of the flare.  Our estimate of the number of $>$500 MeV protons at the Sun in Table \ref{tab:event} was made by assuming that the emission began just after the impulsive phase, peaked at 10 UT, and fell to zero by 12 UT.  We could not estimate the number of protons in the impulsive phase.

\subsection{SOL2011-06-07T06:16} \label{subsec:20110607}

{\bf Is the SGRE time history distinct from that of the impulsive flare? Uncertain.} $>$100 MeV $\gamma$-rays were only observed during one $\sim$30-minute exposure one hour after the flare; there was no evidence for flux variability.  As discussed below, the upper limit on the number of protons in the impulsive phase was a factor of five smaller than observed in the sustained phase.  Therefore, it is unlikely that the SGRE was the tail of the flare emission.

\begin{figure}
\epsscale{.80}
\plotone{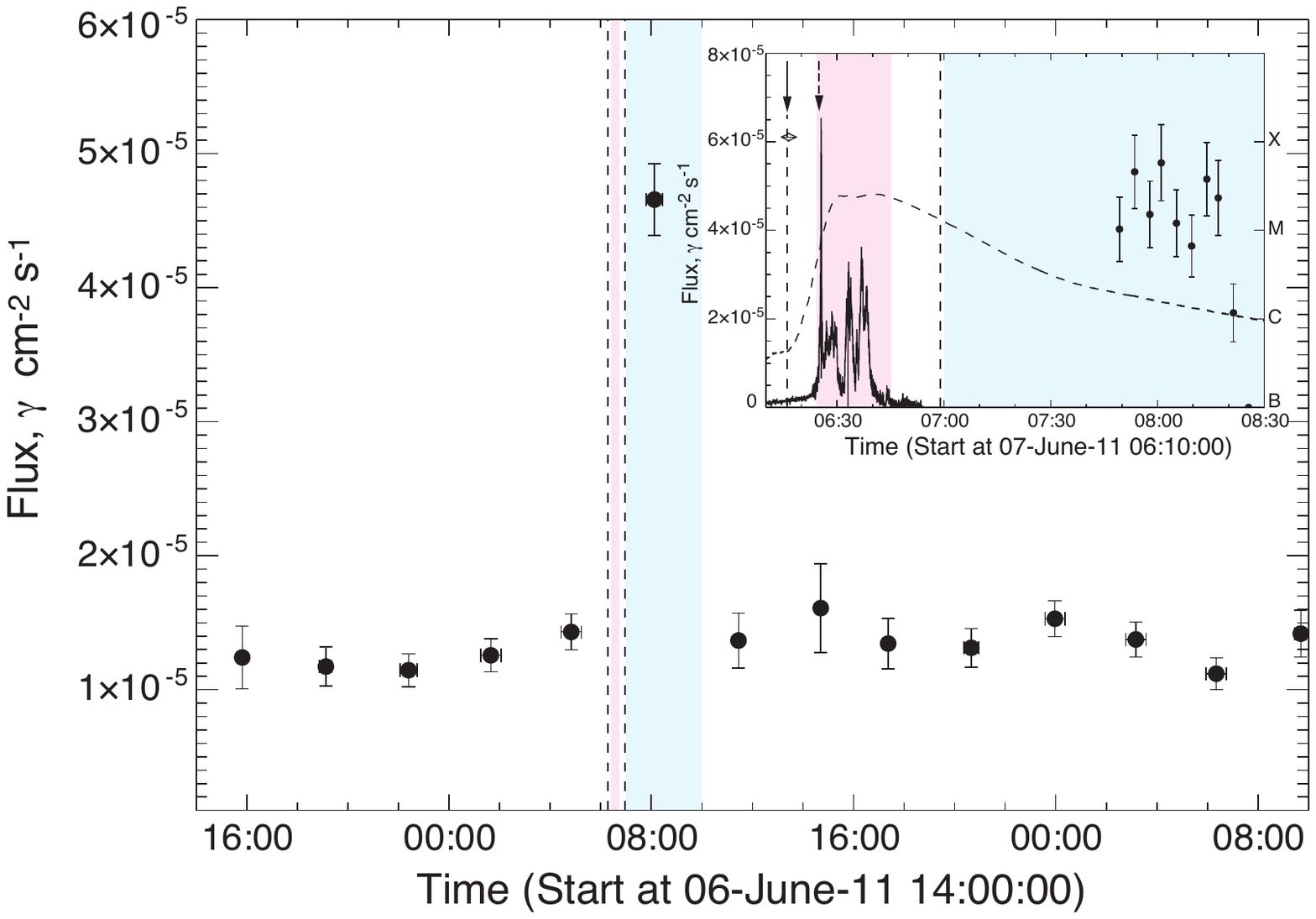}
\caption{Time history of the 2011 June 7 SGRE event observed by LAT observed about an hour after the flare.   The inset shows 4-minute accumulation LAT $>$100 MeV fluxes with $\pm1\sigma$ statistical errors derived from source-class data.  The solid line plots the arbitrarily scaled 100-300 keV count rates observed by GBM during the impulsive flare.  See caption for Figure \ref{110307Ath} for more details.}
\label{110607th}
\end{figure}

{\bf Details}  Plotted in Figure \ref{110607th} (additional details in \citet{acke14}).  M2.5 class flare at W54 lasting $\sim$40 minutes;  $\sim$1200 km s$^{-1}$ CME with onset $\sim$7 minutes before the 100-300 keV X-ray onset; M (metric) and DH (decameter-hectometric) Type II emissions observed with the M onset $\sim$9 minutes after the CME launch;  moderate solar energetic proton event with emission observed $>$100 MeV; impulsive hard X-ray emission up to 300--800 keV observed by both GBM (100--300 keV time history plotted in inset) and {\it RHESSI} with no evidence for the 2.223 MeV neutron-capture line (limit on line flux is used to estimate a limit on number of $>$500 MeV protons during impulsive phase of flare).  There was no LAT solar exposure during the impulsive phase;  the first good LAT solar exposure flare was 07:48--08:22 UT, an hour after the flare.    There  is a significant flux of $>$100 MeV emission during this interval with no evidence for temporal variability; the peak flux in Table \ref{tab:event} is consistent with that reported by \citet{acke14} in their Table 2.  During this time interval prominence material was observed to fall back to the solar surface.  There was no evidence for $>$100 MeV during the next LAT exposure three hours later.   The best fitting pion-decay spectra suggest that the $>$300 MeV proton followed a power law with an index of $\sim$4.5 (Table \ref{tab:event}); this is consistent with that measured by \citet{acke14}.  As LAT had no exposure to the impulsive phase, we used upper limits on the 2.223 neutron-capture line observed in the front {\it RHESSI} detectors to estimate the number of impulsive protons. The number of protons $>$500 MeV in the sustained emission phase exceeded the number in the impulsive phase by at least a factor of five but it is only $\sim$4\% of the number of $>$500 MeV protons observed in space (Table \ref{tab:event}).  The SGRE lasted no more than about three hours; in contrast, the $>$100 MeV proton event at Earth lasted close to two days.

\subsection{SOL2011-08-04T03:41} \label{subsec:20110804}

{\bf Is the SGRE time history distinct from that of the impulsive flare? Uncertain.} The LAT observation began one hour after the impulsive phase and showed no clear temporal variation, and no emission was observed in the next exposure three hours later.  As discussed below, the upper limit on the number of protons in the impulsive phase was a factor of ten smaller than observed in the sustained phase.  Therefore, it is unlikely that the SGRE was the tail of the flare emission.

\begin{figure}
\epsscale{.80}
\plotone{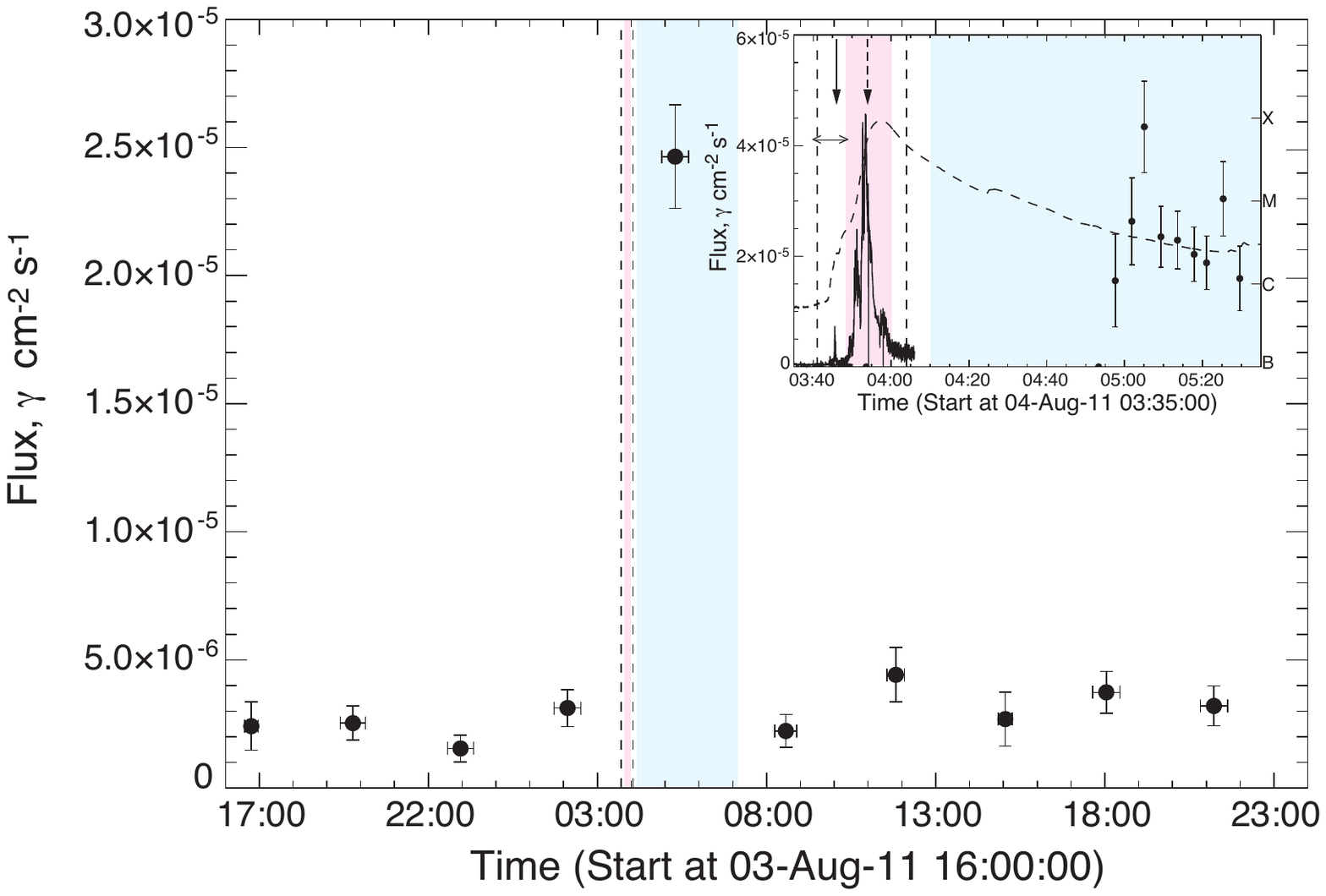}
\caption{Time history of the 2011 August 4 SGRE event observed by LAT about an hour after the flare, plotted at 4-minute resolution in the inset.  Arbitrarily scaled GBM 100-300 keV rates are plotted during the flare.  See caption for Figure \ref{110307Ath} for more details. }
\label{110804th}
\end{figure}

{\bf Details}  Plotted in Figure \ref{110804th}.  M9.3 class flare at W46 lasting $\sim$25 minutes.;  $\sim$1300 km s$^{-1}$ CME with estimated onset $\sim$4 minutes before the 100--300 keV X-ray onset observed by GBM and plotted in inset; M (metric) and DH (decameter-hectometric) Type II emissions observed with the M onset $\sim$8 minutes after the CME launch;  moderate solar energetic proton event with emission observed $>$100 MeV; impulsive hard X-ray emission observed up to $\sim$400 keV by GBM with no evidence for the 2.223 MeV neutron-capture line.  LAT solar exposure between 03:30--03:50 UT during the impulsive phase was poor and \citet{acke14} only classify this as a SGRE event with no impulsive component observable using their LLE data. The recently released Pass8 solar-impulsive class data had only marginal exposure to the rising phase of the hard X-ray emission between 03:43 and 03:49 UT and only upper limits on the $>$100 MeV could be estimated during that interval. The first good LAT solar exposure was between 04:56--05:38 UT, about 90 minutes after the flare.    The measured $>$100 MeV flux during this interval (see Table \ref{tab:event} is consistent with \citet{acke14} with no evidence for temporal variability.  There was no evidence for $>$100 MeV during the next good LAT exposure three hours later.  We estimated a limit on the number of $>$500 MeV protons during the impulsive phase in two ways.  First we obtained an upper limit using the limit on $>$100 MeV emission observed between 03:43--03:49 UT scaled to the entirety of the observed hard X-ray time history.  This gave a limit of  0.4 $\times 10^{28}$ protons $>$500 MeV.  From a limit on the neutron capture line obtained by GBM we obtained a proton limit about four times lower with 95\% confidence.   We find that the number of protons $>$500 MeV in the SGRE phase exceeds the number in the impulsive phase by at least a factor of ten but it is only $\sim$2\% of  the number of $>$500 MeV protons observed in space (Table \ref{tab:event}).  The event at the Sun lasted no more than about four hours; in contrast the $>$100 MeV proton event at Earth lasted close to 20 hours.

\subsection{SOL2011-08-09T07:48} \label{subsec:20110809}

\begin{figure}
\epsscale{1.0}
\plotone{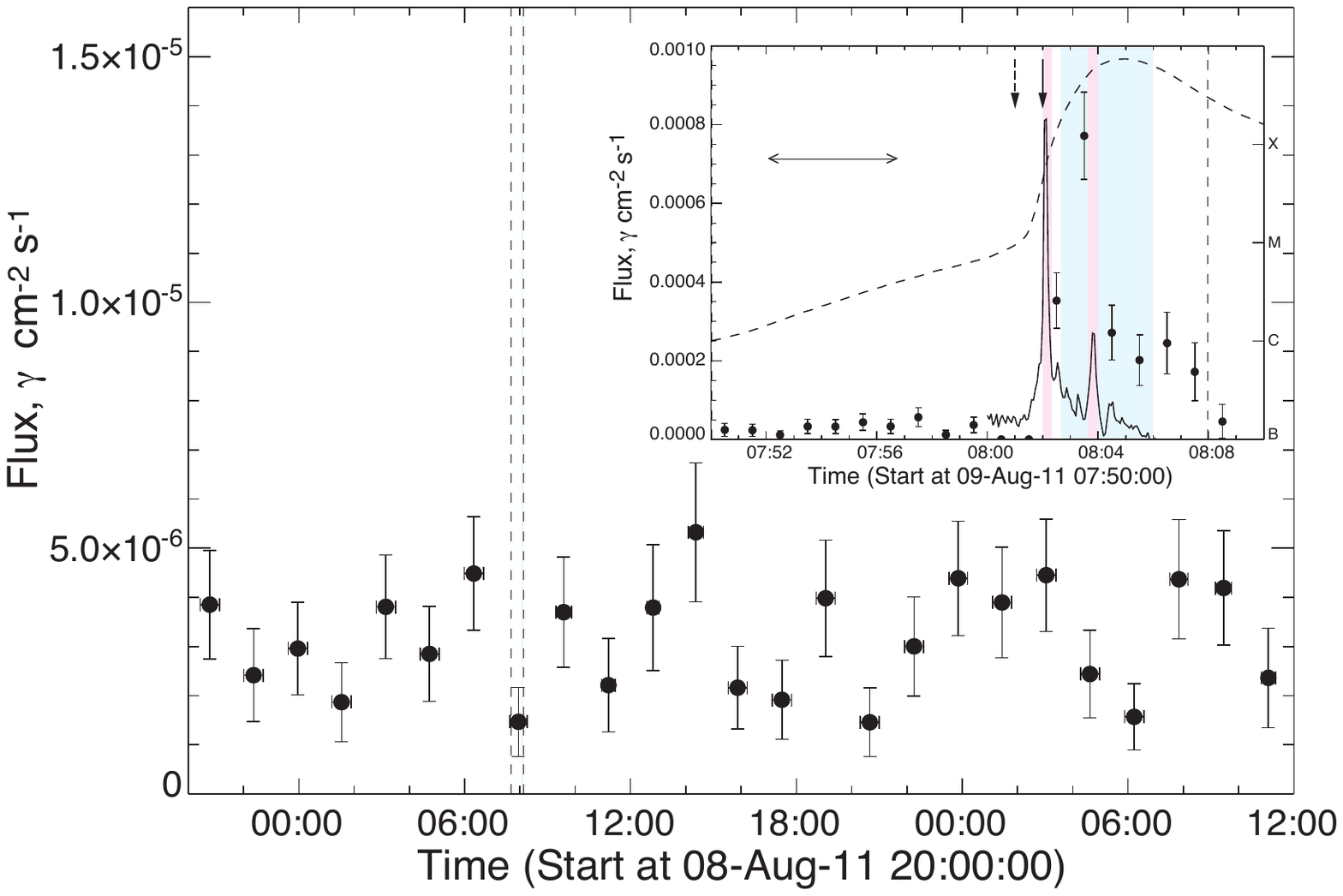}
\caption{Time profile of the 2011 August 9 SGRE event observed by LAT.  Source-class data in the main plot were compromised by high rates in the ACD during the flare resulting in a low flux.  Solar-impulsive-class data were used to plot the $>$100 MeV time history at 1-minute resolution with $\pm1\sigma$ errors in the inset.  The solid trace shows the arbitrarily scaled GBM 100-300 keV rates that differ significantly from the SGRE time history.  See caption for Figure \ref{110307Ath} for more details. }
\label{110809th}
\end{figure}

\begin{figure}
\epsscale{.80}
\plotone{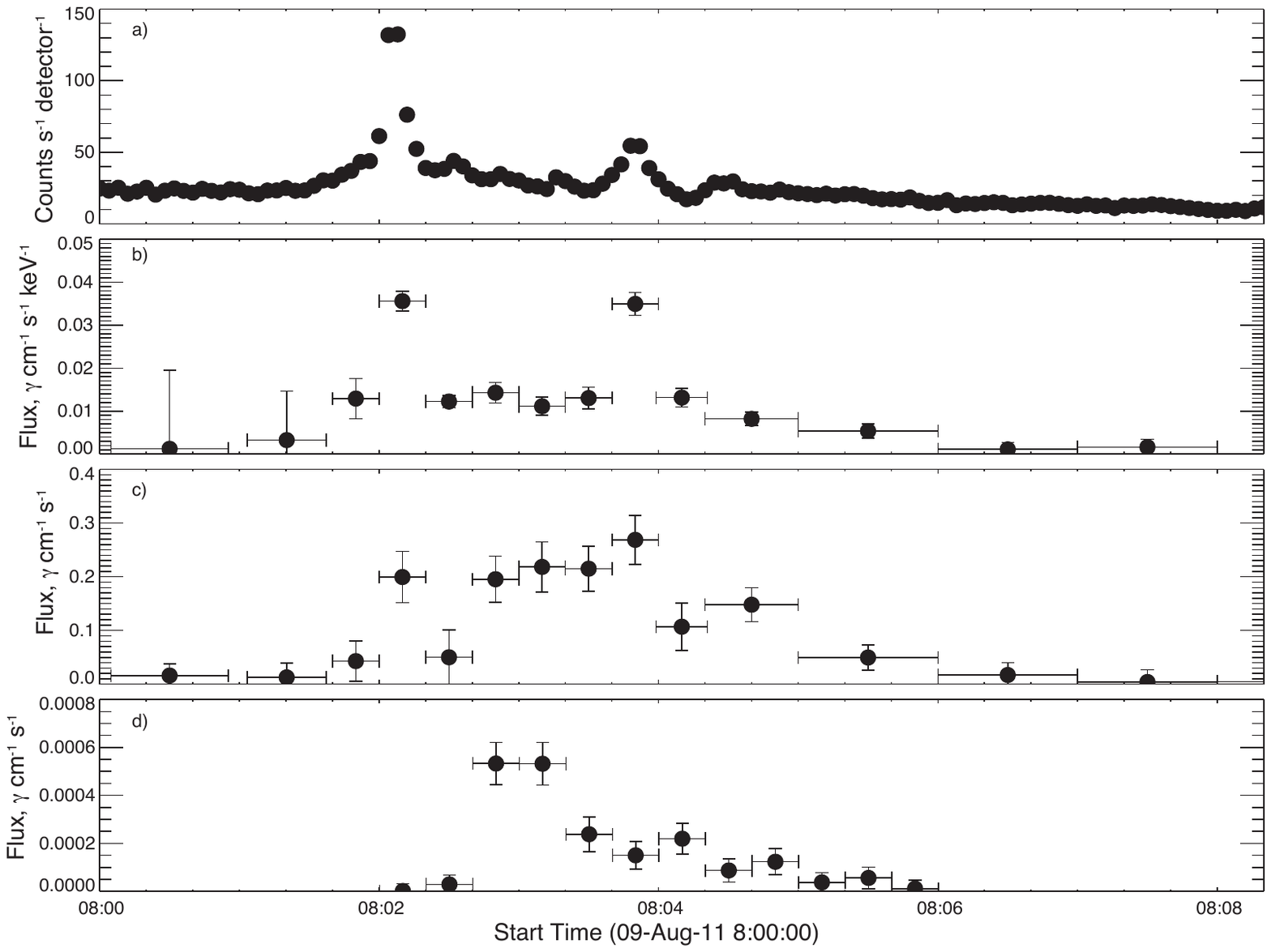}
\caption{Time histories of various hard X-ray and $\gamma$-ray components during the impulsive emission from the 2011 August 9 flare.  a) 100--300 keV X rays observed by {\it RHESSI}, b) flux at 200 keV from power-law electron bremsstrahlung fit to GBM  spectrum $>$200 keV, c) fitted flux of nuclear de-excitation lines in GBM $>$200 keV, and d) flux of $>$100 MeV $\gamma$ rays from fits to LLE data. }
\label{aug9th}
\end{figure}

{\bf Is the SGRE time history distinct from that of the impulsive flare? Yes.}  The event had an impulsive phase dominated by two hard X-ray peaks, with nuclear-line emission detected by GBM and {\it RHESSI}, and $>$100 MeV $\gamma$-ray emission observed by the LAT in both LLE and solar-impulsive-class data. As $>$100 MeV emission was only detected within a minute of the hard X-ray peaks, the event was classified by \citet{acke14} as impulsive.  On closer inspection of the time history, we believe that there are two distinct phases of emission.  This is seen in the inset of Figure \ref{110809th} where the $>$100 MeV flux measured in the solar-impulsive-class data appears to rise just after the hard X-ray peak and reaches a maximum about one minute later.  The $\sim$1-minute high-energy delay from the hard X-ray onset contrasts with the shorter $\sim$10-second  delays observed in the nuclear and $>$100 MeV emission observed in the impulsive 2010 June 12 flare (Event 3 in Table \ref{tab:full}) \citep{acke12a, acke12b}.   The distinct components of the 2011 August 9 flare are revealed in Figure \ref{aug9th} where the impulsive component is reflected in the two peaks at about 08:02:10 and 08:03:50 UT visible in a few hundred keV hard X-rays and in the nuclear emission. The two peaks appear to be of comparable strength, with higher-energy electrons producing the bremsstrahlung (panel b) and tens of MeV protons producing the $\gamma$-ray lines (panel c).  There is an additional temporal component in the $\gamma$-ray line time profile that rose just after the first peak and lasted for about four minutes.  This rise and duration is similar to that observed in the $>$100 MeV flux plotted in panel d), obtained from spectral fits to the LLE data.  It is interesting to note that the CME and Type II radio emission onsets occurred within about one minute of each other and near the time of the first hard X-ray peak, suggesting the formation of a shock deep in the corona that could be responsible for this distinct component of tens to hundreds of MeV protons.  Even though this event lasts only a few minutes, we include this event in our list of SGRE events because of its distinct temporal history.   There is no evidence for SGRE in the two good LAT exposures at 09:20--09:45 and 10:58--11:20 UT following the flare. 

{\bf Details}  Plotted in Figure \ref{110809th}.  X6.9 class flare at W70 lasting $\sim$20 minutes; $\sim$1600 km s$^{-1}$ CME with onset coincident with the rise of the 100--300 keV X-ray emission; M (metric) and DH (decameter-hectometric) Type II emissions observed with the M onset within a minute of the CME launch;  small solar energetic proton event with emission observed $>$100 MeV;  $>$100 MeV $\gamma$-ray emission was observed in LAT solar-impulsive-class data plotted at 1-minute resolution in the inset of the figure beginning just after the prominent hard X-ray peak; GBM spectrum showed clear evidence for the 2.223 MeV neutron-capture line and nuclear de-excitation lines between 08:00--08:08 UT.  The event was classified by \citet{acke14} as impulsive but on closer inspection there are two distinct phases of emission as can be seen in the inset of Figure \ref{110809th}) where the $>$100 MeV flux appears to rise just after the hard X-ray peak and reaches a maximum about one minute later.  This behavior is different than observed in the impulsive 2010 June 12 flare (Event 3 in Table \ref{tab:full}) \citep{acke12a, acke12b}, where the hard X-ray and $\gamma$-ray time histories follow one another closely. 

Comparing the flux observed in nuclear lines with that observed in pion-decay emission from 08:02:40--08:06:00 UT (after removing the contribution from the second impulsive peak), we estimate that the proton spectrum between 5 and 300 MeV was consistent with a power law having an index of 4.3 $\pm$ 0.3.  This compares with a much steeper proton spectrum above 300 MeV, power-law index of 5.8 $\pm$ 0.9 that we measured from LLE spectral fits.  Thus the proton spectrum of the SGRE steepened significantly above a few hundred MeV. There is no evidence for SGRE in the two good LAT exposures between 09:20--09:45 and 10:58--11:20 UT following the flare.  We have also compared the upper limit on $>$100 MeV emission with that observed in nuclear lines during the impulsive phase peaks and conclude that the proton spectrum between 5 and 300 MeV during the flare was steeper than a power law with index of 4.2.  It is interesting that the peak SEP flux is about a factor of 10 smaller than what we might expect for such a fast CME which was  preceded by a broad CME a few hours earlier \citep{gopa04}.

\subsection{SOL2011-09-06T22:12} \label{subsec:20110906}

{\bf Is the SGRE time history distinct from that of the impulsive flare? Yes.} The $>$100 MeV emission was only observed in one $\sim$30-minute exposure during and just after the associated impulsive flare.  Because of possible saturation effects, we used LLE data to obtain the 1-minute $>$100 MeV time history shown in the upper inset of the figure up until 22:34 UT; after this time we used Pass7 source-class data.  The $>$100 MeV time history plotted in the insets reveals two peaks: one delayed by about eight seconds from the peak in 100--300 keV hard X rays, and a second peak with onset within one minute following the $>$100 MeV impulsive peak that reached a maximum at about 22:27 UT and then gradually fell until the end of the LAT solar exposure.  It is similar to what was observed in the 1982 June 3 event, but lasted about twice as long.  

\begin{figure}
\epsscale{.80}
\plotone{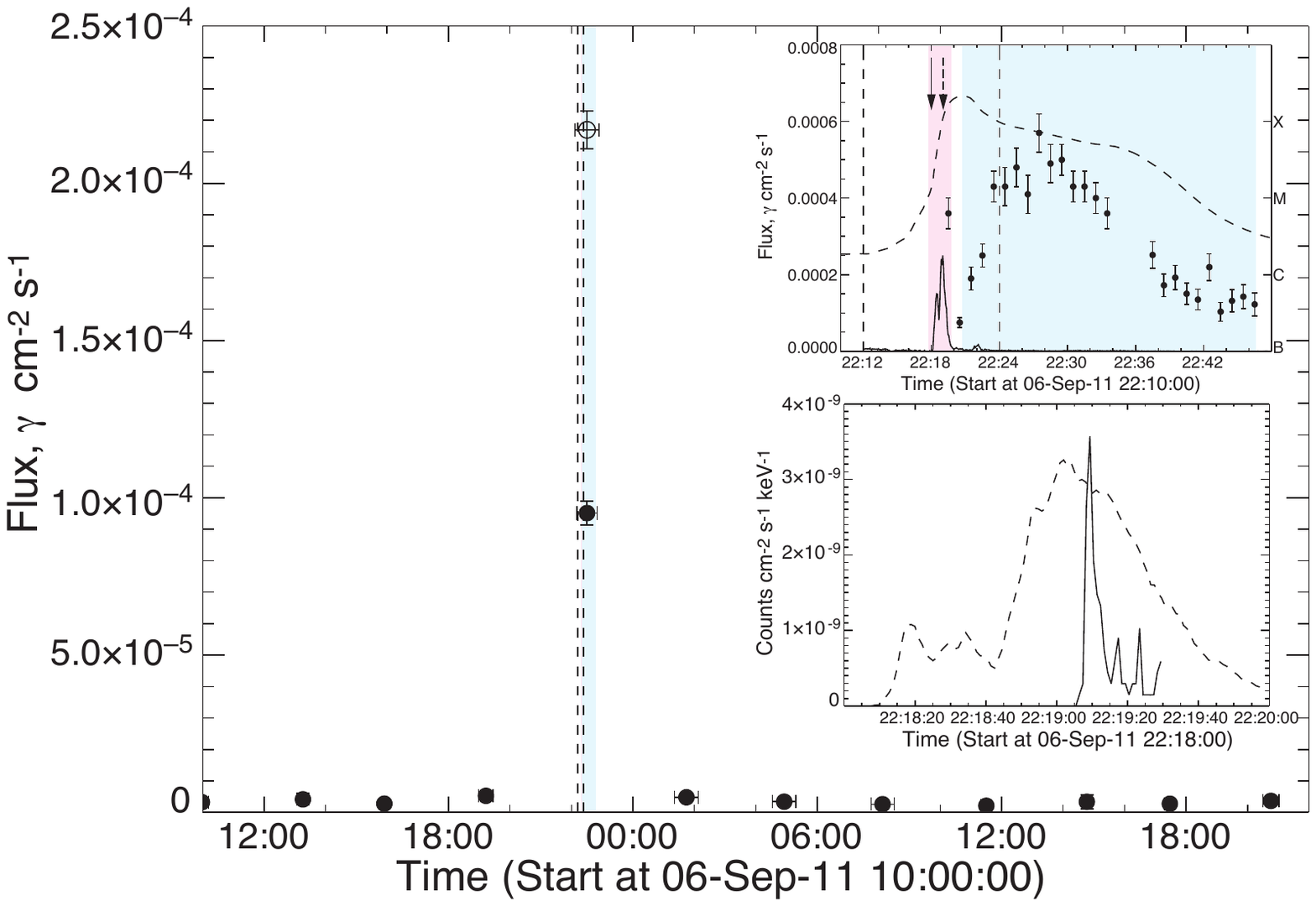}
\caption{Time profile of the 2011 September 6 SGRE event observed by LAT.  Top inset shows the 100--300 keV GBM count rate during the flare along with 1-minute resolution $>$100 MeV $\gamma$-ray fluxes derived from fits to LLE data until 22:33 UT and from source-class data after 22:37 UT.  The bottom inset shows a blowup of the flare region comparing the 100-300 keV rates observed by GBM (dashed trace) with $>$80 MeV fluxes derived from LLE data (solid trace).  The open circle in the main plot is the weighted mean of the $>$100 MeV fluxes plotted in the top inset and derived from solar-impulsive-class data; it is larger than the the source-class data flux, plotted as a filled circle, because of dead time effects in those data.  See caption for Figure \ref{110307Ath} for more details.}
\label{110906th}
\end{figure}

{\bf Details}  Plotted in Figure \ref{110906th}.  X2.1 class flare at W18 lasting $\sim$12 minutes; $\sim$575 km s$^{-1}$ CME in CDAW catalog but listed as $\sim$1000 km s$^{-1}$ for both {\it STEREO A and B} in the CACTUS catalog with onset coincident with rise of 100-300 keV X-rays; only M (metric) Type II emission with onset time within about one minute of CME onset;  small solar energetic proton event but emission observed $>$100 MeV; both {\it RHESSI} and GBM (100--300 keV time history plotted in top inset) observed the impulsive phase of the flare up to energies in excess of 1 MeV with evidence for nuclear-line emission between about 22:18 and 22:20 UT. 

LAT had good solar exposure between 22:12--22:46 UT, covering the entire impulsive phase; there is clear evidence for $>$100 MeV emission in the source-class data, but the measured flux was compromised because of the high rates in the ACD.  We fit the publicly available LLE data for this flare to obtain the 1-minute $>$100 MeV time history plotted in the upper inset up until 22:34 UT; after this time we fit the source-class data. The LLE time history plotted in the top inset reveals a sharp peak during the latter part of the impulsive phase. Studying the LLE emission $>$80 MeV at higher time resolution we find that the peak lasts about 10 seconds is about eight seconds after the peak in the broader 100--300 keV X-ray time history (see bottom inset).  The sustained $>$100 MeV flux plotted in the top inset of Figure \ref{110906th} rose after 20:20 UT, peaking about 22:27 UT, and then falling until the end of the exposure. We have also studied the time history of this event using the recently available solar-impulsive-class data.  The time history of the extended-phase emission agrees well LLE data plotted in the Figure. There is no evidence for $>$100 MeV emission during the next good solar exposure beginning on September 7 at 01:23 UT.  We have fit LAT/LLE SGRE photon spectra after 22:21UT during the rise to and the fall from the peak (Table \ref{tab:event}).  The fitted proton spectrum before the peak is significantly softer (power-law index 5.3) than the spectrum after the peak (power-law index 3.5).  Our fits to the source-class data after 22:37 UT indicate that the proton spectrum also followed a power-law with index of about 3.5.  We've obtained information on the SGRE proton spectrum below 300 MeV by searching for the neutron-capture line in GBM spectrum accumulated between 22:21 and 22:47 UT.   Comparing the 95\% confidence upper limit 2.223 MeV flux with the average flux observed $>$100 MeV by LAT enables us to determine that the proton spectrum from 30--300 MeV was harder than a power-law with index $\sim$4.0 during the rising portion of the emission.  As the measured spectral index above 300 MeV was 5.3 $\pm$ 0.4, the proton spectrum steepened above 300 MeV.

We also obtained spectral information during the impulsive phase of the flare, 22:18--22:20 UT using both GBM and LLE data.  Nuclear line emission was distributed over this full time interval while $>$100 MeV emission was mostly concentrated between 22:19:00 and 22:19:30 UT, as seen by the solid trace plotted in the lower inset with a sharp peak delayed from about eight seconds from a peak in the 100--300 keV X-rays (dashed trace). Our spectral fit to the $>$100 MeV LLE data yields a very soft accelerated proton spectrum $>$300 MeV with power law index steeper than $\sim$6, much softer than observed in the sustained emission phase.  Our fits also rule out a power-law bremsstrahlung origin for the impulsive emission with a confidence of 95\%.  We have fit GBM spectra over the same 22:18--22:20 UT time interval and detected nuclear de-excitation lines and the neutron-capture line.  By comparing the measured pion-decay, neutron-capture line, and de-excitation line fluxes we estimate proton power-law spectral indices from $\sim$4.0 between 5 and 40 MeV and $\sim$4.5 between 40 and 300 MeV.  Thus the impulsive phase proton spectrum steepened with energy above 5 MeV and rapidly above 300 MeV.   

We have estimated the numbers of protons $>$500 MeV during the impulsive and time-extended phases of the event.  We see from Table \ref{tab:event} that the number of impulsive phase protons $>$500 MeV was less than 10\% of the number in the time-extended phase. This fact, along with the vastly different spectra of protons $>$300 MeV indicates that impulsive phase could not have been the major energy source for the SGRE phase.  There is only an upper limit of the number of $>$500 MeV protons in space; it is thirty times larger than the number in the SGRE phase.

\subsection{SOL2011-09-07T22:32} \label{subsec:20110907}

{\bf Is the SGRE time history distinct from that of the impulsive flare? Uncertain.} SGRE was observed in only one LAT exposure about an hour after the impulsive phase and there is no evidence for flux variability during that time interval, suggesting that the measurement may have been made at the peak of the emission.  Thus, we only know that the onset was within one hour of the impulsive phase of the associated flare and that the emission lasted no more than 2.5 hours.  As discussed below, the upper limit on the number of protons in the impulsive phase was within a factor two of the number in the SGRE.  Therefore, it is not possible on energetic grounds to rule out the tail of the flare as the source SGRE.  

\begin{figure}
\epsscale{.80}
\plotone{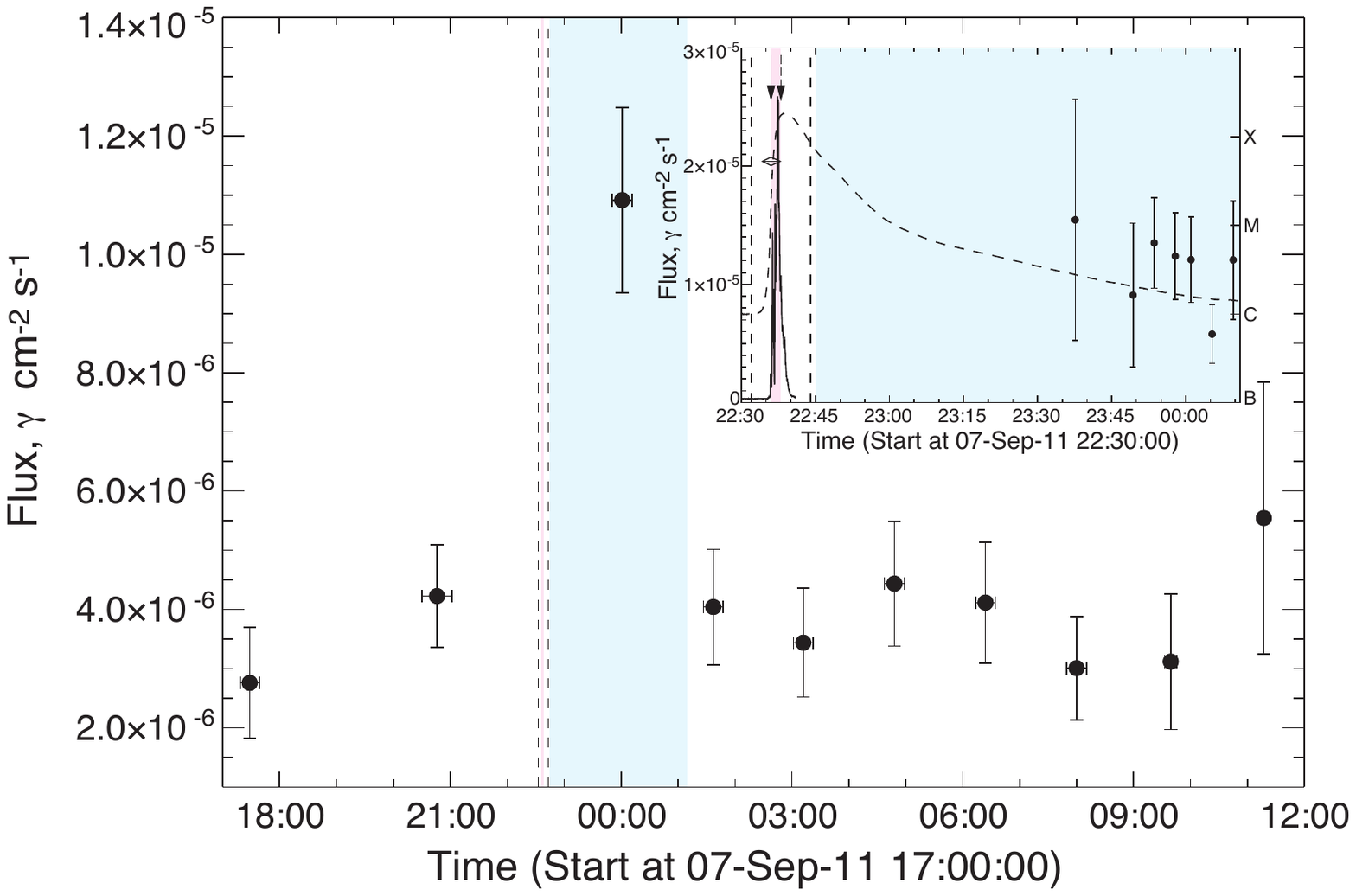}
\caption{Time profile of the 2011 September 7 SGRE event observed by LAT.  The solid trace in the inset is the arbitrarily scaled GBM 100--300 keV rate and the data points with uncertainties are LAT $>$100 MeV fluxes.  See caption for Figure \ref{110307Ath} for more details. }
\label{110907th}
\end{figure}

{\bf Details}  Plotted in Figure \ref{110907th}.  X1.8 class flare at W32 lasting $\sim$12 minutes; $\sim$800 km s$^{-1}$ CME with estimated onset coincident with the 100--300 keV hard X-ray rise; only M (metric) Type II emission observed $\sim$2 minutes after the CME launch; no evidence for solar energetic protons; impulsive hard X-ray emission up to at least 1 MeV observed by GBM (100--300 keV time history plotted in inset).  No LAT solar exposure during the impulsive phase; first good LAT solar exposures at 23:36 (1 minute duration) and 23:52--00:10 UT (4-minute resolution) during a target of opportunity, an hour after the flare. $>$100 MeV flux consistent with \citet{acke14} with no evidence for temporal variability.   There was no evidence for $>$100 MeV during the next LAT exposure 90 minutes later.   There is weak evidence for both nuclear de-excitation and 2.223 MeV line emission observed by GBM during the impulsive phase of the flare.  We obtained a limit on the number of $>$500 MeV protons at the Sun during the flare from the 95\% confidence limit on the 2.223 MeV flux, assuming a proton spectra $>$40 MeV that follows an unbroken power law with an index of 4.5.  The limit on the number of protons during the flare is a factor of two smaller than the number of SGRE protons (Table \ref{tab:event}).  

\subsection{SOL2011-09-24T09:21} \label{subsec:20110924}

{\bf Is the SGRE time history distinct from that of the impulsive flare? Yes.}   LAT had good solar exposure between 09:18--09:45 UT covering the entire flare.  SGRE began about six minutes after the 100--300 keV impulsive hard X-ray peak and lasted about four minutes.

\begin{figure}
\epsscale{.80}
\plotone{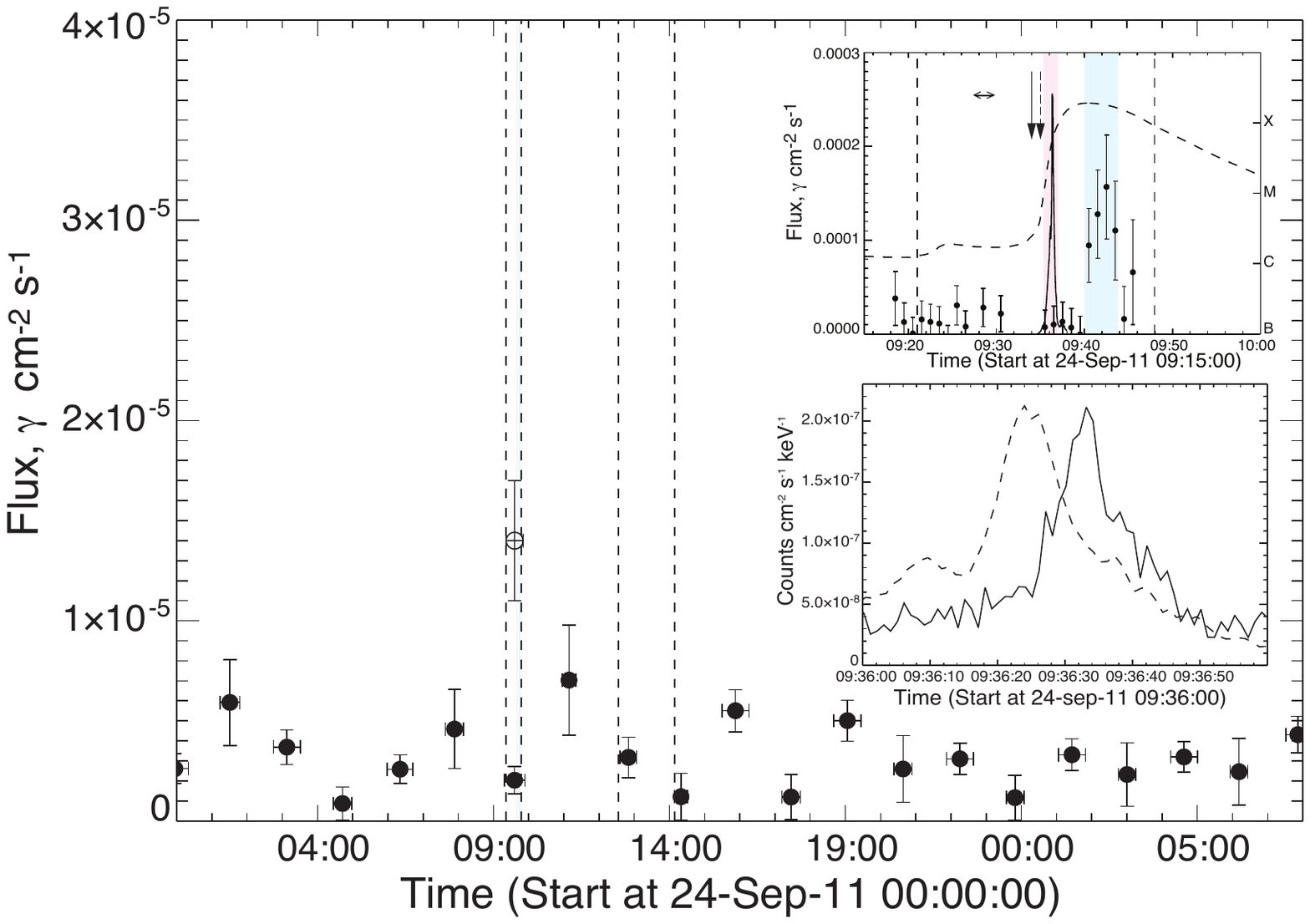}
\caption{Time profile of the SGRE event during the first flare at 09:20 UT on 2011 September 24. The filled circles in the main plot denote the $>$100 MeV fluxes and uncertainties derived from source-class data.  The open circle shows the flux obtained from solar-impulsive-class data revealing SGRE during the flare when the source-class data were compromised.  The upper inset plots the GBM 100-300 keV count rates on the same scale as the solar-impulsive-class $>$100 MeV fluxes.  The short SGRE event began about six minutes after the impulsive hard X-ray peak.  The impulsive flare 10--60 MeV count rates in LLE data (solid trace) and arbitrarily scaled GBM 100-300 keV rates (dashed trace) are plotted in the lower inset at higher time resolution. Otherwise the same caption as Figure \ref{110224th}}
\label{110924th}
\end{figure} 

{\bf Details}  Plotted in Figure \ref{110924th}.  X1.9 class flare at E61 lasting $\sim$27 minutes; the CME speed is uncertain although the CDAW catalog lists it as $\sim$1936 km s$^{-1}$ (both CACTUS and DONKI list lower speeds) with onset $\sim$1 minute before the 100--300 keV X-ray narrow peak; questionable M (metric) Type II emission with onset $\sim$1 minute after the CME launch; only an upper limit on the SEP proton flux was obtained because of the large particle event on September 22.   Both {\it RHESSI} and GBM observed emission up to about 10 MeV during the impulsive flare with evidence for nuclear de-excitation and neutron-capture lines.  LAT had good solar exposure between 09:18--09:45 UT, covering the entire impulsive phase.  The GBM 100--300 keV time history is plotted in the top inset along with the $>$100 MeV flux derived solar-impulsive-class data.  The $>$100 MeV fluxes derived from source class data were affected by the high ACD rates during the flare.  This can be seen by the difference between the open-circle (impulsive class data) and filled-circle (source-class data) fluxes.  There is no clear evidence for $>$100 MeV emission during the impulsive X-ray peak.  The SGRE began about six minutes later and lasted only about five minutes.    LAT had poor solar exposure between 10:58--11:18 UT and good exposure between 12:37--13:00 UT and no SGRE was observed during those time intervals.  The spectrum of the SGRE is consistent with emission from pion-decays produced by protons following a power-law index of 3.4 $\pm$ 1.4.  The 95\% confidence upper limit on the number of $>$500 MeV protons during the impulsive peak is less than 25\% of the number observed in the SGRE. 
  
As shown in the bottom inset, LAT/LLE data reveal a striking peak in $\sim$10--60 MeV $\gamma$-ray emission delayed by about eight seconds from the 100-300 keV hard X-ray peak in the impulsive phase.  This delay is similar to that observed in by LAT in SOL2010-06-12T00:57 (event A3) \citep{acke12a, acke12b}, but in this case the emission is clearly dominated by electron bremsstrahlung; thus the delay measures the time to accelerate electrons to energies $>$10 MeV and is not due to delays associated with accelerating ions to hundreds of MeV as may have been the case for the 2010 June 12 flare.  \citet{acke14} also report this event as impulsive.  LAT LLE photon spectrum appears to follow a power-law in energy visible up to just above 100 MeV with index $\sim$4.8; there is only marginal evidence for pion decay emission at the 67\% confidence level.  If this steep bremsstrahlung spectrum extended to energies below 10 MeV it should have been detectable by GBM; but our studies indicate that it was not detected. GBM measured 2.223 MeV and de-excitation line fluxes indicating that the 4--40 MeV proton spectrum in the impulsive phase had an index of about 4.5.  If this proton spectrum extended without steepening to energies of 300 MeV, the $>$100 MeV pion-decay flux would have been about a factor of two higher than the limits on emission that we set.  This indicates that the impulsive proton spectrum steepened above $\sim$40 MeV.    

\subsection{SOL2012-01-23T03:38} \label{subsec:20120123}

{\bf Is the SGRE time history distinct from that of the impulsive flare? YES.} LAT observation began just after the flare and shows evidence for a decreasing flux from the peak of the impulsive phase followed by the clear onset of the sustained emission about 15 minutes later.  

\begin{figure}
\epsscale{.80}
\plotone{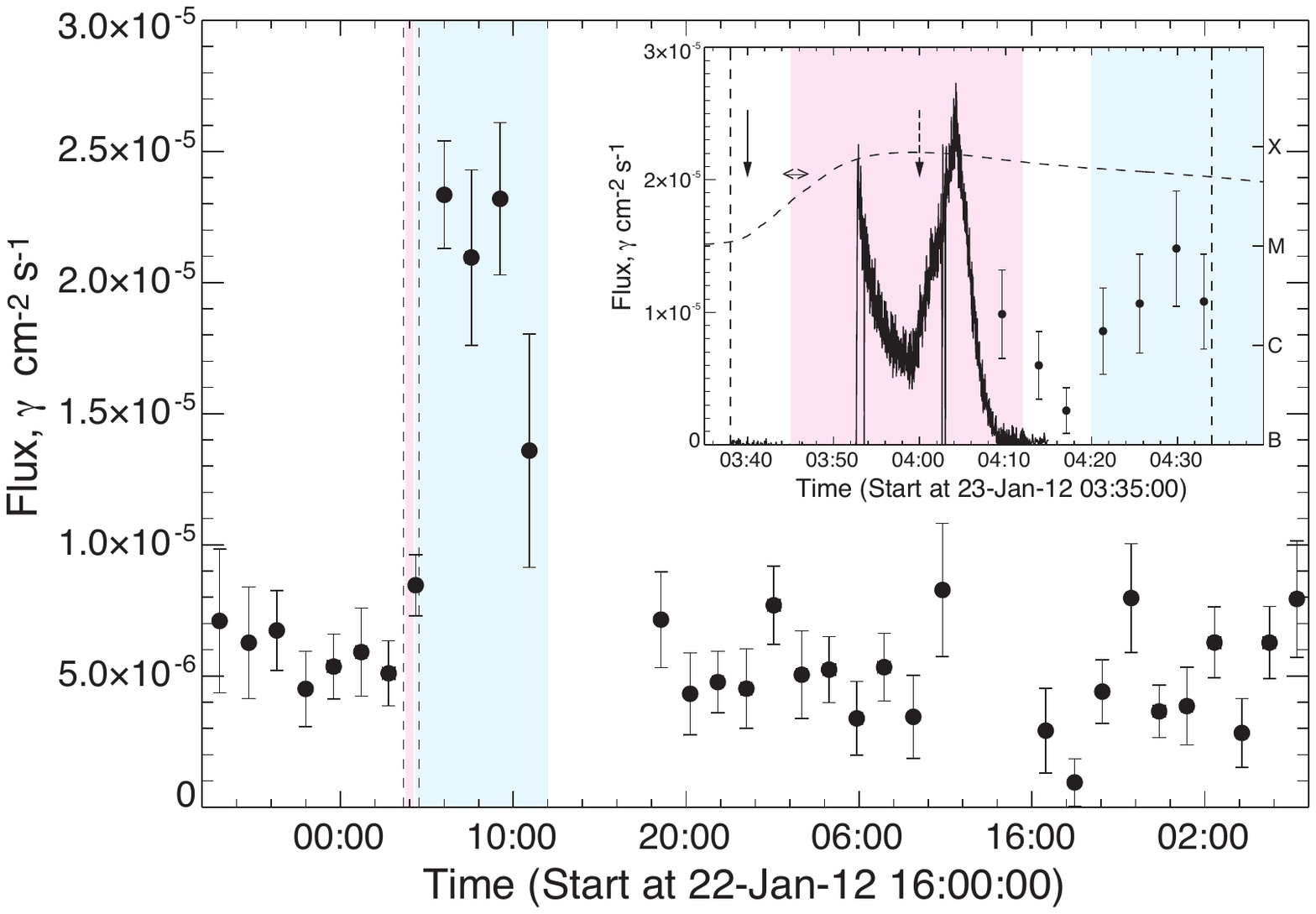}
\caption{Time profile of the 2012 January 23 SGRE event observed by LAT.  The solid trace in the inset shows the arbitrarily scaled GBM 100--300 keV hard X-ray time history compared with $>$100 MeV fluxes and uncertainties plotted at 4-minute resolution.  Plotted by the dashed downward arrow is the DH Type II onset time; there were no metric Type II observations.  See caption for Figure \ref{110307Ath} for more details. }
\label{120123th}
\end{figure}

{\bf Details}  Plotted in Figure \ref{120123th}.  M8.7 class flare at W21 lasting $\sim$55 minutes; $\sim$2000 km s$^{-1}$ CME with onset about the time of soft X-ray event, before {\it Fermi} entered sunlight at 03:53 UT; only DH (decameter-hectometric) Type II emission;  intense solar energetic proton event with emission observed $>$100 MeV; impulsive hard X-ray emission up to 100--300 keV observed by GBM (time history plotted in inset) with no evidence for the 2.223 MeV neutron-capture line.   LAT had good source-class data exposures to the Sun $\sim$04:08--04:36 and 05:46--06:10 UT in response to an autonomous repoint to the Sun, and at 07:18--07:48, 08:58--09:28, and 10:48--11:00 UT (truncated by the SAA). $>$100 MeV emission was observed in each of these intervals.  The source-class data plotted at 4-minute resolution in the inset suggestss that there were $>$100 MeV $\gamma$-rays associated with the impulsive hard X-rays and that the SGRE phase began $\sim$04:20 UT, $\sim$15 minutes after the observed hard X-ray peak.  We reach the same conclusion using the solar-impulsive-class data.   \citet {acke14} report a $>$100 MeV flux of $(0.8 \pm 0.1) \times 10^{-5} \gamma$ cm$^{-2}$ s$^{-1}$ that is consistent with our measurement between 04:08 and 04:34 UT.  We have fit the background-subtracted $\gamma$-ray spectra between 05:46 and 09:29 UT assuming that the emission is due to pion-decay radiation.  The results of these fits are given in Table \ref{tab:event}.   Our best fitting $>$100 MeV $\gamma$-ray fluxes are consistent with those listed by \citet{acke14} with the possible exception of the 07:18--07:48 UT interval that had relatively poor solar exposure and large flux uncertainty. The data between 05:46 and 06:10 UT were good enough to provide an estimate of the accelerated proton power-law spectral index, 5.1.  Power-law spectra with indices of $\sim$5 also fit the weaker fluxes in the next two intervals reasonably well and were used in our determination of the numbers of protons.  Based on these spectral fits we estimate that there were 3$\times 10^{28}$ $>$500 MeV protons accelerated at the Sun during the six-hour period of the SGRE.   

We used two methods for estimating the numbers of protons at the Sun during the impulsive phase of the flare:  1) the 0.023 $\gamma$ cm$^{-2}$ s$^{-1}$ upper limit on the 2.223 MeV line from 03:53--04:09 UT measured by GBM leads to an upper limit of 0.4$\times 10^{28}$ $>$500 MeV; 2) assuming that the $>$100 MeV flux measured between 04:08--04:12 UT followed the normalized hard X-ray time history during the flare, we estimated an average $\gamma$-ray flux of 0.7$\times10^{-4}\gamma$ cm$^{-2}$ s$^{-1}$ during the flare that we treat as an upper limit, resulting in an upper limit of 0.7$\times 10^{28}$ protons $>$500 MeV (Table \ref{tab:event}).   Thus, the number of protons at the Sun during the impulsive phase of the flare is no more than about 20\% of the number in the SGRE.  There were $\sim$70 times more $>$500 MeV SEP protons in interplanetary space than at the Sun.

\subsection{SOL2012-01-27T17:37}  \label{subsec:20120127}

\begin{figure}
\epsscale{.80}
\plotone{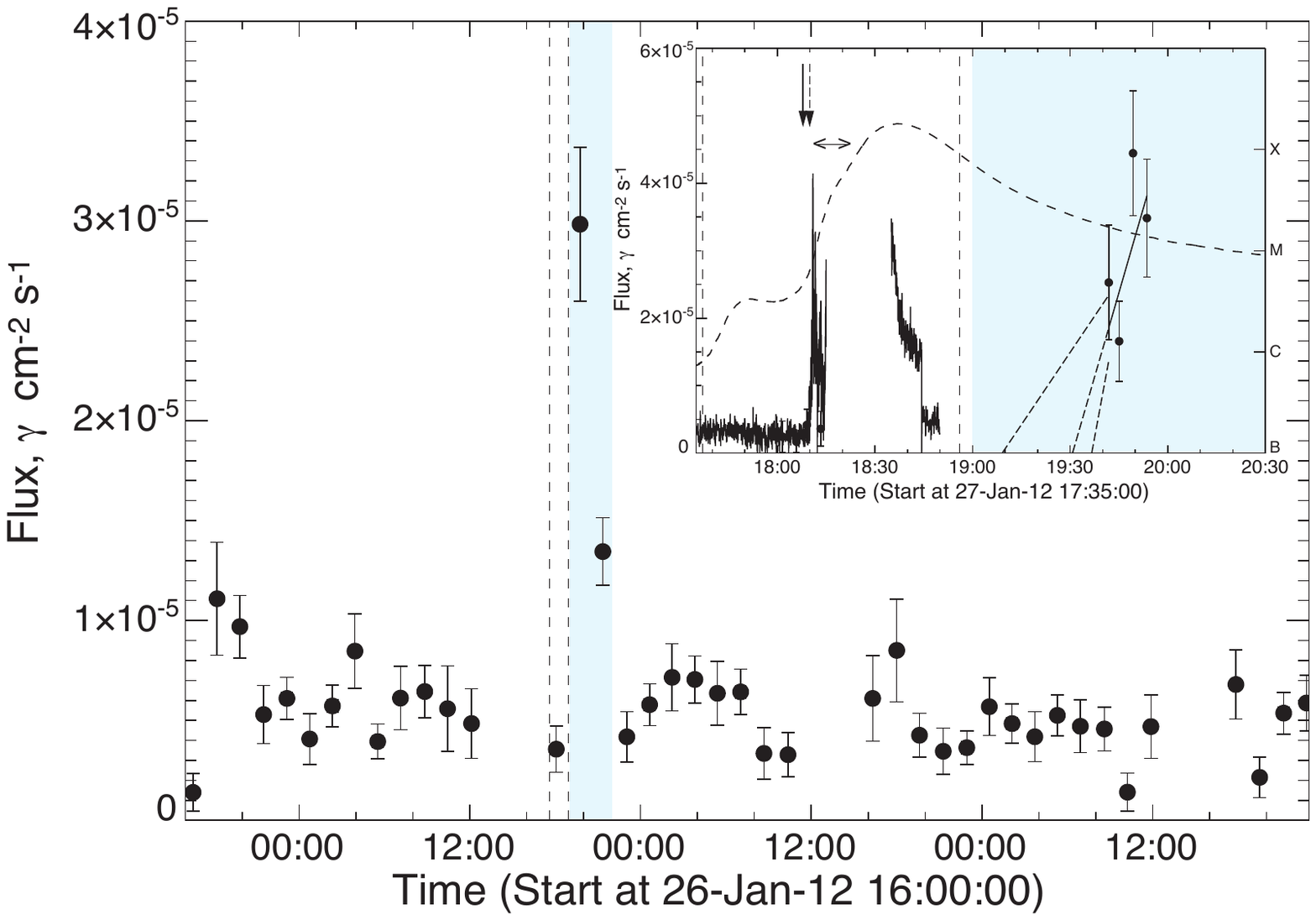}
\caption{Time profile of the 2012 January 27 SGRE event observed by LAT. The solid trace in the inset shows the arbitrarily scaled GBM 100-300 keV rates during the flare; missing data are due to an SAA passage.  The best fit to an increasing $>$100 MeV flux after 19:40 UT is shown by the solid line.  Extrapolations to determine $\gamma$-ray onset and $\pm1\sigma$ deviations are shown by the dashed lines.  See caption for Figure \ref{110307Ath} for more details. }
\label{120127th}
\end{figure}

{\bf Is the SGRE time history distinct from that of the impulsive flare? Yes}  There is evidence that the onset occurred about an hour after the start of the CME as shown in the inset of Figure \ref{120127th}. Much of the impulsive flare was not observed because {\it Fermi} was in the SAA.   This event appears to have lasted about three hours.

{\bf Details}  Plotted in Figure \ref{120127th} using Pass8 data.  X1.7 class flare at about W81(using {\it STEREO A} observations because {\it RHESSI} was being annealed at the time) lasting 79 minutes; $\sim$2500 km s$^{-1}$ CME with estimated onset coincident with the 100-300 keV X-ray rise; M (metric) and DH (decameter-hectometric) Type II emissions observed with the M onset $\sim$5 minutes after the CME launch;  strong SEP event with proton energies not exceeding 100 MeV;  {\it Fermi} was in SAA between 18:15--18:34 UT; 100--300 keV hard X-rays observed by GBM at beginning and end of impulsive phase (see time history plotted in inset).  Poor GBM exposure and large background after SAA prevents estimate of the neutron capture line flux during the impulsive phase, although any line emission would have been strongly attenuated because the flare was near the limb.   Using solar-impulsive-class data, we found no evidence for $>$100 MeV emission during the first part of the flare from 18:10--18:15 UT.   Because the flare was at a heliocentric angle of $\sim 81^{\circ}$, the $>$100 MeV flux at that location would have been attenuated by about 40\% relative to disk center. 

SGRE was observed during the next two exposures 19:36--19:56 and 21:06--21:37 UT.  Within the large uncertainties the $\gamma$-ray fluxes, determined using Pass8 data and listed in Table \ref{tab:event} are consistent with those reported by \citet{acke14}. The increasing fluxes with time {98\% confidence), plotted in the inset of Figure \ref{120127th}, suggest that the onset of the SGRE occurred about an hour after the impulsive flare.  The derived sustained-emission proton spectrum during the first exposure followed a power law with index 4.2 $\pm$ 0.7 while the spectrum taken 90 minutes later appears to be much harder power-law index 2.8 $\pm$ 0.6.  We have estimated the number of $>$500 protons producing the SGRE assuming that they interact at the location of the flare (Table \ref{tab:event}).  The number of $>$500 MeV protons in space was about two orders of magnitude larger those producing the SGRE event (Table \ref{tab:event}).   Due to the limited exposure of GBM and LAT, we have no information on the number of $>$500 MeV protons at the Sun during the flare. 

\subsection{SOL2012-03-05T02:30}  \label{subsec:20120305}

{\bf Is the SGRE time history distinct from that of the impulsive flare? YES.}  The latter part of the impulsive phase of this event was well observed by LAT.  Spectral fits to background-subtracted solar-impulsive-class data were used to obtain the 1-minute resolution time history plotted in the right inset of Figure \ref{120305th}.  There is no evidence for impulsive $>$100 MeV emission associated with the hard X-ray peak.  The SGRE emission began within about five minutes of the prominent hard X-ray peak.  SGRE was observed for about five hours. 

{\bf Details}  Plotted in Figure \ref{120305th}.  X1.1 class flare at E54 lasting $\sim$210 minutes.;  $\sim$1500 km s$^{-1}$ CME with onset an hour after the {\it GOES} start time and $\sim$17 minutes before GBM hard X-ray data become available (note that this was the last of series of three CMEs occurring over a 3-hour period \citep{cola15}) ; only DH (decameter-hectometric) Type II emission;  only upper limit on solar energetic proton flux due to preceding March 4 SEP event; impulsive hard X-ray emission observed up to 100--300 keV by GBM (see time history in right inset) after 03:55 UT when {\it Fermi} moved into daylight; this initiated an autonomous repoint to the Sun that lasted until 06:15 UT; no evidence for the 2.223 MeV neutron-capture line in GBM spectra.  LAT solar exposures in Pass7 source class data: 02:35--03:04, 04:06--04:38, 05:46--06:12, 07:18--07:56, 08:56--09:26 UT.  There is clear evidence in Figure \ref{120305th} for SGRE in the two orbits following the flare, begining at 05:46 UT.   The $>$100 MeV fluxes that we derived for these these two orbits are consistent with those reported by \citet{acke14}.  Our fits also suggest that the proton spectrum may have hardened with time, although the uncertainties are large. \citet{acke14} report a $>$100 MeV flux of 5 $\times 10^{-6} \gamma$ cm$^{-2}$ s$^{-1}$ in the flare exposure between 04:12 and 05:01 having both impulsive and SGRE components.

The inset shows time history of $>$100 MeV emission around the time of the hard X-ray peak near 04:30 UT that we derive using spectral fits to the solar-impulsive-class data.  There is no evidence for $>$100 MeV $\gamma$-rays during the hard X-ray peak between 04:28 and 04:35 UT with a upper limit of 1.5 $\times 10^{-5} \gamma$ cm$^{-2}$ s$^{-1}$. There is evidence for an increase (85\% confidence) in flux after that time, with a peak of $\sim$ 5 $\times 10^{-6} \gamma$ cm$^{-2}$ s$^{-1}$, that  is likely the onset of the SGRE.   The upper limit on the number of $>$500 MeV protons at the Sun during the entire flare using the upper limit of the neutron-capture line flux is significantly larger than the limit obtained using LAT $>$100 MeV observations during the second X-ray peak using  (Table \ref{tab:event}).  We use the conservative limit obtained from the neutron capture line in our studies.    The number of high-energy protons at the Sun during the impulsive flare was therefore probably much less than half of the number in the SGRE.  We have no estimate of the number in space because of the preceding SEP event.

\begin{figure}
\epsscale{.80}
\plotone{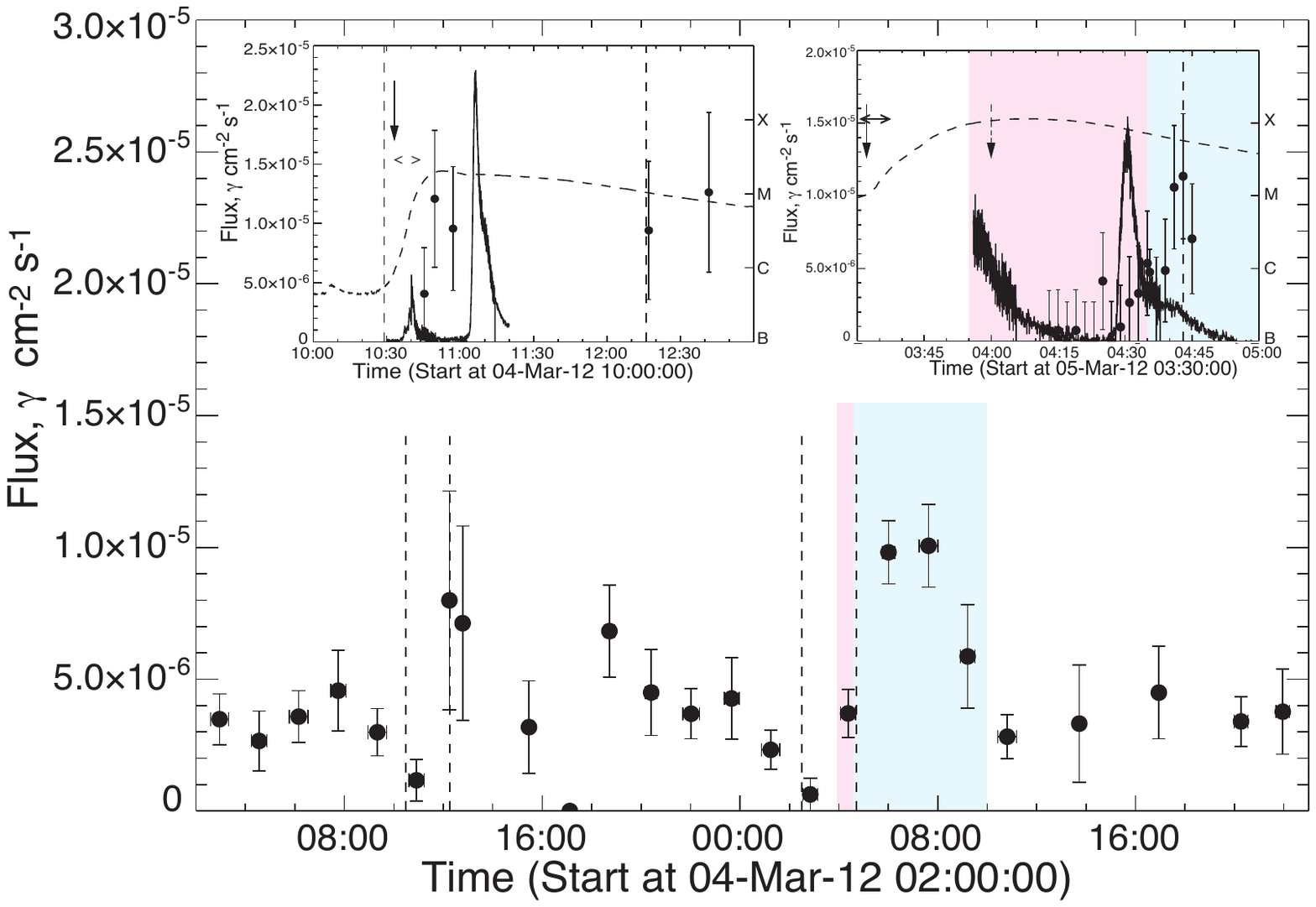}
\caption{Time histories of the 2012 March 4 event with no significant SGRE detected by LAT and the 2012 March 5 event for which sustained emission was observed in three exposures after the flare.  The left inset shows a blowup of the March 4 flare with GBM 50--100 keV rates scaled to the $>$100 MeV flux derived from fits to solar-impulsive-class data.  There is no evidence $>$100 MeV $\gamma$ rays during the flare.   The solid trace in the right inset shows the GBM 100--300 keV rates plotted on the same scale as the $>$100 MeV background subtracted fluxes and $\pm1\sigma$ uncertainties obtained from fits to the solar-impulsive-class data. Plotted by the dashed downward arrow is the DH Type II onset time; there were no metric Type II observations. See caption for Figure \ref{110307Ath} for more details.}
\label{120305th}
\end{figure}

\subsection{SOL2012-03-07T00:02/T01:05}  \label{subsec:20120307}

{\bf Is the SGRE time history distinct from that of the impulsive flare? YES.}  There were two intense sustained-emission events  associated with two flares within about one hour of each other.  LLE data \citep{ajel14} reveal $>$100 emission about 20 minutes after the first flare, a {\it  GOES}-class X5.4 that lasted for at least one hour.  Our study of Pass8 solar-impulsive-class data plotted in the upper right inset reveals that this is a short SGRE event similar to what was observed on SOL2011-09-06T22:12.  The longer SGRE event shown in the lower-left inset began about 02 UT, 45 minutes after an M7 flare, and lasted about 16 hours.  

\begin{figure}
\epsscale{.80}
\plotone{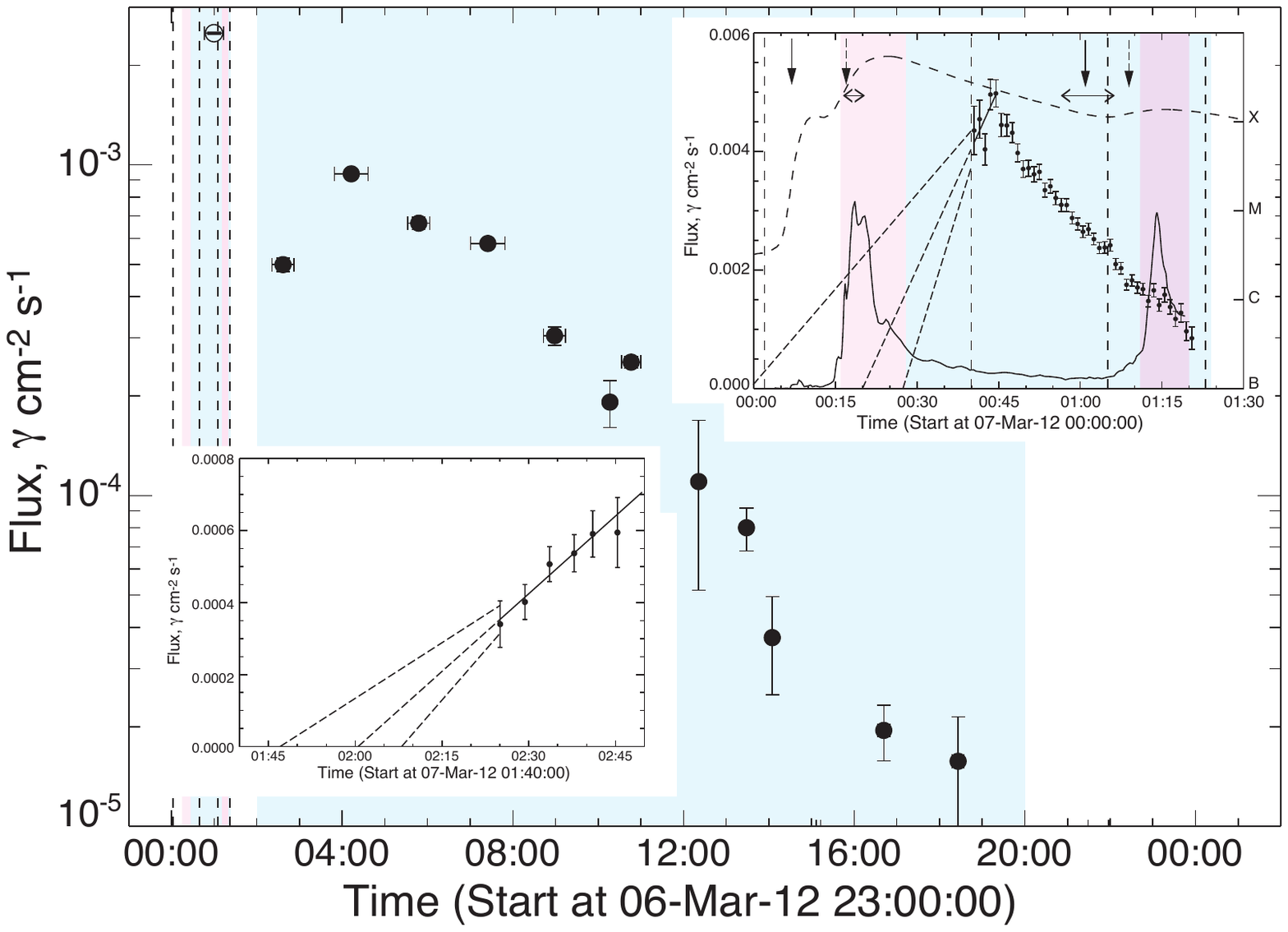}
\caption{Time history 2012 March 7 event when two episodes of SGRE were detected by LAT.  Inset on upper right shows the $>$100 keV time histories (solid trace) derived from the {\it INTEGRAL}/SPI ACD plotted on the same scale as $>$100 MeV fluxes and $\pm1\sigma$ uncertainties derived from fits to one-minute solar-impulsive-class data. The observed rise to a peak indicates that the $>$100 MeV emission is distinct from the X5.4 impulsive flare. The open circle on the main plot is the weighted mean of the $>$100 MeV solar-impulsive-class fluxes plotted in the top inset. The solid line is a fit to the rise in $>$100 MeV flux and the dashed lines are its extrapolation and $\pm1\sigma$ deviations used to estimate the SGRE onset.  The inset on lower left shows four-minute resolution $>$100 MeV fluxes derived from source class data showing the onset of 18-hour duration SGRE. The solid line is a fit to the rise in $>$100 MeV flux and the dashed lines are its extrapolation and $\pm1\sigma$ deviations used to estimate the SGRE onset. See caption for Figure \ref{110307Ath} for more details.}
\label{120307th}
\end{figure}

{\bf Details}  Plotted in Figure \ref{120307th}.  X5.4 and M7\footnote{The flare is listed as an X1.3 class flare because of the high X-ray background from the X5.4 flare} class flares at E27 lasting $\sim$38 and 18 minutes, respectively;  $\sim$2700 and 1800 km s$^{-1}$ CMEs with onsets several minutes before hard X-ray peaks; M (metric) and DH (decameter-hectometric) Type II emissions observed with M onsets delayed by $\sim$17 and eight minutes after the respective CME launches; intense solar energetic proton event with emission $>$100 MeV; impulsive hard X-ray and $\gamma$-ray emission from both flares observed to MeV energies \citep{zhan12}.  Detailed temporal, spectroscopic, and location studies of this event are discussed by \citet{ajel14}.   The $>$100 keV time history in the upper right inset was derived from the {\it INTEGRAL}/SPI anticoincidence detector \citep{zhan12} and covers the entire period of the two flares.  SGRE from both events was observed for up to 18 hours. Because the fluxes from this event were so large it was possible to make LAT observations every orbit, even when the peak solar exposure was about 20\% of the maximum.  {\it Fermi} came into sunlight at 00:30 UT and GBM observed the decay of the X5.4 and most of the M7 flare.  Due to the high hard X-ray rates there were no source-class data during the flares but there were LLE data from 00:35--01:25 UT analyzed by \citet{ajel14}.  We have analyzed data from the new solar-impulsive-class data and plot  $>$100 MeV flux histories in the upper right inset at 1-minute resolution.  The plot shows that the flux peaked at 00:44 UT, suggesting that this emission was not from the impulsive phase but from a short SGRE event similar to that observed associated with SOL2011-09-06T22:12 (event 6).  Good LAT source-class exposures were obtained between: 02:18--02:48, 04:06--04:34, 05:34--06:01, 07:02--07:46, 08:42--09:12,10:33--10:58, 13:23--13:33, 16:35--16:49, and 19:46--20:14 UT.  The fluxes for these observations were derived from Pass8 source-class data. The rising flux between 02:18--02:45 UT, plotted along with an extrapolation to earlier times, suggests that the onset of the 16-hour long SGRE began about 02:00 UT, about 45 minutes after the peak of the M7 flare.   

Table \ref{tab:event} presents the results of our spectroscopic analysis of the LAT data.    As discussed above, there were two discrete SGRE events: SGRE(A) began $\sim$20 minutes after the hard X-ray peak associated with the X5.4 flare and lasted through M7-class flare, and SGRE(B) began about 45 minutes after the M7-class flare.  Assuming that SGRE(A) began at 00:28 UT we estimate that there 4.0 $\times 10^{29}$ $>$500 MeV protons producing the emission.  In contrast there were 1.3$\times 10^{30}$ protons $>$500 MeV responsible for producing SGRE(B).   We confirm the finding of \citet{ajel14} that the spectrum of protons producing the pion-decay $\gamma$ rays appeared to be variable from 00:39 to 01:21 UT.   The spectrum initially softened from an index, s$_p$ $\simeq$3.6 to $\simeq$3.9 at about 01:00 UT and then hardened to an index s$_p$ $\simeq$3.3 about the time of the M7 flare.  This hardening might have been caused by  protons from the M7 flare.   The $\gamma$-ray fluxes and proton spectral indices derived for SGRE(B)  are consistent with those reported by \citet{ajel14}.  The $>$300 MeV proton spectrum exhibited spectral softening through at least the first eight hours of the event.   

We estimated the numbers of protons $>$500 MeV during X5.4 and M7 flares flares using the neutron-capture line fluences of 85 and 65 $\gamma$ cm$^{-2}$, respectively, measured by the {\it INTEGRAL} high-resolution gamma-ray spectrometer \citet{zhan12}.  Assuming a proton power-law index of 4.5 above 40 MeV, we estimated that both flares accelerated only about 2\% of the number of $>$500 MeV protons produced in SGRE(A) and SGRE(B) (Table \ref{tab:event}).   We can also set an upper upper limit on the number of $>$500 MeV protons accelerated during the M7 flare by estimating the excess $>$100 MeV $\gamma$-ray flux over the decay from SGRE(A).  The excess is only significant at the 1-$\sigma$ level. The 95\% confidence limit is about three times lower than what we estimated using the neutron-capture line, suggesting that the proton spectrum above 40 MeV was steeper than the power-law with index 4.5 that we assumed.   We also estimated that the number of $>$500 MeV protons in space was about 100 times the number producing the two SGRE events.  

When the SGRE fluxes exceed $\sim1.0 \times 10^{-4} \gamma$ cm$^{-2}$ s$^{-1}$ it is possible to use measurements of the 2.223 MeV neutron-capture line to constrain the proton spectrum below 300 MeV during these exposures.  We were able to use {\it RHESSI} rear detector observations of the neutron-capture line because the flare occurred just after an anneal of the detectors when the spectral resolution was about 20 keV FWHM at 2.223 MeV.  We obtained background-corrected spectra on orbital time-frames using background $\pm$15 orbits when the environmental conditions were comparable to that during the observing period.  We found no evidence for a significant flux in the neutron-capture line in any exposure.  We then estimated limits on the index of the power-law proton spectrum between about 40 and 300 MeV using our 95\% confidence limits on the 2.223 MeV line flux from {\it RHESSI} compared with the {\it Fermi}/LAT measured $>$100 MeV $\gamma$-ray flux.  The upper limits on the power indices between 40 and 300 MeV are shown in Table \ref{tab:event} in the row below the indices measured above 300 MeV.   All of the limits on the 40--300 MeV index are smaller than those measured above 300 MeV indicating that the proton spectrum producing the SGRE events steepened above about 300 MeV. 

\subsection{SOL2012-03-09T03:22} \label{subsec:20120309}

{\bf Is the SGRE time history distinct from that of the impulsive flare? YES.}   The sustained emission has a clear onset within an hour of the flare, a peak flux about four hours later, and a relatively short decay phase.  As shown in the inset, there is no evidence for impulsive $>$100 MeV emission. 

{\bf Details}  Plotted in Figure \ref{120309th}.  M6.3 class flare at W02 lasting $\sim$56 minutes;  $\sim$850 km s$^{-1}$ CME with onset $\sim$8 minutes after start of 100-300 keV emission; M (metric) and DH (decameter-hectometric) Type II emissions observed with the M onset $\sim$3 minutes after the CME launch; only an upper limit on solar energetic proton flux due to continuing March 7 event; impulsive hard X-ray emission weakly observed up to $>$100 keV by GBM (we plot GBM 50-100 keV time history in the inset); no evidence for the 2.223 MeV neutron-capture line in GBM spectra.  There are good LAT exposures each orbit due ToO in response to the March 7 SEP event.  A LAT solar exposure occurred during the impulsive phase of the flare but there were no source-class data due to high ACD rates; no LLE data are publicly available, but \citet{acke14} indicate that no $>$100 MeV emission was detected. We have analyzed the solar-impulsive-class data to study $>$100 MeV time history during the impulsive phase.  Because of the higher background in the impulsive class of data we have normalized the fluxes to the LAT source-class background rate.  This time history is plotted in the inset.  There is no evidence for impulsive $>$100 MeV emission associated with the hard X-ray peaks.  Other LAT exposures were at 05:10--05:58, 06:46--07:32, and 08:22--09:08  UT.  SGRE commenced about 04:00 UT, after the impulsive flare and consistent with the delayed nature of the event reported by \citet{acke14}, and lasted until about 10:30 UT.   SGRE fluxes in the three exposures are consistent with \cite{acke14} (Table \ref{tab:event}); the spectra of the protons producing pion-decay radiation in the exposures appear to be steep with power-law indices $\geq$6. We estimate that 1.5 $\times 10^{28}$ $>$500 MeV produced the SGRE and that there were fewer than 10\% of this number during the impulsive flare.  Due to large SEP event on March 7 we have no estimate of the number of protons in space.

\begin{figure}
\epsscale{.80}
\plotone{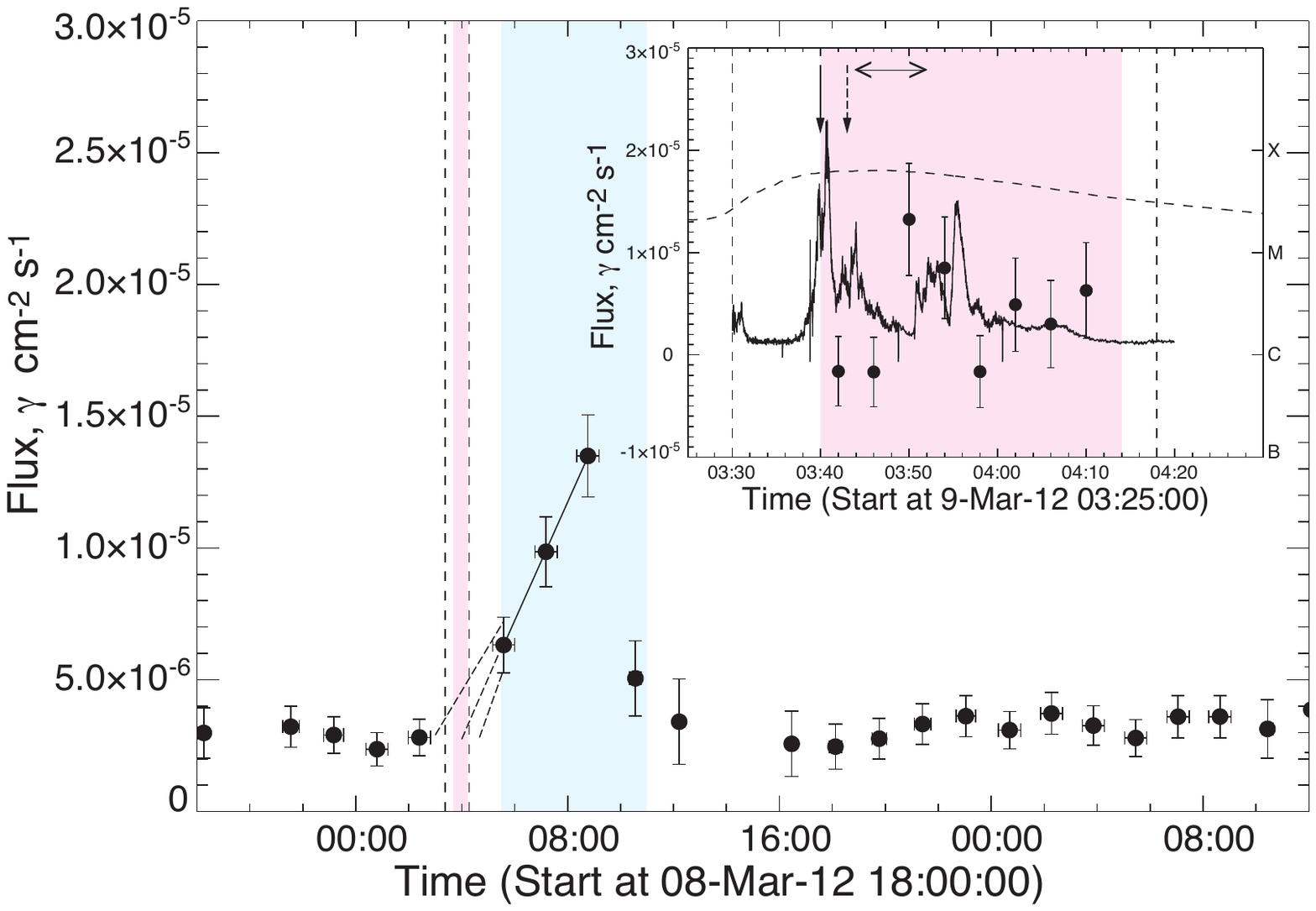}
\caption{Time profile of the 2012 March 9 SGRE event observed by LAT. The best fit to an increasing $>$100 MeV flux after 05:00 UT is shown by the solid line.  Extrapolations to determine $\gamma$-ray onset and $\pm1\sigma$ deviations are shown by the dashed lines. Inset shows GBM 50--100 keV rates plotted on the same scale as $>$100 MeV fluxes derived from fits to solar-impulsive-class data after normalization to the LAT background rate observed in source-class data.  See caption for Figure \ref{110307Ath} for more details.}
\label{120309th}
\end{figure}

\subsection{SOL2012-03-10T17:15} \label{subsec:20120310}

{\bf Is the SGRE time history distinct from that of the impulsive flare? YES.}   The SGRE onset occurred within about two hours of the associated flare and emission lasted about six hours.  The time profile was similar to that observed in the SGRE event on March 9, but the flux was weaker. 

\begin{figure}
\epsscale{.80}
\plotone{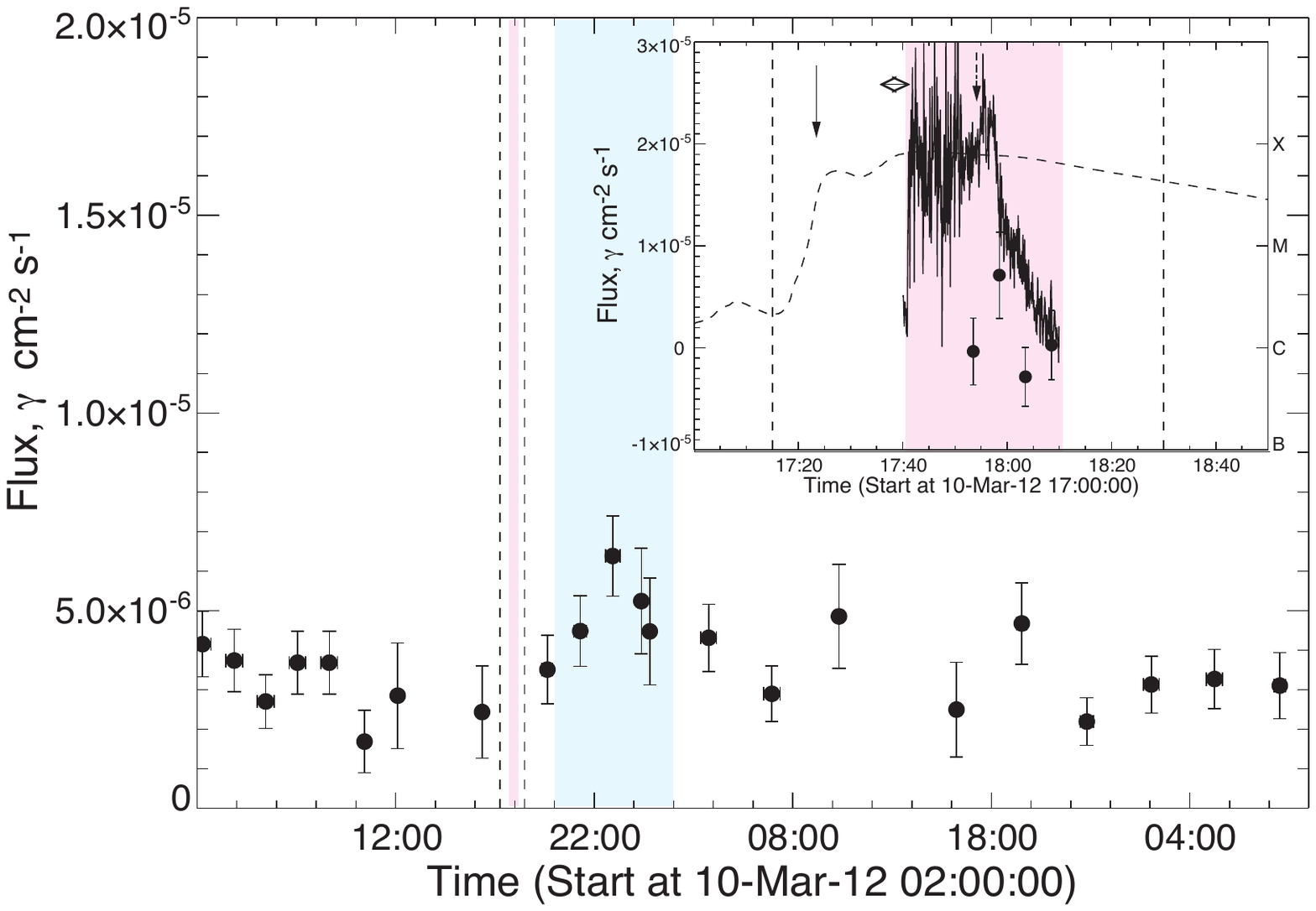}
\caption{Time profile of the 2012 March 10 SGRE event observed by LAT.  Inset shows GBM 50--100 keV rates plotted on the same scale as $>$100 MeV fluxes derived from fits to solar-impulsive-class data after normalization to the LAT background rate observed in source-class data. Plotted by the dashed downward arrow is the DH Type II onset time; no metric Type II emission was detected.  See caption for Figure \ref{110307Ath} for more details. }
\label{120310th}
\end{figure}

{\bf Details}  Plotted in Figure \ref{120310th};  M8.4 class flare at W26 lasting $\sim$75 minutes;  $\sim$1300 km s$^{-1}$ CME with onset $\sim$8 minutes after the {\it GOES} start time and 15 minutes before GBM hard X-ray data became available; only DH (decameter-hectometric) Type II emission detected; only upper limit on solar energetic proton flux due to background from earlier events; impulsive hard X-ray emission observed up to 100--300 keV by GBM after entering daylight (see time history in inset); no evidence for the 2.223 MeV neutron-capture line in GBM spectra.  There are good LAT exposures each orbit due the continuing ToO in response to the March 7 event.  A LAT exposure occurred during the impulsive phase of the flare but there were no source-class data due to high ACD rates. No LLE data are publicly available but \citet{acke14} indicate that no $>$100 MeV emission was detected during the flare. We analyzed the solar-impulsive-class data to study the $>$100 MeV time history during the impulsive phase.  Because of the higher background in the impulsive class of data we have normalized the fluxes to the LAT source-class background rate.  We plot the time history in the inset.  There is no evidence for $>$100 MeV emission associated with the hard X-ray peaks.  Other LAT exposures: 19:24--19:52, 21:02--21:34, 22:34--23:14, 00:10--00:55 (with data gap) UT.  The SGRE was weak, had an uncertain onset time, and lasted about six hours.  Our spectral fits suggest weak 2 $\sigma$ detections in each of three orbits with $>$300 MeV proton spectra having power-law indices steeper than 6 (Table \ref{tab:event}).  We estimate that $\sim$0.5 $\times 10^{28}$ $>$500 MeV protons were responsible for the SGRE.  We obtained a limit on the number of protons during the impulsive flare from a limit on the number of 2.223 MeV photons observed by {\it RHESSI} at the end of the impulsive flare.  This limit is less than the number observed in the sustained emission.  We have also searched for impulsive $>$100 MeV $\gamma$-ray emission using the solar-impulsive-class data in four 5-minute intervals starting at 17:51 UT (see inset of figure).  There is no evidence for emission and we set a limit of 0.03 $\times 10^{28}$ $>$500 MeV protons.  As this observation only covers about 60\% of the impulsive phase we list the upper limit as 0.1 $\times 10^{28}$ in the Table.   Due to the large SEP event on March 7, we only obtained an upper limit on the number of $>$500 MeV protons.

\subsection{SOL2012-05-17T01:25} \label{subsec:20120517}

{\bf Is the SGRE time history distinct from that of the impulsive flare?  Uncertain.}  There is no clear evidence for temporal variability $>$100 MeV in the first exposure taken about 30 min after the last hard X-ray peak and plotted in the inset.  As the flux in the next exposure was not significantly lower, it is likely that the SGRE peaked between the two exposures suggesting that the sustained emission began less than 30 minutes after the last impulsive phase peak. The event lasted about four hours.  As discussed below, the upper limit on the number of protons in the impulsive phase was a factor of four larger than the observed number of protons responsible for the observed emission after the flare.  Therefore, it is not possible on energetic grounds to rule out the tail of the flare as the source of the $>$100 MeV $\gamma$ rays.

\begin{figure}
\epsscale{.80}
\plotone{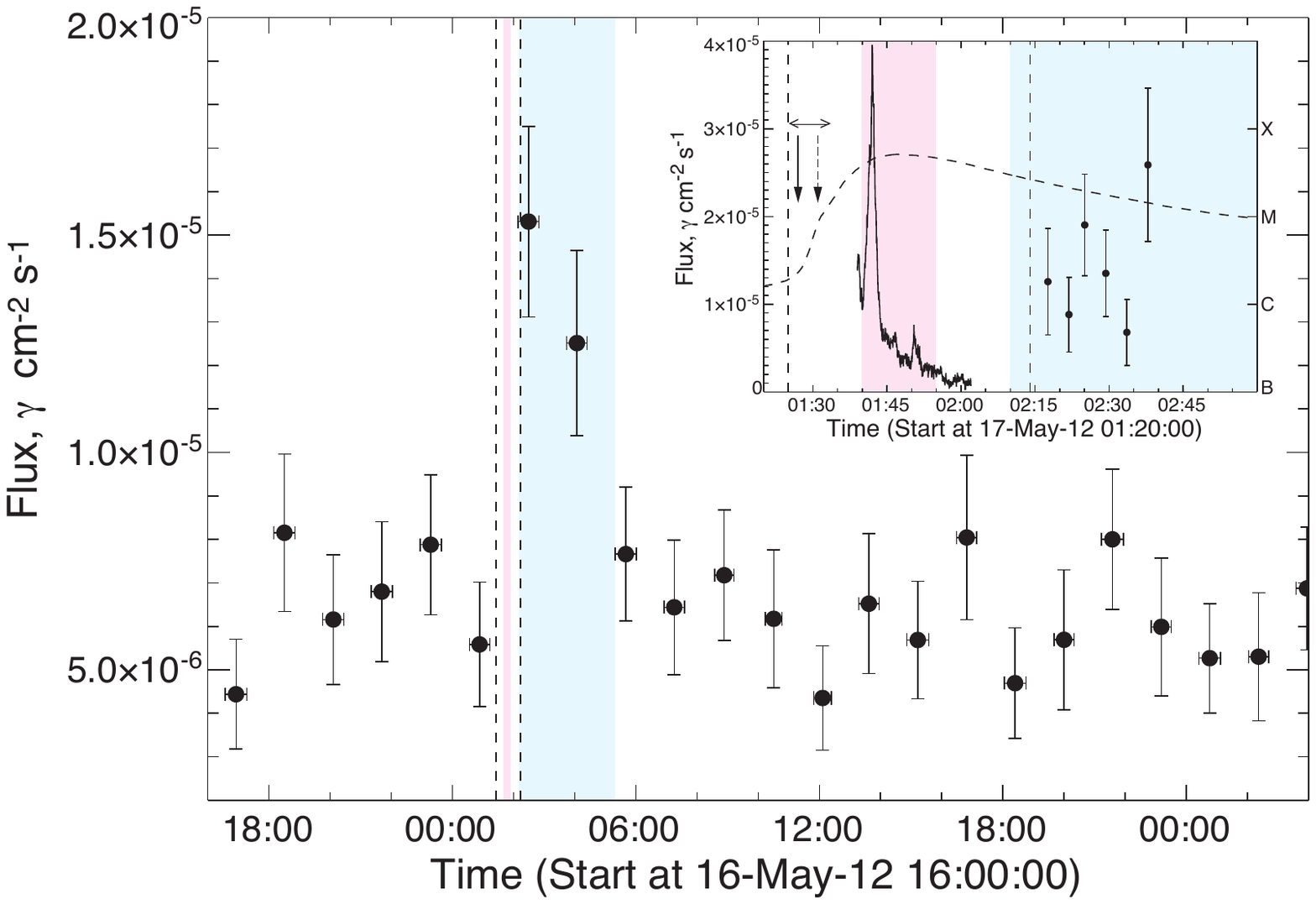}
\caption{Time profile of the 2012 May 17 SGRE event observed by LAT.  The solid trace shows the {\it RHESSI} 50-100 keV rates plotted on the same scale as the $>$100 MeV fluxes.  See caption for Figure \ref{110307Ath} for more details. }
\label{120517th}
\end{figure}

{\bf Details}  Plotted in Figure \ref{120517th}.  M5.1 class flare at W77 lasting $\sim$50 minutes;  $\sim$1600 km s$^{-1}$ CME with onset $\sim$2 minutes after the {\it GOES} start time and 10 minutes before {\it RHESSI} hard X-ray data became available; M (metric) and DH (decameter-hectometric) Type II emissions observed with the M onset $\sim$4 minutes after the CME launch; strong SEP event with proton emission $>$100 MeV and observed as the first ground level event of Cycle 24; impulsive hard X-rays observed by {\it RHESSI} up to 100--300 keV after entering daylight (see 50-100 keV time history plotted in inset); no evidence for 2.223 MeV line in the {\it RHESSI} spectrum during the flare.  

There were good LAT solar exposures every orbit including one just before the flare at 00:34--01:10 and at 02:10--02:48, 03:46--04:22, and 05:22-05:56 UT.  There was significant SGRE in the first two exposures after the flare.  The measured flux (Table \ref{tab:event} in the first exposure is consistent with that reported by \citet{acke14}.  It is not clear why they classify this event as being both impulsive and sustained as there are no $>$100 MeV observations during the flare. The $\gamma$-ray flux just after the flare is plotted at 4-minute resolution in the inset of the Figure and shows no clear evidence for variation in time.  Therefore, there is no clear evidence for an onset of the SGRE but assuming the $>$100 MeV came from a distinct sustained phase the onset had to be $<$30 minutes after the hard X-ray peak.  It is also possible that the falling flux observed in the two orbits after the flare could be due the decay of the impulsive phase.  The proton spectrum producing the $\pi$-decay emission was one of the hardest observed in any SGRE. The number of $>$500 MeV protons producing the SGRE is less than the upper limit on the number during the impulsive phase derived from the limit on the solar 2.223 MeV line.  What is surprising is that the $>$100 MeV flux is so weak that the inferred number of protons in the SGRE phase is about 1000 times less than observed in the SEP in space.  These comparisons of proton numbers suggests that a significant fraction of the SGRE may have been radiated behind the west limb of the Sun.

\subsection{SOL2012-06-03T17:48} \label{subsec:20120603}

{\bf Is the SGRE time history distinct from that of the impulsive flare? YES.}  Solar-impulsive-class data, plotted in the inset, reveal a one-minute $>$100 MeV peak flux coincident with the impulsive hard X-ray peak near 17:53:30 UT and emission following the peak until the end of the exposure at 18:02 UT.   The $>$100 MeV spectrum during the one-minute impulsive flare is steep and can be fit by pion-decay produced by protons following a power-law spectrum with index 6.5 $\pm$ 1.0 (see Table \ref{tab:event}).  In contrast the protons producing the eight-minute SGRE emission following the peak has an average power-law index of 4.3 $\pm$0.7.  There is evidence that the proton spectrum softened during this eight-minute period because the power-law index was 3.6 $\pm$ 0.9 between 17:54--17:55 UT, just after flare peak.  There is also weak evidence (3 $\sigma$)  that the SGRE began two minutes before the peak and its associated proton spectrum was even harder than the eight minute average.  The significant spectral difference between the protons producing the flare and those producing the time-extended emission requires the presence of an additional acceleration process.  It is puzzling that the Type II radio emission began six minutes after the CME onset. 

\begin{figure}
\epsscale{.80}
\plotone{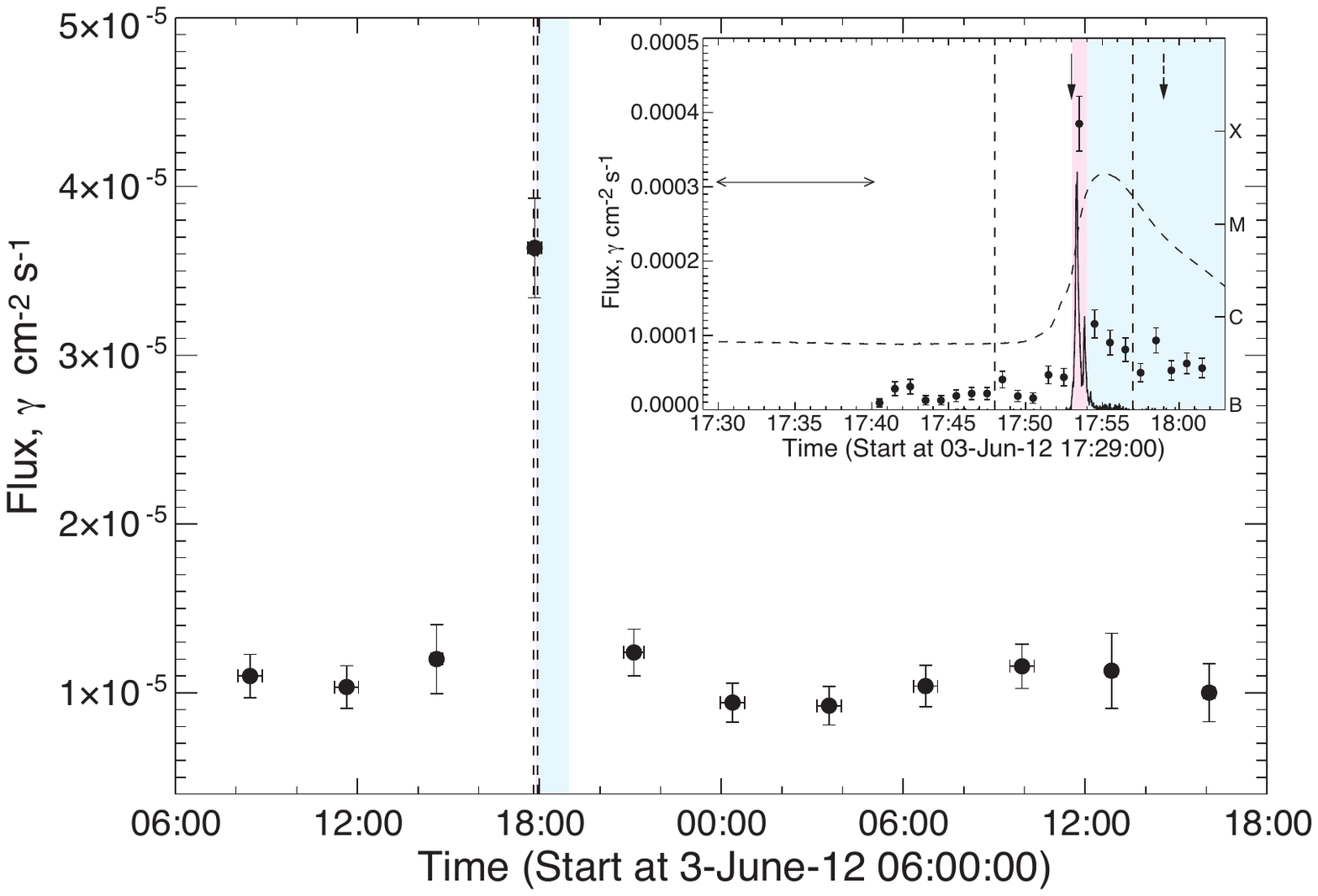}
\caption{Time profile of the 2012 June 03 SGRE event observed by LAT.  Source-class data fluxes are plotted in the main section; the average flux at 18 UT is smaller but consistent with that derived from solar-impulsive-class data plotted in the inset.   Inset compares 100--300 keV rates observed by GBM plotted on the same scale as $>$100 MeV fluxes derived from solar-impulsive-class data.  See caption for Figure \ref{110307Ath} for more details.}
\label{120603th}
\end{figure}

{\bf Details}  Plotted in Figure \ref{120603th}.  M3.3 class flare at E38 lasting $\sim$9 minutes; $\sim$605 km s$^{-1}$ CME with estimated onset coincident with the 100--300 keV X-ray rise; only M (metric) Type II emission observed with onset $\sim$6 minutes after the CME launch; weak SEP event; impulsive hard X-rays observed up to 300-800 keV by {\it RHESSI} and GBM (100--300 keV time history plotted in inset).  This event would not have been included in 4-yr study because of the low CDAW CME speed, but this may have been due to its viewing angle as CACTUS reported a maximum speed of 892 km s$^{-1}$ observed from {\it STEREO B}.  Gamma-ray emission was detected in our automated search of source-class data during a single solar exposure (17:38--18:03 UT) and listed by \citet{acke14} as both an impulsive and SGRE event; our 1-minute resolution plot of solar-impulsive-class $>$100 MeV fluxes in the inset of the figure clearly shows both characteristics. We note that the time profile derived from source-class data is similar, although the fluxes are lower due to ACD live-time effects from the high rate of impulsive hard X-rays.  There is no evidence for $>$100 MeV $\gamma$-ray emission during the next LAT exposure between 20:50 and 21:26 UT.  

There is weak evidence for enhanced $>$100 MeV emission between 17:51 and 17:53 UT in both source and solar-impulsive-class data, suggesting that the sustained emission began as early as two minutes before the impulsive hard X-rays peak.   We fit the flare (17:53-17:54 UT) and post flare (17:54-18:02 UT) solar-impulsive-class data using pion-decay templates.   The proton spectrum during the flare is significantly softer that during the extended phase, power law index of 6.5 vs 4.3, respectively.   The average flux from this study from 17:50--18:02 UT is 30\% higher than reported by \citet{acke14}.  We estimated the number of protons in the time-extended phase by assuming that the emission lasted until 19:00 UT.  If this is the case, then the SGRE was produced by about four-times the number of $>$500 MeV protons as those in the impulsive flare.  We have no information on the number of protons in space.  There is no evidence for the neutron-capture line or for nuclear-deexcitation lines in the GBM spectrum during the one-minute impulsive phase of the flare.  Comparing these limits with the observed $>$100 MeV flux we estimate that the spectrum of flare accelerated protons between about 10 MeV and 300 MeV is consistent with a power law with index of 4 or harder.  This indicates that the flare proton spectrum steepened significantly above a few hundred MeV.

In Figure \ref{jun3ht} we compare the GBM 100-300 keV and $>$80 MeV LLE rates during the flare.  There are two peaks visible $>$80 MeV.  The first peak is coincident with the hard X-ray peak.  The $\gamma$-ray emission is not delayed by about 10 seconds from the hard X-ray emission as has been found in other studies of impulsive flares on 2010 June 12 \citep{acke12a, acke12b}, 2011 September 6 (Appendix \ref{subsec:20110906}, and 2011 September 24 (Appendix \ref{subsec:20110924}.  The second $\gamma$-ray peak is delayed by about 20 seconds, but with no accompanying hard X-ray emission.

\begin{figure}
\epsscale{.80}
\plotone{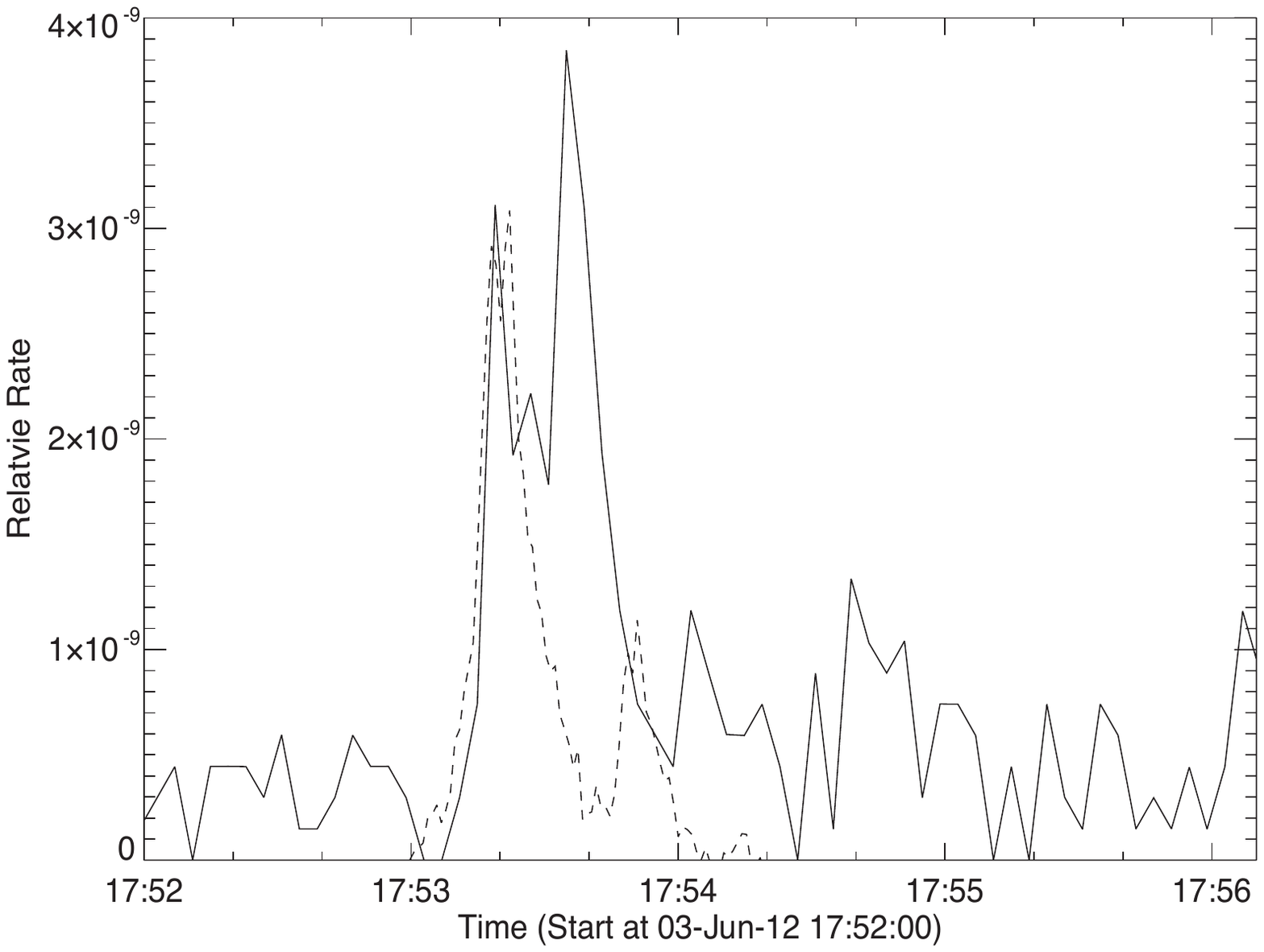}
\caption{Comparison of 100--300 keV rates observed by GBM (dashed trace) and $>$80 MeV rates observed in LLE data (solid trace) during the peak of 2012 June 3 flare. }
\label{jun3ht}
\end{figure}

\subsection{SOL2012-07-06T23:01} \label{subsec:20120706}

{\bf Is the SGRE time history distinct from that of the impulsive flare? Uncertain.}  Neither {\it RHESSI} nor GBM observed the Sun during the impulsive flare.  There is only one LAT exposure in which $>$100 MeV emission was observed; this began about 15 minutes after the impulsive phase and our 4-minute resolution plot in the inset indicates that the flux was falling during this time period.  Thus, we cannot tell whether the emission came from a separate sustained phase beginning in the 15 minutes after the hard X-ray peak or was the tail of the impulsive phase.  As there were no observations during the flare, we have no comparison of the relative numbers of protons in the sustained and impulsive phases.

\begin{figure}
\epsscale{.80}
\plotone{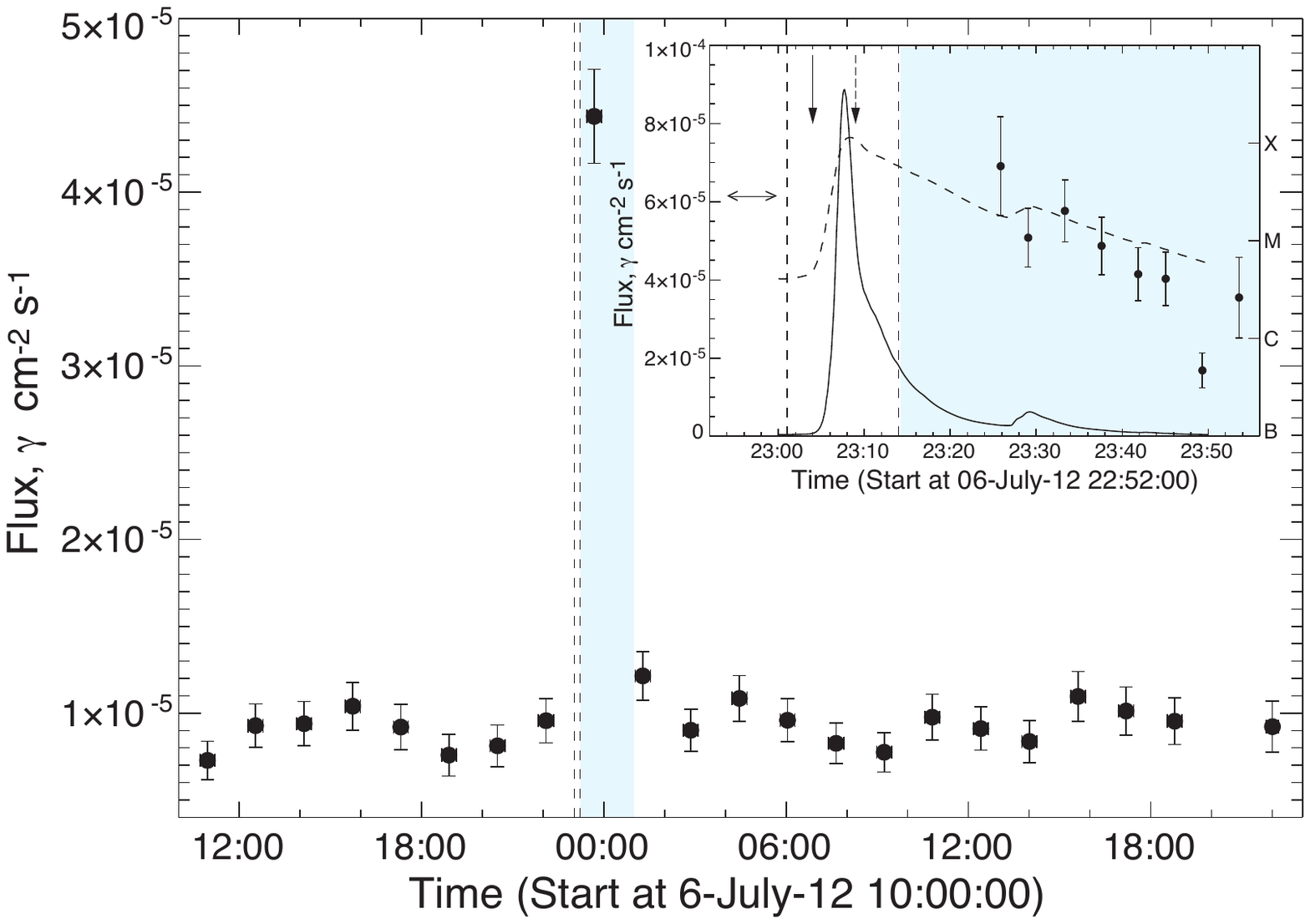}
\caption{Time profile of the 2012 July 6 SGRE event observed by LAT.  As both {\it RHESSI} were in nighttime during the flare, we plot the {\it GOES} 0.5--4{\AA} time history in the inset.  See caption for Figure \ref{110307Ath} for more details. }
\label{120706th}
\end{figure}

{\bf Details}  Plotted in Figure \ref{120706th}.  X1.1 class flare at W52 lasting $\sim$13 minutes;  $\sim$1800 km s$^{-1}$ CME with onset coincident with the rise of the {\it GOES} 0.5--4{\AA} power plotted in the inset; there were no hard X-ray observations during the peak of the flare as both {\it RHESSI} and {\it Fermi} were in nighttime (in lieu of a 100-300 keV X-ray time history, we plot the {\it GOES} 0.5-4{\AA} time history ); M (metric) and DH (decameter-hectometric) Type II emissions observed with the M onset $\sim$5 minutes after the CME launch; moderate SEP event with emission $>$100 MeV.     Listed by \citet{acke14} as both an impulsive and SGRE event but there were no $>$100 MeV observations during the impulsive phase.  Good LAT solar exposures each orbit due to Crab Nebula ToO.  $>$100 MeV $\gamma$ rays observed during the 23:27--23:54 UT solar exposure with a flux (Table \ref{tab:event}) that is consistent with that listed by \citet{acke14}. The inset of the Figure shows the $>$100 MeV emission accumulated in 4-minute intervals over this time period;  the flux appears to be falling. This suggests that the emission may be the decay phase of the impulsive flare, but it could also be a separate component with onset time in the 15 minutes after the impulsive peak.  The best fitting pion-decay spectrum over the full 27-minute exposure is consistent with a proton spectrum with index 5.1, but the spectrum appears to have softened in time: 4.7 $\pm$ 0.8 from 23:27--23:40 UT, and was steeper than 6.0 from 23:40--23:54 UT.  There were $\sim 1 \times 10^{28}$ associated with SGRE assuming an onset just after the impulsive phase; we only have an upper limit on the number of protons in the SEP (Table \ref{tab:event}).

\subsection{SOL2012-10-23T0313}  \label{subsec:20121023}

{\bf Is the SGRE time history distinct from that of the impulsive flare?  Uncertain.}   LAT observed sustained $\gamma$-ray emission during a single 30 minute exposure about 50 minutes after the 5-minute duration impulsive flare on 2012 October 23. There is no evidence for variability in the flux during the exposure and thus we have no information about when the emission began.  The flare itself was observed by both {\it RHESSI} and GBM and the right inset of the Figure \ref{121023th} a shows a high-time resolution plot of the impulsive phase in two GBM energy channels, 100--300 keV (dashed) and $>$9 MeV (solid) revealing two peaks separated by $\sim$15s.  The spectrum of the first peak can be fit by bremsstrahlung from the sum of two power-law electron spectra, one having a steep index 5.1 and the second a hard index of 3; only bremsstrahlung from a steep electron spectrum could fit the second peak.   Although the Sun had just left the LAT aperture, $\gamma$ rays from the first peak scattered in the spacecraft were in LLE data up to $\sim$50 MeV and had a relatively steep spectrum.  This suggests that the impulsive flare was dominated by electron bremsstrahlung and that it was not the source of SGRE.  Thus it is not likely that the there were a sufficient number of impulsive phase protons to produce the SGRE.    This event differed from all of the previous events in that it was not accompanied by a CME.  As discussed below, the upper limit on the number of protons in the impulsive flare was smaller than the number in the SGRE.  Therefore, it is not likely that the tail of the flare was the source of the $>$100 MeV $\gamma$ rays.

\begin{figure}
\epsscale{.80}
\plotone{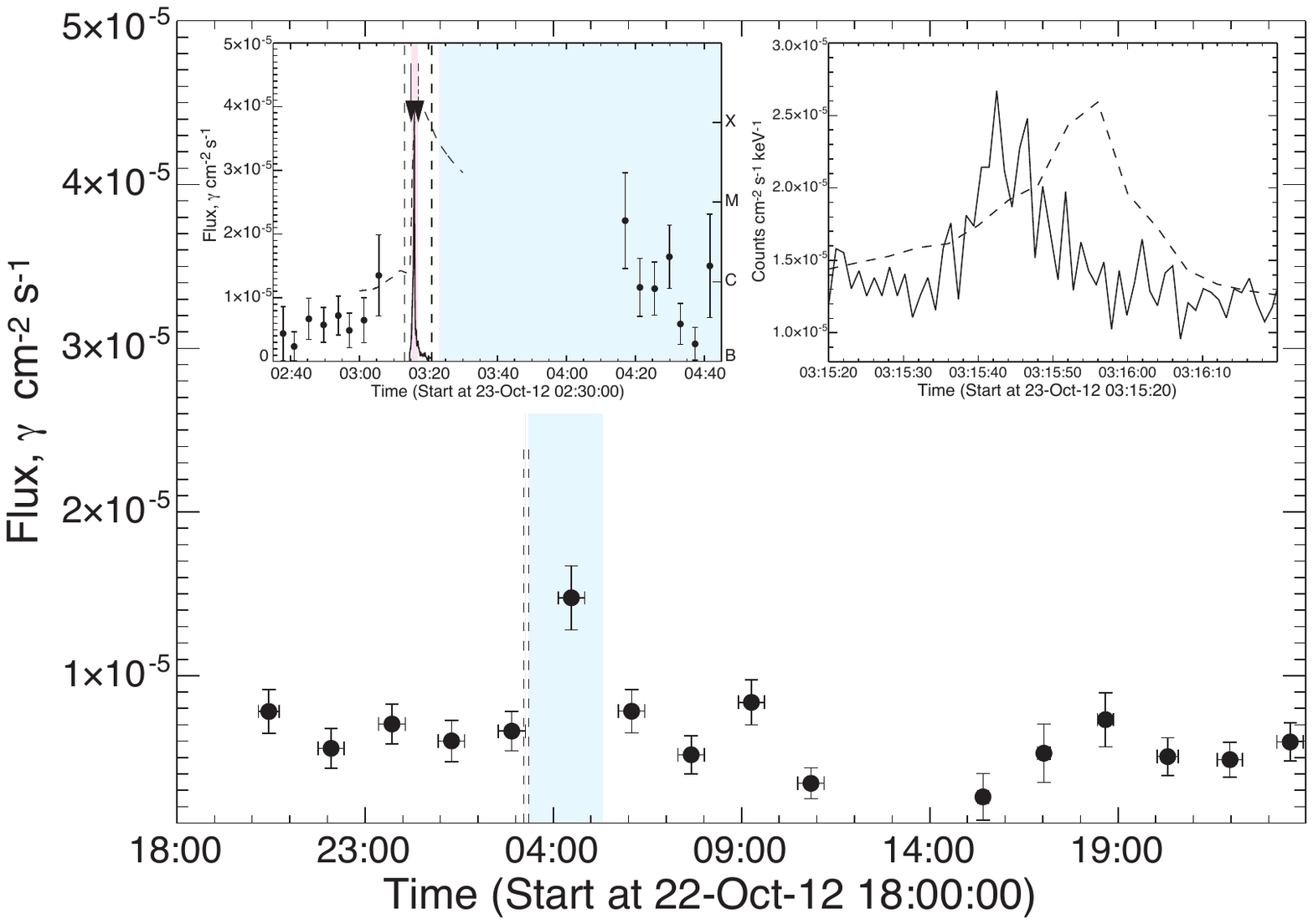}
\caption{Time profile of the 2012 October 23 SGRE event observed by LAT.  Left inset shows GBM 100-300 keV rates plotted on the same scale as $>$100 MeV fluxes derived from source class data.  Right inset compares GBM 100-300 keV (dashed trace) and $>$9 MeV (solid trace) rates.  See caption for Figure \ref{110307Ath} for more details. }
\label{121023th}
\end{figure}

{\bf Details}  Plotted in \ref{121023th}. X1.8 class flare at E57 lasting eight minutes; no CME or SEP detected but Metric Type II radio emission was observed; impulsive hard X-ray and $\gamma$-ray emission observed to $>$9 MeV (see 100--300 keV time history in the left inset).   There is evidence for the eruption of a magnetic loop in AIA 94{\AA} and 131{\AA} images at 03:15 UT (time denoted by the downward arrow in the left inset) along with material moving away from the flare region suggesting that this might have been a failed CME \citep{ji03}.  The SGRE event was only revealed in the Pass8 source class data; it only appeared at 2$\sigma$ significance in the Pass7 source-class data.  LAT had good solar exposures each orbit: 02:34--03:12 (just before the impulsive peak), 04:10--04:40, and  05:45--06:22 UT.  SGRE was only observed between 04:10--04:40 UT and there is no clear evidence for variability during 4-minute integrations.  The onset delay of the $>$100 MeV emission was estimated from the time of the erupting magnetic loop.  The right inset shows a high-time resolution plot of the impulsive phase in two GBM energy channels, 100-300 keV (dashed) and $>$9 MeV (solid) that reveals two peaks separated by $\sim$15s.  The first peak can be fit by the sum of an electron spectra made up of two power-law components, one having a steep spectral index, $\sim$5, and the second a harder spectral index,  $\sim$3; only the steep spectral component of electrons was observed in the second peak and there is no evidence for emission $>$9 MeV.  Although the Sun had just left the aperture of LAT, scattered radiation from the hard first flare peak was observed in LLE data up to about 50 MeV.  Thus, the emission in the impulsive flare was dominated by electron bremsstrahlung.  We estimated the number of $>$500 MeV protons in the SGRE  assuming that is began just after the impulsive flare and peaked in intensity between 04:10 and 04:40 UT.  We estimated the number of protons in the impulsive flare by using the upper limit on the 2.223 MeV capture line.  The results of this study are presented in Table \ref{tab:event} and indicate that the number of $>$500 MeV protons in the flare was only about 20\% of the number producing the SGRE, but the uncertainties are large.

\subsection{SOL2012-11-27T1552}  \label{subsec:20121127}

{\bf Is the SGRE time history distinct from that of the impulsive flare? YES.}  LAT had good exposure throughout the impulsive flare, which was not accompanied by a CME.  $>$100 MeV emission was observed only during this exposure and the fluxes are plotted at 4-minute resolution in the inset along with GBM 100-300 keV rates.  Plots at higher resolution indicate that the SGRE began within a minute of the the narrow impulsive peak observed in hard X-rays.  It lasted for at least 16 minutes, suggesting that it was produced by a distinctly different acceleration process than the flare.  The flare spectrum can be fit with bremsstrahlung from a power-law electron spectrum with a break near 1 MeV.  The upper limit on the number of protons in the flare was more than a factor of ten less than the number in the SGRE based on pion-decay fits to the LAT spectrum.  Therefore, it is not likely that the tail of the flare was the source of the $>$100 MeV $\gamma$ rays.

\begin{figure}
\epsscale{.80}
\plotone{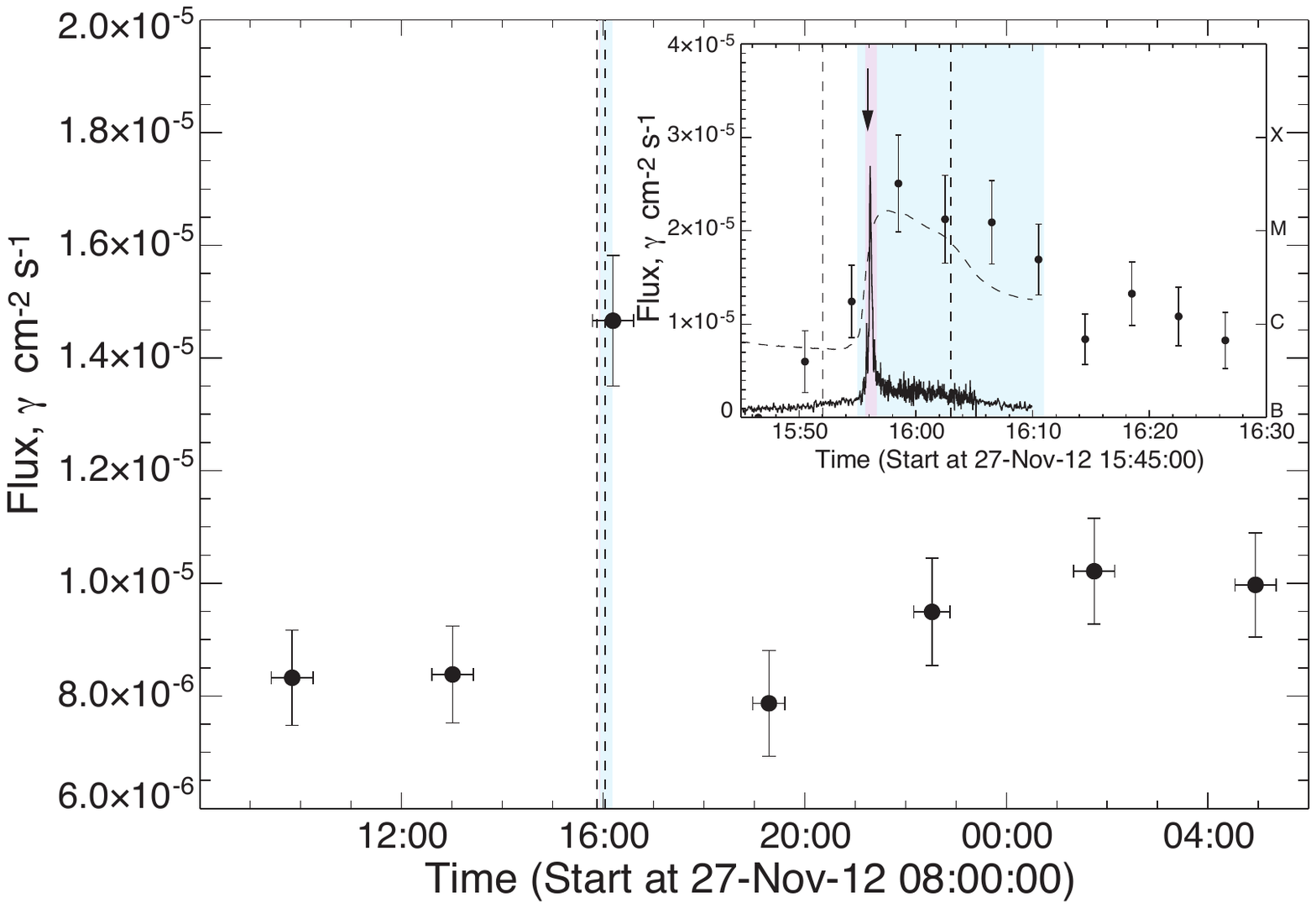}
\caption{Time profile of the 2012 November 27 SGRE event observed by LAT.  Inset displays the expanded flare region with GBM 100--300 keV rates plotted on the same scale as the $>$100 MeV source-class fluxes and $\pm1\sigma$ uncertainties plotted at 4-minute resolution.  See caption for Figure \ref{110307Ath} for more details. }
\label{121127th}
\end{figure}

{\bf Details}  Plotted in Figure \ref{121127th}; M1.6 class flare at W73 lasting $\sim$35 minutes; no CME, Type II emission, or SEP observed; one-minute long impulsive hard X-ray peak reaching energies $>$300 keV observed by GBM; 100--300 keV rates plotted in inset.  There is evidence for the eruption (time denoted by downward arrow in the inset) of a magnetic loop in AIA 171{\AA} at 15:55:50 UT, along with material moving away from the flare region, suggesting that this was a failed CME \citep{ji03}. The Sun was in LAT's field of view during the impulsive phase of the flare. Source-class data could be used for the study because the rates in the ACD from hard X-rays were not too high.  The SGRE flux between 04:10 and 04:40 UT measured using the solar-impulsive-class data is consistent with the flux derived using the source-class data.  We plot the $>$100 MeV $\gamma$ rays at four-minute resolution in the inset to improve the statistical significance of each point.  Plots at one and two minute resolution indicate that the SGRE began within a minute of the X-ray peak and that there no peak in $>$100 MeV $\gamma$-rays coincident with the X-rays.  The SGRE lasted about 16 minutes with marginal evidence for more rapid time variations when plotted at one- and two-minute resolution.  The $>$100 MeV emission appears to have been produced by an acceleration process distinctly different from the flare.  Our fits to the spectra indicate that the protons producing the SGRE followed a power-law with index $\sim$3 that was constant over the observation period. In contrast, the spectrum of electrons producing the flare bremsstrahlung, observed up to 800 keV, followed a power law with index of $\sim$3.7.  There is no evidence for either nuclear de-excitation or 2.223 MeV neutron-capture $\gamma$-rays during the flare in the GBM data.   The number of protons $>$500 MeV in the SGRE was the smallest observed by LAT, but still exceeded the upper limit on the number of protons in the impulsive phase by at least a factor of ten. The onset delay of the $>$100 MeV emission for this event is estimated from the time of the erupting magnetic loop.

\subsection{SOL2013-04-11T06:55}  \label{subsec:20130411}

{\bf Is the SGRE time history distinct from that of the impulsive flare? YES.}  LAT had a good exposure to most of the impulsive flare.   There is no evidence for $>$100 MeV $\gamma$-ray emission during the four-minute long impulsive X-ray peak. The onset of the SGRE was clearly observed at 07:10 UT, about one minute after this peak. It reached a maximum about three minutes later, and lasted only about 20 minutes. 

\begin{figure}
\epsscale{.80}
\plotone{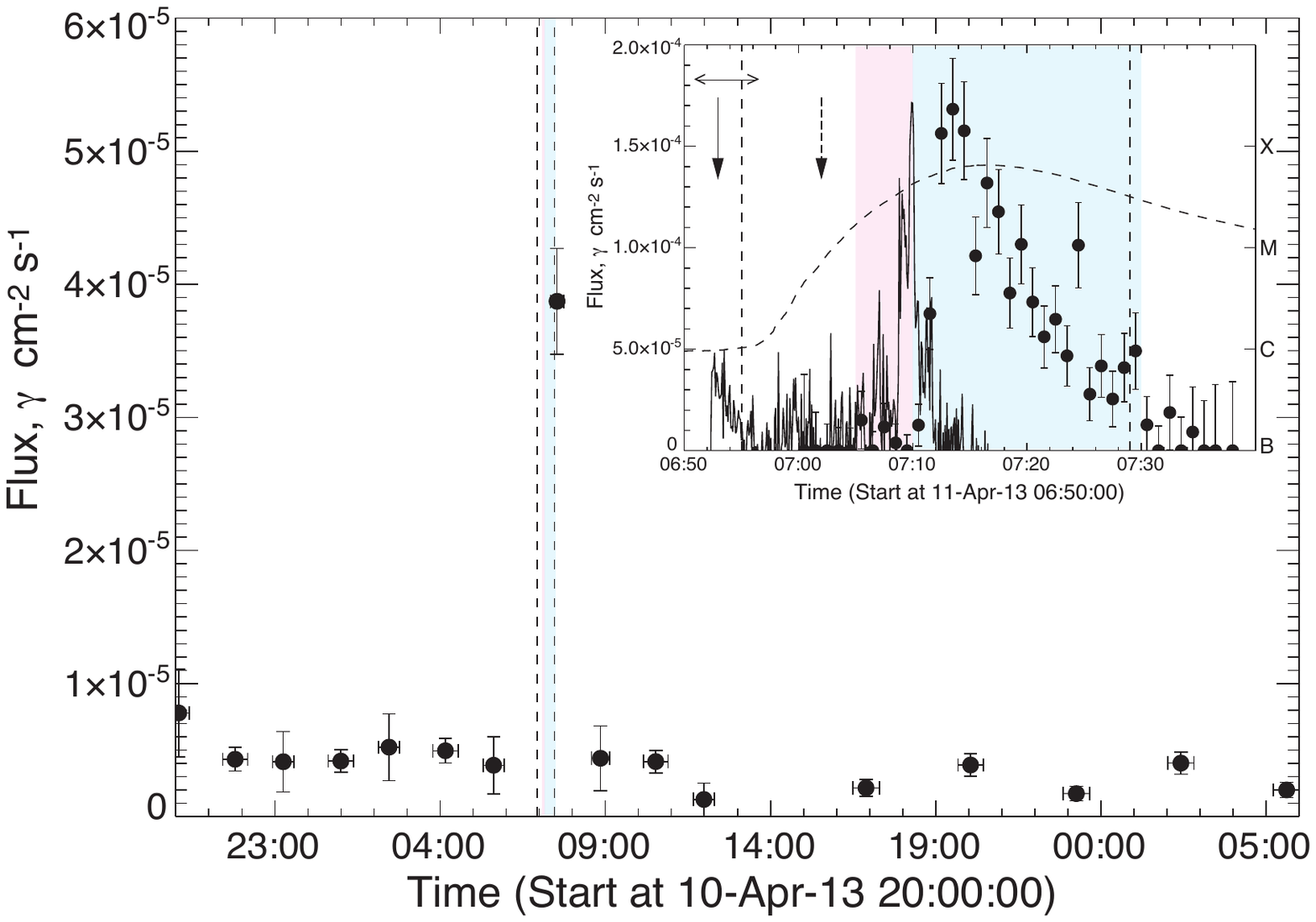}
\caption{Time profile of the 2013 April 11 SGRE event observed by LAT.  Inset shows blowup of flare region with GBM 100-300 keV rates plotted on the same scale as $>$100 MeV fluxes and $\pm1\sigma$ uncertainties derived from fits to solar-impulsive-class data. See caption for Figure \ref{110307Ath} for more details.}
\label{130411th}
\end{figure}

{\bf Details}  Plotted in Figure \ref{130411th}.  M6.5 class flare at W13 lasting $\sim$34 minutes;  $\sim$850 km s$^{-1}$ CME with onset $\sim$15 minutes before the main peak in 100--300 keV flux observed by GBM (plotted in inset); M (metric) and DH (decameter-hectometric) Type II emissions observed with the M onset $\sim$9 minutes after the CME launch; strong SEP event with emission $>$60 MeV.  The Sun was in LAT's field of view during most of the flare but source-class data are only available  between 07:21--07:40 UT due to high rates in the ACD.  However, solar-impulsive data are available beginning at 07:00 UT. The $>$100 MeV fluxes derived from fits to background-subtracted spectra at 1-minute resolution, show no detectable $\gamma$-ray emission during the prominent X-ray peak.  The SGRE began about one-minute later, peaked three minutes after onset, and lasted $\sim$20 minutes.  From fits to spectra in three time intervals during the event, we estimate that $\sim 0.7 \times 10^{28}$ $>$500 MeV protons produced the sustained emission. The proton spectrum $>$300 MeV had an average power-law index of $\sim$5.5 and may have hardened during the event (see Table \ref{tab:event}).  Comparing the 0.05 $\gamma$ cm$^{-2}$ s$^{-1}$ upper limit on the flux in the 2.223 MeV line from 07:10--07:30 UT with the observed flux $>$100 MeV, we estimate that the 40--300 MeV SGRE protons followed a power-law with index harder than 4.5.  This suggests that the SGRE proton spectrum steepened above 300 MeV.  An upper limit on the flux of 2.223 MeV $\gamma$ rays during the impulsive phase of the flare provided a limit on the number of $>$500 MeV protons that is a factor of ten below the number in the SGRE.  An even more constraining limit comes from the upper limit on the $>$100 MeV flux during the hard X-ray peak listed in Table \ref{tab:event}.  The number of $>$500 MeV protons in the SEP was at 2 to 3 orders of magnitude larger than the number the Sun.

\subsection{SOL2013-05-13T01:53} \label{subsec:20130513}

{\bf Is the SGRE time history distinct from that of the impulsive flare? Uncertain.}  LAT had no exposure to the impulsive phase and it only observed $>$100 MeV emission during its first solar exposure two hours later.  We found no evidence for time variation in the flux during this exposure suggesting that the observation may have been made near the peak of the emission.  We note that if the $\gamma$ rays came from the footpoints of the flare at 89E, the flux would have been attenuated by about a factor of 6.  Due to the flare's location near the solar limb, we could not use the 2.223 MeV line to estimate the number of protons in the impulsive flare that could be compared with the number producing the sustained emission. 

{\bf Details}  Plotted in Figure \ref{130513th}.  X1.7 class flare at E89 lasting $\sim$39 minutes;  $\sim$1300 km s$^{-1}$ CME with onset $\sim$ coincident with the 50-100 keV X-ray rise and 10 minutes before the 100--300 keV X-ray rise observed by {\it RHESSI} and GBM; M (metric) and DH (decameter-hectometric) Type II emissions observed with the M onset $\sim$12 minutes after the CME launch; small SEP event with emission $>$60 MeV; 100--300 keV hard X-ray emission observed by {\it RHESSI} (left inset of the figure).  A good LAT exposure ended about the start time of the X-class flare; the next exposure was 04:30--05:15 UT, two hours after the flare and weak $>$100 MeV flux was observed with a flux of $\sim1 \times 10^{-5}$ cm$^{-2}$ s$^{-1}$ produced by a proton spectrum with power-law index $\sim$5.5 (Table \ref{tab:event}).  There was no observable flux variability during the exposure.  If the emission came from the active region it would have been attenuated by about a factor of 6.  There was no significant hard X-ray emission above 300 keV.  Due to the extreme attenuation of the neutron-capture line at E89, we cannot obtain a limit on the numbers of protons at the Sun during the impulsive phase of the flare. 

\begin{figure}
\epsscale{.80}
\plotone{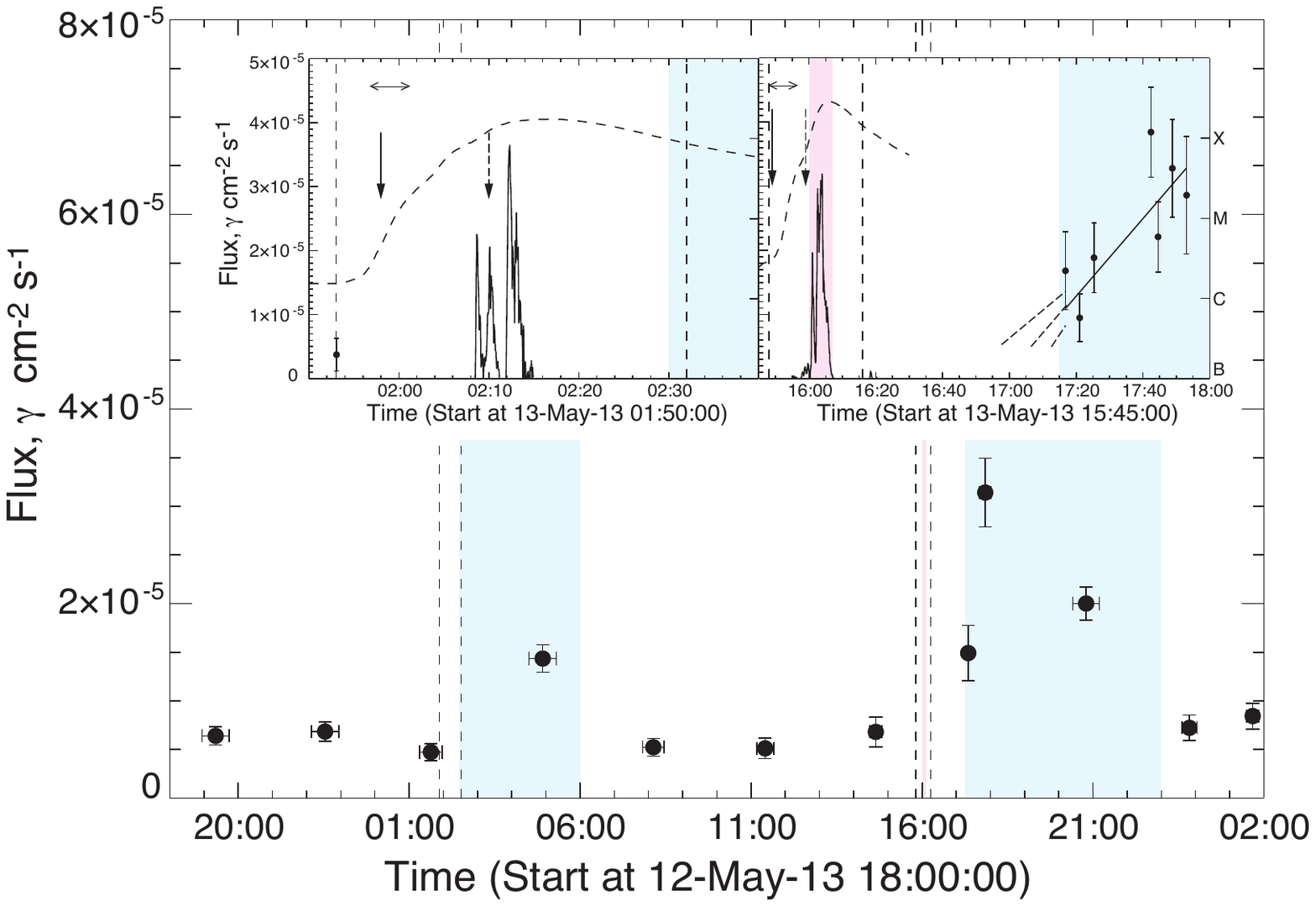}
\caption{Time profiles of two SGRE events on 2013 May 13 observed by LAT.  Blowups of the regions around both flares are shown in the two insets with {\it RHESSI} 100--300 keV time histories scaled to the $>$100 Mev $\gamma$-ray fluxes and $\pm1\sigma$ uncertainties. The best fit to an increasing $>$100 MeV flux after 17:20 UT is shown by the solid line in the right inset.  Extrapolations to determine $\gamma$-ray onset and $\pm1\sigma$ deviations are shown by the dashed lines. See caption for Figure \ref{110307Ath} for more details. }
\label{130513th}
\end{figure}  

\subsection{SOL2013-05-13T15:48}\label{subsec:20130513a}

{\bf Is the SGRE time history distinct from that of the impulsive flare? YES.}   LAT had no exposure to the impulsive flare but it did observe sustained emission in its first solar exposure. The clearly increasing flux in this exposure measured at 4-minute resolution and plotted in the right inset indicates that the SGRE began about an hour after the impulsive flare.  

{\bf Details}  Plotted in Figure \ref{130513th}.  X2.8 class flare at E80 lasting $\sim$28 minutes;  $\sim$1500 km s$^{-1}$ CME with estimated onset $\sim$10 minutes before the 100--300 keV X-ray peaks; M (metric) and DH (decameter-hectometric) Type II emissions observed with the M onset $\sim$12 minutes after the CME launch; strong SEP event with emission $>$60 MeV; hard X-ray and $\gamma$-ray emission observed $>$1 MeV by {\it RHESSI} ({\it RHESSI} 100--300 keV time history plotted in right inset).  Good LAT solar exposures were obtained every other orbit with $>$100 MeV $\gamma$-ray emission observed during the first orbit after the impulsive phase, between 17:15--17:28 and 17:41--17:59 UT (broken by an SAA passage), and during the third orbit between 20:25--21:10 UT.  The emission appears to be rising during the first exposure (see right inset of Figure) with an apparent onset time near 17:00 UT, about one hour after the impulsive flare.  The emission lasted at most five hours.  Fits to the LAT spectra with pion-decay models indicate that the proton spectrum softened between 17:50 and 20:30 UT.  About 3.3 $\times 10^{29} >$500 MeV protons would have been needed to produce the $\gamma$ rays if the interactions took place at a heliocentric angle of 80$^{\circ}$.  {\it RHESSI} had the best exposure to the impulsive flare while GBM began observations in the middle of the flare when its MeV spectrum was dominated by SAA-produced radioactivity.  Only the {\it RHESSI} front detectors could be used to search for line radiation (below 2.4 MeV) because of radiation damage in its rear detectors; the front detector spectral resolution was $\sim$ 30 keV FWHM at 2 MeV.   The impulsive bremsstrahlung spectrum had a spectral index of 2.8 and extended to above 1 MeV; there was also evidence for nuclear de-excitation line emission with 6 $\sigma$ significance, but no evidence for the neutron capture line with a 95\% confidence upper limit on the flux of 0.035 $ \gamma$ cm$^{-2}$ s$^{-1}$.  From this 2.223 MeV line limit the number of protons were at least a factor of five below the number observed during the SGRE phase assuming they all interacted at a heliocentric angle of 80$^{\circ}$ (Table \ref{tab:event}).  This also assumed that the proton spectrum followed a power law with index 4.5 $>$40 MeV.  However, the proton spectrum was likely to be steeper than that because the power-law index between 4 and 40 MeV, determined by comparing the nuclear de-excitation and 2.223 MeV line, was steeper than 5. 

\subsection{SOL2013-05-14T00:00} \label{subsec:20130514}

{\bf Is the SGRE time history distinct from that of the impulsive flare? YES.}   The $>$100 MeV SGRE clearly began about five minutes after the impulsive 100--300 keV X-ray peak.  The SGRE peaked in about one hour and had a more gradual four to five hour decay.   

\begin{figure}
\epsscale{.80}
\plotone{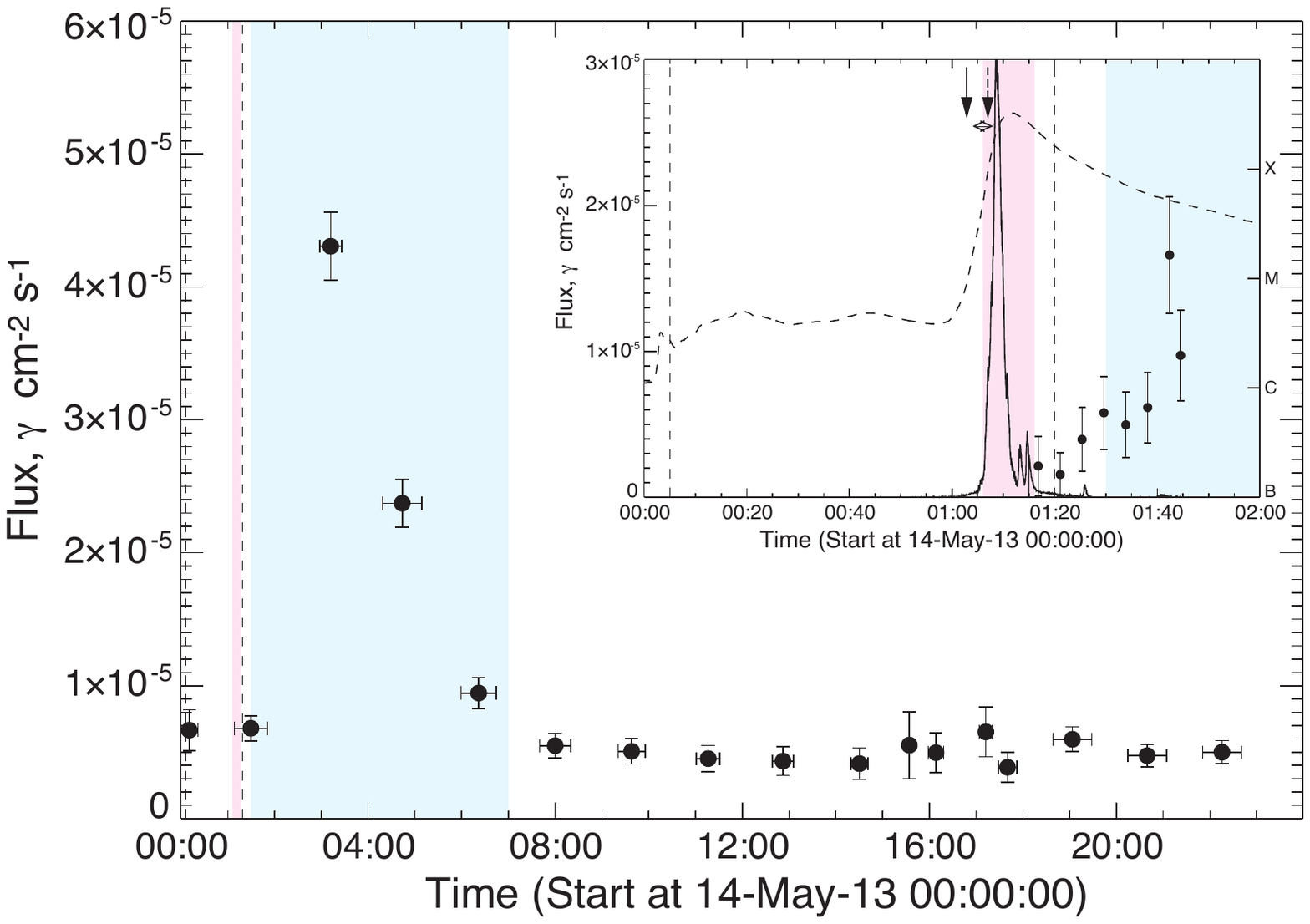}
\caption{Time profile of the 2013 May 14 SGRE event observed by LAT.  Inset shows blowup of flare region with GBM 100--300 keV rates plotted on the same scale as $>$100 MeV fluxes and $\pm1\sigma$ uncertainties derived from source-class data after the flare.  See caption for Figure \ref{110307Ath} for more details. }
\label{130514th}
\end{figure}

{\bf Details}  Plotted in Figure \ref{130514th}.  X3.2-class flare at E77 lasting $\sim$80 minutes (strongest emission began after 01:00 UT and lasted 20 minutes);  $\sim$2600 km s$^{-1}$ CME with estimated onset about one minute before the hard X-ray peak observed by GBM; M (metric) and DH (decameter-hectometric) Type II emissions observed with the M onset $\sim$3 minutes after the CME launch; strong SEP event with emission $>$60 MeV; hard X-ray emission $>$1000 keV observed by GBM; the inset shows the 100--300 keV time history (weak impulsive emission barely visible in the plot continued until 02:00 UT). There were good LAT solar exposures each orbit in response to a ToO on May 13 and even better exposure for two orbits due to an ARR solar pointing in response to the flare.  Pass7 source-class data were originally available from 01:10 -- 01:47 UT, but later versions of it and Pass8 data only began after 01:30 due to concerns about ACD rates.  However, these rates do not appear to be high enough to have seriously affected LAT time histories after 01:15 UT.  The $>$100 MeV flux in the Pass 7 data appears to be rising after 01:15 UT (95\% confidence).  Solar-impulsive-class data are available after 01:11 UT.  We found no evidence for $>$100 MeV emission between 01:11--01:16 UT just after the large X-ray peak.  However, plots of four-minute resolution data are consistent with the rise in flux observed in source-class data.    Other solar exposures were made  02:58--03:23, 04:20--05:06, and 06:01--06:41 UT.   The SGRE peaked in the exposure from 02:58 to 03:23 UT as evidenced by the relatively constant 4-minute resolution fluxes in that exposure.  Our 4-minute resolution plots clearly show that the flux was falling in the next exposure between 04:40--05:06 UT.  From the weak flux observed  between 06:01--06:41 UT, we infer that the emission lasted until about 07:00 UT.  Fits to the background subtracted spectra suggest that the spectrum of protons producing the sustained emission had a steep power-law spectrum (Table \ref{tab:event}).  For an interaction site at 77$^{\circ}$ it required about 5 $\times 10^{28}$ $>$500 MeV protons to produce the SGRE.  The impulsive flare spectrum measured by GBM can be fit by a power-law with index $\sim$2.5 and a weak contribution from nuclear lines.  There is only an upper limit on the neutron-capture line flux of  0.01 $ \gamma$ cm$^{-2}$ s$^{-1}$; this value suggests that the number of $>$500 MeV protons at the Sun during the impulsive phase was at most a few percent of the number in the SGRE phase.

\subsection{SOL2013-05-15T01:25} \label{subsec:20130515}

{\bf Is the SGRE time history distinct from that of the impulsive flare?  YES.}   The sustained emission was relatively weak and below the 4.2 $\sigma$ threshold in our light-bucket analysis; it was identified in the LAT Team's more sensitive Maximum Likelihood study and lasted up to seven hours.   The time history in the main plot suggests that the SGRE phase began no more than an hour after the hard X-ray peak. 

{\bf Details}  Plotted in Figure \ref{130515th}.  X1.2 class flare at E65 lasting $\sim$33 minutes;  $\sim$1350 km s$^{-1}$ CME with estimated onset $\sim$15 minutes before the hard X-ray peak; M (metric) and DH (decameter-hectometric) Type II emissions observed with the M onset $\sim$8 minutes after the CME launch; upper limit on SEP protons due to ongoing event from previous eruption; hard X-ray emission reached 100--300 keV in {\it RHESSI} and GBM (time history in the inset).  Good LAT solar exposures each orbit due to ongoing ToO; source-class solar exposure 01:01--01:33 UT truncated before flare due to high ACD rate. Good source-class exposures 02:37--03:23,  04:13--04:58, 05:52--06:34, 07:33--08:09 UT. This marginal event was noted in the LAT Team's Maximum Likelihood study plotted in the RHESSI browser and was not significant enough to be noted in the Light Bucket study even using the PASS8 source-class data.  As plotted in Figure \ref{130515th} the SGRE was relatively weak and appears to last about six hours.  $>$100 MeV emission appears to have begun before 02:40 UT and there is no evidence for temporal variability during that observing period.  Fit to background-subtracted spectra for the four observations spanning 02:37 to 08:09 UT with pion-decay spectral templates (Table \ref{tab:event}) indicate significant flux levels above 3 $\sigma$ in three time intervals; the proton spectra are typically harder than power laws with indices of --5.  The total number of protons producing the SGRE was $\sim5 \times 10^{27}$.  The upper limit on the number of protons during the impulsive phase derived from the limit on the neutron-capture line is comparable to this.  With solar-impulsive-class data we obtained an upper limit on the $>$100 MeV flux covering the impulsive phase from 01:36--01:46 UT and a limit on the number of protons $>$500 MeV fewer than 5\% of that producing the SGRE.

\begin{figure}
\epsscale{.80}
\plotone{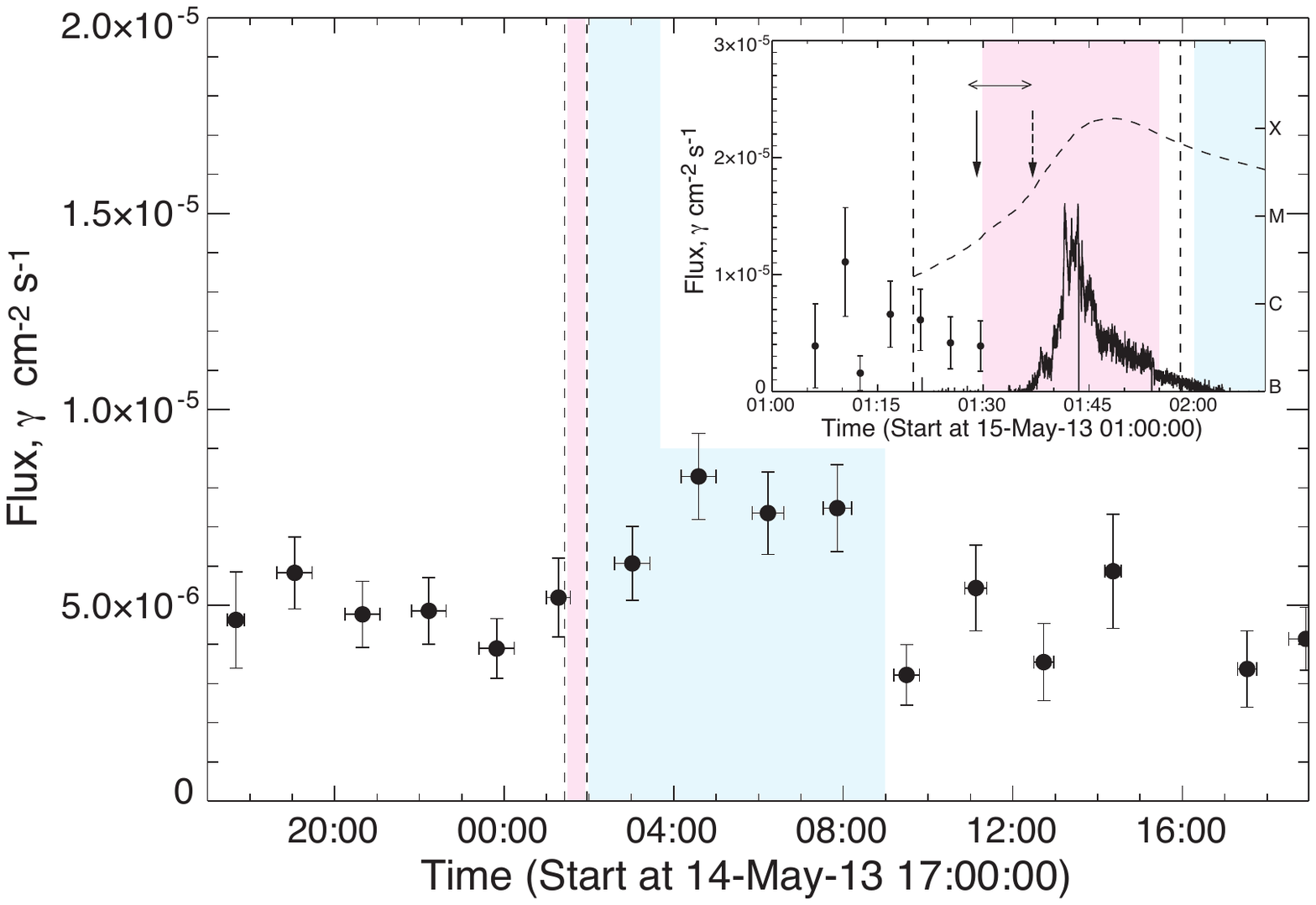}
\caption{Time profile of the 2013 May 15 SGRE event observed by LAT.  Inset shows blowup of flare region with GBM 100--300 keV rates plotted on the same scale as $>$100 MeV fluxes and $\pm1\sigma$ uncertainties derived from source-class data before the flare. See caption for Figure \ref{110307Ath} for more details. }
\label{130515th}
\end{figure}

\subsection{SOL2013-10-11T07:01} \label{subsec:20131011}

{\bf Is the SGRE time history distinct from that of the impulsive flare?  YES.}   This was the first SGRE event detected from a flare behind the solar limb \citep{pesc15,acke17}. The hard X-rays from the flare at E103 peaked about three minutes before the onset of the SGRE.  This can be seen in Figure \ref{131011th} by comparing hard X-ray time profile (green dots) estimated from the derivative of the soft X-ray rates observed by {\it MESSENGER} and the $>$100 MeV flux observed by LAT (data points with uncertainties). 

\begin{figure}
\epsscale{.80}
\plotone{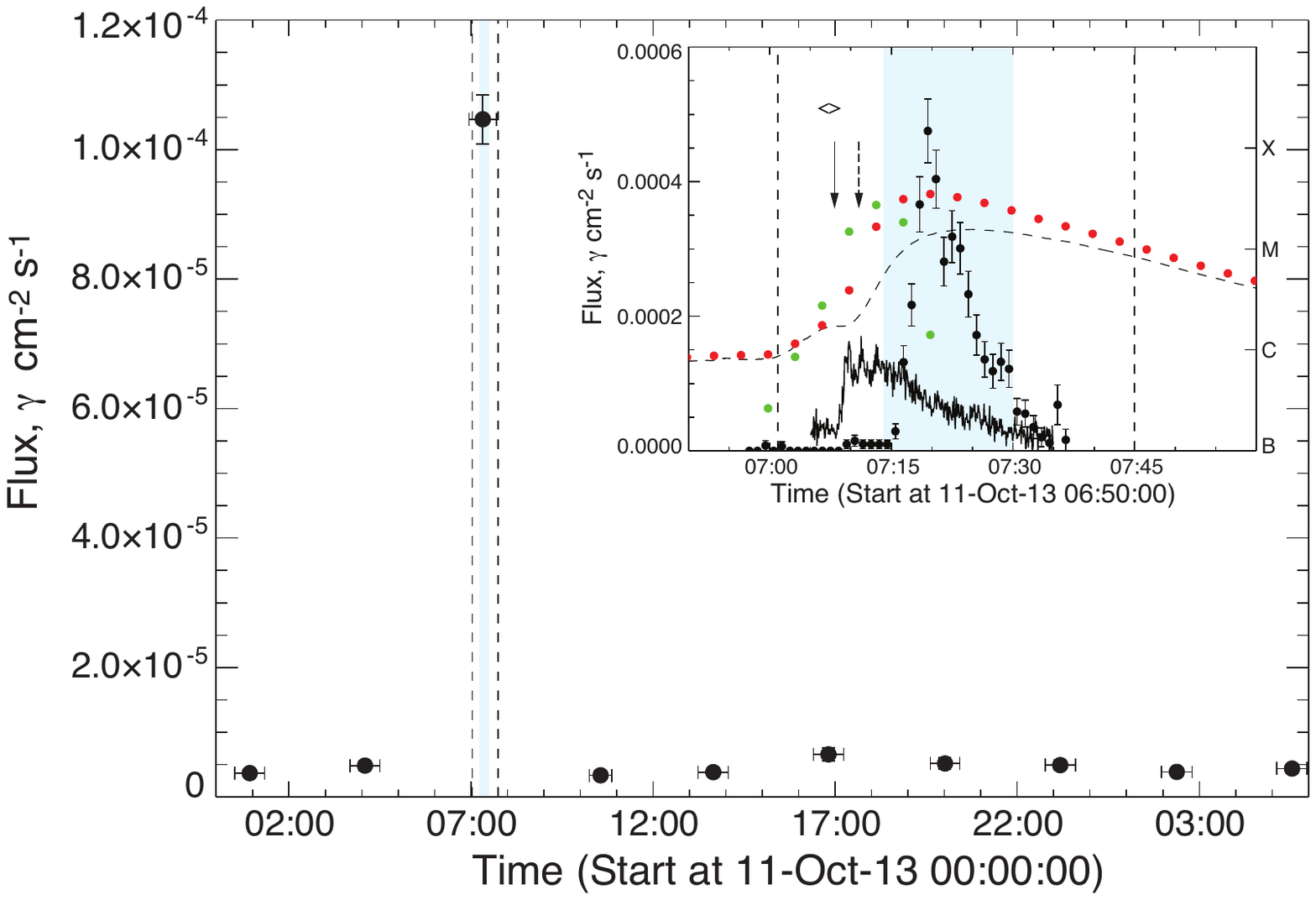}
\caption{Time profile of the 2013 October 11 SGRE event observed by LAT.  The inset shows GBM 100-300 keV rates plotted on the same scale as $>$100 MeV $\gamma$-ray fluxes $\pm1\sigma$ uncertainties derived from source-class data.  In addition to the {\it GOES} X-ray plot we also show as red dots soft X-ray rates from the {\it MESSENGER} SAX instrument that observed the flare region.  The time derivative of this flux is a good representation of the flare hard X-ray emission that we plot as green dots.  See caption for Figure \ref{110307Ath} for more details. }
\label{131011th}
\end{figure}

{\bf Details}  Plotted in Figure \ref{131011th}.  Estimated \citep{pesc15,pesc15c} M4.9 class flare beyond the limb at E106 lasting $\sim$44 minutes;  $\sim$1200 km s$^{-1}$ CME with onset within two minutes of the rise of hard X-rays observed by GBM; M (metric) and DH (decameter-hectometric) Type II emissions observed with the M onset $\sim$3 minutes after the CME launch; strong SEP emission extending to energies $>$60 MeV; hard X-ray emission only observed to 50--100 keV.  Good LAT solar exposures every other orbit: 06:58--0740, 10:16--10:50 UT.  Emission $>$100 MeV was only observed during the first orbit. The 50--100 keV GBM time history (also observed by RHESSI) plotted in the inset of the figure shows an abrupt increase at about 07:08 UT followed by a slower decay lasting until about 07:35 UT.  The red dots plotted at 3.5-minute resolution follow the 1--4 keV time history observed by the Solar Assembly for X-rays (SAX) instrument on {\it MESSENGER} \citep{schl07} that directly observed the flare. This emission preceded the occulted soft X-ray emission observed by {\it GOES} by $\sim$2--3 minutes.  The derivative of the SAX 1--4 keV emission (shown by the green dots) reflects the hard X-ray time history of the flare and has an onset near 07:00 UT.   Thus, the sharp rise in hard X-rays observed by GBM at 07:08 UT is likely to be due to the emission region rising above the solar limb.  { \it RHESSI} hard X-ray images up to 50 keV indicate that the source of the emission was above the Sun's limb \citep{pesc15, acke17}.  The background-subtracted hard X-ray spectrum observed by GBM from 38 to 200 keV between 07:09 and 07:12 UT, before the start of the SGRE, can be fit acceptably (probability 10\%) by an electron spectrum with power-law index 5.0 $\pm$ 0.1 interacting in a thick-target or by an electron spectrum with index 3.3 $\pm$ 0.06 interacting in a thin target.   We note that there is an artifact in GBM NaI spectrum that prevents us from fitting $<$38 keV. 

The time profile of the $>$100 MeV $\gamma$-ray flux, plotted at 1-minute resolution, reveals an increase  beginning at 07:15 UT that peaks in five minutes and falls back to background by about 07:35 UT.  The $\gamma$-ray onset occurred about 15 minutes after the inferred onset of hard X-ray emission observed by {\it MESSENGER} SAX.   This difference in temporal structure suggests that the protons producing the $>$100 MeV emission were accelerated by a second process and were transported to the visible disk where they interacted deep in the solar atmosphere.  \citet{plot17} studied the timing in detail and concluded that the $\gamma$-ray onset occurred just after protons accelerated by the CME shock reached magnetic field lines that reached the visible disk of the Sun.  The centroid of the $>$100 MeV emission by LAT is consistent with a location near the East limb of the Sun at N03E62 with a 1$\sigma$ range in longitude from E39 to just above the Eastern limb \citep{pesc15}.  The background-subtracted $\gamma$-ray spectrum from 07:14--07:30 UT can be fit with a pion-decay template for a power-law proton spectrum with an index of $\sim$ 3.8 (Table \ref{tab:event}) with no evidence for spectral variation during the rising and falling phases.  There is evidence that a better fit would be achieved with a proton spectrum rolling over at energies above 500 MeV.  Assuming that the protons impacted at E85, we estimate that the their total number was $\sim 3.5 \times 10^{28}$; because we do not know the true interaction location, this number is probably uncertain by a factor of 5.  As the flare site was beyond the limb we have no estimate of the number of protons during the impulsive phase.  There is also no evidence for the presence of a solar 2.223 MeV neutron capture line in GBM spectra between 07:15-- 07:30 UT.  Because the $\gamma$ rays in this line would be highly attenuated near the solar limb, we could not obtain information on the proton spectrum below 300 MeV.

\subsection{SOL2013-10-25T07:53} \label{subsec:20131025}

{\bf Is the SGRE time history distinct from that of the impulsive flare? Uncertain.}  LAT had good solar exposure beginning about 15 minutes after the impulsive hard X-ray peak and significant $>$100 MeV emission was detected.  There is a suggestion (90\% confidence) that the $>$100 MeV flux, plotted at 4-minute resolution, was decreasing during that time interval, and there is no evidence for emission in the next good exposure three hours later.  We are, therefore, not able to determine whether the observed emission comes from the tail of the impulsive flare or from SGRE beginning after the flare.  The 95\% confidence limit on the number of impulsive flare protons $>$500 MeV is comparable to the number protons in the SGRE phase.  Therefore, there is no significant constraint on whether the impulsive flare was the source of the sustained emission. 

\begin{figure}
\epsscale{.80}
\plotone{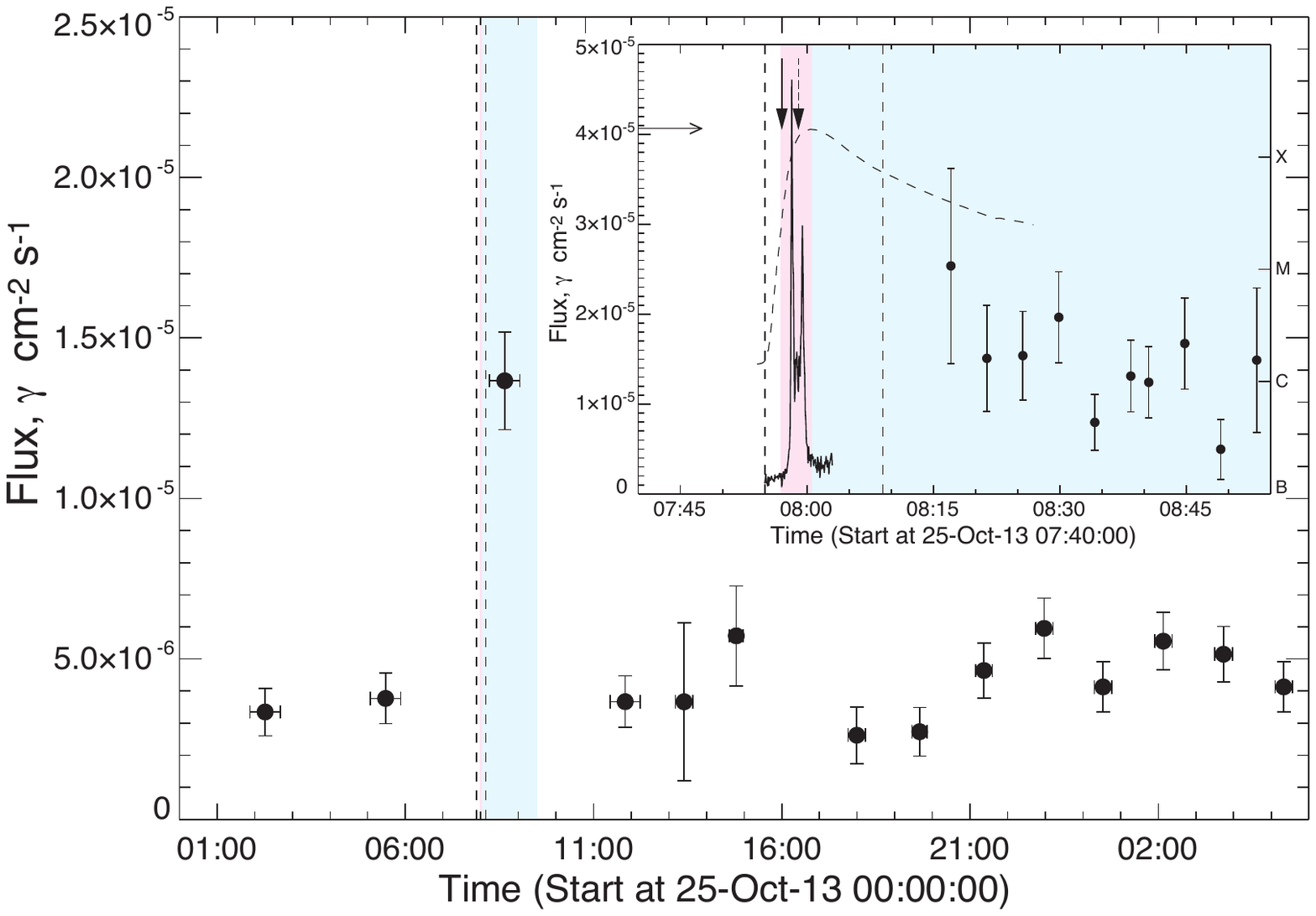}
\caption{Time profile of the 2013 October 25 SGRE event observed by LAT.  The inset shows {\it RHESSI} 100-300 keV rates plotted on the same scale as $>$100 MeV $\gamma$-ray fluxes and $\pm1\sigma$ uncertainties after the flare, derived from source-class data.  See caption for Figure \ref{110307Ath} for more details. }
\label{131025th}
\end{figure}

{\bf Details}  Plotted in Figure \ref{131025th}.  X1.7 class flare at E71 lasting $\sim$16 minutes; a relatively slow $\sim$590 km s$^{-1}$ halo CME with onset coincident with the rise in 100--300 keV X-rays; M (metric) and DH (decameter-hectometric) Type II emissions observed with the M onset $\sim$2 minutes after the CME launch; moderate SEP emission extending to energies $>$60 MeV; hard X-ray emission observed up to in excess of 800 keV by {\it RHESSI} (100--300 keV time history plotted in inset).  Good LAT solar exposures every other orbit: 08:14--08.59. 11:26--12:10 UT with a weak $<$20\% exposure 10:00--10:24 UT.  Emission $>$100 MeV was only observed during the first solar exposure, beginning $\sim$15 minutes after the hard X-ray peak; 4-minute accumulations of LAT source-class data suggest (95\% confidence) that the $>$100 MeV emission was decreasing during the 08:14--08.59 UT exposure; therefore, it is not clear whether the emission is just the tail of flare or an associated event beginning after the impulsive hard X-rays. The spectrum of protons producing the pion-decay $\gamma$ rays was relatively steep with a power-law index between 4 and 7 (Table \ref{tab:event}).  We estimate that $\sim 3 \times 10^{27} >$500 MeV protons were required to produce the SGRE, assuming that it began just after the impulsive flare and lasted until 09:30 UT.  {\it RHESSI} front detectors were used to study the impulsive phase (note that due to radiation damage the spectral resolution was $\sim$20 keV at 511 keV); emission was observed up to about 1500 keV.  There is a hint of nuclear-line emission in the spectrum with about 70\% confidence and only an upper limit on the neutron-capture line flux.  This 95\% limit was used to set an upper limit on the number of $>$500 MeV protons that is comparable to the number observed in the SGRE phase.

\subsection{SOL2013-10-28T15:07} \label{subsec:20131028}

{\bf Is the SGRE time history distinct from that of the impulsive flare?  Uncertain.}   LAT had good exposure about 30 minutes after the impulsive flare and significant $>$100 MeV emission was detected.  There is a weak suggestion (80\% confidence) that the $>$100 MeV flux, plotted at 4-minute resolution, was decreasing during that time interval, and there is no evidence for emission in the next good exposure 90 minutes later.  We are therefore not able to determine whether the observed emission is just the tail of impulsive phase radiation or a sustained-emission component.  We have no information on the number of flare-produced protons that could constrain the flare contribution to the SGRE proton numbers. 

\begin{figure}
\epsscale{.80}
\plotone{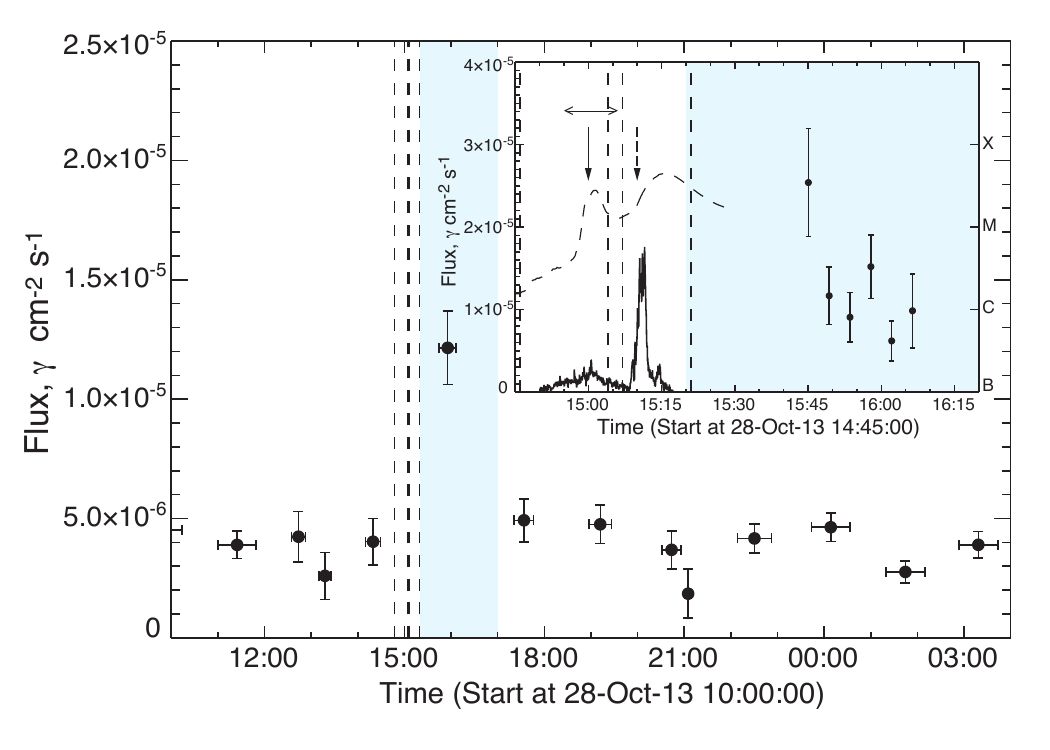}
\caption{Time profile of the 2013 October 28 SGRE event observed by LAT.  The inset shows {\it RHESSI} 100-300 keV rates plotted on the same scale as $>$100 MeV $\gamma$-ray fluxes and $\pm1\sigma$ uncertainties after the flare, derived from source-class data. See caption for Figure \ref{110307Ath} for more details. }
\label{131028th}
\end{figure}

{\bf Details}  Plotted in Figure \ref{131028th}.  M4.4 class flare at E28 lasting $\sim$14 minutes preceded by a weaker flare;  $\sim$800 km s$^{-1}$ CME with onset $\sim$7 minutes before rise of prominent 50-100 keV X-ray peak observed by {\it RHESSI} and GBM; M (metric) and DH (decameter-hectometric) Type II emissions observed with the M onset $\sim$10 minutes after the CME launch; weak SEP emission; hard X-ray emission observed up to just above 100 keV by {\it RHESSI} ({\it RHESSI} 50--100 keV time history plotted in inset).  {\it Fermi} was performing a solar ToO but the exposures were shortened by SAA passages; good solar exposures between 15:46 and 16:06 UT and between 17:21 and 17:40 UT. Emission $>$100 MeV was only observed during the first exposure, beginning $\sim$35 minutes after the hard X-ray peak associated with the second M-class flare; inset shows 4-minute $>$100 MeV accumulations with evidence that the flux was falling (80\% confidence).  Fit to time-integrated $\gamma$-ray spectrum indicates that accelerated proton spectrum was hard (Table \ref{tab:event}). We estimated the number of $>$500 MeV protons at the Sun by assuming that the sustained emission flux peaked at the time of the observation and decreased after that time.  {\it RHESSI} hard X-ray spectrum follows a power law up to about 140 keV. The poor quality of the $\gamma$-ray spectrum due to radiation damage prevented measurement of the neutron-capture line flux and ability to place a constraint on the number of protons during the impulsive flare.  Therefore, we have no information on the number of flare-produced protons that could constrain the flare contribution to the SGRE proton numbers.  We note that there was an impulsive flare on the same day beginning at 01:41 UT with hard X-ray emission extending up to about 1 MeV, but not associated with any SGRE.

\subsection{SOL2014-02-25T00:39} \label{subsec:20140225}

{\bf Is the SGRE time history distinct from that of the impulsive flare? YES.}   LAT began observing intense $>$100 MeV solar $\gamma$-ray emission $\sim$20 minutes after the impulsive flare.  The emission had an onset before 01:00 UT and increased to a maximum near 01:22 UT, after which time it began to fall.  The fluxes during the next exposure three hours later continued to decrease.   

\begin{figure}
\epsscale{.80}
\plotone{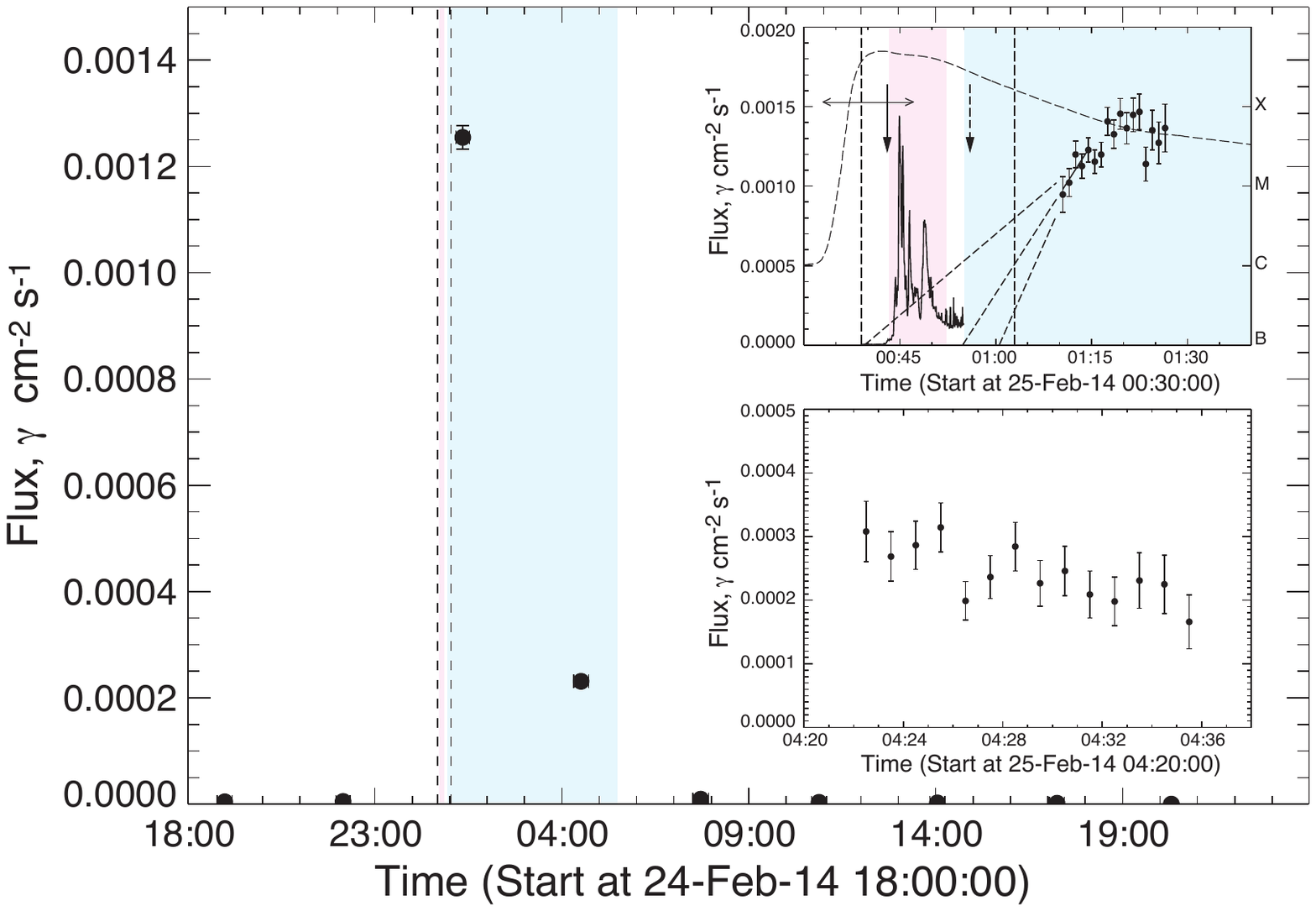}
\caption{Time profile of the 2014 February 25 SGRE event observed by LAT.  Top inset shows a blowup of the flare region with {\it RHESSI} 100-300 keV rates plotted on the same scale as $>$100 MeV $\gamma$-ray fluxes derived from source-class data.  The best fit to an increasing $>$100 MeV flux after 01:10 UT is shown by the solid line in the right inset.  Extrapolations to determine $\gamma$-ray onset and $\pm1\sigma$ deviations are shown by the dashed lines. Bottom inset is a blowup of the 2$^{nd}$ LAT solar exposure showing a falling $\gamma$-ray flux.  See caption for Figure \ref{110307Ath} for more details. }
\label{140225th}
\end{figure}

{\bf Details}  Plotted in Figure \ref{140225th}.  X4.9 class flare at E78 lasting $\sim$24 minutes;  $\sim$2150 km s$^{-1}$ halo CME with onset coincident with the rise of 100-300 keV X-ray emission; M (metric) and DH (decameter-hectometric) Type II emissions observed with the M onset $\sim$13 minutes after the CME launch; strong SEP emission extending to energies $>$700 MeV; hard X-ray/$\gamma$-ray line emission observed into the MeV range by both {\it RHESSI} (100--300 keV time history plotted in inset) and GBM.  {\it Fermi} came into daylight at about 00:39 UT.  Both GBM and {\it RHESSI} observed the entire impulsive phase of the flare. LAT was observing the Galactic Center at the time of the flare but had a good solar exposure from 01:11--01:30 UT.  It also had a good exposure three hours later from 04:21--04:40 UT.  LAT observed emission $>$100 MeV during both these exposures, but not three hours later.  For both the 01:25 and 04:36 UT observations the exposure rapidly increased as Fermi slewed from the Galactic Center to its normal rocking position, peaked for about two minutes and then decreased over the next 10-15 minutes as the Sun left the FoV.   The upper inset shows the 100-300 keV time history observed by {\it RHESSI} and 1-minute resolution time history observed by LAT.  The $>$100 MeV emission rose to a peak near 01:25 UT.  The lower inset shows the falling $>$100 MeV intensity during the exposure beginning 04:20 UT.  In $\S$\ref{subsec:latspect} we discussed fits to the background-subtracted spectrum $>$100 MeV at the peak of the LAT exposure from 01:13:30--01:17:30 where instrumental effects are minimum.  We showed that the data require a pion-decay spectrum produced by a power-law proton spectrum with a break at about 1.3 GeV.  We used this fit and a single power-law fit to data between the peak exposure between 04:24--04:30 UT to estimate the numbers of $>$500 MeV protons at the Sun (Table \ref{tab:event}).  The spectrum softened significantly in the three hours between the observations.

GBM observed clear $\gamma$-ray lines above 2 MeV, including the neutron capture, carbon, and oxygen lines.  The nuclear spectrum dropped precipitously above 7.5 MeV and there was no evidence for any emission above 10 MeV. We fit the background-subtracted GBM spectrum with a power-law and exponentiated power law, along with 2.2 and 0.511 MeV lines, narrow and broad nuclear line and pion-decay  templates.  It concerns us that the fitted width of the solar 2.2 MeV line is $\sim$85 keV. From the measured 2.2 MeV flux of 0.14 $\gamma$ cm$^{-2}$ s$^{-1}$ we estimate that there were 3 $\times 10^{28}$ protons $>$500 MeV protons at the Sun during the impulsive phase assuming the proton spectrum followed a power law with index 4.5 above 50 MeV.   This value is less than 10\% of the number during the sustained-emission phase.

\subsection{SOL2014-09-01T10:54} \label{subsec:20140901}

{\bf Is the SGRE time history distinct from that of the impulsive flare? YES.}   $>$100 MeV emission began about seven minutes after the onset of hard X-rays from the flare located nearly 50$^{\circ}$ beyond the East Limb of the Sun \citep{pesc15c}.  The emission continued for about six hours. 

\begin{figure}
\epsscale{.80}
\plotone{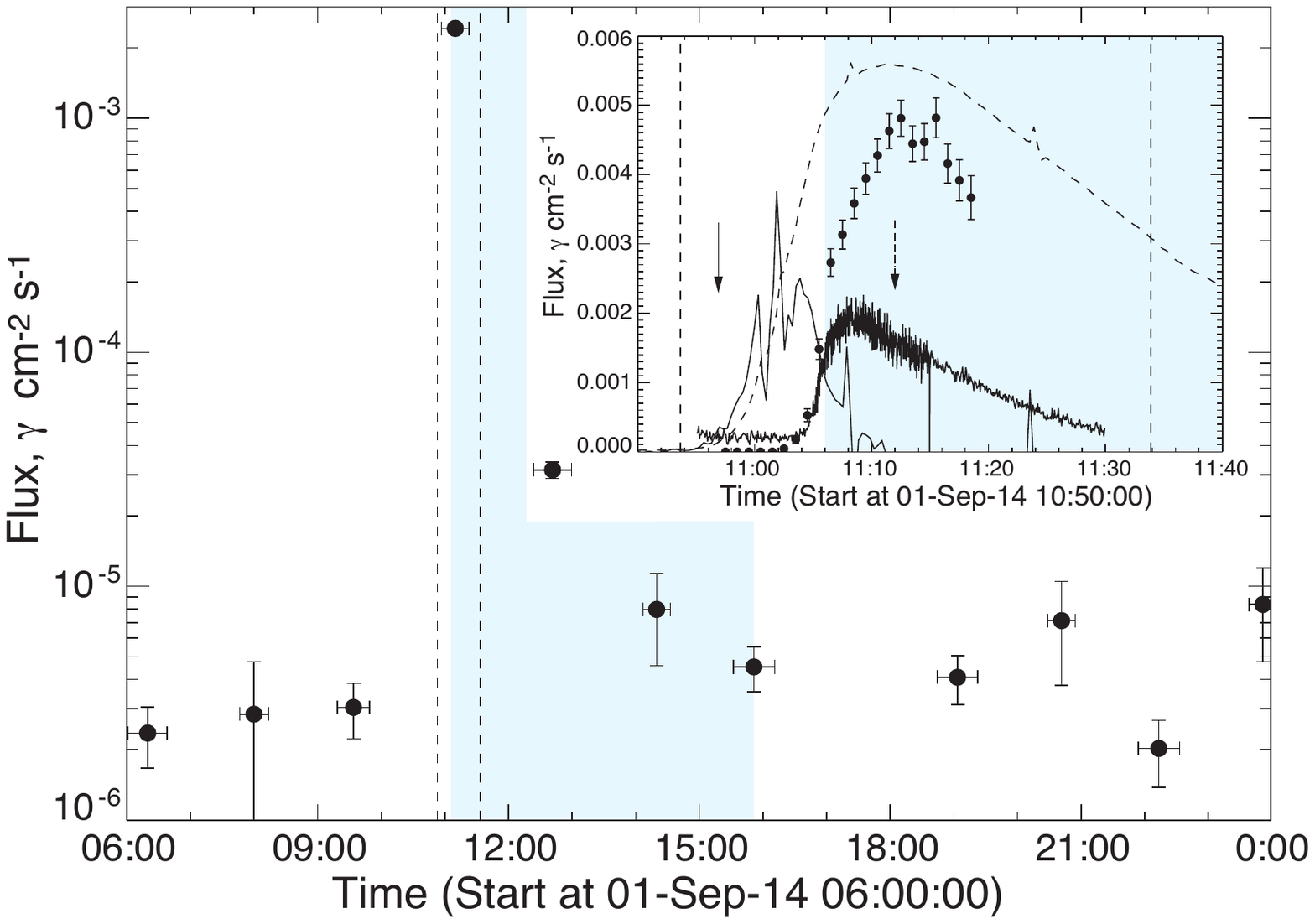}
\caption{Time profile of the 2014 September 1 SGRE event observed by LAT.  The inset shows GBM 100-300 keV rates plotted on the same scale as $>$100 MeV $\gamma$-ray fluxes derived from source-class data.   The dashed curve shows the soft X-ray rates from the {\it MESSENGER} SAX instrument that observed the flare region. The dashed vertical lines are our estimate of the equivalent {\it GOES} start and stop time of this behind-the-limb flare.  The solid trace is a good representation of the flare hard X-ray time history estimated by taking the time derivative of the soft X-ray rates.  See caption for Figure \ref{110307Ath} for more details.}
\label{140901th}
\end{figure}

{\bf Details}  Plotted in Figure \ref{140901th}. Estimated X2.1 {\it GOES} soft X-ray class flare \citep{acke17} at E138 that was observed by the Solar Assembly for X-rays (SAX) instrument on {\it MESSENGER} \citep{schl07} and lasted $\sim$40 minutes (dashed curve in inset of figure); $\sim$1500 km s$^{-1}$ CME with estimated onset time of 10:57 UT, from {\it SDO} 193{\AA}, 211{\AA} images, just at the rise of the inferred flare hard X-ray emission plotted as a solid trace in the inset (estimated by taking the derivative of the SAX soft X-ray time history); M (metric) and DH (decameter-hectometric) Type II emissions observed with the late M onset at 11:13 UT likely the time when the shock was first visible from Earth; very intense SEP emission with comparable peak fluxes of 0.7--4.0 MeV electrons and $>$13 MeV protons consistent with what has been observed in other CME/shock gradual SEP events. 

LAT had good solar exposures every other orbit on September 1 and had exposures four-times smaller in the intervening orbits.  Such a 25\% exposure occurred between 11:06 and 11:20 UT during the behind-the-limb flare when the Sun was at a large angle with respect the LAT telescope axis.  Source-class data could be used to study $>$100 MeV $\gamma$-ray emission because the intense hard X-ray emission from the flare did not reach {\it Fermi}.  At these large solar viewing angles the detector response is small and not as accurately determined.  There were also two good exposures between 12:26--12:58 and 15:36--16:08 UT during which LAT had significantly higher sensitivity to search for delayed high-energy emission.  The $>$100 MeV flux just after the flare was the largest observed by LAT with the exception of the first peak observed on 2012 March 7 just after the X5.6 flare.  The hourly fluxes are plotted logarithmically in the figure in order to reveal the large range in intensity of SGRE that lasted up to six hours,  In the inset we plot $>$100 MeV fluxes at 1-minute resolution along with 100--300 keV rates observed by GBM.  Both the $\gamma$-ray and 100--300 keV X-ray emissions, as viewed from Earth, appear to rise at $\sim$11:04 UT, about seven minutes after the onset of the hard X-ray emission observed from the flare as viewed by {\it MESSENGER}.  The hard X-rays observed by GBM peaked by 11:08 UT while the $>$100 MeV $\gamma$-ray flux peaked about five minutes later. 

We have fit the background-subtracted $>$100 MeV $\gamma$-ray spectrum with a pion-decay spectrum produced by $>$300 MeV protons following a differential power-law spectrum and interacting in a thick target.  Our fits indicate that the spectrum hardened over the duration of the event with spectral indices of 4.25 $\pm$ 0.15, 3.85 $\pm$ 0.1, and 3.45 $\pm$ 0.35, at 11:06--11:12, 11:12--11:20, 12:26--12:58 UT, respectively.  

The NaI detectors on GBM observed a sharp rise in flux from $\sim$10--30 keV about two minutes after the flare onset detected by SAX on {\it MESSENGER}.  This increase was similar to that observed in hard X-rays during the 2013 October 11 event and appears to be due to the appearance of the flare's coronal hard X-ray source above the solar limb.  Higher energy emission observed into the MeV range began as the hard X-ray emission from the flare decreased in intensity.   Our fit to the GBM NaI detector spectrum with the best view of the Sun between 42 and 900 keV from 11:06 to 11:15 UT was acceptable (27\% probability) for thick target  bremsstrahlung model from electrons following a power law spectrum with index 3.2 $\pm$ 0.1 and having low-energy cutoff of $\sim$130 keV.  There is no evidence for spectral variability.  The fact that our fit to the NaI spectrum with thin target bremsstrahlung was not acceptable ($<10^{-6}$ probability) indicates that the electrons interacted deep in the solar atmosphere. This is confirmed by our fit to GBM BGO data up to 40 MeV which was also consistent with thick target bremsstrahlung from electrons following a power-law with index 3.2 $\pm$ 0.1.  Observation of such an energetic population of electrons at the same time as the SGRE suggests that they were accelerated by the same process responsible for the $>$300 MeV protons  (see $\S$\ref{subsec:electrons}).

There is no evidence for the presence of a 2.23 MeV line in the background-subtracted GBM spectrum from 11:04 to 11:30 UT with a 95\% confidence upper limit of 0.016 $\gamma$ cm$^{-2}$ s$^{-1}$.  Three of the rear {\it RHESSI} detectors had moderate spectral resolution at that time, due to a recent anneal, allowing us to search for solar nuclear de-excitation and neutron-capture lines.   Because of contamination from a preceding SAA passage, we could not perform a sensitive search for the de-excitation lines, but we were able to set a 95\% confidence upper limit of 0.028 cm$^{-2}$ s$^{-1}$ from 11:11 to 11:31 UT on the flux in the neutron-capture line, consistent with the GBM result.  Comparing the $>$100 MeV $\gamma$-ray fluence with the upper limit on the 2.223 MeV line fluence we estimate with 95\% confidence that the proton power-law spectral index between 40 and 300 MeV was harder than 3.4, assuming that the protons interacted at a heliocentric angle of 85$^{\circ}$.  For smaller heliocentric angles the index would be even harder.  Because the spectral index for protons $>$300 MeV was $\sim$4 at the same time interval we conclude that the SGRE proton spectrum steepened above a few hundred MeV.   Our estimate of the number of $>$500 MeV protons was also made assuming that the interactions occurred at 85$^{\circ}$.  As there are no measurements of $\gamma$-rays during the flare, we cannot compare this number with number of protons at the flare site.

\subsection{SOL2015-06-21T02:03} \label{subsec:20150621}

{\bf Is the SGRE time history distinct from that of the impulsive flare? YES.}    SGRE began after the most intense portion of the impulsive flare and lasted 10 hours. 

\begin{figure}
\epsscale{.80}
\plotone{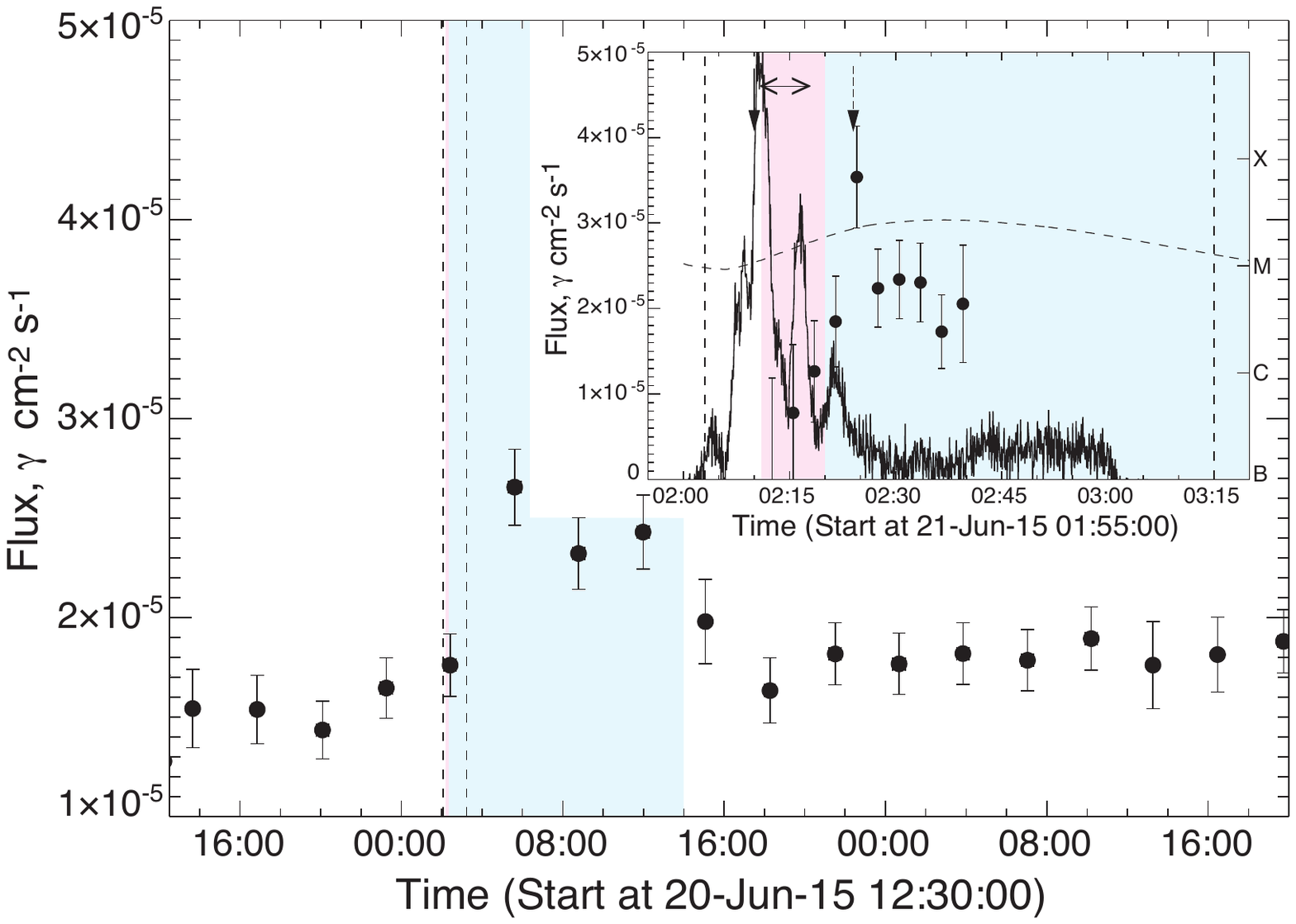}
\caption{Time profile of the 2015 June 21 SGRE event observed by LAT.  The inset shows GBM 50--100 keV rates plotted on the same scale as $>$100 MeV fluxes and $\pm1\sigma$ uncertainties derived from fits to solar impulsive-class data.  These impulsive-class fluxes are not affected by ACD rates that affect the source-class flux plotted in the main figure.  See caption for Figure \ref{110307Ath} for more details. }
\label{150621th}
\end{figure}

{\bf Details}  Plotted in Figure \ref{150621th}.  M2.6 class flare at E16 lasting $\sim$72 minutes;  $\sim$1500 km s$^{-1}$ halo CME with estimated onset time 02:10 UT based on SDO 211{\AA} images coincident with the peak in 100-300 keV X-rays; M (metric) Type II emission observed with onset $\sim$14 minutes after the CME launch but no data on DH (decameter-hectometric) emission; slowly rising SEP emission not observable above 50 MeV by {\it GOES}.  LAT had a good exposure to the Sun between 02:11--02:41 UT, overlapping much of the impulsive flare. The main plot shows Pass 8 source-class data.  As the Sun was near the Crab Nebula in mid-June the celestial background is high. SGRE was detected during the next good LAT solar exposures between 05:22--05:52, 08:32--09:03, and 11:42--12:14 UT.  LAT source-class data during the impulsive phase were compromised by the hard X-ray flux in the ACD at that time.  We have fit the solar-impulsive-class spectra accumulated in 3-minute intervals to obtain the time history of the $>$100 MeV emission during the flare.  This is shown in the inset of the figure and reveals an increase near the end of the impulsive HXR peaks in the 50-100 keV time history from GBM (solid trace).   The $>$100 MeV spectral data between 02:20 and 02:41 UT were not sufficient to determine a proton power-law index.  We estimated the number of $>$500 MeV protons at the Sun in the SGRE using the four exposures.  We also obtained upper limits on the number of >500 MeV protons in the impulsive phase using two different methods.  We first obtained an upper limit on the neutron-capture line flux in the {\it RHESSI} front detectors (15 keV FWHM resolution at 1275 keV at that time in the Mission) after 02:08 UT.  From this limit we estimated that there were no more than half the number of protons founds in the SGRE phase.  Integrating the LAT flux observed in the solar-impulsive-class data after 02:11 UT, including what appears to be SGRE, we obtained a more constraining upper limit  20\% of the number in the sustained emission phase.

\section{Table of Physical Characteristics of Sustained $\gamma$-Ray Emission} \label{sec:resultstable}

Table \ref{tab:event} provides the results of our spectroscopic study of the 30 events listed in Table \ref{tab:latlist}.   The first column lists the date of the event and the event number in parentheses.  In column two we list the type of emission providing the information in each row. There are three types: 1) the sustained $\gamma$-ray emission (SGRE), 2) impulsive flare, and 3) the solar energetic particles (SEP).  If there are more than LAT for each event, we put the number of the exposure in parenthesis, e.g. SGRE(2).   The third column gives the time intervals of the LAT solar exposures providing the information on the SGRE.  The estimated $>$100 MeV $\gamma$-ray flux and uncertainty, based on our spectral fits with pion-decay templates, is listed in column 4.\footnote{The flux may be different than that plotted in the time history figures in Appendix \ref{sec:append} because the latter included background and were estimated assuming a harder power-law photon spectrum.}  The best-fit spectral index and uncertainty of the protons producing this flux, assuming a differential power-law spectrum at energies above the 300 MeV proton, are listed in column 5.   The fifth column gives the indices of the proton power-law spectra for the different solar exposures (with exception of the 2014 February 25 event where a power-law times an exponential was necessary to fit the data as discussed in $\S$\ref{subsec:latspect}).  From the $>$100 MeV $\gamma$-ray flux, proton spectral index, and heliocentric angle of the flare site and assumptions about the temporal structure of the SGRE, we estimated the required numbers of $>$500 MeV protons at the Sun in column 7 for each emission interval given in column 6. These calculations were all done using the results of \citet{murp87} and taking into account absorption of the photons in a spherical solar geometry; the details will be discussed in a future publication.  We choose 500 MeV for specifying the proton number because the number above that energy is not sensitive to  the proton spectral index. This is primarily due to the 300 MeV threshold energy for pion production.  Because of the limited duty cycle of the LAT observations, our knowledge of the temporal evolution of the SGRE is limited.  We generally estimate the number of accelerated protons by assuming that the flux increased linearly from onset of the $>$100 MeV emission to its peak and decayed linearly from the peak to a time before the first null LAT observation.  For events where more measurements are available we estimated the total number of protons by assuming that the flux changed linearly between measurements.   In the row labelled 'SGRE Total' we list our estimate of the total number of $>$500 MeV protons and uncertainty, based on both statistical errors and our confidence in the time history and duration of the event.

For intense SGREs the fluxes below 100 MeV may be detectable by {\it RHESSI} and GBM.  For these exposures, we provide estimates of the proton spectral index between 30 and 300 MeV during the sustained-emission exposures based on a comparison of measured fluxes of the 2.223 MeV neutron-capture line and $>$100 MeV emission, using the procedure outlined in $\S$\ref{subsec:below100MeV}.  This information is provided in column 5 of the rows listed as 'SGRE $<$300 MeV'.   In the rows containing the flare {\it GOES} class, we include in column 7 our estimates for the number of $>$500 MeV protons accelerated during the impulsive phase of the flares, using either available LAT measurements ($\S$\ref{subsec:pionfits}) or observations of the 2.223 MeV neutron capture line by {\it RHESSI} or GBM (see $\S$\ref{subsec:below100MeV}).  In column 7, in the rows labelled `SEP' we list estimates of the integrated number of $>$500 MeV protons in space.  The asymmetric uncertainties are 1$\sigma$.

\begin{deluxetable}{ccccccc}
\tablecaption{Spectral Characteristics of Sustained-Emission Events \label{tab:event}}
\tablehead{
\colhead{Date (Event)} & \colhead{Type} & \colhead{Observing} & \colhead{Flux $>$100 MeV}& \colhead{Proton PL Index, s} &\colhead{Emission} & \colhead{$10^{28}$ Protons } \\
\colhead{yyyy/mm/dd} & \colhead{}  &\colhead{Interval, UT} & \colhead{10$^{-4} \gamma$ cm$^{-2}$ s$^{-1}$ }& $>$300 MeV &\colhead{Interval, UT}  &\colhead{$>$500 MeV}
}
\colnumbers
\startdata
2011/03/07 (1) & SGRE(1)  & 20:10--20:39 &  0.17 $\pm$ 0.03 & 3.3 $\pm$ 0.45        &  20:00--20:39  & 0.1   \\
& SGRE(2) & 23:21--00:03 &  0.29 $\pm$ 0.03 & 4.1 $\pm$ 0.4    & 20:39--00:03  &  1.6  \\
 & SGRE(3) & 02:32--03:13 &  0.35 $\pm$ 0.03 & 4.3 $\pm$ 0.4    & 00:03-03:13    &   2.4 \\
 & SGRE(4) & 05:43--06:25 &  0.14 $\pm$ 0.02 & 6.7 $\pm$ 1.4  & 03:13--06:25 	&  2.0\\
   & SGRE(5) &  &   &  & 06:25--11:00 	&  1.1\\
 & SGRE Total  &  &  & & 20:00--11:00 &  7.2 $\pm$ 2.1 \\
 & M3.7 flare &  & &  & 19:58--20:06&   $<$0.07\tablenotemark{b}\\
 & SEP &  & & & & $<$10\\
 \\
 2011/06/02 (2) & SGRE   & 09:41--10:28 &  0.07 $\pm$ 0.02   & 4.2 $\pm$ 2.0   & 08:10 -- 12:00  & 0.03 $\pm$ 0.02   \\
 \\
 2011/06/07 (3) & SGRE(1)   & 07:48--08:19 &  0.29 $\pm$ 0.03   & 4.5 $\pm$ 0.6   & 07:00 -- 08:20  & 0.5   \\
 &SGRE(2) &  & &   & 08:20--10:00 &   0.6 \\
 & SGRE Total &  & &   &  07:00--10:00 &  1.1 $\pm$ 0.4 \\
 & M2.5 flare &  & &   & 06:24--06:45   &   $<$0.2\tablenotemark{b}\\
& SEP &  & & & &    25.2 ${^{+22.6}_{-11.9}}$ \\
\\
2011/08/04 (4) &SGRE(1)  & 04:56--05:37 &  0.28 $\pm$ 0.03   & 4.6 $\pm$ 0.6   & 04:10--05:10  &   0.4    \\
 & SGRE(2) & & &  & 05:10--07:10  & 0.8  \\
 & SGRE Total & & &  & 04:10--07:10  & 1.2 $\pm$ 0.3  \\
 & M9.3 flare &   & &  & 03:48--04:00   & $<$0.1\tablenotemark{b} \\
 & M9.3 & & &  &    & $<$0.4 \\
 & SEP &  & &   &  &   41.4 ${^{+ 26.4}_{-16.1}}$\\
\\
2011/08/09 (5) & SGRE  & 08:02:40--08:06:00 &   2.0 $\pm$ 0.2   & 5.8 $\pm$ 0.9 & 08:02:40--08:06:00   & 0.4 +- 0.1    \\
 & SGRE $<$300 MeV  &   &  & 4.3 $\pm$ 0.3\tablenotemark{c} &  &    \\
& X6.9 flare &  &  &  & 08:02:00--08:02:20    & $<$0.01  \\
& X6.9 flare & &  &  & 08:03:40--08:04:00    & $<$0.01 \\
\\
2011/09/06 (6) & SGRE(1)  & 22:21--22:28 &  3.9 $\pm$ 0.2   & 5.3 $\pm$ 0.4  & 22:21--22:28  &  1.2  \\
 & SGRE(1) $<$300 MeV  & 22:21--22:28  &   &  $<$4.0\tablenotemark{c}   &   &  \\
 & SGRE(2) & 22:28--22:34 &  4.4 $\pm$ 0.2  &  3.5 $\pm$ 0.3  & 22:28--22:34   & 0.6 \\
  & SGRE(2) $<$300 MeV & 22:28--22:34  &   &  $<$3.6\tablenotemark{c}   &   &  \\
 & SGRE(3)   & 22:37--22:47 &  1.5 $\pm$ 0.1 &  3.5 $\pm$ 0.2  & 22:37--22:47  & 0.4 \\
 & SGRE(3) $<$300 MeV & 22:37--22:44  &   &  $<$3.9\tablenotemark{c}   &   &  \\
& SGRE Total   &     &   &  & 22:21--23:20   & 2.2 $\pm$ 0.4  \\
& X2.1 flare &  & 6.6 $\pm$ 1.0 & $>$6  & 22:18--22:20   & 0.13 $\pm$ 0.05 \\
 & SEP &  & &   &  &   $<$62\\
\\
2011/09/07 (7) &  SGRE(1)& 23:51--00:09 &  0.07 $\pm$ 0.02   & 4.4 $\pm$ 1.4   & 22:45--00:00  &   0.1 \\
& SGRE(2) & &  &  &  00:00--01:10  &    0.1\\
& SGRE Total & &  &  &  22:45--01:10  &    0.2 $\pm$ 0.1\\
&X1.8 flare &  &  &  &  22:36--22:38  &   $<$0.1\tablenotemark{b}\\
 & SEP &  & &   &  &   $<$68
\\
\\
2011/09/24 (8) & SGRE &09:40--09:44 & 0.3 $\pm$ 0.09 & 3.4 $\pm$ 1.4 & 09:40--09:44 & 0.03 $\pm$ 0.01 \\
  & X1.9 flare & & & & 09:35:30--09:37:00 &$<$0.007\\
\\ 
2012/01/23 (9) & SGRE(1)  & 05:46 -- 06:10 &  0.19  $\pm$ 0.02  &    5.1 $\pm$ 0.8 & 04:20--05:59   &  0.8  \\
 & SGRE(2)  & 07:18 -- 07:48 &  0.16  $\pm$  0.05   &   &  05:59--07:34  & 0.7   \\
 & SGRE(3)   & 08:58 -- 09:28&  0.19  $\pm$ 0.03   &   & 07:34--09:14  &  0.8  \\
 & SGRE(4)   & &     &   & 09:14--12:00  &  0.7  \\
& SGRE Total  & &    &   & 04:20--12:00    &  3.0  $\pm$ 0.6\\
 & M8.7 flare &  & &  &  03:53--04:09  & $<$0.4\tablenotemark{b}\\
 & M8.7 flare &  &   &   & 03:45--04:09   &    $<$0.7 \\
& SEP &  & &  &   &    212 $\pm$ 88
\\
\\
2012/01/27 (10) & SGRE(1) & 19:36--19:56 & 0.26 $\pm$ 0.05    & 4.2 $\pm$ 0.7  & 19:00--19:50   &   0.5   \\
 & SGRE(2)   & 21:06--21:37 & 0.05 $\pm$ 0.02    & 2.8 $\pm$ 0.6  & 19:50--21:21  & 1.1  \\
& SGRE(3) &  &  & & 21:21--22:00  & 0.08 \\
 & SGRE Total  &  &  & &  19:00-22:00 &1.7 $\pm$ 1.0 \\
& SEP &  & &  &   &   656  $\pm$ 260
\\
\\
2012/03/05 (11)& SGRE(1)   & 05:46--06:12 & 0.10 $\pm$ 0.015    & 4.9 $\pm$ 0.9  & 04:30--05:59  &   0.12   \\
 & SGRE(2)   & 07:18--07:56 & 0.075 $\pm$ 0.019    & 3.6 $\pm$ 0.8  & 05:59--07:36   &  0.25 \\
 & SGRE(3)  & &       &    & 07:36--10:00  & 0.16 \\
  & SGRE Total   & &       &    &  04:30--10:00 & 0.53 $\pm$ 0.15 \\
& X1.1 flare &  &   &   & 03:55--04:35  & $<$0.3\tablenotemark{b}\\
&                  & 04:28--04:35  & $<$0.15  &   & 04:28--04:35  & $<$0.04\\
\\
2012/03/07 (12) & SGRE(A) Total & 00:39--01:24 &   28.7 $\pm$ 0.4 & 3.6 +- 0.3  & 00:28--01:24   &   40 $\pm$ 15 \\
 &  SGRE(B1)   &02:18--02:48 &  5.8 $\pm$ 0.3  & 3.5 $\pm$ 0.2   &  02:00--02:34   & 3   \\
 & SGRE(B1) $<$300 MeV  &   &  & $<$3.3\tablenotemark{c} &  &    \\
 & SGRE(B2) & 03:50--04:34 &  10.0 $\pm$ 0.2  & 3.85 $\pm$ 0.1   & 02:34--04:12   & 23   \\
  & SGRE(B2) $<$300 MeV  &   &  & $<$3.3\tablenotemark{c}   &  &   \\
 & SGRE(B3) & 05:34--06:01 &  8.7 $\pm$ 0.4  & 4.25 $\pm$ 0.2    & 04:12--05:46    & 30   \\
 & SGRE(B4) & 07:02--07:46 &  6.2 $\pm$ 0.2  & 4.5 $\pm$ 0.15  &  05:46-07:24   & 27   \\
    & SGRE(B4) $<$300 MeV &   &  & $<$3.3\tablenotemark{c} &  &    \\
 & SGRE(B5)  & 08:42--09:12 &  4.1 $\pm$ 0.3  & 4.8 $\pm$ 0.5   & 07:24-08:48    & 18   \\
    & SGRE(B5) $<$300 MeV  &   &  & $<$3.7\tablenotemark{c} &   &    \\
 & SGRE(B6)   & 10:33--10:58 &  2.5 $\pm$ 0.2  & 5.2 $\pm$ 0.4   & 8:48--10:46    & 17   \\
    & SGRE(B6) $<$300 MeV  &   &  & $<$3.7\tablenotemark{c} &    &   \\
 & SGRE(B7)  & 13:23--13:33 &  0.6 $\pm$ 0.2  &  & 10:46--13:27   & 9   \\
 & SGRE(B8)  & 16:35--16:49 &  0.22 $\pm$ 0.06 &  & 13:27--16:41  &   3 \\
 & SGRE(B9) & 19:46--20:14 &  0.07 $\pm$ 0.02 &   & 16:41-20:01  &   1 \\
 & SGRE(B) Total &   &   &  & 02:00--20:01 &   131 $\pm$ 15 \\
 & X5.4 flare &  &  &   & 00:16--00:28   & 1.4\tablenotemark{b}  \\
 & M7 flare &   &  &   & 01:11--01:20  &    1.1\tablenotemark{b} \\
 & M7 flare &   &  &  &  01:12--01:17  &  $<$0.4 \\
 & SEP &  & &  &   &   13300 ${^{+31800}_{-9360}}$
\\
\\
2012/03/09 (13) & SGRE(1)  & 05:10--05:58 & 0.06 $\pm$ 0.03 & $>$6  & 04:30--06:00  &      0.1 \\
& SGRE(2)  &06:46--07:32 &  0.11 $\pm$ 0.03 & 6 $\pm$ 1.5    &  06:00--07:09  &     0.3  \\
& SGRE(3)  &08:22--09:08 &  0.15 $\pm$ 0.03 & 6.7 $\pm$ 1.5   & 07:09--08:46  &     0.7  \\
 & SGRE(4)&  &  &  & 08:46--10:30   & 0.4 \\
  & SGRE Total &  &  &  & 04:30--10:30   & 1.5 $\pm$ 0.6\\
& M6.3 flare  & &  &   &  03:40--04:14&$<$0.1 \\
& M6.3 flare  & &  &   &  03:30--04:06&$<$0.6\tablenotemark{b} 
\\
\\
2012/03/10 (14) &  SGRE(1)   & 20:59--21:33 & 0.02 $\pm$ 0.01 & $\geq$6 & 20:00--21:15   &     0.05   \\
&  SGRE(2) &22:35--23:15 &  0.043 $\pm$ 0.027 &$\geq$6     & 21:15--22:55 &    0.18  \\
& SGRE(3)  &00:10--00:56 &  0.038 $\pm$ 0.015 & $\geq$6    & 22:55--00:33  &    0.2  \\
& SGRE(4) & &   &     & 00:33-02:00  &    0.09  \\
& SGRE Total & &   &      &20:00-02:00  &    0.5 $\pm$ 0.3  \\
& M8.4 flare &  &  &   & 17:51--18:11   &$<$0.1\\
& M8.4 flare &  &  &   & 17:41--18:05   &$<$0.2\tablenotemark{b}\\
& SEP &  & &  &   &   $<$8.8
\\
\\
2012/05/17 (15) & SGRE(1)  &  02:10--02:48 &  0.08 $\pm$ 0.03  & 2.6 $\pm$ 0.6   &02:05--02:29   & 0.02   \\
 & SGRE(2) & 03:46--04:22 &  0.05 $\pm$ 0.025  & 2.2 $\pm$ 1.0  &  02:29--04:02  & 0.1  \\
  & SGRE(3)  &   &    &   & 04:02--05:20   & 0.03   \\
 & SGRE Total &  &    &   & 02:10--05:20  & 0.15 $\pm$ 0.1   \\
  & M5.1 flare & & & &01:40--01:55   & $<$0.6\tablenotemark{b}  \\
 & SEP  &   &    &    &   & 850 ${^{+416}_{-278}}$
\\
\\
2012/06/03 (16) & SGRE(1)  &  17:51--17:53 &  0.24 $\pm$ 0.08?  & 2.5 ${_{-?}^{+1.5}}$   & 17:51--17:53   & 0.01?   \\
& SGRE(2) & 17:54--18:02 &  0.54 $\pm$ 0.06  & 4.3 $\pm$ 0.7  &   17:54--18:02    &  0.15  \\
& SGRE(3)  &  &    &    &   18:02--19:00    & 0.58   \\
& SGRE Total &  &    &    &   17:54--19:00    & 0.74 $\pm$ 0.35  \\
& M3.3 flare &  &  3.3 $\pm$ 0.4  & 6.4 $\pm$ 1.0  & 17:53--17:54   & 0.19 $\pm$ 0.05
\\
\\
2012/07/06 (17) & SGRE(1)  &  23:27--23:54 &  0.37 $\pm$ 0.04  &5.1 $\pm$ 0.6 &23:14--23:40    & 0.2   \\
 & SGRE(2)   &  23:27 to 23:40 &    &4.7 $\pm$ 0.8 &    &    \\
 & SGRE(3)  &  23:40 to 23:54 &    & $>$6.0 &   &    \\
& SGRE(4) &  &    &    &   23:40--01:00    & 0.7   \\
& SGRE Total &  &    &    &   23:14--01:00    & 0.9 $\pm$ 0.3   \\
& SEP  &   &    &    &   & $<$29
\\
\\
2012/10/23 (18) & SGRE(1)  &  04:10--04:40 &  0.13 $\pm$ 0.03  &4.6 $\pm$ 1.2 & 03:20--04:25   &  0.18  \\
& SGRE(2)  &  &    &    &   04:25--05:20    & 0.15  \\
& SGRE Total   &  &    &    &   03:20--05:20    & 0.33 $\pm$ 0.2   \\
& X1.8 flare &  &  &   &03:15--03:17   & $<$0.06\tablenotemark{b}
\\
\\
2012/11/27 (19) & SGRE Total    &  15:55--16:11 &  0.13 $\pm$ 0.03   & 2.9 $\pm$ 0.6 & 15:55--16:11   &  0.04 $\pm$ 0.02  \\
& M1.6 flare &   &  &   &15:55:40--15:56:44  & $<$0.002\\
\\
2013/04/11 (20) &  SGRE(1)  &  07:10--07:14 &  1.0 $\pm$ 0.1  &5.8 $\pm$ 0.8    & 7:10--07:14   & 0.21   \\
& SGRE(1) $<$300 MeV &07:10--07:30  &  & $<$4.5\tablenotemark{c} &   &   \\
& SGRE(2) &07:14--07:20  & 1.1 $\pm$ 0.09  & 5.2 $\pm$ 0.5  & 07:14--07:20  &  0.30 \\
& SGRE(3) &07:20--07:30  & 0.53 $\pm$ 0.06  & 5.0 $\pm$ 0.7  & 07:20--07:30  &  0.23 \\
& SGRE Total &  &  &   &07:10--07:30   &  0.74 $\pm$ 0.3  \\
& M6.5 flare &  &  &   &07:07--07:11   & $<$0.03  \\
& SEP  &   &    &    &   & 660  $\pm$ 500\\
\\
2013/05/13 (21) & SGRE Total   &  04:30--05:15 &  0.10 $\pm$ 0.02  &5.3 $\pm$ 1.7   & 02:30--06:00   &  2.2 $\pm$ 1.5   \\
\\
2013/05/13 (22) &  SGRE(1) &  17:15--17:28 &  0.11 $\pm$ 0.05  &4 $\pm$ 2  & 17:00--17:22    & 0.06   \\
& SGRE(2)  & 17:41--17:59 &  0.25 $\pm$ 0.05  & 2.9 $\pm$ 0.6 & 17:22--17:50  & 0.2  \\
& SGRE(3)  & 20:25--21:10 &  0.2 $\pm$ 0.04  & 6.2 $\pm$ 1.0   &  17:50--20:45  &  2.0  \\
& SGRE(4)  &  &  &   & 20:45--23:00  &  1.0 \\
& SGRE Total   &  &  &   & 17:00--23:00  &  3.3 $\pm$ 1.8  \\
&  X2.8 flare & &  &   &16:00--16:07   & $<$0.6\tablenotemark{b}
\\
\\
2013/05/14 (23) &  SGRE(1) & 01:30 -- 01:47   &  0.06 $\pm$ 0.02  &  6 $\pm$2   &  01:20--01:40   & 0.05  \\
  & SGRE(2)   &  02:58--03:23 &  0.42 $\pm$ 0.03  & 5.0 $\pm$ 0.5   & 01:40--03:10    & 1.3   \\
 & SGRE(3)  &04:20--05:06  & 0.26 $\pm$ 0.03   & 6.6 $\pm$ 0.9  &  03:10--04:43  &  2.1 \\
 & SGRE(4)  &06:01--06:41  & 0.08 $\pm$ 0.02   & $>$6.5  & 04:43--06:20    &  1.2 \\
 & SGRE(5)  &  &   &     & 06:20--07:00   &  0.1 \\
 & SGRE Total & &   &     & 01:20--07:00   &  4.8 $\pm$ 2.0\\
 & X3.2 flare &  &  &  & 01:06--01:16 & $<$0.2\tablenotemark{b}  \\
\\
2013/05/15 (24) & SGRE(1)  & 02:37--03:23   &  0.02 $\pm$ 0.01  &  4.5$\pm$2  &  02:00--03:00    & 0.02   \\
  & SGRE(2)  &  04:13--04:58 &  0.06 $\pm$ 0.02  & 5$\pm$1.5    & 03:00--04:35   & 0.15   \\
 &SGRE(3) &  05:52--06:34  & 0.03 $\pm$ 0.01   & $<$2.5   & 04:35--06:13   &  0.12 \\
 & SGRE(4)  & 07:33--08:09  & 0.03 $\pm$ 0.01   & $<$2.5  & 06:13--07:51  &  0.12\\
 & SGRE(5)  &  &   &    & 07:51--09:00    &  0.04 \\
 & SGRE Total  &  &   &    & 02:00-09:00    &  0.45 $\pm$ 0.25\\
 &  X1.2 flare & &  &  &01:36--01:46   & $<$0.01\\
 &  X1.2 flare & &  &  &01:30--01:55   & $<$0.36\tablenotemark{b}\\
\\
2013/10/11 (25) & SGRE Total   & 07:14--07:30   &  2.0 $\pm$ 0.1  &  3.75 $\pm$ 0.2   & 07:14--07:30    & 1.9 $\pm$ 1.0  \tablenotemark{e}\\
 & SGRE $<$300 MeV   & 07:14--07:30   &   &  $<$4.2 \tablenotemark{c}   &     & \\
\\
2013/10/25 (26)& SGRE(1)   & 08:14 -- 08.59  &  0.13 $\pm$ 0.02  & 5.6 $\pm$ 1.4    & 08:02--08:30    & 0.11  \\
 & SGRE(2)  &   &    &    & 08:30--09:30     &  0.21   \\
 & SGRE Total   &   &    &    & 08:02--09:30     &  0.32 $\pm$ 0.15   \\
 & X1.7 flare   &  &    &    & 07:58--08:02     &  $<$0.32\tablenotemark{b}
\\
\\
2013/10/28 (27) & SGRE(1)  &15:46--16:06   & 0.07 $\pm$ 0.02    &  2.5 $\pm$ 0.5   & 15:20--15:55    & 0.014   \\
 & SGRE(2)   &   &    &    & 15:55--17:00     &  0.026\\
& SGRE Total  &   &    &    & 15:20--17:00     &  0.04 $\pm$ 0.02\\
\\
\\
2014/02/25 (28) & SGRE(1)  & 01:13--01:17  &  14.1 $\pm$ 0.5  & $\sim$3.5 \tablenotemark{d}   &  00:55--01:15    &  6 \\
  & SGRE(1) $<$300 MeV  & 01:10--01:26  &   & $<$3.5 \tablenotemark{c}   &      &   \\
  & SGRE(2)  & 04:21--04:40  & 3.5  $\pm$ 0.2  & $\sim$7   &  01:15--04:30    &  73 \\
    & SGRE(2) $<$300 MeV  & 04:21--04:40  &  & $<$3.8 \tablenotemark{c}  &      &   \\
 & SGRE(3)  &  &   &    &  04:30--05:30    &  9  \\
& SGRE Total  &  &   &    &  00:50--05:30    &  88 $\pm$ 40  \\
& X4.9 flare   &   &    &    & 00:43--00:52     &  3 $\pm$1.5\tablenotemark{b}   \\
 & SEP &   &    &    &   & 96000  $\pm$ 96000
\\
\\
2014/09/01 (29) & SGRE(1)  & 11:06--11:12  &  42 $\pm$ 2  & 4.25 $\pm$ 0.15   &  11:06--11:12    &  14.6\\
  & SGRE(2)  & 11:12 -- 11:20  &  50 $\pm$ 2  & 3.85 $\pm$ 0.1  &  11:12--11:20    &  38  \\
  & SGRE $<$300 MeV  &  11:04 -- 11:30   &  & $<$3.4\tablenotemark{c,e}   &  &
\\
  & SGRE(3)  & 12:26 -- 12:58  &  0.33 $\pm$ 0.03  & 3.45 $\pm$ 0.35  &  11:20--12:42    &  91  \\
  & SGRE(4)  &   &    &  &  12:42--15:52    &  3.6  \\
& SGRE Total &  &   &    &  11:02--15:52    &  146 $\pm$ 80\tablenotemark{e}
\\
\\
2015/06/21 (30) & SGRE(1)  & 02:20--02:41  &  0.12 $\pm$ 0.04  & ?  &  02:20--02:41    &  0.07\\
 & SGRE(2)   & 05:20--05:53  &  0.13 $\pm$ 0.03  & 3.2 $\pm$ 0.7   &  02:41--05:37    &  0.57\\
  & SGRE(3)  &08:30--09:02   & 0.09 $\pm$ 0.02 & 2.7 $\pm$ 0.7 &  05:37--08:46    & 0.38  \\
& SGRE(4)  &11:41--12:14   & 0.07 $\pm$ 0.02 & 3.3 $\pm$ 1.3 &  08:46--11:56    & 0.27  \\
  & SGRE(5) &   &    &  &  11:56--14:00    & 0.08 \\
    & SGRE Total  &   &    &  &  02:20--14:00    & 1.4 $\pm$ 0.7 \\
   & M2.6 flare  &   &    &    & 02:11--02:20     &  $<$0.2   \\
& M2.6 flare  &   &    &    & 02:09--02:25     &  $<$0.6\tablenotemark{b}\\
\enddata

\tablenotetext{b}{from 2.223 MeV line flux assuming protons follow a power-law spectrum with index s=4.5 above $\sim$30 MeV see $\S$\ref{subsec:below100MeV}}
\tablenotetext{c}{95\% confidence limit on index between 30 and 300 MeV based on comparing 2.223 MeV line flux upper limit and $>$100 MeV $\gamma$-ray flux, see $\S$\ref{subsec:below100MeV}}
\tablenotetext{d}{Better fit E$^{-2}$*exp-(E/1300 MeV) }
\tablenotetext{e}{Heliocentric angle of 85$^{\circ}$}

\end{deluxetable}

\section{Estimate of Number of SEP Protons} \label{sec:sep}

For comparison with the number of protons interacting in the solar atmosphere in SGRE events, we made an estimate of the number of $>$500 MeV SEP protons emitted from the Sun and escaping to at least 1 AU.  We used the event-integrated fluences recorded by the High-Energy Proton and Alpha Detector (HEPAD) on GOES-13 and GOES-15\footnote{\url{https://ngdc.noaa.gov/stp/satellite/goes/datanotes.html}}.  HEPAD records the proton flux in three differential bins covering 330--700 MeV and one integral bin at $>$700 MeV.  We corrected the observed fluxes for both background and the incorrect geometry factors used in the routine processing (H. Sauer, 2007, private communication).  Because we integrated the fluxes over the $\sim$1--2-day duration of the particle event, we treat the derived fluence as an omnidirectional average.  We validated the HEPAD proton fluences by comparison with both earlier GLE measurements \citep{tylk09} and the recent PAMELA measurements of the 2012 May 17 event \citep{brun16}.  To extract the number of $>$500 MeV protons from the HEPAD fluence, we used the heuristic method outlined by \citet{mewa05} in an analogous study at lower energies.  This method requires the estimation of two correction factors, one related to interplanetary transport and a second related to the large scale distribution of the proton fluence in interplanetary space at 1 AU.
  
The transport factor is called a crossing-correction because it takes into account that some protons may cross back and forth across the 1-AU boundary multiple times during the event.  We estimated the crossing factor in three ways.   First, we analyzed the time-dependent front-back asymmetry observed by the world-wide neutron monitor network in the GLE of 2012 May 17 and the particularly well-observed and modeled GLE of 15 April 2001 (Shea and Smart, private communication, 2013).  We also modeled the crossing-factor using both a Monte Carlo method \citep{chol10} (Chollet, private communication 2011) and an analytic formulation of the transport process (C.K. Ng, private communication, 2011).  In both implementations, we treated the release of protons from the Sun as of sufficiently short duration to be modeled as a delta-function in time.  We adopt the scattering law used by \citet{bieb04} in their modeling of the 2001 April 15 GLE.  We used both their best-fit values for the scattering parameters extracted for that event, as well as modest levels of variation about the best-fit values.  Both the GLE analysis and the modeling efforts indicate that the crossing-factor for $>$500 MeV protons observed over a ~1-day period is a factor of two, with systematic uncertainly of +/- 50\%.  The measured GOES fluence was therefore reduced by this factor.
  
Because the fluence of $>$500 MeV protons is measured only at Earth, we used two methods to estimate the large-scale spatial distribution of the particles. First, we examined the historical record of 57 GLE fluences observed between 1956 and 2006 versus solar longitude of the source region.  That distribution peaks at the nominal best-connected longitude of W58 and falls off exponentially in both directions with an e-folding width of 23$^{\circ}$.  Second, two of the events in this study were recorded by both STEREO spacecraft, giving two additional widely-spaced measurements at 1 AU.  The STEREO HED instrument, however, records protons only up to 100 MeV.  We therefore used a power-law fit to the measurements in the two highest-energy bins (30--60 and 60--100 MeV) to estimate the event-integrated fluence at $>$100 MeV,  Given these two measurements and that of GOES, we fit the longitude distribution with a Gaussian.  Both the GLE-derived exponential and the STEREO Gaussians yield large-scale fluence multipliers between ~1 and ~4 for events with solar source regions located between W20 and W90.  Events with source locations outside this range were excluded from the analysis because the extrapolations are larger and may not be reliable.  Finally, we used the same angular distribution to estimate the latitudinal extent of the proton fluence, but with two modifications.  First, we assumed that the fluence is highest for measurements in the ecliptic, regardless of the latitude of the source region.  Second, we assumed that there is no fluence above 65$^{\circ}$ latitude, the highest latitude at which Ulysses observed solar protons.  Note that both of these assumption serve to place a lower-bound on the  interplanetary proton fluence: if the latitude distribution peaked at a non-zero latitude or the fluence extended above 65 degrees latitude, the estimated number of interplanetary protons would be larger.

\section{Solar Radio Bursts} \label{sec:solarradio}

Solar radio bursts provide valuable diagnostics of flare and eruptive phenomena in the solar  corona. We focus on two particular types of radio burst commonly seen at metric and longer wavelengths (e.g. \citet{wild63,mcle85}):

\begin{itemize}

\item Type III bursts, which are attributed to beams of energetic electrons
propagating on field lines that extend into the outer corona. The 
durations of individual beams are short, of order seconds or less, although the
waves they generate can last longer, and on a
frequency-time plot they are seen to drift very rapidly from high to low frequency.
They have to drift over a significant frequency range in order to be identified as
Type IIIs.  In most cases, the electrons in the beams are thought to have
typical energies of order 10 keV (e.g. \citet{lin85}).

\item Type II bursts have slower frequency-time drift rates (also from high
to low frequency), with
instantaneously narrow bandwidth and usually striking fundamental-harmonic 
structure (at least at higher frequencies). They are attributed to electrons accelerated at
a shock and radiating in its vicinity as it moves through the corona: 
the speed of a coronal shock is of order of (but larger than) the Alfv{\'e}n
speed, v$_{\rm A}$ (typically 500 km s$^{-1}$), which is much slower than the velocity of 
an electron in a Type III-producing beam.

\end{itemize}

Both burst types are believed to radiate at the local plasma frequency,
$f_{\rm p}\,=\,9000\,\sqrt{n_{\rm e}}$, where $n_{\rm e}$ is the ambient
electron density (cm$^{-3}$). The high-to-low frequency drift corresponds to the 
motion of the radiating source outwards through a decreasing density gradient.
There is therefore a mapping between the frequency of emission and height in
the solar atmosphere: thus, emission at 100\,MHz (metric wavelengths) 
occurs around 0.5\,R$_\odot$
above the photosphere, 10\,MHz (decametric wavelengths) is several R$_\odot$ above the photosphere,
and 1\,MHz (hectometric wavelengths) is 10-20\,R$_\odot$ out (these heights are crude estimates,
varying greatly from one atmospheric density model to another).

The diagnostic value of Type III bursts is that they indicate that electrons
accelerated low in the corona have access to magnetic field lines that carry
them far out into the solar wind, if they are seen to emit down to low
frequency ($\sim$\,1\,MHz). Type III bursts often occur early in the
impulsive phase of solar flares, but they can also occur for extended
periods at low frequencies later in an event (e.g. \citet{cane02}).

Type II bursts require the presence of a shock that can accelerate electrons, 
and CMEs provide a natural driver for
this process. Shock formation in a magnetized plasma such as the solar
corona requires that the driver exceed the local MHD fast-mode speed, which
is close to the Alfv{\'e}n speed, v$_{\rm A}$.  v$_{\rm A}$
varies with density and magnetic field strength in the solar corona. High
values of the Alfv{\'e}n speed occur in strong magnetic-field regions low
in the atmosphere, generally decreasing initially outwards with height, 
with values of order 100-200\,km\,s$^{-1}$, before increasing again to a
local peak in the Alfv{\'e}n speed at a height of order several R$_\odot$,
with a value of order 500\,km\,s$^{-1}$. \citet{gopa01} suggest that
metric Type II bursts form at heights below the local peak in v$_{\rm A}$,
while Type II bursts observed below 10 MHz may form at heights above the local peak
in v$_{\rm A}$. This picture is consistent with the finding of \citet{cane05} that
there are in fact two classes of Type II burst: the coronal Type II bursts,
typically observed at metric wavelengths below the height of the local peak 
in v$_{\rm A}$ and never observed to propagate to low frequencies characteristic
of the interplanetary (IP) medium; and IP Type II bursts, which can occur at low
frequencies at the same time as coronal Type IIs at higher frequencies, and are
capable of drifting well out into the solar wind to frequencies below 1 MHz.

The significance of Type II bursts is that, if they are indeed a reliable
indicator of shock formation, then no shock acceleration of protons or electrons can 
take place before the onset of Type II radio emission. Type II bursts occur in
conjunction with an accompanying flare, and the emission can start at any time 
after the onset of the flare's impulsive phase. \citet{make15} find that at
the onset of metric Type II burst radio emission, the 
average radial height of an associated CME is of order 1.7 R$_\odot$.
The implicit assumption in this
analysis is that the Type II burst occurs at the nose of the CME, i.e., the
location on the CME with the greatest height above the solar surface.
The complication in this scenario is that Type II radio emission might not
occur at this location. The sources of Type II radio emission must be spatially 
localized, since a
very large source would necessarily extend over a large range of electron
densities and therefore have a large instantaneous frequency bandwidth,
which, by definition, is not observed (i.e., such broadband emission would not have the
usual characteristics of Type II bursts in dynamic spectra). When radio imaging observations
of Type II bursts have been able to identify the location of Type II
emission relative to a CME, they generally do not show that the
Type II emission occurs at the nose of the CME, but rather is often located
on the flanks of the CME \citep{chen14,feng15} or is associated with some 
feature distinct from the CME (e.g.
\citet{gary84,magd12,bain12,zimo12}.
This complication implies that conditions other than simple shock formation might
be required in order to see Type II radio emission, and therefore the
presence of a Type II may be {\em sufficient} but not {\em necessary} evidence for the
presence of a shock. If this is the case, the onset time of Type II emission
does not constrain the start of shock acceleration.

In addition, the possible distinction between coronal Type II bursts, occurring
while CMEs are still below a radius of order 2\,R$_\odot$, and IP Type IIs is a
further complication, since acceleration of energetic particles can in
principle occur at either shock. The onset time of an IP Type II can be
difficult to determine from low-frequency radio spectra because of the presence of
bright Type III emission in the same frequency and time range.

The following table indicates the presence or absence of Type II and Type
III radio emission for each of the flares associated with the 30 sustained
FERMI LAT $>$\,100\,MeV events. Radio dynamic spectra were available for all 30
events, both from ground-based observations of metric frequencies (25-180 MHz from
the four stations of the US Air Force Radio Solar Telescope Network, as well as
the radio spectrograph at Culgoora operated by the Australian Space Weather
Services section), and from space-based observations of decametric-hectometric
frequencies (DH, 1-10 MHz) provided by the WAVES receivers on the WIND and STEREO
spacecraft. Identification of Type II bursts relied on the NOAA event reports
provided by RSTN for the metric data and the WIND/WAVES burst list for the DH
data. In cases where the authors are not entirely convinced that a Type II is 
present in the dynamic spectra, question-marks qualify the report.

For the Type III emission, we attempt to determine whether it is
present in the impulsive phase (identified as the initial rise in soft X-rays) as
well as in the later phase, at least several minutes later \citep[see discussion in
][]{duff15}. These two phases generally have different interpretations:
acceleration of electrons to high energies with resulting hard X-ray production is
generally most prolific in the impulsive phase, while there may also be late-phase
energy release which is typically more evident in heating and soft X-rays. In either case,
the presence of Type III emission out to low frequencies implies that accelerated
electrons have ready access to open field lines in the acceleration region.

\begin{deluxetable}{rrccc}[t]
\floattable
\tabletypesize{\scriptsize}
\tablecaption{Radio Bursts from LAT Sustained $>$100 MeV Events\label{tab:radio}}
\tablehead{
\colhead{Number} & \colhead{Date, {\it GOES} class} & \colhead{Type II} &
\colhead{Type III metric}&\colhead{Type III DH\tablenotemark{a}} \\
\colhead{} & \colhead{} & \colhead{Metric\tablenotemark{b} | DH\tablenotemark{a} }& \colhead{Impulsive | Late} &
\colhead{Impulsive | Late} 
} 
\colnumbers
\startdata
1 & 2011/03/07, M3.7 &    Y? | Y & 	    N | Y   &     N | Y  \\
2 & 2011/06/02, C3.7 &     N | Y & 	    Y | N   &     Y | N  \\
3 & 2011/06/07, M2.5 &    Y? | Y & 	    Y | Y   &     Y | Y  \\
4 & 2011/08/04, M9.3 &    Y | Y & 	    Y | Y   &     Y | Y  \\
5 & 2011/08/09, X6.9 &    Y? | Y & 	    Y | Y   &     Y | Y  \\
6 & 2011/09/06, X2.1 &    Y | Y & 	    N | Y   &     N | Y  \\
7 & 2011/09/07, X1.8 &    Y | N & 	    Y | Y   &    Y? | Y  \\
8 & 2011/09/24, X1.9 &    Y? | N &	    Y | Y   &     Y | Y  \\
9 & 2012/01/23, M8.7 &    N | Y & 	    Y | Y   &     Y | Y  \\
10 & 2012/01/27, X1.7 &    Y | Y & 	    N | N   &     Y | Y  \\
11 & 2012/03/05, X1.1 &    N | Y & 	    Y | Y   &     Y | Y  \\
12 & 2012/03/07, X5.4 &   Y? | Y & 	    Y | Y   &     Y | Y  \\
   & 2012/03/07, M3\ \ \ & Y? | Y &	 Y & Y  \\
13 & 2012/03/09, M6.3 &    Y | Y & 	    N | Y   &     N | Y  \\
14 & 2012/03/10, M8.4 &    N | Y & 	    N | Y   &     N | Y  \\
15 & 2012/05/17, M5.1 &    Y | Y & 	    Y | Y   &     Y | Y  \\
16 & 2012/06/03, M3.3 &    Y | N & 	    Y | N   &     Y | Y  \\
17 & 2012/07/06, X1.1 &    Y | Y & 	    Y | N   &     Y | N  \\
18 & 2012/10/23, X1.8 &    Y | N & 	    N | N   &     N | N  \\
19 & 2012/11/27, M1.6 &    N | N & 	    N | N   &     N | N  \\
20 & 2013/04/11, M6.5 &    Y | Y & 	    Y | Y   &     Y | Y  \\
21 & 2013/05/13, X1.7 &    Y | Y & 	    Y | Y   &     Y | Y  \\
22 & 2013/05/13, X2.8 &    Y | Y & 	    N | Y   &     Y | Y  \\
23 & 2013/05/14, X3.2 &    Y | Y & 	    N | Y   &     Y | Y  \\
24 & 2013/05/15, X1.2 &    Y | Y & 	    N | N   &     N | Y  \\
25 & 2013/10/11, M4.9 &    Y | Y & 	    N | Y   &     Y | Y  \\
26 & 2013/10/25, X1.7 &    Y | Y & 	    N | Y   &     N | Y  \\
27 & 2013/10/28, M4.4 &    Y | Y & 	    Y | Y   &     Y | Y  \\
28 & 2014/02/25, X4.9 &    Y | Y & 	    Y | Y   &     Y | Y  \\
29 & 2014/09/01, X2.1 &   Y? | Y & 	    Y | Y   &     Y | Y  \\
30 & 2015/06/21, M2.6 &   Y | Y & 	    N | N   &     Y | Y  \\
\enddata
\tablenotetext{a}{DH = decametric/hectometric: this refers to observations at
frequencies below 10 MHz that can only be carried out from space due to the
ionospheric cutoff. The WIND and both STEREO spacecraft carry WAVES receivers
operating in this frequency range, and observations in the 1-10 MHz range were
used for the identifications indicated here.}
\tablenotetext{b}{Metric dynamic-spectra radio observations are obtained in the 
frequency range 25-180 MHz by the four stations of the US Air Force Radio Solar
Telescope Network (RSTN), and by the radio spectrograph at Culgoora operated by
Australia's Space Weather Services section.}
\end{deluxetable}

In all 30 events except one, there is evidence for the presence of Type II emission 
at metric or DH wavelengths, or both. In all except two events, there is evidence
for Type III emission at metric or DH wavelengths, and those are the two events
(2012-10-23, 2012-11-27) without reported CMEs.

\bibliography{references}

\end{document}